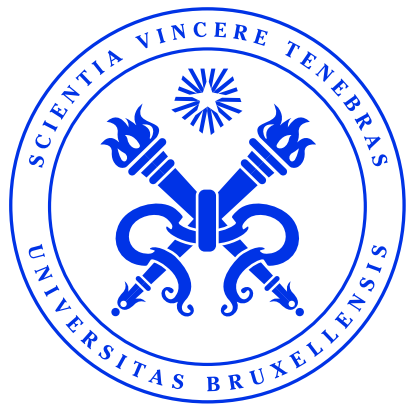

# VISCOSITY AND MICROSCOPIC CHAOS :
# THE HELFAND-MOMENT APPROACH

## Sébastien Viscardy

Thèse présentée en vue de l'obtention
du grade de Docteur en Sciences

Réalisé sous la direction de
**Pierre Gaspard**

**Interdisciplinary Center for Nonlinear Phenomena and Complex Systems**

**Faculté des Sciences**

**Service de Chimie-Physique**

September 2005

**ULB** UNIVERSITÉ LIBRE DE BRUXELLES, UNIVERSITÉ D'EUROPE

Thesis defended at the **Université Libre de Bruxelles** on the 21th of September, 2005.

Composition of the Jury:

| | | |
|---|---|---|
| Claudine Buess-Herman | Université Libre de Bruxelles | President |
| J. Robert Dorfman | University of Maryland | External member |
| Pierre Gaspard | Université Libre de Bruxelles | PhD director |
| Michel Mareschal | Université Libre de Bruxelles | Effective member |
| Grégoire Nicolis | Université Libre de Bruxelles | Expert |
| John W. Turner | Université Libre de Bruxelles | Effective member |

# Table of Contents

# List of Figures

# List of Tables

# Résumé de la thèse

Depuis les premiers développements de la physique statistique réalisés au 19<sup>ème</sup> siècle, nombreux ont été les travaux dédiés à la relation entre les processus macroscopiques *irréversibles* – tels que les phénomènes de transport – et les propriétés de la dynamique *réversible* des atomes et des molécules. Depuis deux décennies, l'*hypothèse du chaos microscopique* nous en apporte une plus grande compréhension. Dans cette thèse, nous nous intéressons plus particulièrement aux propriétés de *viscosité*.

Dans ce travail, nous considérons des systèmes périodiques de particules en interaction. Nous proposons une nouvelle méthode de calcul de la viscosité valable pour tous systèmes périodiques, quel que soit le potentiel d'interaction considéré. Cette méthode est basée sur la formule dérivée par Helfand exprimant la viscosité en fonction de la variance du *moment de Helfand* croissant linéairement dans le temps.

Dans les années nonante, il a été démontré qu'un système composé de seulement deux particules présente déjà de la viscosité. Les deux disques *durs* interagissent en collisions élastiques dans un domaine carré ou hexagonal avec des conditions aux bords périodiques. Nous appliquons notre méthode de calcul des propriétés de viscosité dans les deux réseaux. Nous donnons également une explication qualitative des résultats obtenus.

L'étude de la relation entre les propriétés de viscosité et les grandeurs du chaos microscopique représente l'une des principales tâches de cette thèse. Dans ce contexte, le *formalisme du taux d'échappement* joue un rôle majeur. Ce formalisme établit une relation directe entre cette grandeur et la viscosité. Nous étudions numériquement cette relation et la comparaison avec les résultats obtenus par notre méthode sont excellents.

D'autre part, le formalisme du taux d'échappement suppose l'existence d'un *répulseur fractal*. Après avoir mis en évidence son existence, nous appliquons le formalisme proposant une formule exprimant la viscosité en termes de l'exposant de Lyapunov du système – mesurant le caractère chaotique de la dynamique – et de la dimension fractale du répulseur. L'étude numérique de cette relation dans le modèle à deux disques durs est réalisée avec succès et sont en excellent accord avec les relations obtenus précédemment.

Enfin, nous nous penchons sur les systèmes composés de $N$ disques durs ou sphères dures. Après une étude de l'équation d'état et des propriétés chaotiques, nous avons exploré les propriétés de viscosité dans ces systèmes. Les données numériques obtenues sont en très bon accord avec les prévisions théoriques d'Enskog. D'autre part, nous avons utilisé notre méthode de calcul de la viscosité dans des systèmes de Lennard-Jones. De plus, nous avons proposé une méthode analogue pour le calcul numérique de la *conduction thermique*. Nos résultats sont en très bon accord avec ceux obtenus par la méthode de Green-Kubo.

# Acknowledgements

First I would like to deeply thank Prof. Grégoire NICOLIS for having passed on to me his great interest for statistical physics since my undergraduate studies, and for his continuous support and encouragement during this thesis. I also wish to express my sincere gratitude to Prof. René LEFEVER as the chief of the *Service de Chimie-Physique* and for the high quality of his teaching of thermodynamics having exerted a deep influence on me.

My deepest thanks go to my PhD supervisor Prof. Pierre GASPARD who has welcomed me in his group since my master thesis in Chemistry. His proficiency, his availability as well as his kindness are such that I cannot imagine better conditions for achieving this dissertation.

I gratefully acknowledge everyone I interacted with during the realization of this thesis, especially Profs. J. Robert DORFMAN, Henk VAN BEIJEREN, and William G. HOOVER, Drs. Isabelle CLAUS, Thomas GILBERT, David SANDERS, Renaud LAMBIOTTE and Rainer KLAGES.

All scientists using computers know that many problems may occur and prevent one to go on with its research which is why it is a pleasure to thank Gerald HOUART for his essential help and devotion as a computer support.

Through these last years, I have gathered a lot of excellent memories and shared great companionship with all my colleagues. I would like to thank them all, especially Isabelle CLAUS, Ines DE VEGA, Massimiliano ESPOSITO, Thomas GILBERT, Cem SERVANTIE, Olivier LEJEUNE, David ANDRIEUX, Eric GERRITSMA, Nathan GOLDMAN, and all the others that I might have forgotten to thank.

Finally, I would like to thank all my family and my friends, especially Guillaume GODART and Olivier LEJEUNE, for their essential support. Without them, my thesis would never have been achieved. Furthermore, this list would not be complete without including my music teachers and all my friends of the music Academy of Schaerbeek. Music has taken an important and single role in my life. I would like to thank all of them by quoting Nietzsche: *Without music, [my] life would be a mistake*.

# Chapter 1

# Introduction

## 1.1   Motivation

This thesis is dedicated to the study of the transport property of viscosity from the viewpoint of dynamical systems theory. Viscosity – just like diffusion – is among the so-called irreversible properties that are studied in nonequilibrium thermodynamics.

To understand their description, we can easily consider a familiar phenomenon. If you were to put a drop of ink in a glass of water, we would know that the drop would diffuse throughout the whole vessel. We could not imagine the diffused ink coming back to the starting point. Furthermore, in the case of viscosity, a steel ball moving in a liquid will be subjected to a force that will limit its velocity. The ball will never regain its energy that is lost to the surrounding liquid. These are considered as two typical examples of macroscopic irreversible phenomena.

On the other hand, at the microscopic level, we observe another description. As a matter of fact, you view the atoms as very small spheres colliding with each other. According to Newton's equations of motion, the trajectory of these spheres can in principle go in one direction of time as well as the other one (back in time) after reversing their velocities.

This apparent paradox between microscopic and macroscopic descriptions is a known fundamental problem in statistical mechanics that has been studied since the 19th century with the pioneering work of Boltzmann. The question "How does the microscopic reversible dynamics produce irreversible phenomena at our scale?" remains an important issue. This question is the central issue of this thesis.

For fifteen years, the possibility that the chaotic behavior of the microscopic dynamics plays a role to understand the irreversible character of phenomena such as diffusion or viscosity has been investigated. Successive works in this direction have presented successful results and have established clear connections between the two scales. Among them are found the escape-rate formalism and the microscopic constructions of the hydrodynamic modes (Gaspard, 1998; Dorfman, 1999; Gaspard and Nicolis, 1990; Dorfman and Gaspard, 1995; Gaspard and Dorfman, 1995; Gaspard, 1993; Gaspard, 1996). Therefore in this work, we aim to study the viscosity coefficients in the framework of this approach. We will also consider periodic systems composed of only two hard disks. As a matter of fact, Bunimovich and Spohn have proved that these systems already present the property of viscosity (Bunimovich and Spohn, 1996). By using the method developed by Helfand (1960) , we will study the properties of viscosity. Fractal objects testifying the chaotic character of the microscopic dynamics will be put in evidence and, in this context, we will apply the escape-rate formalism. The results will be compared with those obtained by the Helfand-moment method. Furthermore, an extension to many-particle systems will also be studied.

The development of sciences is a long process. The problems that we are currently facing are the heirs to the scientific development in the past and their resolution will be the starting point of new questions in the future. In other words, *the being is nothing but a moment in the becoming*. It is precisely the reason why we choose to situate this thesis in the long history of the areas of physics having given rise to the problems in which we are interested (Viscardy, 2005). We will start with the presentation of the discovery of the concept of irreversibility as a fundamental property in nature. The development of hydrodynamics as the domain of physics being particularly concerned by the process of viscosity will be considered in section 1.3. Further, in section 1.4, we focus on the kinetic theory that has a long history and has played an important role in the establishment of the statistical mechanics, that is to say, in the understanding of the relation between the microscopic and macroscopic behaviors. In section 1.5, we cover the brief history of the development of a new revolutionary field of the science of the twentieth century. Sometimes compared to the scientific revolutions implied by quantum mechanics and the relativity, the phenomenon of chaos has been discovered in most of the natural sciences such as physics, chemistry, meteorology and geophysics. Our intent is to show the role of microscopic chaos in statistical mechanics of irreversible processes. Finally, section 1.6 will briefly demonstrate the benefits of the use of hard-sphere systems.

## 1.2   Irreversibility

The history of modern science holds its richness in the variety of conceptions which were developed successively throughout the years[1]. In the 17th century, it was the doctrine of the *clockmaker God* or *mechanical philosophy* which prevailed above the others.

The development of these ideas began with the work of Boyle (1627-1691) and others. According to his doctrine, Boyle stated that the world works like a clock: once created and energized, it can run forever in a deterministic way and without any need for a divine intervention. To ensure that this "clockwork universe" never runs down, Descartes (1596-1650) more or less intuitively introduced the statement that the total amount of motion in the world must remain constant. He defined this quantity as the *scalar* momentum $mv$ (quantity of matter multiplied by its scalar speed[2]) (Blackwell, 1966). However, as Descartes himself later observed, experiments did not confirm his enunciation of the conservation law of motion. Huygens (1629-1695) corrected it in 1668 by modifying it into the *vector* form. Huygens worked also on the problem of collisions. This revised law allowed him to claim that the vector sum of the product of the mass by the vector velocity $m\mathbf{v}$ remains unchanged after a collision even if it was inelastic and with dissipation of energy after the collision itself. Introducing this modification, it then appears that the world could stop after a certain amount of time. This was totally contradictory with the clockworld concept.

---

[1]For general informations on this section, see for example Holton and Brush (1985) and Brush (1983) .

The only way to avoid this possibility was to postulate that the matter is composed of elastic particles. If the macroscopic objects lose some motion after a collision, it is only in appearance because the motion is transferred to the invisible particles of the objects. Therefore, the idea emerged that heat is related to a rapid motion of the invisible parts inside the macroscopic bodies.

In the 18th century, the study of the motion of bodies and its change after collisions was of great importance. Moreover, in addition to this study, another quantity so-called *vis viva*, $mv^2$, was introduced for the first time. Besides, its conservation law in elastic collisions was remarkably proved by the same man who helped formulate the conservation of momentum: Huygens. This major contribution induced the first step to the central quantity acknowledged in mechanics: the energy. The concept of *vis viva* was used up to the 19th century, when the factor $\frac{1}{2}$ was added and it then became "kinetic energy".

The world-machine concept is often attributed to Newton (1642-1727) because of the importance played by his brilliant *Principia* published in 1687 (Newton, 1999). In his masterpiece, the Newtonian dynamics is depicted for the first time although it clearly appeared that the author was really opposed to it as he pointed out in his *Opticks* (Newton, 1952) published in 1704

> By reason of the [...] weakness of elasticity in solids, motion is much more apt to be lost than got, and is always upon the decay.

Therefore, in a certain sense, he suggested the existence of the dissipation of motion in the world. However, it should be specified that this viewpoint has to be considered in the light of the polemic between Newton[3] and Leibniz concerning the role of God (Koyré, 1957).

During the 18th century the concept of the "Newtonian world-machine" continued to dominate the thought of scientists. This century is also the one of the birth of Geology. It began with the question of the temperature of Earth's interior (Brush, 1994). Yet in 1693 Leibniz thought that Earth was originally hotter and cooled down, at least on the outside. Later Buffon (1707-1788) studied this question by leading some experiments on the cooling of heated spheres of iron. By considering a molten Earth at the origin he hence found that our planet was about 75,000 years (de Buffon, 1774). At the opposite, the founder of Geology James Hutton (1726-1797) disagreed with this theory of cooling. Defending his "Uniformitarism" (principle telling that the geological processes in the past have to be explained by using only the laws and physical processes that can now be observed) and accepting the hypothesis that the interior of the Earth is much hotter than its surface, he thought this situation had been like that forever. For him the geological processes are cyclic: alternance of periods of erosion and denudation implying the destruction of the mountains, and periods of uplifts of new continents (thanks to the subterranean fires). Actually one of his disciples Playfair promoted the uniformitarist viewpoint by citing the mathematical works by Lagrange and Laplace showing the cyclic movement of the planets around the sun. This position was perpetuated by Lyell (1797-1875) during the 19th century.

---

[2]More precisely, Descartes considered the volume of the bodies instead of their mass, the volume being for him the real measure of quantity of matter to be considered, as pointed out by Blackwell (1966) .

[3]More precisely, his student and faithful friend Clarke (1675-1729).

In the early 19th century, the first mathematical theory describing the propagation of heat was developed. Indeed, the French mathematician Fourier (1768-1830) was interested by the problem of the cooling of the Earth. This was the main motivation that lead him to study the heat conduction in solids. In 1819, in his *Mémoire sur le refroidissement séculaire du globe terrestre* (Fourier, 1819), he came up with an equation which presented a significant feature: unlike Newton's laws of motion, Fourier's equation is *irreversible*. Any system in which a temperature difference exists presents a heat flow from the high to low temperature. This was the discovery of an equation describing processes in which the past does not play the same role as the future. His theory therefore presents an important turning point in the history of physics due to the powerful analysis he developed (what we now call Fourier's analysis), and because it is explicitly based on a postulate of irreversibility. In the first half of the 19th century the eventual contradiction between the reversible Newtonian dynamics and the irreversible heat conduction did not appear because the most studied problem at this time was at the phenomenological level and the Newtonian mechanics had already been successfully applied to problems with dissipative forces (friction, etc.). The contradiction appeared during the first attempt to explain the macroscopic processes in terms of the assumed reversible Newtonian dynamics of the particles composing the system.

A few years after Fourier's work on the propagation of heat, Carnot (1796-1832) published in 1824 his essay *Réflexions sur la force motrice du feu* (Carnot, 1824). Driven by engineering ideas, he focused his research on the question of the limited efficiency of the steam engine. By using the *caloric theory*[4] appearing during the 18th century and developed by Lavoisier (1743-1794), Carnot concluded that a difference of temperature existing between two bodies gives the possibility of doing work by allowing heat to expand a gas as the heat flows from the hot body to the cold one[5]. However, an engine wrongly designed results in loosing the chance of doing work that might have been done. Therefore, proving that heat always flows from hot to cold bodies in the engine as it does anywhere in nature, and discovering that this equalization represents a loss of the opportunity to produce mechanical work, Carnot pointed out the existence of *dissipation*. More important, the discovery made first by Carnot in 1831, then Mayer (1814-1878) in 1842 (Mayer, 1842) and later by Joule (1818-1889) (Joule, 1847), brought up a fundamental statement: work and heat are actually two different expressions of the same quantity: the energy. Indeed, although heat and mechanical work seem to be two independent quantities, we now understand that heat can be transformed into work and vice versa. In other terms, the energy of an isolated system remains constant. A science of transformations was born: *thermodynamics*. This equivalence between heat $Q$ and work $W$ consists in the so-called *first law of thermodynamics*: the conservation of the internal energy $E$ of a closed system: $dE = dQ + dW$.

As we have seen, the original problem giving birth to thermodynamics is the separation between the concepts of conservation and reversibility. In mechanical transformations, the conservation of energy (at the beginning the *vis viva*) coincides with the idea of reversibility. On the other hand, the physico-chemical transformations can preserve the energy while they cannot be reversed. Hence it became necessary to define a quantity characterizing this irreversibility. In 1854 Clausius (1822-1888) introduced the quantity $dQ/T$ as a measure of the quantity of work lost during the transfer of heat from a hot to a cold body (Clausius, 1854; Clausius, 1856). Later, in 1865, he gave it the famous name *entropy*[6] (Clausius, 1865). Thanks to this new quantity, the *second law of thermodynamics* can now be stated very simply: the entropy of an isolated system always tends to increase. We thus have a well-established principle regarding the irreversible processes. We may therefore state that the two laws of thermodynamics combine the two concepts of conservation of energy and of nonreversibility of macroscopic phenomena.

---

[4]The caloric theory supposed that heat is a fluid composed of particles independent of the rest of matter. These particles repel each other but are attracted to the particles of ordinary matter. According to this theory it has to be an eventual material substance and thus should be limited. Hence the increase of the temperature of water by rotating a drill should be due to the transfer of the so-called *caloric* from the drill to the water. But, in 1798, Thompson (1753-1814) realized that one can produce heat without any limit. He then concluded that heat is not a chemical substance or a material substance but is the expression of a movement. However, since he did not propose any alternative theory of heat, the caloric theory had an important influence until the 1830's and the works done in particular by Joule.

[5]Actually later, Carnot abandoned the caloric theory.

[6]The ethymology of this word expressed in the most clearly way the essence of thermodynamics: indeed it comes from the Greek words $\varepsilon\nu\varepsilon\rho\gamma\varepsilon\iota\alpha$ (energy) and $\tau\rho\sigma\pi\eta$ (transformation).

As we will later see in section 1.4, the atomic theory increased its influence during the 19th century. The development of the kinetic theory of gases played an important role to prove the discrete character of matter. One of the most important leaders in this area was Boltzmann (1844-1906). He particularly proposed a more general definition of entropy, in terms of the probabilities of molecular arrangements, that can change even when there is no heat flow (e.g. the system become more disordered). His theory based on a statistical description was a first understanding of the link between the atomic level and the macroscopic phenomena, that is between the reversible microscopic dynamics and the macroscopic irreversible processes.

In the nineteenth-century society, in which social and economical activities, science and technology induced a progress never observed before in the history of Humanity, arose the idea of the evolution of species with Lamark (1744-1829) and later with Darwin (1809-1882). According to Darwin the species that we know today are the result of a long process of natural selection. This mechanism has induced the production of increasingly complex and organized beings, from the unicellular bacteries to the mammals such as the human beings. Whereas thermodynamics introduced a quantity measuring the continuous growth of disorder, biology put in evidence the continuous growth of order, of organization in the biological world. As a result, these two points of view influenced some philosophers: on the one hand, Adams (1838-1918), influenced by the discovery of constant dissipation of energy in the universe, expressed a pessimistic vision concerning the future of the society, as argued in *The Degradation of the Democratic Dogma* (Adams, 2004), and made the second law of thermodynamics an explicit basis for the tendency of history (Burich, 1987); on the other hand, the Darwinism strongly influenced philosophers like Spencer (1819-1903) who thought that all in universe goes gradually from a state of confused homogeneity to a differentiated heterogeneous state (Freeman, 1974; Rousseau, 1945). These two points seemed to be contradictory. One had yet to wait for the development of the nonequilibrium thermodynamics, especially attributed to Prigogine (1917-2003) and his coworkers, to resolve this apparent contradiction. Indeed they showed that, out of thermodynamic equilibrium, i.e. in open systems, matter is able to exhibit self-organization, although the internal production of entropy is absorbed by the environment (Glansdorff and Prigogine, 1971;

Nicolis and Prigogine, 1977). These recent works have pointed out the importance of nonequilibrium phenomena and have led to important advances on the properties of nonequilibrium systems. It is in this general context that this thesis is written.

## 1.3   Hydrodynamics and viscosity

The vital importance of water throughout all civilizations induced a special interest for the study of the properties and behaviors of fluids (in particular of water). The earliest quantitative application of "real fluid" or viscosity effects was by the ancient Egyptian Amenemhet (~1600 BC). He made a 7 degree correction to the drainage angle of a water clock to account for the viscosity change of water with temperature (which can be significant between day and night in this region). And in general the early centres of civilization in Egypt, Mesopotamia, India and China systematically used various machines for irrigation and water supply (Blair, 1944). In particular, Archimedes (287-212 BC), is not only considered as the father of hydrostatics for his famous law, but has also developed the so-called Archimedes screw, a water elevating machine, which has been used for different purposes (e.g. to extract water from rivers). Notice also that Heron of Alexandria (~ 10?-75) treated in *Mechanica* the problem of friction between a body and a plane and, considering a horizontal, frictionless plane, he said

We demonstrate that a weight in this situation can be moved by a force less than any given force[7].

The earliest extensive treatise devoted to practical hydraulics is a manual on urban water supply written by Roman soldier and engineer Sextus Julius Frontius (first century BC), inspector of the aqueducts and the public fontains in Rome. He noted that the amount of water discharged through an orifice in a given interval of time depends, not only on the size of the orifice, but also on its depth $h$ below the surface of the water in the reservoir from which the water is supplied. Starting upon the results of Galileo's experiments with falling bodies, Torricelli (1608-1647) came to the conclusion that the velocity $v$ of the water exiting out of the reservoir is proportional to the square root of the $h$. This is *Torricelli's theorem* which has been mathematically expressed as $v = \sqrt{2gh}$ later by Daniel Bernoulli (1700-1782). On the other hand, the centuries since the *Renaissance* have been characterized by the major influence of the engineerings on the development of physics. More specifically, the construction of bridges and canals induced a lot of theoretical studies on the flow of fluids. By observing the behavior of the flow of water in rivers, da Vinci (1425-1519) already came to the conclusion that, when the river becomes shallower or narrower, the water flows faster. Later Castelli (1577-1644) confirmed this result and gave the so-called *velocity-area law*: $vA = $ const, where $v$ is the velocity of water and $A$ the cross-sectional area of the flow. It consists of the first idea of continuity of flow (for an incompressible fluid), which was later developed by Euler.

---

[7]Cited by Russo (2004).

The term *hydrodynamics* was first used by Daniel Bernoulli as the title of his book *hydrodynamica* published in 1738 (Bernoulli, 1738). His theory was original because he was the first scientist to combine the sciences of hydrostatics (considering the pressure) and hydraulics (considering the motion of fluids). From the conservation of *vis viva* he reached the famous so-called *Bernoulli's principle* relating the velocity of flow at a point in a pipe to the pressure there: $\frac{\rho v^2}{2} + P = $ const. But in the modern sense, hydrodynamics began with the work of d'Alembert (1717-1783) and especially Euler (1707-1783).

During his Berlin period as an engineer, Euler was in charge of the construction of canals, the supply of water for King's Sans-Soucis Palace, and the improvement the water turbine. However, his strong interest for mathematics, mechanics and physics was much greater. Hydrodynamics was among his numerous topics of interest. In 1755, he derived the fundamental equations of hydrodynamics (Euler, 1755) by introducing the new and main concept of *fluid particle*[8]. First he proposed the modern and general form for the equation of continuity (1.1), which expresses the conservation of matter[9]

$$\frac{d\rho}{dt} + \rho \frac{\partial v_j}{\partial r_j} = 0 . \tag{1.1}$$

where $\rho$ is the mass density and $v_i$ the *i*-component of the velocity of the considered fluid particle. Furthermore, considering a small parallelepiped and pressure $P$ acting on its different faces depicted in Fig.1.1, by starting from the Newton's second law, he derived the general equations for the motion of an ideal fluid, the so-called *Euler's equations of motion*

$$\rho \frac{dv_i}{dt} = \rho \left( \frac{\partial v_i}{\partial t} + v_j \frac{\partial v_i}{\partial r_j} \right) = -\frac{\partial P}{\partial r_i} + f_i , \tag{1.2}$$

where $f_i$ is the *i*-component of the external force (gravity,....).

But these equations do not present the property of fluids which was more and more of interest at the beginning of the nineteenth century, that is the internal friction, or the *viscosity*[10] in modern terminology. Indeed, the dissipative feature of the fluid dynamics does not appear in Euler's equations and it was not until then considered as a fundamental property in nature. The first studies on this property go back already to Newton's Book II of his *Principia* (Newton, 1999). His theory stated that, if a portion of fluid is kept moving, the motion gradually communicates itself to the rest of the fluid. He ascribed to this effect the term *defectus lubricitatis*, that is, a *lack of slipperiness*. It is nothing but viscosity. Considering two particles of a viscous flow sliding one over another, then there exists friction, or viscous resistance along the surface of contact. The force of this resistance per unit area is nowadays known as the *shear stress* $\tau$. The hypothesis proposed by Newton[11] is that the shear stress depends on the speed with which the two layers slide one over another. By extension, we obtain that the shear stress is proportional to the velocity gradient in the perpendicular direction of the movement of the fluid layers: $\tau \sim \frac{\partial v_x}{\partial y}$.

---

[8]A fluid particle is imagined as an infinitesimal body, small enough to be treated mathematically as a point, but large enough to possess such physical properties as volume, mass, density, and so on.

[9]We use the Einstein's convention of summation over repeated indices: $v_j \frac{\partial v_i}{\partial r_j}$ has to be understood as $\sum_j v_j \frac{\partial v_i}{\partial r_j}$.

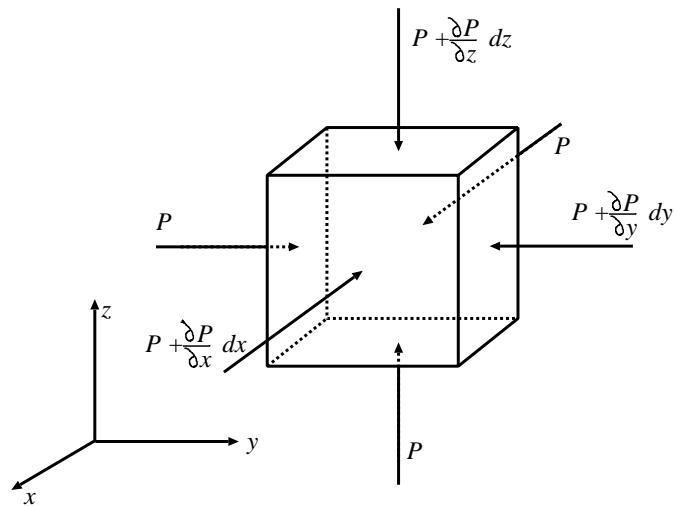

**Figure 1.1**. Pressures acting ont Euler's fluid infinitesimal parallelepiped.

At the end of the 18th century and the beginning of the 19th century, numerous and extensive investigations on the flow of pipes and open channels started to be carried out by scientists and engineers such as Du Buat, Girard, Prony, Coulomb, and especially Poiseuille (1797-1869). Being interested

---

[10]The word viscosity derives from the latin word *viscum* for mistletoe. The mistletoe berries produce a viscous glue which was used to lime birds. Viscosity is expressed in *Pa.s* or in *Poise* ($10^{-1}Pa.s$) in the honor of Poiseuille (1797-1869).

[11]In his Principia, Newton made precisely the following hypothesis:

The lack of slipperiness of the parts of a fluid is, other things being equal, proportional to the velocity with which the parts of the fluid are separated from one another (Book II, Section 9 in Newton (1999) ).

in the circulation of blood in capillary vessels, he then used glass capillaries of very much smaller bore that any of his predecessors. Long series of measurements of the quantity $Q$ of liquid discharged in unit time were carried out in function of the different factors. He then obtained in the forties the relation $Q = K\frac{PD^4}{L}$ where $P$ is the pressure, $D$ and $L$ respectively the diameter and the length of the capillary (Poiseuille, 1840a; Poiseuille, 1840b; Poiseuille, 1846). And in 1856, Wiedemann (1826-1899) as well as Hagenbach in 1860, deduced mathematically, by making use of Newton's hypothesis, that the constant $K$ is such that (Wiedemann, 1856; Hagenbach, 1860)

$$Q = \frac{\pi P R^4}{8\eta L} \tag{1.3}$$

where $R$ is the radius of the capillary. A new factor $\eta$ appears which Wiedemann proposed to call *Zähigkeitconstante der Flüssigkeiten*, that is, the *viscosity coefficient of the liquids* (Wiedemann, 1856). So for the first time the term viscosity was used in a technical sense.

In parallel, the first scientist having theoretically used the property of viscosity in the fundamental equations of hydrodynamics was the French civil engineer Navier (1785-1836) in a mémoire read in 1822 to the "Académie des Sciences" (Navier, 1823). His idea was that any pressure tends to reduce the distances between the molecules of the fluid. Taking into account the intermolecular repulsive forces produced by this action, which gave him the explanation of the Newton's *lack of slipperiness*, he added to the pressure acting on a small volume the contribution of this effect ($P\delta_{ij} \to P\delta_{ij} - \eta\frac{\partial v_i}{\partial r_j}$), that is a dissipative term implying the presence of the viscosity coefficient $\eta$. Being rediscovered by Stokes (1819-1903) in 1845 without introducing any molecular behavior (Stokes, 1845), the history has given the name of these two persons to the famous *Navier-Stokes equations*

$$\rho\frac{dv_i}{dt} = -\frac{\partial P}{\partial r_i} + \eta\left(\frac{\partial^2 v_i}{\partial r_j\,\partial r_j}\right) + f_i \ . \tag{1.4}$$

On the other hand, in 1829, Poisson (1781-1836) derived an equation in a mémoire devoted to the motion of solid bodies and fluids in which appeared a second constant (Poisson, 1831), in addition to the viscosity introduced by Navier. These two constants measure the response of the substance to two different types of forces: the first tending to shear it, and the second tending to expand or contract it. We saw above that the first type is characterized by the viscosity coefficient $\eta$ which is more precisely called *shear viscosity*. But until sixty years ago, most of the standard works did not consider the other constant $\eta'$ called the *second viscosity coefficient* by making it simply proportional to the shear viscosity. Since Stokes (1845) , it was usually admitted that

$$\eta' + \frac{2}{3}\eta = 0 \ , \tag{1.5}$$

which is generally called *Stokes' relation*. Stokes used this relation by arguing that in most cases of practical interest the rate of expansion would be so small that the value assigned to this sum would be unimportant. And before sixty years ago no direct determination of $\eta'$ has yet been made. Indeed it is only in 1942 that Tisza made this determination by evaluating the ratio $\frac{\eta'}{\eta}$ when ultrasonic vibrations are passed through a liquid (Tisza, 1942). He then showed that this vanishing relation is not justified in general and that the re-introducion of the second viscosity coefficient is necessary in order to get an agreement between the theory and experiments (Markham et al., 1951; Karim and Rosenhead, 1952; Brush, 1962b). Consequently, a new coefficient $\zeta$ can be defined as

$$\zeta = \eta' + \frac{2}{3}\eta \tag{1.6}$$

and is the so-called *bulk viscosity coefficient*.

The bulk viscosity is more difficult to measure experimentally than the shear viscosity. Moreover, its measure is generally less known. The different ways to measure it imply the phenomenon of absorption of sound in liquids and gases. Before Tisza (1942) , it was well-known that the absorption of sound in polyatomic gases and in liquids obtained experimentally exhibited quite great disagreements with the absorption predicted by the classical theory based on viscosity. Previously, it was usually admitted that the absorption is given as

$$\alpha_{\mathrm{cl}} = \frac{2\pi^2\nu^2}{\rho v_s^3}\left[\left(\frac{1}{c_V} - \frac{1}{c_P}\right)\kappa + \frac{4}{3}\eta\right] \ , \tag{1.7}$$

where $\nu$ is the acoustic frequency, $\rho$ the mass density, $v_s$ the acoustic velocity, $c_V$ and $c_P$ the specific heats respectively at constant volume and constant pressure, and $\kappa$ the thermal conductivity. Hence, only the shear viscosity was considered. In his paper (Tisza, 1942), he pointed out that Stokes' assumption (1.5) is not justified except in dilute monoatomic gases. However, for polyatomic gases and liquids, the bulk viscosity should be quantitatively quite important, sometimes more than the shear viscosity (see Table 1.1). Consequently, Tisza obtained a modified expression for the absorption

| fluid | $\eta \times 10^5$ Pa s | $\zeta/\eta$ |
|---|---|---|
| He | 1.98 | 0 |
| Ar | 2.27 | 0 |
| $H_2$ | 0.887 | 32 |
| $N_2$ | 1.66 | 0.8 |
| $O_2$ | 2.07 | 0.4 |
| $CO_2$ | 1.50 | 1.000 |
| Air | 1.85 | 0.6 |
| $H_2O$ (liquid) | 85.7 | 3.1 |
| Ethyl alcohol | 110 | 4.5 |
| Glycerine | 134.000 | 0.4 |

**Table 1.1**. Shear and bulk viscosities obtained experimentally at 300 K and 1 Atm in monoatomic, diatomic and polyatomic gases, and in liquids; values given by Thompson (1972).

coefficient in which the bulk viscosity is added in the expression (1.7) as follows

$$\alpha_{\text{obs}} = \frac{2\pi^2\nu^2}{\rho v_s^3} \left[ \left( \frac{1}{c_V} - \frac{1}{c_P} \right) \kappa + \frac{4}{3}\eta + \zeta \right] . \tag{1.8}$$

or by considering the ratio of the viscosity coefficients $\frac{\zeta}{\eta}$

$$\frac{\zeta}{\eta} = \frac{4}{3} \frac{\alpha_{\text{obs}} - \alpha_{\text{cl}}}{\alpha_{\text{cl}}} \left( 1 + \frac{3}{4} \left( \frac{1}{c_V} - \frac{1}{c_P} \right) \frac{\kappa}{\eta} \right) . \tag{1.9}$$

On the other hand, the tangential force acting for example on the plane $xy$ should be related to the sliding in the two perpendicular directions composing this plane, that is $\frac{\partial v_x}{\partial z} + \frac{\partial v_z}{\partial x}$ in the direction of $x$, and $\frac{\partial v_y}{\partial z} + \frac{\partial v_z}{\partial y}$ in the direction of $y$. This is why Barré de Saint-Venant (1797-1886) combined both of them in 1843 (de Saint-Venant, 1843) and obtained more precise expressions for the pressure tensor, e.g. $P_{xx} = P + 2\eta \frac{\partial v_x}{\partial x}$ and $P_{xy} = \eta \left( \frac{\partial v_x}{\partial y} + \frac{\partial v_y}{\partial x} \right)$. It allows one to write the Navier-Stokes equations in a more complete form as[12]

$$\rho \frac{dv_i}{dt} = -\frac{\partial}{\partial r_j} \left[ P \, \delta_{ij} - \eta \left( \frac{\partial v_i}{\partial r_j} + \frac{\partial v_j}{\partial r_i} - \frac{2}{3}\delta_{ij} \, \frac{\partial v_l}{\partial r_l} \right) - \zeta \, \delta_{ij} \frac{\partial v_l}{\partial r_l} \right] + f_j . \tag{1.10}$$

When the fluid is considered incompressible ($\frac{\partial v_l}{\partial r_l} = 0$), one recovers the original Navier-Stokes equation (1.4). Let us also mention that these equations are valid in so far as Newton's assumption is reasonable. A wide range of fluids (water, air, etc.) follows the linear relation between the shear stress $\tau$ and the velocity gradient. But liquids composed by lengthened molecules (polymers, etc.) obey to different viscosity laws, which are called non-Newtonian.

---

[12]For further information on the history of hydrodynamics, see Rouse and Ince (1957), Tokaty (1994), Mikhailov (1994) and Stokes (1846). A more rigorous derivations of Navier-Stokes equations can be found for example in (Landau and Lifshitz, 1959).

As we will see in the next section, the 19th century and the beginning of the 20th century before the first world war has seen the decisive development of the scientific atomistic conception of matter and the will to derive macroscopic phenomena like viscosity in terms of the Newtonian microscopic dynamics. These works gave rise to the powerful kinetic theory of gases.

## 1.4    Kinetic theory of gases

The idea of atomism goes back to the fifth century BC with the ancient Greek philosophers Leucippus and Democritus. However, the history of the kinetic theory of gases does not really begin until the seventeenth century when Torricelli, Pascal (1623-1662) and Boyle first established the physical nature of the "sea" of air that exerts mechanical pressure on surfaces in contact with it. This development of the concept of air pressure can be considered as part of the change in scientific attitudes which led to the explanations based simply on matter and motion instead of postulating "occult forces" or teleological principles. Consequently, Boyle (1662) discovered the law $PV = $ const now called *Boyle law*.

Usually we consider Daniel Bernoulli as the first scientist having proposed a kinetic theory of gases. Actually, in his famous treatise *Hydrodynamica* (Bernoulli, 1738), he gave a derivation of the gas law of Boyle and Mariotte from "billiard ball" model, assuming that the gas consists of a very large number of small particles in rapid motion. Moreover, using the principle of conservation of *vis viva* he concluded that pressure is proportional to the square of the velocities of gas particles, and thus proportional to the temperature. In other terms heat is nothing but atomic motion. However, in spite of these results, Bernoulli's model was forgotten. He was about a century ahead of his time with his theory. The latter success of Lavoisier's caloric theory buried it by proposing another conception of the matter based on the equilibrium between the caloric repulsion (due to the atmosphere of caloric whose density increases with temperature) and the gravitational attraction between matter particles.

In chemistry, the atomic theory emerged at the beginning of the 19th century with Dalton (1766-

1844). Thanks to the laws of chemical discontinuities, that is the law of *constant proportions*[13] discovered in 1806 by Proust (1754-1822) and the law of *multiple proportions*[14] in 1802-1804 by Dalton himself, he proposed in 1808 a discrete view of matter, which should be composed by indivisible entities: the atoms which are characterized by different *atomic weights*. But these laws and experiences at that time did not reject a concept of *equivalents* first proposed in 1792 by Richter (1762-1807). Because of different ambiguities amongst the partisans of the atomic theory, the equivalentism had a great success during the 19th century. However, many new phenomena and laws were discovered that only the atomic hypothesis could explain. Among these, the law of Gay-Lussac (1778-1850) discovered in 1809[15],which allowed to Avogadro (1776-1850) to emit his famous hypothesis (1811) saying that all gases considered in the same conditions of temperature and pressure contain, for equal volumes, the same number of molecules[16]. We can also cite the law of Dulong (1785-1838) and Petit (1791-1820)[17] (1819); the emergence of structural chemistry, in particular the development of the theory of valence by Kekulé (1829-1896), organic chemistry and the concept of isomerism, and stereochemistry. But one of the most important works having played a fundamental role in the development of the atomic theory and its acceptance is the establishment of the periodic table of the elements in 1869 by Mendeleev (1834-1907) (Mendeleev, 1869a; Mendeleev, 1869b). Indeed, by assuming his periodic law in the properties such as the valence of elements when these are listed in order of atomic weight, he predicted the existence of unknown elements corresponding to the gaps in his famous table (see Fig.1.2) and to which a given atomic weight is associated (Brush, 1996). The discovery of some of them in the following years confirmed the periodic property of elements. In the polemic between atomists and equivalentists, Mendeleev's brilliant idea heavily contributed to the acceptance of the atomic conception of the matter amongst the chemists[18].

While the caloric theory was being brought to its final stage of perfection by Laplace, Poisson, Carnot and others, Herapath (1790-1868) in 1820 and 1821, as well as Waterston (1811-1883) in

---

[13]The proportion between two combining elements cannot vary continuously.

[14]When two elements can combine according to different ratios, the ratios of their weight in the different cases are simple, that is integer numbers.

[15]The ratios of volumes of reacting gases are integer numbers.

[16]Let us notice that Avogadro was the first scientist having used the term *molecule* in the modern sense. The word is derived from the french word *mol'ecule* meaning "extremely minute particle" which comes from the Latin word *molecula*. This term is the diminutive of *mole* meaning "mass, cluster, great number, multitude". On the other hand, the word *atom* is derived from the Latin term *atomus* used by Lucretius, and from the Greek word $\alpha\tau o\mu o\varsigma$, which means uncut.

[17]The atomic weight multiplied by the specific heat of an element is independent of the nature of the element.

[18]For further information on the history of atomic theory in chemistry, see for example Pullman (1995) and Leicester (1956).

```
                                    Ti =  50      Zr =  90      ? = 180
                                    V  =  51      Nb =  94      Ta = 182
                                    Cr =  52      Mo =  96      W  = 186
                                    Mn =  55      Rh = 104,4    Pt = 197,4
                                    Fe =  56      Ru = 104,4    Ir = 198
                             Ni = Co = 59         Pd = 106,6    Os = 199
  H = 1                             Cu = 63,4     Ag = 108      Hg = 200
          Be =  9,4      Mg = 24    Zn = 65,2     Cd = 112
          B  = 11        Al = 27,4  ?  = 68       Ur = 116      Au = 197?
          C  = 12        Si = 28    ?  = 70       Sn = 118
          N  = 14        P  = 31    As = 75       Sb = 122      Bi = 210?
          O  = 16        S  = 32    Se = 79,4     Te = 128?
          F  = 19        Cl = 35,5  Br = 80       J  = 127
  Li = 7  Na = 23        K  = 39    Rb = 85,4     Cs = 133      Tl = 204
                         Ca = 40    Sr = 87,6     Ba = 137      Pb = 207
                         ?  = 45    Ce = 92
                         ?Er = 56   La = 94
                         ?Yt = 60   Di = 95
                         ?In = 75,6 Th = 118?
```

**Figure 1.2**. Mendeleev's periodic table having suggested the existence of unknown elements (picture from (Mendeleev, 1869b)).

1845 proposed their kinetic theory[19]. However both of them were rejected by the Royal Society. It must be emphasized that what needed to be established was not simply a connection between heat and molecular motion, for that was already admitted by many scientists and was not considered incompatible with the caloric theory; it was rather the notion that heat is *nothing but* molecular motion, and the idea that molecules *move freely through space* in gases rather than simply vibrating around fixed positions. This statement could not yet be accepted.

During the period between 1842 and 1847 the general scientific and intellectual climate implied quite simultaneous works by different scientists on a concept which is one of the most important reasons of the revival of kinetic theory: the *conservation of energy*. Indeed, the influence of the German romanticism and *Naturphilosophie*, especially the idea that there must be a single unifying principle underlying all natural phenomena (Brush, 1967). Furthermore, this period inherited different discoveries of various conversion processes[20] and gave birth to this principle usually attributed to Mayer (1814-1878) (Mayer, 1842) and Joule (1818-1889) (Joule, 1847). Mayer emphasized the philosophical generality of the principle while Joule provided the experimental verification in particular cases. From that moment heat, mechanical work, electricity and other apparently different entities are considered as different forms of the same thing, now called *energy*. And in 1847 Joule and Helmholtz (1821-1894) indicated quite clearly that *mechanical* energy is regarded as the basic entity. It was this prejudice toward mechanical explanations that made the kinetic theory appear to be an obvious consequence of the principle of conservation of energy. If we are convinced that heat and mechanical energy are interconvertible, what is more natural than to conclude that heat is a mechanical energy?

---

[19]For further informations on their theories, see respectively Brush (1957a, 1957b).

[20]Oersted's discovery of electromagnetism (1820), Seebeck's discovery of thermo-electricity (1822), Faraday's many discoveries in electricity and magnetism, and many others.

The real breakthrough for the kinetic theory took place when Krönig (1822-1879) assumed in 1856 (Krönig, 1856) that the molecules of gas move with constant velocity in straight line until they strike against other molecules, or against the surface of the container[21]. But the kinetic theory was still confronted to objections like those supported by Buys-Ballot (1817-1890). Since the kinetic theory claimed that the velocities of molecules were of the order of several hundred meters per second, he pointed out that we would expect two gases to mix with each other very rapidly. However, the experience shows that the process takes a certain time, of the order of several minutes. In order to answer this objection he showed that, in real gases, the molecules could not travel without colliding with other molecules. Consequently, Clausius (1858) introduced a new concept: the so-called *mean free path* of a molecule between two successive collisions . This concept was major not only for the further developments, but also because it establishes in concrete terms one of the most fundamental ideas of the kinetic theory of gases rejected in the past: that molecules can move freely through space and yet collide with each other.

The early kinetic theorists assumed that molecules tending to equilibrium move all at the same velocity. Maxwell (1831-1879) was the first scientist who introduced the idea of random motion for the molecules – hence statistical considerations[22]. In 1860, in his first paper on kinetic theory entitled *Illustrations of the dynamical theory of gases* (Maxwell, 1860), he suggested that, instead of tending to equalize the velocities of all molecules, the successive collisions would produce a statistical distribution of velocities in which all might occur, with a known probability. For thermal equilibrium he could then derive from symmetry considerations his famous distribution function which in modern notation is given $f(v) = 4\pi(m/2\pi k_B T)^{3/2} v^2 \exp(-mv^2/2k_B T)$. Hence statistical character appeared to be a fundamental element of kinetic theory.

In 1859, Maxwell came to the kinetic theory as an exercise in mechanics involving the motions of systems of particles acting on each other only by impact. Being interested in viscosity he proposed a mechanism which allowed him to establish a relation between the (shear) viscosity and the mean free path. By considering gas divided into parallel layers and supposing that the motion is uniform in each layer but varying from one to another, he showed that the viscosity $\eta$ should have to be proportional to the mean free path $\langle l \rangle$, the mass density $\rho$ and the molecular mean velocity $\langle v \rangle$, relation written as

$$\eta_M = \frac{1}{3} \langle v \rangle \ \rho \ \langle l \rangle \ . \tag{1.11}$$

---

[21]Actually his publication did not represent a real advance compared to previous works by Bernoulli and Herapath. But his influence in the physics community was important and induced an special interest among the physicists for the kinetic theory.

[22]Actually Maxwell was influenced by works on statistics, in particular the works realized by Quetelet (1796-1874) on the height distribution of a population of soldiers. Quetelet stated the idea of the 'average man' with 'average height' etc., as the result of an experimentally established normal distribution function. Maxwell worked out the idea of an 'average molecule' in analogy with that of the 'average man'.

He also published this result in his first article on the kinetic theory (Maxwell, 1860). However, as Clausius and himself concluded, the mean free path is inversely proportional to the density, which implied the surprising result that the viscosity in a gas does not depend on its density. Whereas it was known that this dependence does exist in liquid, and according to the opinion of Stokes, Maxwell thought his predictions were absurd and therefore that the kinetic theory is wrong, or at least inadequate. But at that time a few experiments on the gas viscosity had already been realized. The first studies probably began in the 1840's with e.g. Graham (1846, 1849). However, accurate experiments had not yet been done on the viscosity of *gases*, therefore, in 1866, Maxwell himself carried out his own experiment and found out that viscosity remains constant over a large range of pressure (Maxwell, 1866). This work played an important role in the development of the kinetic theory and its acceptance by most of scientists who so far had been doubting it. For instance, in 1865, Loschmidt (1821-1895) gave the first convincing estimate of the diameter of an air molecule (about 10 $\overset{\circ}{\text{A}}$ which is about four times too large) as well as the Avogadro number (about $6.025 \times 10^{23}$) (Loschmidt, 1865). On the other hand, Maxwell formula, combined with the equation of state for real gases derived in 1873 by van der Waals (1837-1923)[23] (van der Waals, 1873), belongs to the list of thirteen different phenomena mentioned in 1913 by Perrin (1870-1942) and allowing one to evaluate the atomic magnitudes as well as the Avogadro number (Perrin, 1991).

As the kinetic theory establishing a microscopic mechanical basis to the macroscopic processes such as viscosity was attracting more and more attention by physicists, the idea to reconcile the second law of thermodynamics with the principles of mechanics emerged and the first scientist having been concerned by it was Boltzmann (1844-1906). His first major achievement (Boltzmann, 1868) was to extend Maxwell's distribution law to the case of an external force field (field deriving from the potential energy $V$) is present. He obtained the so-called *Boltzmann factor* which, combined with Maxwell's velocity-distribution law, constitutes the basic principle of statistical mechanics (Maxwell-Boltzmann distribution law).

---

[23]Work based on the *virial theorem* introduced by Clausius (1870): 'the mean kinetic energy of the system is equal to the mean of its virial multiplied by $-\frac{1}{2}$: $\overline{E} = -\frac{1}{2} \sum_i \overline{\mathbf{r}_i \cdot \mathbf{F}_i}$', where $\mathbf{r}_i$ and $\mathbf{F}_i$ are respectively the position of the $i$th particle and the force acting on it.

In 1872, Boltzmann attempted to establish an equation describing the changes in the distribution resulting from collisions between molecules. Indeed he considered the single particle distribution function $f(\mathbf{r}, \mathbf{v}, t)$ so that $f(\mathbf{r}, \mathbf{v}, t)\delta\mathbf{r}\delta\mathbf{v}$ gives the average number of molecules in the infinitesimal volume $\delta\mathbf{r}\delta\mathbf{v}$ around the position $\mathbf{r}$ and the velocity $\mathbf{v}$. The evolution of this distribution function is governed by the integro-differential equation which is now called the *Boltzmann equation* (Boltzmann, 1872; Boltzmann, 1995)

$$\frac{\partial f}{\partial t} = -\mathbf{v} \cdot \frac{\partial f}{\partial \mathbf{r}} + J_B(f, f) \; . \tag{1.12}$$

Here $J_B$ is the binary collision term[24] taking only two particle collisions into account, which is a good approximation for dilute gases. An important assumption made by Boltzmann is the so-called *Stosszahlansatz*[25], which assumes that the velocities of colliding particles must be uncorrelated. Later Jeans (1877-1946) developed this statement and introduced the assumption of *molecular chaos*[26] (Jeans, 1903; Jeans, 1904). Boltzmann showed that collisions always push $f$ toward the equilibrium Maxwell distribution. In particular the quantity $H = \int f \log f$ always decreases with time unless $f$ is the Maxwell distribution in which case $H$ remains constant. This is *Boltzmann's H-theorem* $\frac{dH}{dt} \leq 0$ (Boltzmann, 1872). Boltzmann suggested that $H$ could be considered as a generalized entropy having a value for any state, contrary to the thermodynamic entropy defined only for equilibrium states. On the other hand, he explained the reason why the Maxwell distribution law is the one corresponding to the thermal equilibrium by showing that this distribution is the one most likely to be found, because it corresponds to the largest number of microstates. By comparing the two approaches (the kinetic approach and the one based on probabilities), Boltzmann could conclude that the process of irreversible approach to equilibrium, which is a typical example of entropy-increasing process, corresponds to a transition from less probable to more probable microstates. Entropy itself can therefore be interpreted as a measure of probability. By defining $\Omega$ the probability of a macrostate, which is proportional to the number of associated microstates, Boltzmann proposed a generalized entropy of a macrostate $S = k_B \log \Omega$ (see Fig.1.3), now called the *Boltzmann entropy*.

---

[24] $J_B(f, f) = \int \sigma_{\text{diff}} |\mathbf{v}_1 - \mathbf{v}| \left( f' f_1' - f f_1 \right) d\mathbf{v}_1 d\Omega$ where $f$ and $f_1$, $f'$ and $f_1'$ are the functions of the two particles respectively before and after the collision. $\sigma_{\text{diff}}$ represents differential cross section and $\omega$ the angular variables of integration. For a detailed derivation of the Boltzmann equation, see . Boltzmann (1995) and Dorfman (1999).

[25]"Assumption about the number of collisions".

[26]For some informations on the distinctions between both assumptions, see (Ehrenfest and Ehrenfest, 1990).

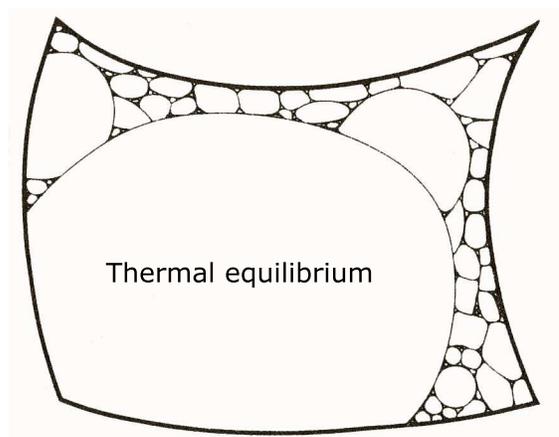

**Figure 1.3**. Division of the phase space of microstates associated with the same macrostate. The volume of the phase space associated with the thermal equilibrium appears to be so much greater than the others that the system, out of equilibrium at the origin, evolves naturally to this macrostate. The entropy $S$, proportional to the logarithm of the corresponding phase-space volume, then takes its maximal value. Picture from Penrose (1991).

With the turning of the century Gibbs (1839-1903) transformed the kinetic theory into a more abstract mathematical theory based on the concept of ensembles of similar systems. Actually Boltzmann (1884) himself was working in this way by introducing the so-called *Ergoden*, which corresponds to the *microcanonical ensemble* of Gibbs (1839-1903). In his definitive treatise, Gibbs (1902) presented this ensemble as one of a series of ensembles. This new formalism appeared to be a powerful technique for calculating and predicting the physical properties of bulk matter *at equilibrium* from the properties of atoms.

Despite all the brillant discoveries due to the atomic hypothesis of matter accumulated at the end of the 19th century, a lot of scientists (amongst the most important) continued to reject the atomic theory. In general, they belonged to movements called *equivalentism* represented by especially Berthelot (1827-1907), *energetics* with Ostwald (1853-1932) and Duhem (1861-1916), or *empiriocritism*, the doctrine of Avenarius (1843-1896) and Mach (1838-1916)[27]. The debate then were still vigorous between anti-atomists and the partisans of the kinetic theory, especially Boltzmann. Fortunately, during the first decade of the 20th century, the triumph of the kinetic theory and the atomic hypothesis arrived with the work done in 1905 by Einstein (1879-1955) on the *Brownian motion* (Einstein, 1905). Being aware of the attacks on kinetic theory by Mach, Ostwald and others, Einstein started his article by contrasting the predictions made by thermodynamics and by the kinetic theory (in particular by pointing out that thermodynamics distinguishes the Brownian particle and the "hypothetic" molecules composing the liquid, whereas kinetic theory does not and says that the colloidal particle should follow the Clausius' *equipartition theorem*). By combining Stokes' formula for the force on a sphere moving through a viscous fluid, and the formula for the osmotic pressure of dissolved molecules, he derived an expression for the mean-squared displacement of the Brownian particle submitted to the molecular agitation, which gave a way to evaluate the Avogadro's number $N_{Av}$. In 1908, Perrin (1870 - 1942) made different experiments in application to the Einstein's Brownian-motion theory and found $N_{Av} = 6.7 \times 10^{23}$ (Perrin, 1908). Later, in 1913, he published his now famous book *Les atomes* (Perrin, 1991) in which he proposed thirteen different experiments based on different phenomena for evaluating Avogadro's number. The quite narrow range $6.0 - 7.5 \times 10^{23}$ thus obtained is such that we now attribute the demonstration of the discrete feature of the matter to Perrin, especially due to his efforts to convince the opponents of atoms' existence[28]. Whereas Ostwald and others recognized it, some of them like Mach and Duhem continued to reject them (Pullman, 1995).

---

[27]For further informations, see for example Pullman (1995), Brush (1967) and Kubbinga (1990).

If the performances made by the kinetic theory were already numerous, new works and predictions were added and thus confirmed the depth of this theory. In 1911, Enskog (1884-1947) predicted that diffusion would appear in a mixture when one imposes a temperature gradient, phenomenon now called *thermal diffusion*. One year later, Chapman (1888-1970) derived an expression for the thermal diffusion coefficient.

Around the end of the 19th century and the beginning of the 20th century different works based on the mean-free-path method were achieved by a lot of physicists (Maxwell, Boltzmann, Tait, Rayleigh,...)[29] to improve the expression for the viscosity in low-density gases. But one has to wait for the almost simultaneous works carried out by Chapman in 1916 and Enskog in 1917 to obtain an expression for the viscosity derived from the kinetic equation[30] (Chapman, 1917; Enskog, 1917). The derivation known today is that of Enskog, and is based on a series solution of Eq.(1.12) for the distribution function by introducing a parameter $\lambda$. He therefore derived analytical expressions for the transport coefficients for general potential of interactions. In the case of the hard-sphere potential, the

---

[28]Let us mention that the recent measurements gives us $6.02214199(47) \times 10^{23}$ for the current value of the Avogadro's number (Mohr and Taylor, 1998).

[29]Especially the dependence in temperature and the factor $\frac{1}{3}$ in Eq.(1.11).

[30]Notice that Chapman started from the Maxwell's transport theory whereas the Enskog's derivation was based on Boltzmann's. For a brief discussion on both methods, see Brush (1962a).

viscosity coefficient is written as

$$\eta_B = 1.0162 \, \frac{5}{16 \, \sigma^2} \, \sqrt{\frac{mk_BT}{\pi}} \tag{1.13}$$

where $\sigma$ and $m$ are respectively the diameter and the mass of the particles. This expression for the viscosity is now called the *Boltzmann viscosity*. This new expression confirms the temperature-dependence and the absence of density-dependence predicted by earlier theory, especially that of Maxwell in Eq.(1.11)[31]. Furthermore, being available for any potential interaction, the Chapman-Enskog's developement turned out to be a method for determining the actual force law of real molecules by comparing the experimental and theoretical results for transport properties as viscosity and thermal conductivity. This was the way followed by Lennard-Jones (1894-1954) until the early thirties when he proposed the now called *Lennard-Jones 6-12 potential* which is still considered today as the most realistic potential existing between particles (Lennard-Jones, 1931)[32].

Maxwell and others showed both theoretically and experimentally that the viscosity coefficient is independent of the density in dilute gases. However, even in moderately dense gases, it appears that this property is no longer observed. In 1922, Enskog proposed an extension of his previous method in order to predict this dependence (Enskog, 1922). Considering a hard-sphere system he assumed that the collision rate in a dense gas would be changed by a factor $\chi$ which could be related to the equation of state. He modified the Boltzmann equation (1.12) by replacing $f f_1$ in which both $f$ and $f_1$ are evaluated in the same point in space, by $f(x, y, z, ...) f_1(x_1, y_1, z_1, ...)$, where the point points $(x, y, z)$ and $(x_1, y_1, z_1)$ must be separated by the distance $\sigma$. In such a way, he obtained[33]

$$\eta_E = \eta_B \left( \frac{1}{\chi} + \frac{4}{5} \, b_0 n + 0.7614 \, b_0^2 n^2 \chi \right) \tag{1.14}$$

in which the factor $\chi = 1 + \frac{5}{8} b_0 n + 0.2869 \, (b_0 n)^2$, $b_0 = \frac{2\pi\sigma^3}{3}$ and $n$ the number density. A density-dependent expression for the viscosity is therefore obtained[34]. One had to wait for a decade to compare Enskog's theoretical predictions to experimental data. Indeed, Michels and Gibson (1931) measured the viscosity of nitrogen gas at pressure up to 1000 atm (1013250 hPa) and obtained a good agreement with the semi-empirical Enskog theory presented below. Using this method, they also evaluated the dimensions of the gas molecule at different temperatures. Let us point out that the only work allowing Enskog to test his theory was the one accomplished by Warburg and von Babo (1882) for carbon dioxide. The agreement is quite good (see Table 1.2).

---

[31] By substituting the mean velocity $\overline{v}$ and the mean free path $\overline{l}$ by their expression in terms of temperature and atomic properties, one obtains $\eta = \frac{2}{3\,\sigma^2} \sqrt{\frac{mk_BT}{\pi^3}}$.

[32] For more historical elements of the development of expression for interatomic forces, see for example Brush (1970).

[33] For the development of Enskog's theory, see for example Chapman and Cowling (1970) and Hirschfelder et al. (1954).

[34] Let us mention that, in 1899-1900, by considering the analogy with the modification of the equation of state for ideal gas when the effect of finite molecular size is taken into account, and the correction the mean free path due to the effect of

| Mass density $[g/cm^3]$ | Kinematic viscosity (experimental) $\frac{\eta}{\rho}\,[10^3\,poises/(g/cm^3)]$ | Kinematic viscosity (calculated) $\frac{\eta}{\rho}\,[10^3\,poises/(g/cm^3)]$ |
|---|---|---|
| 0.100 | 1.80 | 1.91 |
| 0.170 | 1.12 | 1.17 |
| 0.240 | 0.908 | 0.899 |
| 0.310 | 0.784 | 0.774 |
| 0.380 | 0.724 | 0.719 |
| 0.450 | 0.702 | 0.701 |
| 0.520 | 0.704 | 0.705 |
| 0.590 | 0.722 | 0.724 |
| 0.660 | 0.756 | 0.754 |
| 0.730 | 0.795 | 0.792 |

**Table 1.2.** Kinematic viscosity $\frac{\eta}{\rho}$ at temperature $T = 313.45K$, near the critical point. Comparison between the experimental results obtained by Warburg and von Babo (1882) for carbon dioxide and Enskog's predictions. As the density increases, he kinematic viscosity goes through a minimum after which it increases. Enskog's predictions confirm this property (data given in Enskog (1922)'s paper).

In the fifties and sixties, new experimental data for noble gases (for viscosity as well as for thermal conductivity) were added to allow a comparison and a test of theoretical predictions[35]. The particular advantage of Enskog's theory compared to other theories for dense gases is that it requires the adjustment of only a few parameters like the diameter of the spheres in order to observe an agreement between Enskog's viscosity and the experimental data for a larger range of density[36].

---

excluded volumes on the collision rate, Jäger (1865-1938) modified the dilute gas viscosity formula (1.11) and obtained

$$\eta_J = \eta_M \left( \frac{1}{A} + 8\,b_0 n + 16\,b_0^2 n^2 A \right)$$

where $A = 1 + \frac{5b_0 n}{2} + \cdots$. The similarity with the Enskog viscosity (1.14) is remarkable (Brush, 1976).

[35]see for example Senger (1965, 1968), Hanley et al. (1972).

[36]Indeed effective values may be attributed to the parameters $b$ and $\chi$. It was Michels and Gibson (1931) who introduced such a procedure by identifying the pressure $P$ in the equation of state for rigid spheres with the thermal pressure $T(\partial p/\partial T)_V$ of the real gas

$$bn\chi = \frac{1}{R}\left(\frac{\partial PV}{\partial T}\right)_V - 1\,.$$

For low densities, one should require that the the Enskog viscosity (1.14) reduces to the Chapman-Enskog viscosity (1.13) by requiring $\lim_{n\to 0}\chi = 1$, so that we get (Hirschfelder et al., 1954)

$$b = B + T\frac{dB}{dT}$$

In Table 1.1, it is shown that the bulk viscosity is vanishing in the dilute monoatomic gases (argon and helium) as predicted by the Enskog theory. Indeed Enskog's bulk viscosity (1.15) for hard spheres

$$\zeta = 1.002 \chi \eta_B (b \, n)^2 \tag{1.15}$$

vanishes for very low density. However, a non-vanishing ratio $\zeta/\eta$ has been measured in liquid argon (Naugle and Squire, 1965) as well as in dense gaseous argon (Madigosky, 1967). A comparison with the modified Enskog theory briefly presented above can be found in Hanley and Cohen (1976). Consequently Stokes' relation is not shown to be justified against many cases, not only through theoretical predictions, but also experimentally.

In the sixties it was observed that transport coefficients, in particular the viscosity coefficient $\eta$, cannot be expressed in a power series in terms of the density $n$ ($\eta = \eta_B + \eta_1 n + \eta_2 n^2 + \eta_3 n^3 + \dots$) Indeed it has been shown that correlations between molecules is observed over large distances, larger than the range of the intermolecular interaction. Hence Dorfman and Cohen (1967) obtained theoretically that the coefficient of the quadratic term contains a contribution proportional to the logarithm of the density so that the previous density expansion should be rewritten as

$$\eta = \eta_B + \eta_1 \, n + \eta_2' \, n^2 \ln n + \eta_3 \, n^3 + \dots \, . \tag{1.16}$$

Nevertheless although the importance of the discovery of long-distance correlations is great in the modern kinetic theories, their contributions do not seem to be important, as it was showed by Sengers (1966) for the viscosity in a hard-disk gas , and later by Kamgar-Parsi and Sengers (1983) in a gas of hard spheres . On the other hand, since the early seventies several attempts have been made to detect this logarithmic-density dependence of experimental data, in particular for the viscosity coefficient[37]. Even if the viscosity is generally the transport coefficient measured with the highest precision, it has never been shown that the addition of the logarithmic term was necessary. Although we now know that Enskog viscosity is not exactly correct, the latter may still be used in order to compare numerical results. This is why, despite these facts, we consider Enskog's theory in this thesis[38].

Following Enskog theory proposed in 1922, we have to wait 1946 to observe a revival of interest

---

where $B$ is the second virial coefficient.

[37]See for example Kestin et al. (1971, 1980, 1981).

[38]A historical survey of the discovery of the logarithmic-density dependence is given by (Brush, 1972) Brush (1972), pp.72-80.

for the extension of kinetic theory to dense gases and liquids. Indeed, after an original idea of Yvon (1935), a hierarchy of coupled equations ruling the time evolution of the reduced $n$-particle distribution functions was derived by Bogoliubov, Born, Green, Kirkwood and Yvon such that, by truncating this BBGKY hierarchy, kinetic equations were derived leading to expansions of the transport properties in terms of the particle density (Yvon, 1935; Bogoliubov, 1946; Born and Green, 1946; Kirkwood, 1946).

In 1928, Pauli (1900-1958) derived a *master equation* in order to apply quantum mechanics to irreversible processes (Pauli, 1928). This derivation also contained an assumption analogue to Boltzmann's *Stosszahlansatz* – the repeated random phase assumption. Since then, different works have been completed by means of master equations, such as the generalization of the classical Boltzmann equation to quantum systems by Uehling and Uhlenbeck (1933). In the fifties, master equations were derived for weakly coupled systems by van Hove (1955, 1957, 1959) as well as Brout and Prigogine (1956). Kinetic equations were also obtained for plasmas since thanks to Landau (1936) and Vlassov (1938). In this context, Balescu (1960) and Lenard (1960) derived in the sixties the so-called *Lenard-Balescu* transport equation for plasmas.

In parallel, the relationship between molecular fluctuations in systems due to thermal agitation, and transport coefficients characterizing the irreversible processes was studied starting with Einstein's work on the Brownian motion (Einstein, 1905) to reach *the linear response theory* developed in the fifties by Green (1951,1960), Kubo (1957), and Mori (1958a). This theory expresses the transport coefficients in terms of the integral of autocorrelation function. Moreover, aiming at establishing an Einstein-like relation for transport processes, Helfand (1960) derived formulas relating the transport coefficients to the variance of their associated quantities now called *Helfand moment*.

The transport coefficients are usually calculated based on some kinetic equation. Each kinetic equation involves a stochastic assumption such as Boltzmann's *molecular chaos* which is reached by some truncation of the evolution equation. Such phenomena as the transport processes, playing such an important role in nature, must present a more fundamental justification in terms of the intrinsic properties of the underlying microscopic dynamics. We can conclude that microscopic chaos seems to be the key for a better understanding of the irreversibility of the macroscopic phenomena.

## 1.5   Microscopic chaos

It is interesting to note that the idea of considering a gas as a disordered state is old. During the 16th century Paracelsus (1493-1541) studied gases and observed that the air and vapors did not have a fixed volume or shape. He then gave them the name of *chaos* in analogy with $\chi\alpha o\varsigma$ in the greek mythology, being at the origin of the universe. Later, van Helmont (1577-1644), who is the first to have made the distinction between the different gases and the air, invented the word *gas* according to the flemish pronounciation of chaos (Leicester, 1956).

In the end of the 19th century was generally accepted that a large number of degrees of freedom is a necessary condition to observe unpredictable behavior. This great number was the necessary element in order to introduce a statistical description of phenomena having a deterministic underlying dynamics. However, in 1892, Poincaré (1854-1912) showed that a low-dimensional deterministic system, namely the *three-body problem*, exhibits an unpredictable behavior which is now commonly called *chaos* (Poincaré, 1892). Since his work, a large number of nonlinear dynamical systems, the time evolution of their variables **x** being given by $\frac{dx_i}{dt} = F_i(x_j)$, have been discovered. In these systems, chaos appears to be a generic phenomenon rather than an exotic one (Nicolis, 1995).

As a matter of fact, in 1873, Maxwell himself had already emphasized the sense of unstable behaviors

> When the state of things is such that an infinitely small variation of the present state will alter only by an infinitely small quantity the state at some future time, the condition of the system, whether at rest or in motion, is said to be stable; but when an infinitely small variation in the present state may bring about a finite difference in the state of the system in a finite time, the condition of the system is said to be unstable. It is manifest that the existence of unstable conditions renders impossible the prediction of future events, if our knowledge of the present state is only approximate, and not accurate [39].

On the other hand, in the late years of the same century, Hadamard (1865-1863) studied the geodesic flows on a negative-curve surface and concluded that

> each change of the initial direction, as small as possible, of a geodesic which remains at a finite distance is enough for implying an arbitrary variation of the behavior of the trajectory (Hadamard, 1898).

By considering Hadamard's work in a book, Duhem (1906) also concluded in the chapter entitled *Exemple de déduction mathématique à tout jamais inutilisable* that, in Hadamard's billiard, the trajectory obtained mathematically becomes unusable for a physicist since the an experimental measure realized by any physical procedure as precise as possible is always endowed of an error which grows with time. Furthermore, Poincaré clearly explained that

> A very small cause which escapes our notice determines a considerable effect that we cannot fail to see, and then we say that the effect is due to chance (Poincar´e, 1908), implying that determinism and chance may be combined thanks to the unpredictability. The only mathematician having taken into account what we now call the *sensitivity to the initial conditions*, property put in evidence by Poincaré and the others, was Birkhoff (1884-1944) who developed in the twenties the theory of *billiards* (Birkhoff, 1927). Otherwise, one had to wait for the works done in particular by Lorenz (1963) in meteorology to consider again the *sensitivity to the initial conditions* defining chaos, and the *attractor* he put in evidence. Others like Smale (1967) considered global behavior of phase space rather than a particular trajectory, and invented his *horseshoe* (see Fig.1.4) in order to get a visual analogy to the sensitive dependence to the initial conditions. Finally, Feigenbaum (1978) discovered the universality of the *period-doubling bifurcation cascade*[40].

---

[39]Cited by Hunt and Yorke (1993). The text of the conference of Maxwell may be found in Garnett (1882).

In 1892 Lyapunov (1857-1918) defended his doctoral thesis *The general problem of the stability of motion* (Lyapunov, 1892). In his work he proposed a method which provides ways of determining the stability of sets of ordinary differential equations. He introduced the well-known *Lyapunov exponents*, quantities characterizing the growth ($\lambda_i > 0$) or decay ($\lambda_i < 0$) rate of the distance between two nearby trajectories in the phase space of dynamical systems[41] (Eckmann and Ruelle, 1985). The positivity of such quantities expresses a dynamical instability and induces a *sensitivity to initial conditions*, property defining *chaos*. Two trajectories initially very close separate exponentially in time. This sensitivity to initial conditions limits the possible predictions on the trajectories because they are only known through a given precision $\epsilon_{\text{initial}}$. By considering a rate of separation of very close trajectories given by the maximum Lyapunov exponent $\lambda_{\text{max}}$, the error between the predicted and the actual trajectories grows as $\epsilon_t \simeq \epsilon_{\text{initial}} \exp(\lambda_{\text{max}} t)$. After a finite time, the error becomes larger than the final allowed precision $\epsilon_{\text{final}}$, which then defines the Lyapunov time $t_{\text{Lyap}} \simeq (1/\lambda_{\text{max}}) \ln (\epsilon_{\text{final}}/\epsilon_{\text{initial}})$. Given these initial and final precisions, predictions after the Lyapunov time are no longer relevant. This result is the requirement of going a statistical description.

Inspired by the ideas of Krylov (1979), who argued that trajectories in phase space for simple fluid separate exponentially, Sinai and his coworkers proved the ergodic hypothesis for billiard systems, and characterized the stochastic-like behavior of these deterministic systems (Sinai, 1970a). This type of stochastic-like behavior was then studied extensively for simple systems with few degrees of freedom.

---

[40]For further informations see Gleick (1987) and Ott (1993).
[41]See Chapter 3.

The unstable character of dynamical systems is therefore such that even deterministic systems can generate random behaviors. But on the other hand, this instability of the dynamics produces information in time allowing to reconstruct the system trajectory in phase space. Indeed the separation in time of nearby trajectories gives us the possibility to distinguish the trajectories. In this context, in the late fifties Kolmogorov and Sinai applied the concept of entropy per unit time introduced a decade before in the information theory by Shannon to the dynamical systems. He defined the so-called *Kolmogorov-Sinai entropy* $h_{KS}$ (Eckmann and Ruelle, 1985). This new quantity measures the (expo-nential) rate at which information is obtained in time in random processes[27]. The KS entropy per unit time is the dynamical analogue of the entropy per unit volume defined in equilibrium statistical mechanics. Later, in the seventies, Sinai, Bowen, and Ruelle proposed the mathematical foundations of a new formalism for chaotic systems applying the techniques used in statistical thermodynamics, which is called the *thermodynamic formalism* (Ruelle, 1978; Beck and Schlögl, 1993). In this formalism, a new quantity, the *topological pressure*, was defined and played a role in dynamical systems very similar to that of the free energy for statistical-mechanical systems.

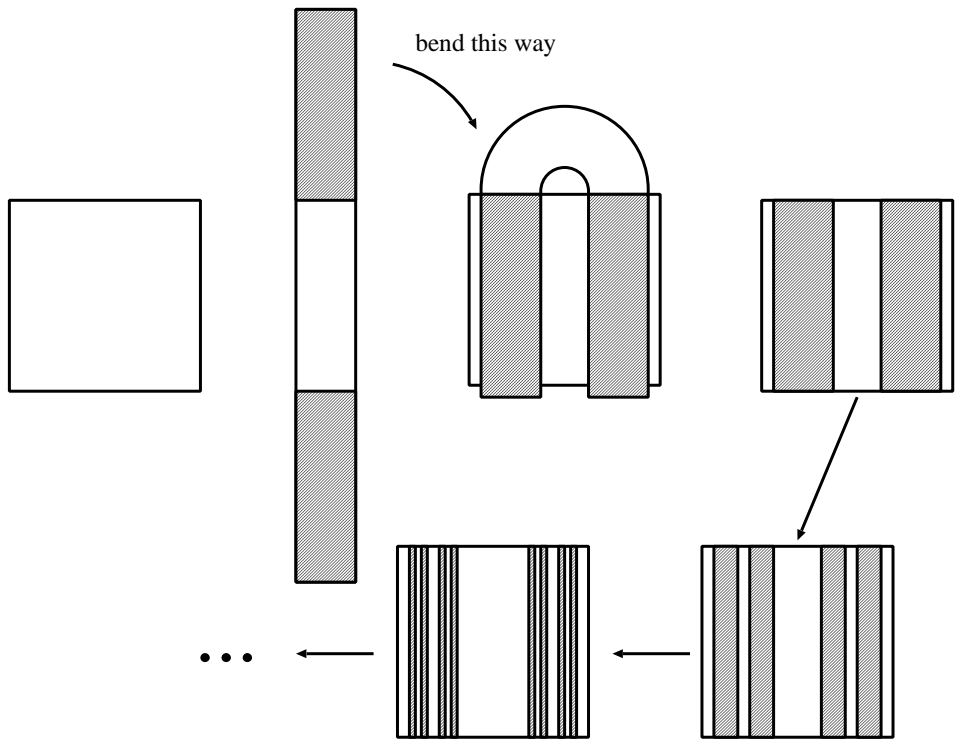

**Figure 1.4.** *Smale's horseshoe*: Succession of stretchings and foldings illustrating the evolution of phase-space due to the chaotic dynamics.

The two aspects of chaos, that is the *dynamical instability* characterized by the Lyapunov exponents, and the *dynamical randomness* by the KS entropy are strongly related, the second one being the consequence of the first one. In 1977, Pesin proposed a theorem for closed systems (systems without any escape of trajectories out of the initial phase space) which is now known under the name of *Pesin's identity*, which relates the KS entropy to the sum of all the positive Lyapunov exponents of the system (Pesin, 1977; Eckmann and Ruelle, 1985)

$$h_{\text{KS}} = \sum_{\lambda_i > 0} \lambda_i \ . \tag{1.17}$$

As we saw, the instability of the dynamics in phase space induces an exponential separation of trajectories characterized by the Lyapunov exponents. Because the phase-space volume accessible to the trajectories is finite (e.g. the volume defined by the energy of the system), they have to fold onto themselves. We then observe successive stretchings and foldings of phase-space volumes, as the Smale's horseshoe illustrates it (see Fig.1.4), creating strange objects (after an infinite number of such operations) nowadays called *fractals*.

Whereas the term *fractal* is associated with Mandelbrot, the history of this intriguing discipline of mathematics began during the 1870's, when continuous functions without derivatives were discovered. For a long time, the idealization of nature implied a smooth and regular representation of real objects (Chabert, 1994). In mathematics, one dealt with continuous functions such that a tangent could be drawn at (almost) each point. However, Riemann (1826-1866) already claimed a contradictory opinion. And in the 1870's Weierstrass (1815-1897) gave an example having no derivative in any point[42]. In geometry, mathematicians, such as Koch (1870-1924) in 1904, proposed continuous curves without a tangent at any point obtained by an elementary geometric construction (Koch, 1904). *Koch's curve* depicted in Fig.1.5 is a clear example of such objects. Moreover, the Japanese mathematician O. Takagi (1875-1960) working at Göttingen proposed in 1903 a simple example of continuous but nondifferentiable function known today as *Takagi's function* (Takagi, 1903). On the other hand, studying the Brownian motion, Perrin observed experimentally that the trajectory drawn by the Brownian particle is highly irregular and he deduced that

---

[42]The paper was read in 1872 in the Royal Prussian Academy of Sciences, but was only published on the original version in 1895 (Weierstrass, 1895).

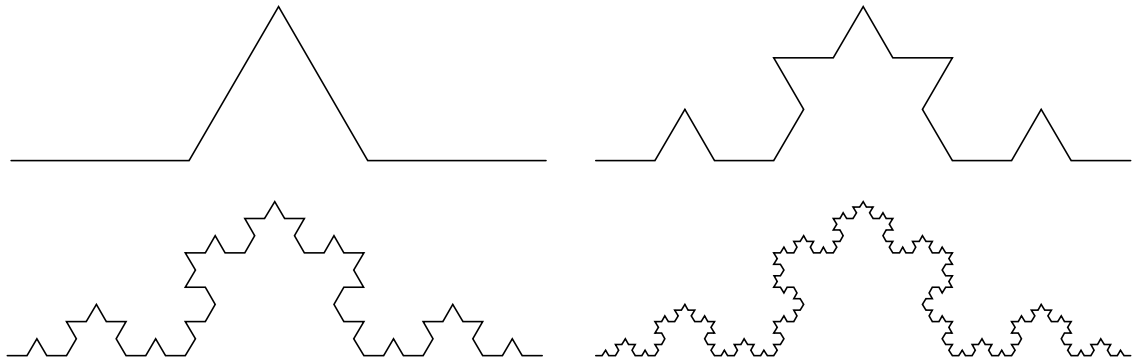

**Figure 1.5**. *Koch's curve*: the first four steps of construction. The operation consists in dividing each segment into 3 parts, and on the one in the medium is constructed an equilateral triangle.

> If the functions with derivative are the simplest ones, the simplest to be treated, functions with derivatives are the exception; or, to use geometric language, curves with no tangent at any point become the rule (from Ref. (Perrin, 1991)).

Indeed according to the precision of the measure, the length of the trajectory is different, the larger the precision is, the larger is the trajectory. The relevance of the notion of length thus vanishes and has to be replaced by a new quantity characterizing such objects, that is the *dimension*. In this context, Hausdorff (1868-1942) proposed in 1919 a new notion of dimension which is no longer a whole number, but can take noninteger values (Hausdorff, 1919). For example, in the case of Koch's curve, its *Hausdorff dimension* is neither equal to one, that is the dimension of a line, nor equal to two, the dimension of the plane, but between both of them, that is 1.26 .

By introducing the term *fractal*[43], Mandelbrot (1975) brought back into fashion all the old works on these strange objects. He used them in a lot of different disciplines such as in physics (turbulence), geography ("How long is the Coast of Britain?"), astronomy (craters of the moon), biology, etc. , in order to show that objects without a tangent at any point are nothing but the rule in nature. The property being common to the different objects he treated, and thus the one implying a certain unity in nature, is the *self-similarity*. A fractal object is such that an enlarged part is similar to the whole object, which implies a scale invariance[44].

In the context of statistical mechanics establishing relationships between the microscopic and macroscopic scales, we deal with microscopic chaos implied by the defocusing character of successive collisions between atoms and molecules. Let us consider the Brownian motion of a colloidal particle. The high-dimensional microscopic chaos of the surrounding fluid, characterized by the spectrum of

---

[43]This word is derived from the Latin word *fractus* meaning *irregular, broken*.
[44]Several fundamental papers on the development of theory of fractals are reprinted in Elgar (1993).

Lyapunov exponents, induces a dynamical randomness given by the KS entropy, which is calculated by Pesin's identity (1.17). This huge dynamical randomness appears to be at the origin of the erratic motion of the Brownian particle. Consequently it gives us a new interpretation[45] of the observed stochastic processes in terms of a high-dimensional chaos in the microscopic Newtonian dynamics. Therefore the statistical mechanics of irreversible processes no longer needs stochastic models in order to describe the macroscopic irreversibility. The microscopic chaos provides deterministic systems which present a stochastic-like behavior without need of any stochastic assumption. Moreover, chaos and the sensitivity to initial conditions implying unpredictability for long times justifies the use of statistical mechanics, even for systems with a low-dimensional phase space.

In this perspective, during the last two decades, different theories have been developed establishing connections between irreversible phenomena (transport and chemical-reaction processes) and the chaotic properties of the microscopic dynamics. Among them we find the *thermostated-system approach* developed in the eighties by Nosé (1984a, 1984b), Hoover (1985, 1991), Evans and Morriss (Evans et al., 1990; Evans and Morriss, 1990), and others; the *escape-rate formalism* (Gaspard, 1998; Dorfman, 1999) proposed in 1990 by Gaspard and Nicolis (1990) for diffusion and extended in 1995 by Dorfman and Gaspard to the other transport processes (Dorfman and Gaspard, 1995; Gaspard and Dorfman, 1995); and the *hydrodynamic-mode method* (Gaspard, 1998) developed during the nineties by Gaspard and coworkers (Gaspard, 1993; Gaspard, 1996).

## 1.6   Hard-sphere systems

In statistical mechanics, we aim to express macroscopic properties through the microscopic dynamics. However, an important assumption is needed on the latter, that is the interatomic force to be considered[46]. In the history of kinetic theory of gases and statistical mechanics, *hard-sphere potential* has played a fundamental role in the success of this powerful branch of physics. Hard-ball system without attractive forces is the system on which one can find the greatest amount of works achieved since the first developments of kinetic theory. The first who assumed that the atoms are very small billiard balls was Daniel Bernoulli in the previously cited book *Hydrodynamica* (1738). To recover

---

[45]An experimental work has been done exhibiting the chaotic character of the microscopic dynamics by Gaspard et al. (1998).

[46]An interesting review and discussion concerning the many potentials of interaction considered in the history of the kinetic theory of gases can be found in Brush (1970, 1976, 1983).

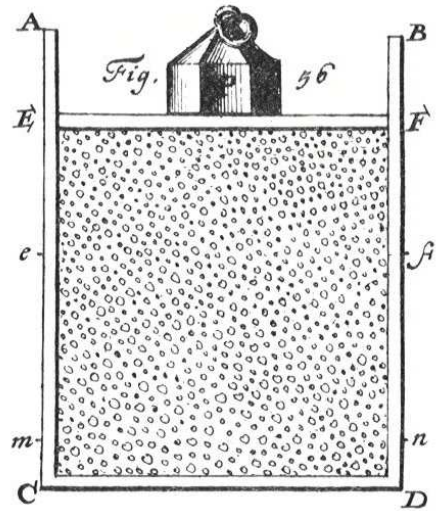

**Figure 1.6**. Bernoulli's gas model. From *Hydrodynamica* (Bernoulli, 1738).

Boyle's law, he expressed the product $PV$ in terms of the velocity of these particles

$$PV = \frac{1}{3} Nm \langle v \rangle^2 \qquad (1.18)$$

by imagining that a gas is composed of infinitely small particles (see Fig. 1.6).

Previously, in section 1.4, we saw that the hard-ball system, the simple model of particles in interaction, allowed the first kinetic-theorists to predict non-intuitive properties or recover experimental data for gases. Maxwell (1860) derived the expression (1.11) predicting the independence of the viscosity of a dilute gas on the density (he observed this phenomenon himself through his own experiments (Maxwell, 1866)). Thanks to this work, Loschmidt (1865) was able to give a first convincing estimation of the size of atoms and of the Avogadro number. Later, Enskog (1922) proposed his theory by considering such models and recovered that the kinematic viscosity $\frac{\eta}{\rho}$ versus the density $\rho$ presents a minimum, as previously experimentally observed by Warburg and von Babo (1882).

Later, Kirkwood and Monroe (1941) predicted that a phase transition of fluid-solid type should be observed in hard-ball systems. This prediction was confirmed during the following decade by Alder and Wainwright who enjoyed the first computers available for numerical simulations. By molecular dynamics in periodic boundary conditions[47], they calculated the equation of state for hard spheres (Alder and Wainwright, 1957) as well as for hard disks (Alder and Wainwright, 1962) and put in evidence the existence of the phase transition. The observation of the fluid-solid phase transition in a system without attractive force between particles suggested that such a phase transition is due to the repulsive part of the potential of interaction, whereas it was well known that the gas-liquid is related to the attractive interatomic force[48].

---

[47]These conditions have been used in order to avoid the effect of surface (Alder et al., 1955). Indeed, since the number of particles considered in their works as well as in ours, an important proportion of these particles would be at the surface of the box.

On the other hand, systems with hard-sphere potential in the dynamical-system theory is often used. Indeed, in the seventies, Sinai (1970b) (Sinai, 1970b) provided a proof of ergodicity in the two-hard-disk system. Szász and Simanyi extended these results to hard-ball systems (Simanyi and Szász, 1995; Szász, 1996). Furthermore, hard-sphere systems have been often considered during the development of dynamical-system theory for the study of transport processes by Sinai, Bunimovich and others. As discussed above, Bunimovich and Sinai showed that the diffusion coefficient in the Lorentz gas is finite and positive (Bunimovich and Sinai, 1980a). More recently, Bunimovich and Spohn proved the existence of the viscosity coefficient in the two-hard-disk system (Bunimovich and Spohn, 1996; Bunimovich, 2000).

Therefore, although systems with such a potential of interaction might seem to be naive and be the source of disagreements with experiments, they have presented, and still present today, a great interest for developing a better understanding of matter as well as irreversible processes.

## 1.7   Outline

In this work, we will consider the viscosity as a process of momentum transport and we will study the link between microscopic and macroscopic properties in the context of the chaotic hypothesis.

Chapter 2 is devoted to the viscosity and its microscopic expression in terms of the dynamical quantity called *Helfand moment $G(t)$*. The Helfand-moment method is based on the linear growth in time of the variance of the $G(t)$ through an Einstein-like equation. We propose a method available for any system of particles with periodic boundary conditions in order to calculate the viscosity.

In chapter 3 we develop three important methods establishing a link between the properties of the microscopic chaos and the transport properties and in particular, viscosity. The first one, known under the name of *thermostated-system approach*, imposes a constraint on the system, maintaining it in a nonequilibrium state. The second one, the well-known *escape-rate formalism*, introduces absorbing boundary conditions inducing an escape process. This latter is related to the viscosity on the one hand, and in the other hand to properties of the microscopic chaos. Finally, the third method, the *hydrodynamic-mode method*, is based on the construction of the hydrodynamic modes in terms of the singular distributions due to the deterministic chaotic character of the microscopic dynamics.

---

[48]Later studies considering soft repulsive forces such as the inverse square showed that crystal solid also appears at low temperatures and high densities.

Chapter 4 presents our main model. It is composed of two hard disks in a periodic lattice. Two geometries are considered: the square and the hexagonal lattices. The system presents a fluid-solid phase transition. First, we have exposed different properties of the model, such as the pressure, the mean free path, and the Lyapunov exponent. The viscosity is evaluated for both geometries thanks to the Helfand-moment method proposed in chapter 2. We then give a qualitative explanation of the behavior of the viscosity in function of the density, especially at the phase transition and in the solid phase. We apply the *escape-rate formalism* in calculating the escape rate, and the characteristic quantities of the chaotic dynamics of particles such as the Lyapunov exponent and fractal dimension, in order to study the relationships between microscopic properties and viscosity as a transport process.

In chapter 5, we extend the model to many-particle systems. We consider systems of $N$ hard disks as well as $N$ hard spheres. We evaluate the equation of state and the Lyapunov spectrum for such systems. The viscosity is studied in two- and three-dimensional systems of hard-ball particles. Furthermore, we use our method proposed in chapter 2 to the computation of the viscosity in systems of particles interacting through the *Lennard-Jones potential*. In the same chapter, a similar method is proposed for the *thermal conductivity* which allows us to calculate this other transport coefficient in periodic systems through an Einstein-like equation.

Finally we develop the main conclusions and perspectives of our work in chapter 6.

# Chapter 2

# Helfand-moment method

This chapter is devoted to the viscosity and its evaluation in terms of microscopic dynamics by means of the fluctuations at equilibrium. First we rewrite the Navier-Stokes equations in which appears the viscosity (see section 1.3). It is then established how the transport processes (such as viscosity) can be related to the fluctuations at equilibrium due to the thermal agitation of atoms and molecules. It is in this context that the Helfand-moment method was developed. In this thesis, we propose a general Helfand-moment method for systems with periodic boundary conditions. This method may be applied by considering any potential of interaction. Moreover, the Helfand-moment method proposed in this thesis is discussed in comparison with the literature on this topic.

## 2.1   Phenomenological approach to viscosity

The Navier-Stokes equations (1.10), which were introduced in section 1.3, are the central equations in hydrodynamics. First, let us write them in the modern way. By developing the total differential $\rho \frac{dv_i}{dt}$ and by using the continuity equation (1.1), we get the general form

$$\frac{\partial \rho v_i}{\partial t} = -\frac{\partial \Pi_{ij}}{\partial r_j} \, , \tag{2.1}$$

where the *momentum flux density tensor* $\Pi_{ij}$ is written as:

$$\Pi_{ij} = \rho \, v_i v_j + P \, \delta_{ij} - \sigma'_{ij} \, . \tag{2.2}$$

In section 1.3 we supposed that the system is isotropic. This is often the case and the viscous stress tensor $\sigma'_{ij}$ is then written as

$$\sigma'_{ij} = \eta \left( \frac{\partial v_i}{\partial r_j} + \frac{\partial v_j}{\partial r_i} - \frac{2}{d} \, \delta_{ij} \, \frac{\partial v_l}{\partial r_l} \right) + \zeta \, \delta_{ij} \, \frac{\partial v_l}{\partial r_l} \, , \tag{2.3}$$

for a $d$-dimensional system. When the fluid is anisotropic, the friction may depend on the selected direction. The viscous stress tensor has to be modified in order to integrate the anisotropy. This last tensor depends linearly on the velocity gradient tensor in the framework of Newtonian fluids. The most general quantity relating two second-order tensors is a fourth-order tensor

$$\sigma'_{ij} = \eta_{ij,kl} \frac{\partial v_k}{\partial r_l} \, .$$
(2.4)

$\eta_{ij,kl}$ is the *viscosity tensor*. This is the most general expression for the viscous stress tensor including anisotropic as well as isotropic systems for Newtonian fluids. The theory of Cartesian tensors shows that the basic isotropic tensor is the Kronecker tensor $\delta_{ij}$ and that all the isotropic tensors of even orders can be written like a sum of products of tensors $\delta_{ij}$ (Aris, 1962)

$$\eta_{ij,kl} = a \, \delta_{ij} \, \delta_{kl} + b \, \delta_{ik} \, \delta_{jl} + c \, \delta_{jk} \, \delta_{il} \, ,$$
(2.5)

where $a, b$ and $c$ are scalars. Since the viscous stress tensor is symmetric $\sigma'_{ij} = \sigma'_{ji}$, we find that $b = c$ so that only two of these coefficients are independent. After a rearrangement we obtain the expression (2.3). The coefficients $\eta = b$ and $\zeta = a + (2/d)b$ are respectively the *shear and bulk viscosities* and they can be expressed in terms of the elements of the fourth-order viscosity tensor as:

$$
\begin{aligned}
\eta &= \eta_{xy,xy} \, , \\
\zeta &= \frac{1}{d} \, \eta_{xx,xx} + \frac{d-1}{d} \, \eta_{xx,yy} \, .
\end{aligned}
$$
(2.6)

Such coefficients must respect thermodynamic laws, especially the second law expressing that the internal entropy production has to be positive or vanishing $\sigma \geq 0$. According to the balance equation for entropy density $\rho s$, the entropy production for a non-reactive system composed of only one component can in general be written as (de Groot and Mazur, 1984)

$$\frac{\partial(\rho s)}{\partial t} + \frac{\partial}{\partial r_j}(\rho s v_j) = \frac{1}{T^2}\kappa(\nabla T)^2 + \frac{1}{T}\sigma'_{ij}\frac{\partial v_i}{\partial r_j} \geq 0$$
(2.7)

Let us consider the second term introducing the viscous stress tensor for isotropic systems. By replacing $\sigma'_{ij}$ by Eq. (2.4), we have

$$\sigma'_{ij}\frac{\partial v_i}{\partial r_j} = \eta \left( \frac{\partial v_i}{\partial r_j}\frac{\partial v_i}{\partial r_j} + \frac{\partial v_j}{\partial r_i}\frac{\partial v_i}{\partial r_j} - \frac{2}{d}\delta_{ij}\frac{\partial v_i}{\partial r_j}\frac{\partial v_l}{\partial r_l} \right) + \zeta\delta_{ij}\frac{\partial v_i}{\partial r_j}\frac{\partial v_l}{\partial r_l}$$
(2.8)

$$\sigma'_{ij}\frac{\partial v_i}{\partial r_j} = \eta \left[ \left(\frac{\partial v_i}{\partial r_j}\right)^2 + \frac{\partial v_i}{\partial r_j}\frac{\partial v_j}{\partial r_i} \right] - \frac{2}{d}\eta\left(\frac{\partial v_l}{\partial r_l}\right)^2 + \zeta\left(\frac{\partial v_l}{\partial r_l}\right)^2$$
(2.9)

where the product $\delta_{ij}\frac{\partial v_i}{\partial r_j}$ gives the divergence of the velocity vector

$$\delta_{ij}\frac{\partial v_i}{\partial r_j} = \frac{\partial v_l}{\partial r_l}. \tag{2.10}$$

By decomposing the first term as follows:

$$\eta\left[\left(\frac{\partial v_i}{\partial r_j}\right)^2 + \frac{\partial v_i}{\partial r_j}\frac{\partial v_j}{\partial r_i}\right] = \frac{\eta}{2}\left(\frac{\partial v_i}{\partial r_j} + \frac{\partial v_j}{\partial r_i}\right)^2, \tag{2.11}$$

we obtain

$$\sigma'_{ij}\frac{\partial v_i}{\partial r_j} = \frac{\eta}{2}\left(\frac{\partial v_i}{\partial r_j} + \frac{\partial v_j}{\partial r_i}\right)^2 - \frac{2}{d}\eta\left(\frac{\partial v_l}{\partial r_l}\right)^2 + \zeta\left(\frac{\partial v_l}{\partial r_l}\right)^2, \tag{2.12}$$

and this allows us to rewrite the balance equation for the entropy density (2.7) as

$$\frac{\partial(\rho s)}{\partial t} + \frac{\partial}{\partial r_j}(\rho s v_j) = \frac{1}{T^2}\kappa\left(\nabla T\right)^2$$

$$+ \frac{\eta}{2T}\left(\frac{\partial v_i}{\partial r_j} + \frac{\partial v_j}{\partial r_i}\right)^2 - \frac{2\eta}{Td}\left(\frac{\partial v_l}{\partial r_l}\right)^2 + \frac{\zeta}{T}\left(\frac{\partial v_l}{\partial r_l}\right)^2 \geq 0. \tag{2.13}$$

where $s$ is the entropy per unit mass.

Let us consider the terms where the shear viscosity $\eta$ appears and let us put $\frac{\eta}{2T}$ in evidence

$$\frac{\eta}{2T}\left[\left(\frac{\partial v_i}{\partial r_j} + \frac{\partial v_j}{\partial r_i}\right)^2 - \frac{4}{d}\left(\frac{\partial v_l}{\partial r_l}\right)^2\right] \equiv \frac{\eta}{2T}\,\mathcal{G} \tag{2.14}$$

We then have

$$\mathcal{G} = \left(\frac{\partial v_i}{\partial r_j} + \frac{\partial v_j}{\partial r_i}\right)^2 - \frac{2}{d}\delta_{ij}\left(\frac{\partial v_i}{\partial r_j} + \frac{\partial v_j}{\partial r_i}\right)\left(\frac{\partial v_l}{\partial r_l}\right) \tag{2.15}$$

$$= \left(\frac{\partial v_i}{\partial r_j} + \frac{\partial v_j}{\partial r_i}\right)^2 - \frac{4}{d}\delta_{ij}\left(\frac{\partial v_i}{\partial r_j} + \frac{\partial v_j}{\partial r_i}\right)\left(\frac{\partial v_l}{\partial r_l}\right)$$

$$+ \left(\frac{2}{d}\,\delta_{ij}\frac{\partial v_l}{\partial r_l}\right)^2 \tag{2.16}$$

knowing that

$$\delta_{ij}\left(\frac{\partial v_i}{\partial r_j} + \frac{\partial v_j}{\partial r_i}\right) = 2\,\frac{\partial v_l}{\partial r_l} \tag{2.17}$$

and

$$\delta_{ij}\,\delta_{ij} = d. \tag{2.18}$$

We may then gather the terms as

$$\mathcal{G} = \left( \frac{\partial v_i}{\partial r_j} + \frac{\partial v_j}{\partial r_i} - \frac{2}{d} \, \delta_{ij} \frac{\partial v_l}{\partial r_l} \right)^2 \tag{2.19}$$

Finally Eq. (2.13) can be rewritten as

$$\frac{\partial (\rho s)}{\partial t} + \frac{\partial}{\partial r_j}(\rho s v_j) = \frac{1}{T^2} \kappa \, (\nabla T)^2$$
$$+ \frac{\eta}{2T} \left( \frac{\partial v_i}{\partial r_j} + \frac{\partial v_j}{\partial r_i} - \frac{2}{d} \, \delta_{ij} \frac{\partial v_l}{\partial r_l} \right)^2 + \frac{\zeta}{T} \left( \frac{\partial v_l}{\partial r_l} \right)^2 \geq 0. \tag{2.20}$$

Hence, we obtain the condition of positivity for $\kappa, \eta$ and $\zeta$

$$\kappa \geq 0, \quad \eta \geq 0, \quad \zeta \geq 0. \tag{2.21}$$

Let us anticipate on chapter 4 by pointing out that, for anisotropic systems, these conditions of positivity for elements of the viscosity tensor no longer hold, and are to be reconsidered.

## 2.2 Irreversibility and microscopic fluctuations at equilibrium

During the 17th century, the development in microscopy was more and more important, in particular due to the progress brought by van Leeuwenhoeck (1632-1723). Thanks to the magnification provided by the microscopes was discovered the microscopic world of cells, that were called the *animalcules*. In this context, the observation of irregular motion of small grains immerged in a fluid has been already done (Gray, 1696). The interpretation of such a phenomenon until the 1820's was that these organic grains are endowed with living force. But in 1828, Brown (1773-1858) observed that inorganic particles also exhibit the same kind of dancing motion (Brown, 1828). He put in evidence that such a behavior has a physical rather than a biological cause, and thus opened the way to a new area in physics. This behavior of particles suspended in fluids is now called *Brownian motion*[1].

During the following decades different interpretations were given to this phenomenon (e.g. the motion would have been due to a local difference of temperature created by the light used to observe the particle). It is only in 1863 that Wiener (1826-1896)[2] refuted these explanations and proposed to look for the origin of the phenomenon in the liquid itself (Wiener, 1863). Consequently, he is sometimes considered as the discoverer of the origin of the Brownian motion (Perrin, 1909). It is to Gouy (1854-1926) however that goes the credit for having really prepared the way for our present point of view, since his experiments established conclusively

> i) that the Brownian movement appears for any particle, and the more viscous the liquid is, and the bigger the particles are, the lower the magnitude of the movement is; ii) that the phenomenon is perfectly regular, appears at constant temperature and in absence of any cause of external movement (Gouy, 1889).

---

[1] In fact similar observations had been recorded in France, in the same year, by Adolphe Brongniart (1801-1876).

[2] Let us point out that we refer to Ludwig Christian Wiener and not to Norbert Wiener (1894-1964).

However it was not before Einstein (1905) and von Smoluchowski (1906) that a successful theory was proposed for the Brownian motion[3]. As we saw in the previous chapter, it played an important role in the proof of the discontinuous feature of matter. Einstein hence expressed the diffusion coefficient of a Brownian particle in terms of the mean-square of its position[4]

$$D = \lim_{t \to \infty} \frac{\left\langle [x(t) - x(0)]^2 \right\rangle}{2\,t} \; . \tag{2.22}$$

More than putting simply in evidence the discrete character of the matter, he established a relationship between the diffusion coefficient of a Brownian particle and the spontaneous fluctuations intrinsic to the medium, which are due to the random collisions of particles of the surrounding fluid. Hence these natural fluctuations, generated by the microscopic dynamics at the thermodynamic equilibrium, define a limit on the accuracy of the measuring instruments (Barnes and Silverman, 1934).

Different phenomena similar to the Brownian movement were discovered later. In particular, in 1918, Schottky observed that the thermionic current in a vacuum tube presents rapid and irregular changes in magnitude, due to the random emission from the cathode. It induces fluctuations of the

---

[3]Actually in his thesis *Théorie de la Spéculation* (Bachelier, 1900), Louis Bachelier (1870-1946), the founder of mathematical finance, arrived at the same 'displacement' law, not for the colloidal diffusion, but for the 'mean displacement' of stock prices over time.

[4]For further informations, see Perrin (1909), Brush (1968) and Haw (2002). In addition, let us quote:

> It is appropriate to examine with greater attention these corpuscles, the disorderly motion of which can be observed in rays of sunshine: such chaotic movements attest to the underlying motion of matter, hidden and imperceptible. You will indeed observe numerous such corpuscles, shaken by invisible collisions, change path, be pushed back, retrace their steps, now here, now there, in all directions. It is clear that this to-and-fro movement is wholly due to atoms. First, the atoms move by themselves, then the smallest of the composite bodies, which are, so to speak, still within the reach of the forces of the atom, jostled by the invisible impulse from the latter, start their own movement; they themselves, in turn, shake slightly larger bodies. That is how, starting from atoms, movement spreads and reaches our senses, in such a way that it is imparted to these particles which we are able to discern in a ray of sunshine, without the collisions themselves which produce them being manifest to us (given by Pullman (1995)).

At first sight, this quotation might be given by a 19th-century scientist. But it is quite amazing to know that it goes back to the famous Roman poet and philosopher Lucretius (99-55 BC)).

voltage in any circuit in which the tube is connected, phenomenon now called *Schottky effect* (Schottky, 1918). But it is another phenomenon to which is attributed a special place in the development of statistical mechanics of irreversible processes. In 1927, Johnson observed experimentally that spontaneous fluctuations of potential difference is produced in any electric conductor and concluded that the thermal agitation of the electric charges in the conductor is the cause of this phenomenon (Johnson, 1927; Johnson, 1928). In 1928, Nyquist obtained theoretically the same results (Nyquist, 1928) and is now considered at the origin of the fluctuation-dissipation theorem. Later, in 1946, Kirkwood studied the Brownian motion in liquids and derived a new formula for the friction constant which involves the autocorrelation of the intermolecular force acting on the Brownian particle at a certain time $t$ with its value a time $t + \tau$ (Kirkwood, 1946). But it is to Callen and Welton that one must attribute the generalization of the Nyquist relation for any dissipative system. In this way, they established the well-known *fluctuation-dissipation theory* (Callen and Welton, 1951). Let us consider a charged Brownian particle in a liquid driven by an external electric field. The random collisions of the molecules of the liquid induce, on one hand, a random driving force on the Brownian particle maintaining this in constant irregular motion (fluctuation). But, on the other hand, they imply a resistance to the driving motion, trying to slow down the charged particle (dissipation). Because of their common origin (the thermal agitation), these two effects are related. This relationship is precisely the aim of the so-called fluctuation-dissipation theory. These different works issued from Nyquist's discovery contributed to the establishment of the *linear-response theory* developed in the fifties by Green (1951,1960), Kubo (1957) and Mori (1958), which relates the transport coefficients to the integral of time auto-correlation functions. In particular, the shear viscosity is expressed in terms of the time correlation of the microscopic expression of the $xy$-component of the stress tensor $J_{xy}$ as

$$\eta = \lim_{V \to \infty} \frac{1}{V k_B T} \int_0^\infty \left\langle J_{xy}(0) J_{xy}(t) \right\rangle \, dt \, , \tag{2.23}$$

while the bulk viscosity $\zeta$ is given by

$$\zeta + \frac{4}{3}\eta = \lim_{\substack{N,V \to \infty \\ n=N/V}} \frac{1}{V k_B T} \int_0^\infty dt \, \langle (J_{xx}(0) - \langle J_{xx} \rangle) (J_{xx}(t) - \langle J_{xx} \rangle) \rangle \tag{2.24}$$

in $d = 3$. The relations (2.23) and (2.24) are nowadays called the *Green-Kubo formulas*.

Notice that Mori has shown that, in the case of dilute gases, this expression reduces to the Chapman-Enskog results (Mori, 1958b).

The simplicity of Eq. (2.22) obtained by Einstein (1905) presents a particular interest. The extension of such a relation to the other transport coefficients could be useful. In this context, Helfand (1960) proposed quantities associated with the different transport processes in order to establish Einstein-like relations such as Eq. (2.22) between the transport coefficients and its associated quantities. In the case of self-diffusion, the associated Helfand moment is nothing but the position of one particle $x_i$. On the other hand, for the shear viscosity coefficient, we have

$$\eta = \lim_{t \to \infty} \frac{1}{2tVk_BT} \left\langle \left[ G_{xy}(t) - G_{xy}(0) \right]^2 \right\rangle , \qquad (2.25)$$

where $G_{xy}(t)$ is the *Helfand moment* associated with the shear viscosity. This technique will play an important role in this thesis and we will develop this in section 2.5.

Hence fluctuations at equilibrium induced by the thermal agitation play a fundamental role in modern statistical mechanics of nonequilibrium processes. The advantages of such approaches to irreversible processes are that irreversible phenomena can be described by the tools of equilibrium statistical mechanics and, hence, the construction of nonequilibrium distribution functions is not necessary. Furthermore, such relations are valid in general and can therefore be applied to dilute gases as well as to dense gases and liquids[5].

The theories presented above imply the necessity to express the stress tensor in terms of the molecular variables. The following section is devoted to this point.

## 2.3   Microscopic expression of the viscosity

At the microscopic level, atoms and molecules evolve in time according to Newton's equation of motion

$$\begin{aligned} \frac{d\mathbf{r}_a}{dt} &= \frac{\mathbf{p}_a}{m} \\ \frac{d\mathbf{p}_a}{dt} &= \sum_{b \neq a} \mathbf{F}(\mathbf{r}_{ab}) \end{aligned} \qquad (2.26)$$

where $a, b = 1, \ldots N$ ($N$ being the number of particles in the system) and $\mathbf{r}_{ab} = \mathbf{r}_a - \mathbf{r}_b$. We may

---

[5]A general overview of these theories is given for example by Kubo (1966).

express the momentum density in terms of microscopic variables as

$$\hat{\mathbf{g}}(\mathbf{r}) = \sum_{a=1}^{N} \mathbf{p}_a \delta(\mathbf{r} - \mathbf{r}_a) \ . \tag{2.27}$$

If we introduce a smooth test function $f(\mathbf{r})$ which is time independent, Eq. (2.27) becomes

$$\int d\mathbf{r} f(\mathbf{r}) \hat{\mathbf{g}}(\mathbf{r}) = \int d\mathbf{r} f(\mathbf{r}) \sum_a \mathbf{p}_a \delta(\mathbf{r} - \mathbf{r}_a) = \sum_a \mathbf{p}_a f(\mathbf{r}_a) \ . \tag{2.28}$$

Let us take the following definition for the *microscopic momentum current density* $\hat{\tau}_{ij}$

$$\frac{\partial \hat{g}_i}{\partial t} + \frac{\partial \hat{\tau}_{ij}}{\partial r_j} = 0 \tag{2.29}$$

appearing in the equation of momentum conservation. By multiplying this equation by $f(\mathbf{r})$ and by integrating over $\mathbf{r}$, one has

$$
\begin{aligned}
\int d\mathbf{r} f(\mathbf{r}) \frac{\partial \hat{g}_i}{\partial t} &= -\int d\mathbf{r} f(\mathbf{r}) \frac{\partial \hat{\tau}_{ij}}{\partial r_j} \\
&= -\int d\mathbf{r} \left[ \frac{\partial}{\partial r_j} \left( f \hat{\tau}_{ij} \right) - \frac{\partial f}{\partial r_j} \hat{\tau}_{ij} \right] \\
&= -\int f \, \hat{\tau}_{ij} \, dA_j + \int d\mathbf{r} \frac{\partial f}{\partial r_j} \hat{\tau}_{ij}
\end{aligned}
\tag{2.30}
$$

where $dA_j$ is an element of area perpendicular to the axis $r_j$. This boundary term vanishes because $f(\mathbf{r}) \to 0$ for $\mathbf{r} \to \infty$. By using Eq. (2.28), we then get

$$
\begin{aligned}
\int d\mathbf{r} f(\mathbf{r}) \frac{\partial \hat{g}_i}{\partial t} &= \int d\mathbf{r} \frac{\partial f}{\partial r_j} \hat{\tau}_{ij} = \frac{d}{dt} \sum_a f(\mathbf{r}_a) p_{ai} \\
&= \sum_a \frac{df(\mathbf{r}_a)}{dt} p_{ai} + \sum_a f(\mathbf{r}_a) \frac{dp_{ai}}{dt} .
\end{aligned}
\tag{2.31}
$$

First, consider the first term

$$\sum_a \frac{df(\mathbf{r}_a)}{dt} p_{ai} = \frac{1}{m} \sum_a \nabla f(\mathbf{r}_a) \cdot \mathbf{p}_a \ p_{ai} \tag{2.32}$$

knowing that $f$ is a time-independent function and that $\frac{d\mathbf{r}_a}{dt} = \frac{1}{m}\mathbf{p}_a$. The second term of Eq. (2.31) is

developed as follows

$$
\begin{aligned}
\sum_a f(\mathbf{r}_a)\frac{dp_{ai}}{dt} &= \sum_a \sum_{b \neq a} f(\mathbf{r}_a) F_i(\mathbf{r}_a - \mathbf{r}_b) \\
&= \frac{1}{2} \sum_a \sum_{b \neq a} \left[ f(\mathbf{r}_a) - f(\mathbf{r}_b) \right] F_i(\mathbf{r}_a - \mathbf{r}_b)
\end{aligned}
\tag{2.33}
$$

where we use the equation of motion Eq. (2.26), and $F_i(\mathbf{r}_a - \mathbf{r}_b) = -F_i(\mathbf{r}_b - \mathbf{r}_a)$ since the forces of interaction between particles are central. Hence Eq. (2.31) becomes

$$
\begin{aligned}
\frac{d}{dt} \sum_a f(\mathbf{r}_a) p_{ai} &= \frac{1}{m} \sum_a \nabla f(\mathbf{r}_a) \cdot \mathbf{p}_a \, p_{ai} \\
&\quad + \frac{1}{2} \sum_a \sum_{b \neq a} \left[ f(\mathbf{r}_a) - f(\mathbf{r}_b) \right] F_i(\mathbf{r}_a - \mathbf{r}_b)
\end{aligned}
\tag{2.34}
$$

Let us consider a pair of particle $a \neq b$. We introduce an arbitrary smooth curve $\lambda \to \mathbf{r}_{ab}(\lambda)$ such that $\mathbf{r}_b = \mathbf{r}_{ab}(0)$ and $\mathbf{r}_a = \mathbf{r}_{ab}(1)$. Thanks to this curve, we can express $f(\mathbf{r}_a) - f(\mathbf{r}_b)$ as

$$
\begin{aligned}
f(\mathbf{r}_a) - f(\mathbf{r}_b) &= f(\mathbf{r}_{ab}(1)) - f(\mathbf{r}_{ab}(0)) \\
&= \int_0^1 \frac{df(\mathbf{r}_{ab}(\lambda))}{d\lambda} d\lambda \\
&= \int_0^1 \nabla f(\mathbf{r}_{ab}(\lambda)) \cdot \frac{d\mathbf{r}_{ab}}{d\lambda} d\lambda \ .
\end{aligned}
\tag{2.35}
$$

Consequently, Eq. (2.34) is rewritten as

$$
\begin{aligned}
\frac{d}{dt} \sum_a f(\mathbf{r}_a) p_{ai} &= \frac{1}{m} \sum_a \nabla f(\mathbf{r}_a) \cdot \mathbf{p}_a \, p_{ai} \\
&\quad + \frac{1}{2} \sum_a \sum_{b \neq a} \int_0^1 d\lambda \frac{d\mathbf{r}_{ab}}{d\lambda} \cdot \nabla f(\mathbf{r}_{ab}(\lambda)) F_i(\mathbf{r}_a - \mathbf{r}_b) \ .
\end{aligned}
\tag{2.36}
$$

Let us introduce the delta function $\delta(\mathbf{r} - \mathbf{r}_a)$ as follows

$$
\begin{aligned}
\frac{d}{dt} \sum_a f(\mathbf{r}_a) p_{ai} &= \frac{1}{m} \int d\mathbf{r} \nabla f(\mathbf{r}) \cdot \sum_a \mathbf{p}_a p_{ai} \, \delta(\mathbf{r} - \mathbf{r}_a) \\
&\quad + \frac{1}{2} \int d\mathbf{r} \sum_a \sum_{b \neq a} \int_0^1 d\lambda \frac{d\mathbf{r}_a}{d\lambda} \cdot \nabla f(\mathbf{r}) F_i(\mathbf{r}_a - \mathbf{r}_b) \, \delta(\mathbf{r} - \mathbf{r}_{ab}(\lambda))
\end{aligned}
\tag{2.37}
$$

so that

$$\frac{d}{dt} \sum_a f(\mathbf{r}_a) p_{ai} = \int d\mathbf{r} \frac{\partial f(\mathbf{r})}{\partial r_j} \left\{ \frac{1}{m} \sum_a p_{ai} p_{aj} \delta(\mathbf{r} - \mathbf{r}_a) \right.$$
$$\left. + \frac{1}{2} \sum_a \sum_{b \neq a} \int_0^1 d\lambda \frac{dr_{abj}}{d\lambda} F_i(\mathbf{r}_a - \mathbf{r}_b) \delta(\mathbf{r} - \mathbf{r}_{ab}(\lambda)) \right\} . \qquad (2.38)$$

With Eq. (2.29), we then find the following expression for the microscopic momentum current density

$$\hat{\tau}_{ij} = \frac{1}{m} \sum_a p_{ai} p_{aj} \delta(\mathbf{r} - \mathbf{r}_a) + \frac{1}{2} \int_0^1 d\lambda \ F_i(\mathbf{r}_a - \mathbf{r}_b) \frac{dr_{abj}}{d\lambda} \delta(\mathbf{r} - \mathbf{r}_{ab}(\lambda)) . \qquad (2.39)$$

Let us take the integral over a volume $V$ of $\tau_{ij}$

$$\begin{aligned}
J_{ij}(t) &= \int_V d\mathbf{r} \tau_{ij}(\mathbf{r}, t) \\
&= \sum_a \frac{1}{m} p_{ai} p_{aj} \int_V d\mathbf{r} \delta(\mathbf{r} - \mathbf{r}_a) \\
&+ \frac{1}{2} \sum_a \sum_{b \neq a} F_i(\mathbf{r}_a - \mathbf{r}_b) \int_0^1 dr_{abj} \int_V \frac{d\mathbf{r}}{d\lambda} \delta(\mathbf{r} - \mathbf{r}_{ab}(\lambda)) .
\end{aligned} \qquad (2.40)$$

Finally we get the microscopic current

$$J_{ij}(t) = \sum_a \frac{1}{m} p_{ai} p_{aj} + \frac{1}{2} \sum_a \sum_{b \neq a} F_i(\mathbf{r}_a - \mathbf{r}_b)(r_{aj} - r_{bj}) \qquad (2.41)$$

which enters in the Green-Kubo formula for the shear viscosity (2.23).

In Appendix A, we show how to obtain the complete viscosity tensor in terms of autocorrelation function of the microscopic momentum density, that is, the Green-Kubo formula

$$\eta_{ij,kl} = \frac{\beta}{V} \int_0^\infty \left[ \langle J_{ij}(0) \ J_{kl}(t) \rangle - \langle J_{ij} \rangle \langle J_{kl} \rangle \right] \ dt . \qquad (2.42)$$

## 2.4 The periodic boundary conditions and their consequences

The main purpose of this work consists in the study of transport processes (mainly the viscosity) which are basically bulk properties of matter. While, in systems of macroscopic size, only a very small fraction of particles is close to the wall of the container, this is no longer the case in molecular-dynamics simulation since the system is necessarily composed of a finite number of particles . Indeed, consider a three-dimensional system with $N = 10^{21}$. Since the number of atoms on the surface is of order $N^{2/3}$, it means that only $10^{14}$ of them are near the walls, that is to say one in $10^7$. On the other hand, in molecular dynamics, a typical number of particles is 1000. Consequently, hundreds of atoms are close to the surface and affect the computation of bulk properties.

The standard way to avoid such problems is to consider a system with periodic boundary conditions (p.b.c.). Considering periodic systems implies that the particles in a "fundamental cell" are reproduced in the neighbouring ones, and so on. Therefore the dynamics of particles of the fully-extended system can be reduced to the one of the particles moving inside the fundamental cell. The opposite boundaries of this cell are identified. It is equivalent to say that the particles are moving on a torus (see Fig. 2.1).

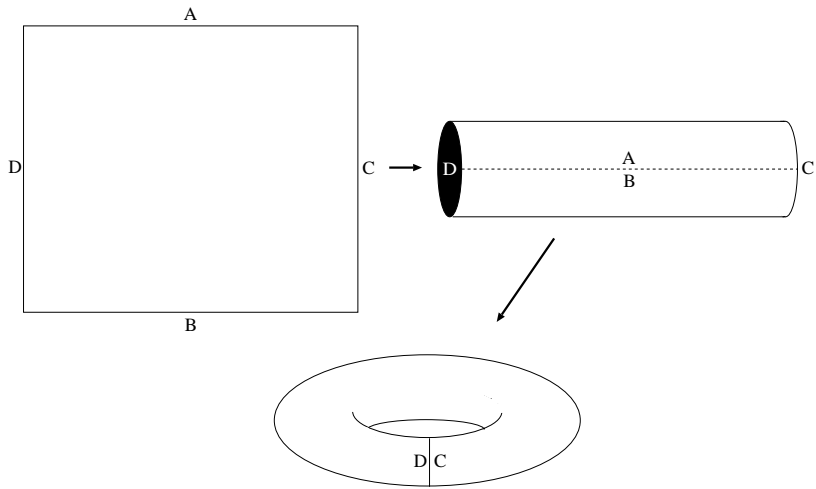

**Figure 2.1**. Construction of the torus from the fundamental cell with periodic boundary conditions. The opposite boundaries are identified with each other.

Due to the periodicity of the system, forces can be exerted by image particles as well as real particles. In consequence, the force $\mathbf{F}_{ab}$ exerted by $b$ on $a$ also contains the contributions of all the images of $b$ (see Fig. 2.2)

$$\mathbf{F}_{ab} = -\sum_{\beta} \frac{\partial V(\mathbf{r}_{ab})}{\partial \mathbf{r}_{ab}} \qquad (2.43)$$

where $\mathbf{r}_{ab} = \mathbf{r}_a - \mathbf{r}_b - \beta L$ and $\beta$ is the cell translation vector (Haile, 1997).

In molecular dynamics, it is usual to consider a short range for the potential of interaction. One possibility is to introduce a *cutoff distance* $r_{\text{cutoff}}$ such that, for larger distances, the potential vanishes.

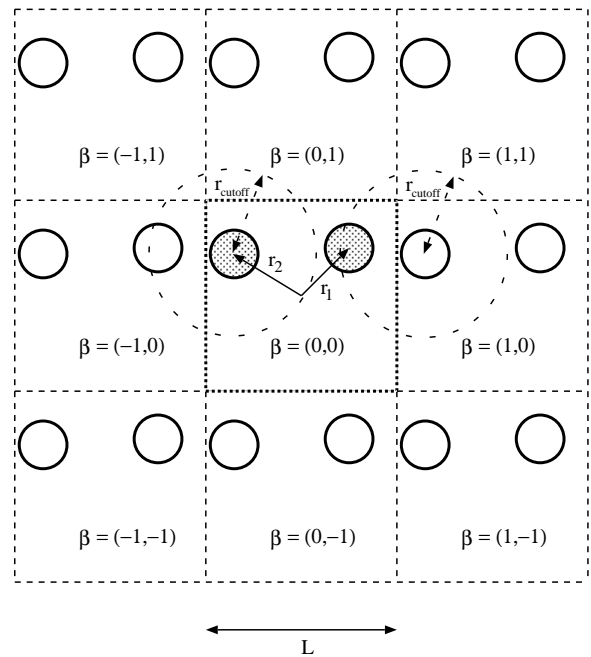

**Figure 2.2**. Illustration of a two-dimensional system with periodic boundary conditions. The fundamental cell, in the center, is represented with its neighbouring cells. In this example, due to the short range of the interparticle interaction, here imposed by the finite value of $r_{\text{cutoff}} \leq \frac{L}{2}$, the only non-vanishing term of the sum over the different $\beta$ in Eq. (2.43) is such that the minimum image convention is satisfied. In other words, the distance taken into account for the calculation of the force exerted on the particle 1 by the particle 2 is $\mathbf{r}_1 - \mathbf{r}_2 - \beta^{(1,2)}L$ with $\beta^{(1,2)} = (1, 0)$.

For example, the Lennard-Jones potential is written as

$$V(r_{ab}) = \begin{cases} 4\epsilon \left[ \left( \frac{\sigma}{r_{ab}} \right)^{12} - \left( \frac{\sigma}{r_{ab}} \right)^6 \right] & r_{ab} \leq r_{\text{cutoff}} \\ 0 & r_{ab} > r_{\text{cutoff}} \end{cases} \tag{2.44}$$

which is illustrated in Fig. 2.3.

When the forces are short ranged compared to the dimension $L$ of the box, we need consider only those image cells that adjoin the fundamental cell. In the two-dimensional case, only eight neighbouring cells must be taken into account (see Fig. 2.2), and 26 in three-dimensional systems. Furthermore, among the 9 or 27 terms in the sum (2.43), only one has a separation $|\mathbf{r}_a - \mathbf{r}_b - \beta L|$ less than $\frac{L}{2}$. Then either atom $b$ or only one of its images can exert a force on particle $a$. Indeed, as shown in Fig. 2.2, particle 1 is only in the range of interaction of the image of particle 2 that the position is $\mathbf{r}_2 + \beta^{(1,2)}L$ with $\beta^{(1,2)} = (1, 0)$. This is the so-called *minimum image convention*. Here we define the quantity $\mathbf{L}_{b|a}$ used this thesis as

$$\mathbf{L}_{b|a} = \beta^{(a,b)}L \tag{2.45}$$

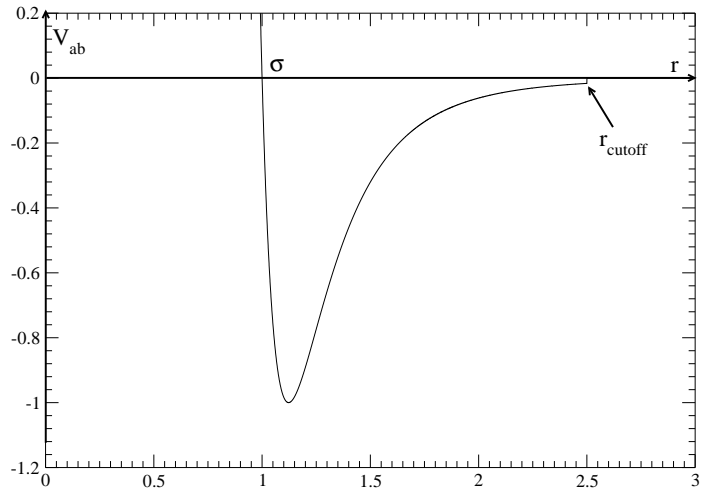

**Figure 2.3**. Lennard-Jones potential with a finite cutoff distance. The parameters $\epsilon$ and $\sigma$ are fixed to unity. The cutoff distance $r_{\text{cutoff}}$ equals 2.5 .

where $\beta^{(a,b)}$ is thus the cell translation vector satisfying the minimum image convention for particles $a$ and $b$. On other words, $\mathbf{L}_{b|a}$ is the vector to be added to $\mathbf{r}_b$ in order to satisfy this latter convention.

It is important to notice that we do not suppose here that the force field $\mathbf{F}(\mathbf{r})$ is periodic but the dynamical system itself is periodic as the consequence of the assumptions that the particles can interact with their images. In order to define a dynamics which is periodic in the box of size $L$ the positions should jump in order to satisfy the minimum image convention. As a consequence of this assumption, the positions and momenta used to calculate the viscosity by the Green-Kubo method actually obey modified Newton's equations

$$\frac{d\mathbf{r}_a}{dt} = \frac{\mathbf{p}_a}{m} + \sum_s \Delta\mathbf{r}_a^{(s)} \, \delta(t - t_s) \, ,$$

$$\frac{d\mathbf{p}_a}{dt} = \sum_{b(\neq a)} \mathbf{F}(\mathbf{r}_a - \mathbf{r}_b - \mathbf{L}_{b|a}) \, , \qquad (2.46)$$

where $\Delta\mathbf{r}_a^{(s)}$ is the jump of the particle $a$ at time $t_s$ with $\left|\Delta\mathbf{r}_a^{(s)}\right| = L$. We notice that modified Newton's equations (2.46) conserve energy, total momentum and preserve phase-space volumes (Liouville's theorem).

## 2.5 Helfand moment

As we have shown above, fluctuations at equilibrium are of interest in the study of irreversible processes, such as transport processes. In this context, Einstein (1905) obtained a relation of the diffusion in terms of the variance of the position of a Brownian particle which is submitted to the thermal agitation. Although Kubo (1957) was the first to claim that the extension of such an expression to the other transport coefficients (the two viscosities and thermal diffusion) should be possible, it is to Helfand that we owe this effort achieved in 1960. In the particular case of shear viscosity, he obtained

$$\eta = \lim_{t \to \infty} \frac{\beta}{2tV} \left\langle \left[ G_{xy}(t) - G_{xy}(0) \right]^2 \right\rangle , \qquad (2.47)$$

where $G_{xy}(t)$ is precisely the dynamical quantity associated with shear viscosity, and is called *Helfand moment*

$$G_{xy}(t) = \sum_{a=1}^{N} p_{ax}(t) y_a(t) . \qquad (2.48)$$

In the case of bulk viscosity, he obtained

$$\zeta + \frac{4}{3}\eta = \lim_{t \to \infty} \frac{\beta}{2tV} \left[ \langle G_{xx}(t) G_{xx}(t) \rangle - \langle G_{xx}(t) \rangle \langle G_{xx}(t) \rangle \right] \qquad (2.49)$$

with $G_{xx}(t)$, the Helfand moment associated with bulk viscosity, written as

$$G_{xx}(t) = \sum_{a=1}^{N} x_a(t) p_{ax}(t). \qquad (2.50)$$

More generally, we can define such a relation for each element of the viscosity tensor:

$$\eta_{ij,kl} = \lim_{t \to \infty} \frac{\beta}{2tV} \left[ \langle G_{ij}(t) G_{kl}(t) \rangle - \langle G_{ij}(t) \rangle \langle G_{kl}(t) \rangle \right] \qquad (2.51)$$

in terms of quantities $G_{ij}(t)$ to be defined by

$$G_{ij}(t) = \sum_{a=1}^{N} p_{ai} r_{aj} \qquad (2.52)$$

if we take $G_{ij}(0) = 0$. In Eqs. (2.47), (2.49) and (2.51), the average $\langle \cdot \rangle$ is performed with respect to the equilibrium state. We notice that, for the microcanonical state (see Appendix A),

$$\beta = \frac{1}{k_{\mathrm{B}}T} \; \frac{N}{N-1} \; .$$  (2.53)

The *Helfand moment* $G_{ij}(t)$ may be defined as the integral of the microscopic current appearing in the Green-Kubo relation:

$$G_{ij}(t) = G_{ij}(0) + \int_0^t J_{ij}(\tau) \, d\tau \; .$$  (2.54)

In order to simplify the terminology we will use Helfand moment for the *Helfand moment associated with viscosity* (shear or bulk viscosities, according to the case).

This quantity can be interpreted as the center of momenta of all the particles in the system. An important difference between diffusion and viscosity is that the dynamical variable $G_{xy}$ evolves with all the particles contrary to $x$. Indeed the Helfand moment is written as a sum over all the particles of the system. This difference can be understood as following: the $x$-component of the arbitrarily selected particle contains all the information of the diffusion. On the other hand, for viscosity, the process of momentum transport in the system implies the intervention of all the particles.

### 2.5.1   Hard-ball systems

In 1970, Alder, Gass and Wainwright calculated the viscosity coefficients of hard-ball systems with Einstein-like formulas based on expressions for Helfand moments which are specific to hard-ball systems (Alder et al., 1970). Their use of Helfand-moment method was motivated by the hard-sphere character of the potential interaction between the particles of the system. Indeed the Green-Kubo method previously introduced presents a disadvantage. Since the force consists of a $\delta$ function at each collision, the autocorrelation form is more difficult to use directly. Their expression takes into account only the elastic collisions between the hard balls. The Helfand moment can be obtained by direct integration of the microscopic current according to Eq. (2.54) with $G_{ij}(0) = 0$

$$G_{ij}(t) \;\; = \;\; \int_0^t d\tau \, J_{ij}(\tau)$$  (2.55)

$$= \;\; \int_0^t d\tau \left[ \sum_{a=1}^{N} \frac{p_{ai} p_{aj}}{m} + \frac{1}{2} \sum_{a \neq b} F_i(\mathbf{r}_a - \mathbf{r}_b) \, (r_{aj} - r_{bj}) \right] ,$$  (2.56)

with $\mathbf{r}_a - \mathbf{r}_b$ satisfying the minimum image convention. Between the collisions, the trajectory is a straight line and the particle velocities change only at each collision. Therefore, the first term in the integral, which is kinetic, is constant during two successive collisions and changes only at the collisions. The second term, i.e. the potential term, vanishes between two successive collisions and contributes only at collisions. Indeed, for a hard-ball potential, the forces between the particles $a$ and $b$ colliding at the time $t_c$ of the collision $c$ can be written in terms of the change $\Delta \mathbf{p}_a^{(c)} = \mathbf{p}_a(t_c + \epsilon) - \mathbf{p}_a(t_c - \epsilon)$ of momentum of the particle $a$ at the collision $c$ as

$$
\begin{aligned}
\mathbf{F}(\mathbf{r}_a - \mathbf{r}_b) &= +\Delta \mathbf{p}_a^{(c)}\, \delta(t - t_c)\,, \\
\mathbf{F}(\mathbf{r}_b - \mathbf{r}_a) &= -\Delta \mathbf{p}_a^{(c)}\, \delta(t - t_c)\,,
\end{aligned}
\tag{2.57}
$$

for $t_c - \epsilon < t < t_c + \epsilon$, because $\Delta \mathbf{p}_b^{(c)} = -\Delta \mathbf{p}_a^{(c)}$. The forces with the other particles which are not engaged in the collision vanish. Therefore, we obtain

$$
G_{ij}(t) = \sum_{(c-1,c)} \left( \sum_{a=1}^{N} \frac{p_{ai} p_{aj}}{m} \right)_{(c-1,c)} \Delta t_{c-1,c} + \sum_c \Delta p_{ai}^{(c)}\, r_{abj}^{(c)}\, \theta(t - t_c)\,,
\tag{2.58}
$$

where, in the first term, $\Delta t_{c-1,c}$ is the time of flight between the collisions $c - 1$ and $c$ during which the momenta remain constant and, in the second term, $a$ and $b$ denote the particles interacting at the collision $c$ and $r_{abj}^{(c)} = r_{aj}(t_c) - r_{bj}(t_c)$. The first sum runs over the intercollisional free flights $(c - 1, c)$ between the initial time $t = 0$ and the current time $t$, while the second sum runs over the collisions occurring between the time $t = 0$ and $t$. If $C$ denotes the last collision before the current time $t$, we notice that the last term of the first sum is $\Delta t_{C,C+1} = t - t_C$.

### 2.5.2 Systems with periodic boundary conditions

In the present work, we propose a more general expression of the Helfand moment for systems with periodic boundary conditions (Viscardy and Gaspard, 2003a). Instead of considering the evolution of the Helfand moment through the successive collisions between hard particles as in the previous method, we take into account the periodicity of the system. In a system of $N$ particles on a torus and satisfying the minimum image convention, the Helfand moment has to be modified in order to take into account the periodic constraints on the motion of the particles. Formally, we can then write the

Helfand moment to which is added a term $I(t)$ gathering the modifications as

$$G_{ij}(t) = \sum_a p_{ai}(t) r_{aj}(t) + I(t) \ . \tag{2.59}$$

As we saw above, the current is the time derivative of the Helfand moment

$$
\begin{aligned}
\frac{dG_{ij}(t)}{dt} &= \sum_a \frac{dp_{ai}(t)}{dt} r_{aj}(t) + \sum_a p_{ai}(t) \frac{dr_{aj}(t)}{dt} + \frac{dI(t)}{dt} \\
&= \sum_a \frac{p_{ai}(t) p_{aj}(t)}{m} + \sum_a \sum_s p_{ai}(t) \Delta r_{aj}^{(s)} \delta(t - t_s) \\
&\quad + \sum_a \sum_{b \neq a} F_i(\mathbf{r}_a - \mathbf{r}_b - \mathbf{L}_{b|a}) r_{aj}(t) + \frac{dI(t)}{dt}
\end{aligned}
\tag{2.60}
$$

where we have introduced the modified Newton equations (2.46). The term implying the interparticle force $\mathbf{F}(\mathbf{r}_a - \mathbf{r}_b - \mathbf{L}_{b|a})$ may be modified as

$$
\begin{aligned}
\sum_a \sum_{b \neq a} F_i(\mathbf{r}_a - \mathbf{r}_b - \mathbf{L}_{b|a}) r_{aj} &= \frac{1}{2} \sum_{a=1}^N \sum_{b \neq a} F_i(\mathbf{r}_a - \mathbf{r}_b - \mathbf{L}_{b|a}) r_{aj} \\
&\quad + \frac{1}{2} \sum_{b=1}^N \sum_{a \neq b} F_i(\mathbf{r}_b - \mathbf{r}_a - \mathbf{L}_{a|b}) r_{bj} \ .
\end{aligned}
\tag{2.61}
$$

Since the force $\mathbf{F}$ is central, we obtain $F_i(\mathbf{r}_a - \mathbf{r}_b - \mathbf{L}_{b|a}) = -F_i(\mathbf{r}_b - \mathbf{r}_a - \mathbf{L}_{a|b})$, which implies that

$$\sum_a \sum_{b \neq a} F_i(\mathbf{r}_a - \mathbf{r}_b - \mathbf{L}_{b|a}) r_{aj} = \frac{1}{2} \sum_{a=1}^N \sum_{b \neq a} F_i(\mathbf{r}_a - \mathbf{r}_b - \mathbf{L}_{b|a}) \ (r_{aj} - r_{bj}) \ . \tag{2.62}$$

In systems with periodic boundary conditions, the current is written as

$$J_{ij}(t) = \sum_{a=1}^N \frac{p_{ai} p_{aj}}{m} + \frac{1}{2} \sum_{a=1}^N \sum_{b \neq a} F_i(\mathbf{r}_a - \mathbf{r}_b - \mathbf{L}_{b|a}) \ (r_{aj} - r_{bj} - L_{b|aj}) \ . \tag{2.63}$$

In consequence, Eq. (2.60) becomes

$$\frac{dG_{ij}(t)}{dt} = J_{ij}(t) + \frac{1}{2} \sum_{a=1}^N \sum_{b \neq a} F_i(\mathbf{r}_a - \mathbf{r}_b - \mathbf{L}_{b|a}) L_{b|aj} + \sum_{a=1}^N \sum_s p_{ai}(t) \Delta r_{aj}^{(s)} \delta(t - t_s) + \frac{dI(t)}{dt} \tag{2.64}$$

By comparison,we obtain

$$\frac{dI(t)}{dt} = -\sum_a \sum_s p_{ai}(t)\Delta r_{aj}^{(s)}\delta(t-t_s) - \frac{1}{2}\sum_{a=1}^{N}\sum_{b\neq a} F_i(\mathbf{r}_a - \mathbf{r}_b - \mathbf{L}_{b|a})L_{b|aj} \qquad (2.65)$$

and, finally, $I(t)$ can be expressed as

$$I(t) = -\sum_a \sum_s p_{ai}^{(s)}\Delta r_{aj}^{(s)}\theta(t-t_s) - \frac{1}{2}\sum_{a=1}^{N}\sum_{b\neq a}\int_o^t d\tau \, F_i(\mathbf{r}_a - \mathbf{r}_b - \mathbf{L}_{b|a}) \, L_{b|aj} \,. \qquad (2.66)$$

We then obtain our general expression for the Helfand moment available for any system submitted to periodic periodic boundary conditions

$$G_{ij}(t) = \sum_{a=1}^{N} p_{ai}(t)\, r_{aj}(t) - \sum_{a=1}^{N}\sum_s p_{ai}^{(s)}\,\Delta r_{aj}^{(s)}\,\theta(t-t_s) - \frac{1}{2}\sum_{a=1}^{N}\sum_{b\neq a}\int_0^t d\tau\, F_i(\mathbf{r}_a - \mathbf{r}_b - \mathbf{L}_{b|a})\, L_{b|aj} \quad (2.67)$$

where $G_{ij}(0) = 0$, $p_{ai}^{(s)} = p_{ai}(t_s)$ and $\theta(t-t_s)$ is the *Heaviside step function* at the time $t_s$ of the jump $s$

$$\theta(t-t_s) = \begin{cases} 1 & \text{for } t > t_s \,, \\ 0 & \text{for } t < t_s \,. \end{cases} \qquad (2.68)$$

We notice that the last two terms of Eq. (2.67) involve the particles near the boundaries of the box. The second term is due to the passage of the particles to or from the neighboring boxes, while the third term concerns the pairs of interacting particles separated by a wall of the box. The expression (2.67) which we propose here can be used to obtain the viscosity coefficients thanks to the Einstein-like formulas (2.51) in a molecular dynamics defined on the torus. We emphasize that the expression (2.67) may apply to systems of particles interacting with a smooth potential under the condition that the range is finite, or to systems of hard balls in elastic collisions. We show in Appendix C that the hydrostatic pressure can also be written in terms of the Helfand moment (2.67).

## 2.6   Validity of our Helfand-moment method

Since the beginning of the nineties, some confusions have been propagated in the literature concerning the use of the mean-squared displacement equation for shear viscosity. First it concerns the unfortunately well-known McQuarrie equation. On the other hand, several works have been done which have wrongly concluded that the mean-square displacement equation for shear viscosity, that is the *Helfand-moment method*, is inapplicable for systems with periodic boundary conditions. These confusions and criticisms are reported in particular by Erpenbeck (1995). Since this method is central in this thesis, a section has to be devoted to such problems in order to avoid any misconception.

### 2.6.1   McQuarrie expression for shear viscosity

In his well-known book *Statistical Mechanics*, McQuarrie (1976) reported the work achieved by Helfand (1960). The derivation he proposed is quite different but he obtained the same relation as Helfand's one, that is[6]

$$\eta = \frac{\beta}{2tV} \left\langle \sum_{a,b=1}^{N} [x_a(t) - x_b(0)]^2 \, p_{ay}(t) p_{by}(0) \right\rangle .$$   (2.69)

Unfortunately, McQuarrie let as an exercise the derivation of the final expression from Eq. (2.69) and wrote "his" mean-squared displacement equation for shear viscosity $\eta_{MQ}$ as follows

$$\eta_{MQ} = \lim_{t\to\infty} \frac{\beta}{2tV} \left\langle \sum_{a=1}^{N} \big[ x_a(t) p_{ay}(t) - x_a(0) p_{ay}(0) \big]^2 \right\rangle ,$$   (2.70)

whereas Helfand obtained

$$\eta_H = \lim_{t\to\infty} \frac{\beta}{2tV} \left\langle \left[ \sum_{a=1}^{N} x_a(t) p_{ay}(t) - x_a(0) p_{ay}(0) \right]^2 \right\rangle .$$   (2.71)

The difference between both expressions is on the position of the sum over particles, and it seems quite obvious that such a difference is simply due to a typing error. Nevertheless, the McQuarrie expression (2.70) at first sight presents a certain interest compared to Helfand's one (2.71), in the sense that the sum over the $N$ particles may be put out of the average. Consequently, one obtains a sum of averages no longer depending on the different particles. Eq. (2.70) can then be rewritten as

$$\eta_{MQ} = \lim_{t\to\infty} \frac{\beta N}{2tV} \left\langle \big[ x_1(t) p_{1y}(t) - x_1(0) p_{1y}(0) \big]^2 \right\rangle .$$   (2.72)

In other words, the McQuarrie relation seems to present the interesting advantage that shear viscosity would be evaluated through a *single-particle* expression whereas Helfand expressed the viscosity by a *collective* approach.

---

[6]Eq. (3.13) in Helfand (1960)'s paper and Eq. (21-304) in McQuarrie (1976)'s book.

The first time that Eq. (2.70) has been considered was in the work by Chialvo and Debenedetti (1991). Without giving a theoretical proof of the validity of the last equation or the equivalence with Eq. (2.71), they provided a numerical comparison between both methods and concluded that the difference between $\eta_H$ and $\eta_{MQ}$ is small. Later Chialvo, Cummings and Evans (1993) tended to prove the McQuarrie expression. But thereafter Allen, Brown and Masters showed (1994), by comparison with their own Green-Kubo results, that the numerical calculations for shear viscosity obtained by Chialvo and Debenedetti (1991) are incorrect, whereas Allen (1994) devoted a comment the paper of Chialvo, Cummings and Evans (1993), and concluded that the McQuarrie expression is not valid and is not able to yield shear viscosity. This conclusion was confirmed later by Erpenbeck (1995). As mentioned above, it is not really surprising since Eq. (2.70) seems quite clearly to be the result of a typing error, and we can therefore emphasize that the viscosity is a collective transport process, implying the intervention of all the particles.

### 2.6.2   Periodic systems and Helfand-moment method

A more important problem concerning our own work is the different criticisms (Allen, 1993; Allen et al., 1994; Erpenbeck, 1995) claiming that a mean-squared displacement equation for shear viscosity, that is the Helfand expression, is unusable for systems submitted to periodic boundary conditions. First, it was pointed out that Alder *et al.* method expressed by Eq. (2.58) is not based on the Helfand expressions, but instead is based on time correlations of the time integrals of the microscopic currents in the Green-Kubo formula (Erpenbeck, 1995).

The main attack about the use of periodic systems for the calculation of viscosity concerns the bounded motion of particles in time because the torus is finite. Consequently, according to the criticisms, the Helfand moment is also bounded in time, what implies that the "pure" Helfand-moment method would be invalid since it would then be rightly expected to obtain a vanishing shear viscosity for long times. By this argument, Allen concluded that the *only* correct way, it seems, to handle $G_{xy}(t)$ is to write it as $\int_0^t \dot{G}_{xy}(\tau) \, d\tau$, and express $\dot{G}_{xy}$ in pairwise, minimum-image form (Allen (1993)).

In other words, the Alder *et al.* method would be the only valid method for studying viscosity, that is, through a method intermediate between the Helfand and Green-Kubo methods. Let us mention that this opinion was recently followed by Hess, Kröger and Evans having considered systems with soft-potential interactions (Hess and Evans, 2001; Hess et al., 2003), as well as by Meier, Laesecke and Kabelac (2004,2005).

However, in order to recover the real microscopic current for infinite systems, the original expression of the Helfand moment (2.48) *must be modified*. It is precisely what we did above for developing our Helfand-moment method by adding the two terms

$$-\sum_{a=1}^{N}\sum_{s} p_{ai}^{(s)} \, \Delta r_{aj}^{(s)} \, \theta(t-t_s) - \frac{1}{2}\sum_{a=1}^{N}\sum_{b\neq a} \int_{o}^{t} d\tau \, F_i(\mathbf{r}_a - \mathbf{r}_b - \mathbf{L}_{b|a}) \, L_{b|aj} \qquad (2.73)$$

to the first one whose the variance is indeed bounded in time. And it is precisely this sum over times of the jumps and the interactions between each particle with the images of the other ones (due to the minimum image convention) that will contribute to the linear growth in time of the variance of the Helfand moment. Hence, the method we propose here is consistent, completely equivalent to the Green-Kubo formula, and presents a certain advantage. The advantage of the Helfand-moment method is that it expresses the transport coefficients by Einstein-like formulas, directly showing their positivity. Moreover, this method is very efficient because it is based on a straightforward accumulation which is numerically robust.

## 2.7   Existence proof of the viscosity

Since computer tools are used in statistical mechanics, a lot of works devoted to the calculations of transport coefficients in different models has been achieved. But in most cases, the transport coefficient of interest has not been rigorously proved to exist, i.e., to be finite, non-vanishing and positive (required by the positivity of the entropy production). The existence of strictly positive coefficients required the establishment of a *central limit theorem*. Only a few systems can claim to have such an advantage. The case of diffusion in *periodic Lorentz gas* considered in 1980 by Bunimovich and Sinai (1980, 1981) appeared as the easiest since it requires a central limit theorem for the position of the point-like particle moving in the physical space. And very recently, it has been suggested by numerical studies that the central limit theorem could also be satisfied for a polygonal billiard channel (Sanders, 2005a; Sanders, 2005b). In this context, Bunimovich and Spohn (1996) proved that a periodic two-hard-disk model (assuming that the diameter of the particles is sufficiently large) satisfies such a theorem for the stress tensor, proving consequently the existence of viscosity coefficients already in this very simple model. Let us point out that Ladd and Hoover (1985) have shown numeri-
cally that viscosity already exists with only two particles. This result is the starting point of our study
of viscosity since it allows us to consider systems with only two particles.

In the same paper (Bunimovich and Spohn, 1996), Fließer devoted a section to the numerical
study of viscosity properties in this two-hard-disk system of square geometry. In this thesis, we study
not only this case, but we also extend this model to the hexagonal geometry. Indeed, the hexagonal
geometry presents some advantages: viscosity is well defined in the fluid phase in which the diffusion
coefficient is non-vanishing (i.e. in the finite-horizon regime; see section 4.2.3). Moreover, as we show
in section 4.4, viscosity in the square lattice is not *isotropic* (contrary to the hexagonal one), which
implies that the viscosity tensor may not be reduced to the shear and bulk viscosity coefficients. A
part of this work is devoted to the comparison between both geometries.

## 2.8  Conclusions

Our Helfand-moment method has several theoretical and numerical advantages: (i) It is strictly
equivalent to the Green-Kubo method. (ii) The Einstein-like formula (2.47) or (2.51) directly show
the positivity of the viscosity coefficient or viscosity tensor because $t$, $\beta$, and $V$ are positive. Moreover,
the Helfand moments directly obey central limit theorems, expressing the Gaussian character of the
dynamical fluctuations of collective variables in systems with finite viscosity. (iii) Thanks to our
expression (2.67) of the Helfand moment, the viscosity coefficients are given by a straightforward
accumulation over the successive jumps $s$. For a given system with $N$ particles, numerical convergence
can be reached in the limit of an arbitrarily large number of jumps $s$, under conditions of existence of
the viscosity coefficients.

By defining the Helfand moment as the integral (2.54) of the microscopic current for a system
with minimum image convention, we obtain the expression (2.67) which can be used to directly
calculate $\Delta G_{ij}(t) = G_{ij}(t) - G_{ij}(0)$ for the Einstein-Helfand relation, remaining consistent with the
requirements imposed by the periodic boundary conditions and with the Green-Kubo formula for a
system satisfying the minimum image convention. Furthermore, this new expression shows that the
Helfand-moment method can be applied to all periodic systems since it is available for any potential
interaction and, therefore, overcome the difficulties previously mentioned in the literature (Erpenbeck,
1995; Allen et al., 1994; Allen, 1993).

**Chapter 3**

# Relationships between chaos and transport

In this chapter, we briefly outline the three different approaches establishing links between transport processes at the macroscopic level and quantities of the underlying microscopic chaotic dynamics. The first method, the so-called *thermostated-system approach*, introduces an external force inducing a nonequilibrium state. In order to keep constant the temperature a *thermostat* has to be introduced to evacuate the excess of energy. Such systems are no longer conserving the volumes in the phase space so that the sum of Lyapunov exponents is not vanishing. Since the method establishes a link between this sum and the transport coefficients, it was not clear whether this link was an artefact of the method or a hint for a more general property. The second approach is the *escape-rate formalism* which introduces absorbing boundary conditions inducing an escape process characterized by an *escape rate* and a fractal repeller. The escape rate can be related on one hand to the transport coefficients and, on the other hand, to chaotic quantities of the microscopic dynamics such as the positive Lyapunov exponents and the fractal dimensions of the repeller. This method provides a relationship which hold for Hamiltonian systems without violation of Liouville's theorem. Finally, the third approach is based on the construction of the hydrodynamic modes at the microscopic level. This method goes behond the stochastic assumption introduced by Boltzmann's *Stosszahlansatz*. This approach has shown that the hydrodynamic modes are no longer smooth but must be considered singular. This property plays a fundamental role for the tranport processes. Indeed, in the case of diffusion, the fractal dimension of the diffusive modes can be related to the diffusion coefficient. Furthermore, the singular character gives the positivity of the entropy production in nonequilibrium systems. The construction of nonequilibrium steady states confirms the singular feature of the hydrodynamic modes.

However, before a presentation of these different methods, we devote a part of this chapter to some generalities and definitions about Liouvillian dynamics and chaotic dynamical systems.

## 3.1   Dynamical systems

A lot of natural systems present a dynamics which can be mathematically described by ordinary differential equations (Ott, 1993; Nicolis, 1995)

$$\frac{d\mathbf{X}}{dt} = \mathbf{F}(\mathbf{X}) \;, \tag{3.1}$$

$\mathbf{X} = x_1, x_2, ..., x_M$ being the $M$-dimensional vector composed by the relevant variables of the system.

If the function $\mathbf{F}(\mathbf{X})$ does not depend explicitly on the time $t$, the system is said to be autonomous. Let us mention that the $M$-dimensional space $\Gamma$ of the variable $\mathbf{X}$ is called the *phase space*.

Phase-space volumes may be preserved or, on the contrary, they may expand or contract under the time evolution. Let us consider the volume $V(0)$ at time $t = 0$ inside the $(M-1)$-dimensional surface $S_0$ in the phase space. Each point of this volume $V(0)$ evolves in time and, after a time $t$, are confined in the volume $V(t)$ limited by the surface $S_t$. The comparison between the last volume and the initial volume brings one an information on the conservative or dissipative character of the dynamical system. If they are equal, the system is said to be *volume-preserving* or *conservative*. Otherwise, because the system has to be confined in a finite volume, one should find $V(t) < V(0)$ for $t > t_0$, and the system is then called *dissipative* (Nicolis, 1995).

More formally, one deals with a conservative when the divergence of the vector field $\mathbf{F}$ vanishes: $\frac{\partial F_i}{\partial X_i} = 0$. Indeed, the time evolution of the volume $V(t)$ is given by

$$
\begin{aligned}
\frac{dV(t)}{dt} &= \frac{d}{dt} \int_{V_t} d\mathbf{X} = \oint_{S_t} \frac{dX_j}{dt} dS_j \\
&= \oint_{S_t} F_j dS_j = \int_{V_t} \frac{\partial F_j}{\partial X_j} d\mathbf{X}
\end{aligned}
\tag{3.2}
$$

so that $\frac{dV}{dt} = 0$ for a conservative dynamics.

The equations (3.1) governing the evolution of dynamical systems induce a so-called *flow*

$$
\mathbf{X} = \mathbf{\Phi}^t \mathbf{X}_0 \; .
\tag{3.3}
$$

In terms of this flow, the evolution of the phase-space volumes can be express by introducing the determinant of (3.3)

$$
J(\mathbf{X}) = \left| \det \frac{\partial \Phi_j^t}{\partial X_j} \right| = \exp \left( \int_0^t \frac{\partial F_j}{\partial X_j} \, d\tau \right) \; .
\tag{3.4}
$$

In a conservative system, $J(\mathbf{X}) = 1$.

## 3.2   Hamiltonian systems

In statistical mechanics, we consider that the dynamics of the atoms and molecules is governed by Newton's laws of motion, of which the formalism has been developed by Hamilton (1805-1865).

The Hamiltonian systems are defined by the Hamilonian function

$$H(\mathbf{X}, t) = H(\mathbf{q}, \mathbf{p}, t) \ ,$$

where $\mathbf{q} = (q_1, q_2, ..., q_N)$ and $\mathbf{p} = (p_1, p_2, ..., p_N)$, $N$ being the *number of degrees of freedom*. Hamiltonian dynamical systems are governed by the *Hamiltonian equation of motion*

$$\begin{aligned}
\frac{d\mathbf{q}}{dt} &= \frac{\partial H}{\partial \mathbf{p}} \\
\frac{d\mathbf{p}}{dt} &= -\frac{\partial H}{\partial \mathbf{q}} \cdot \ .
\end{aligned} \tag{3.5}$$

If the Hamiltonian $H$ is time independent, the system is autonomous. In this case, we have from Eqs.(3.5) that

$$\frac{dH}{dt} = \frac{\partial H}{\partial p_j} \frac{dp_j}{dt} + \frac{\partial H}{\partial q_j} \frac{dq_j}{dt} = 0 \ . \tag{3.6}$$

$H(\mathbf{q}, \mathbf{p})$ is then a *constant of motion* and this constant is equal to the energy $E$. Hence the energy is conserved in such systems.

The energy being fixed, the $2N$-dimensional phase space is reduced to a $(2N - 1)$-dimensional space; the trajectories of the dynamical systems are then constraint to move on this hypersurface at fixed energy.

By writing the Eqs. (3.5) under the general equation (3.1) for dynamical systems with $\mathbf{X}\,(\mathbf{q}, \mathbf{p})$, we have

$$\frac{d\mathbf{X}}{dt} = \mathbf{F}(\mathbf{X}) = \Sigma \cdot \partial_{\mathbf{X}} H \ , \tag{3.7}$$

where the *fundamental matrix* $\Sigma$ of the *symplectic structure* is introduced

$$\Sigma = \begin{pmatrix} \mathbf{0_N} & \mathbf{1_N} \\ -\mathbf{1_N} & \mathbf{0_N} \end{pmatrix} \ , \tag{3.8}$$

in which the unit matrix $\mathbf{1_N}$ is $N$-dimensional, and $\mathbf{0_N}$ is a $(N \times N)$-matrix with vanishing elements.

The divergence of the vector field (3.7) of a Hamiltonian system always vanishes because

$$\nabla \cdot \mathbf{F} = \frac{\partial}{\partial q_j} \frac{dq_j}{dt} + \frac{\partial}{\partial p_j} \frac{dp_j}{dt} = \frac{\partial^2 H}{\partial q_j \partial p_j} - \frac{\partial^2 H}{\partial p_j \partial q_j} = 0 \ . \tag{3.9}$$

It shows that the volumes of the phase space of Hamiltonian systems are incompressible. This is what claims *Liouville's theorem*.

## 3.3 Liouvillian dynamics

In statistical mechanics one introduces statistical ensembles of identical systems described by the variables $\mathbf{X} = (x_1, x_2, ..., x_M)$ and having different initial conditions. An ensemble of initial conditions $\mathbf{X}^{(i)}$ is then considered instead of having only one.

The purpose of statistical mechanics is to establish a bridge between the macroscopic observables (e.g. $A$) and the microscopic variables $\mathbf{X}$. In this context, a fundamental property of statistical mechanics is that the most probable value of an observable is equal to the average of this observable over the statistical ensemble. It allows one to consider the average of the observable instead of the real observable, average defined by

$$\langle A \rangle = \lim_{N \to \infty} \frac{1}{N} \sum_{i=1}^{N} A\left(\mathbf{X}^{(i)}\right) . \tag{3.10}$$

The average value of an observable (3.10) may then be expressed as

$$\langle A \rangle = \int_{\Gamma} A(\mathbf{X}) f(\mathbf{X}) \, d\mathbf{X} . \tag{3.11}$$

if the statistical ensemble of systems can be described by a density distribution in the phase space $\Gamma$

$$f(\mathbf{X}) = \lim_{N \to \infty} \frac{1}{N} \sum_{i=1}^{N} \delta(\mathbf{X} - \mathbf{X}^{(i)}) \tag{3.12}$$

with the normalization $\int_{\Gamma} f(\mathbf{X}) \, d\mathbf{X} = 1$.

The time evolution of the probability density $f$ representing a statistical ensemble obeys the principle of probability conservation and is governed by continuity equation similar to the one of hydrodynamics

$$\frac{\partial f}{\partial t} + \frac{\partial}{\partial x_i}(F_i f) = 0 , \tag{3.13}$$

where $\mathbf{X} = (x_1, x_2, \ldots, x_i, \ldots)$. By introducing the so-called *Liouville operator* $\hat{L}$, this equation can be rewritten as

$$\frac{\partial f}{\partial t} = \hat{L} f \qquad \text{where} \quad \hat{L}(\ldots) = -\frac{\partial}{\partial x_i}(F_i \ldots) \tag{3.14}$$

and is called the *Liouville equation*. In the case of Hamiltonian systems, the Liouville operator is given by the Poisson bracket of the Hamiltonian with $f$

$$\hat{L}f = \{H, f\} = \frac{\partial H}{\partial q_i}\frac{\partial f}{\partial p_i} - \frac{\partial H}{\partial p_i}\frac{\partial f}{\partial q_i} \ . \tag{3.15}$$

A concept closely related to the probability density is the *probability measure $\mu$* which can be defined as

$$\forall A \subset \Gamma \quad : \quad \mu(A) > 0;$$
$$\forall A, B \subset \Gamma | A \cap B = \varnothing \quad : \quad \mu(A + B) = \mu(A) + \mu(B);$$
$$\mu(\Gamma) = 1 \ . \tag{3.16}$$

A measure is said to be *invariant* if it is stationnary under the time evolution of the system

$$\mu_i(\mathbf{\Phi}^{-t}A) = \mu_i(A) \qquad \forall A \subseteq \Gamma \ . \tag{3.17}$$

The relation between a measure $\mu$ and the corresponding density $f$ is given by

$$d\mu = f(\mathbf{X})\,d\mathbf{X} \tag{3.18}$$

and the measure over a finite domain $A$ of the phase space $\Gamma$ by

$$\mu(A) = \int_A d\mu = \int_A f(\mathbf{X})\,d\mathbf{X} \ . \tag{3.19}$$

## 3.4   Chaotic systems

As we saw in section 1.5, the essential property which characterizes the chaotic systems is the *sensibility to initial conditions*. In such systems, two trajectories in phase space, arbitrarily close to each other at initial time, separate from each other exponentially with time. The fundamental quantity of chaotic systems measuring this phenomenon is the *Lyapunov exponent*.

### 3.4.1   Linear stability, tangent space and Lyapunov exponents

An important characterization of trajectories is given by their *linear stability*, which controls how an infinitesimal perturbation of a trajectory evolves in time. This perturbation may be evaluated by integration of the evolution equations (3.1) of both trajectories, the first considered as the reference and passing by the point $\mathbf{X}$, the second being separated by an infinitesimal quantity $\delta\mathbf{X}$. We then have

$$\frac{d\delta\mathbf{X}}{dt} = \mathbf{F}(\mathbf{X} + \delta\mathbf{X}) - \mathbf{F}(\mathbf{X}) = \frac{\partial\mathbf{F}(\mathbf{X})}{\partial\mathbf{X}} \cdot \delta\mathbf{X} \tag{3.20}$$

at the linear order in $\delta\mathbf{X}$. These vectors $\delta\mathbf{X}$ belong to a linear tangent space of the phase space $\Gamma$ in each point $\mathbf{X}$. This space, noted $\mathcal{T}\,\Gamma(\mathbf{X})$ is called *tangent space*. Since (3.20) is linear, all solutions are of the type

$$\delta\mathbf{X}_t = \frac{\partial\boldsymbol{\Phi}^t(\mathbf{X}_0)}{\partial\mathbf{X}_0} \cdot \delta\mathbf{X}_0 = \mathbf{M}(t, \mathbf{X}_0) \cdot \delta\mathbf{X}_0 \ , \tag{3.21}$$

with $\mathbf{X}_0$ and $\delta\mathbf{X}_0$, the values of $\mathbf{X}$ and $\delta\mathbf{X}$ at $t = 0$, and $\mathbf{M}(t, \mathbf{X}_0)$ called *fundamental matrix*.

The infinitesimal perturbation $\delta\mathbf{X}$ can growth exponentially, what we have already mentioned by the property of sensibility to initial conditions. This growth is characterized by the *Lyapunov exponent* associated with a arbitrary tangent vector $\overline{\mathbf{e}}$

$$\lambda(\mathbf{X}, \overline{\mathbf{e}}) = \lim_{t \to \infty} \frac{1}{t} \, \ln \, \| \, \mathbf{M}(t, \mathbf{X}) \cdot \overline{\mathbf{e}} \, \| \ . \tag{3.22}$$

In 1968 Oseledec proposed what he called a *multiplicative ergodic theorem* claiming that, for an ergodic system, the Lyapunov exponent in the direction $\overline{\mathbf{e}}$ is independent of the position $\mathbf{X}$ of the trajectory in the phase space (Oseledec, 1968)

$$\lambda(\mathbf{X}, \overline{\mathbf{e}}) = \lambda(\overline{\mathbf{e}}) \ . \tag{3.23}$$

The tangent space can hence be decomposed into subspaces such that with each subspace is associated a Lyapunov exponent. It implies that the sum of the multiplicities $l^k$ of each exponent equals to the dimension of the tangent space (and of the phase space)

$$\sum_{k=1}^{L} l^k(\mathbf{X}) = dim\mathcal{T}\,\Gamma = dim\Gamma \ . \tag{3.24}$$

In the case of Hamiltonian systems, a theorem concerning the Lyapunov exponents, which is called *pairing rule* (Young, 1982), that is, for a positive exponent, it exists another exponent equal in magnitude, but negative. Consequently, the sum over all the exponents vanishes. This result is not surprising and appears to be intuitive. Indeed, in the Hamiltonian systems conserving the volume of phase space, if the volume is stretched in one direction, it has to be contracted at the same rate in another direction in order to keep constant the volume. On the other hand, one can show that to each quantity conserved by the Hamiltonian equations (the energy, for example), it corresponds a vanishing Lyapunov exponent. Finally, one can show that, if a trajectory has not got any stationnary point, its Lyapunov exponent in the direction of the flow is also equal to zero.

### 3.4.2   Kolmogorov-Sinai entropy

In the previous section was introduced a quantity characterizing quantitatively the property of sensibility to initial conditions, that is the Lyapunov exponents. In other words, the latter measure the dynamical instability, the signature of chaos. This sensibility to initial conditions implies that the trajectories in phase space deviate from each other in the unstable directions. Let us consider an arbitrary error on the measure of initial conditions of a trajectory chosen as the reference. The so-obtained stretching of this phase-space volume induces a gain of information on the real initial conditions since the instability of the dynamics tends to separate the trajectories emerging from this initial phase-space volume, separation which allows us to distinguish more and more the ensemble of trajectories when time goes on. On the other hand the dynamical instability also implies a randomness if we consider the same phenomenon from another viewpoint. If we choose particular initial conditions for the trajectory describing the dynamics of the system, after a certain time, the prediction of the evolution of the system loses its validity because of the dynamical instability and, consequently, *chaotic deterministic systems generate randomness*. The quantity measuring the gain of information in time and the dynamical randomness is the so-called *entropy per unit time*. This entropy characterizes the temporal disorder by analogy to the entropy per unit volume, introduced in thermodynamics and statistical mechanics, which characterizes the spatial disorder. Let us consider a partition $\mathcal{P}$ of phase space $\Gamma$ into $M$ cells (see Fig.3.1)

$$\mathcal{P} = \{C_1, C_2, ..., C_M\} \ . \tag{3.25}$$

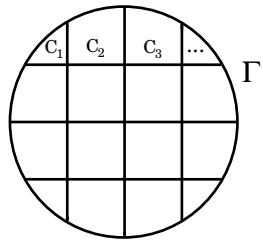

**Figure 3.1**. Partition $\mathcal{P}$ of phase space $\mathcal{M}$ into cells $C_{\omega_i}$ with $\omega_i = 1, \ldots, M$.

Let us furthermore introduce an invariant probability measure $\mu_e$. The probability $\mu_e(\omega_0, \omega_1, ..., \omega_{n-1})$ of having a trajectory which visits successively the cells $(C_{\omega_0}, C_{\omega_1}, ..., C_{\omega_{n-1}})$ at times $t = 0, \Delta t, ..., (n-1)\Delta t$ is given by a $n$-time correlation function (Gaspard, 1998)

$$\mu_e(\omega_0, \omega_1, ..., \omega_{n-1}) = \int_\Gamma \mu_e(d\mathbf{X}) \chi_{\omega_0}(\mathbf{X}) \chi_{\omega_1}(\Phi^{\Delta t}\mathbf{X}) ... \chi_{\omega_{n-1}}(\Phi^{(n-1)\Delta t}\mathbf{X}) \tag{3.26}$$

where $\mathbf{\Phi}$ is the flow introduced at Eq. (3.3) and $\chi_{\omega_i}$ is the indicator function of the cell $C_{\omega_i}$ defined as

$$\chi_{\omega_i}(\mathbf{X}) = \begin{cases} 1 & si\ \mathbf{X} \in C_{\omega_i} \ , \\ 0 & si\ \mathbf{X} \notin C_{\omega_i} \ . \end{cases} \tag{3.27}$$

The entropy per unit time of this partition $\mathcal{P}$ is defined as

$$h(\mathcal{P}) = \lim_{n \to \infty} -\frac{1}{n\Delta t} \sum_{\omega_0, ..., \omega_{n-1}} \mu(\omega_0, \omega_1, ..., \omega_{n-1}) \ \ln \ \mu(\omega_0, \omega_1, ..., \omega_{n-1}) \ . \tag{3.28}$$

However, this entropy depends on the chosen partition. In this context, Kolmogorov have showed how to avoid such a restriction. By definition, one calls *Kolmogorov-Sinai entropy* the *supremum* over all the partitions $\mathcal{P}$

$$h_{\text{KS}} = \text{Sup}_\mathcal{P} \ h(\mathcal{P}) \ , \tag{3.29}$$

which is independent of the partition and defines an intrinsic quantity to the dynamics of the system $\mathbf{\Phi}^t$ and to the invariant measure $\mu$. But the origin of this random property of the system is nothing but the sensibility to initial conditions and the stretching of phase-space volumes. A relationship between the KS entropy and the positive Lyapunov exponents (responsible of the stretching) is then given by the so-called *Pesin's theorem* (1.17)

$$h_{\text{KS}} = \sum_{\lambda_i > 0} \lambda_i \ . \tag{3.30}$$

### 3.4.3   Escape rate and escape-rate formula

As mentionned in section 1.5, the necessity of introducing techniques used in statistical thermody-namics for the study of chaotic dynamical systems. The theory obtained by Bowen and Ruelle is the so-called *thermodynamic formalism* (Ruelle, 1978; Beck and Schlögl, 1993). The idea is to introduce a functional of physical observables which is the generating functional of the average and of the time correlation functions of the given observable $A(\mathbf{X})$. This observable has to be averaged over given orbits of the invariant set $\mathcal{A}$ considered. With this aim, we introduce the notion of separated subsets.

A *separated subset* $\mathcal{S} = \mathbf{Y}_1, \ldots, \mathbf{Y}_s \subset \mathcal{A}$ is composed of points which are separated by a distance $d_T$ larger than $\epsilon$ over a time interval $[-T, +T]$, that is

$$d_T(\mathbf{Y}_i, \mathbf{Y}_j) = \max_{-T \le t \le +T} \left\| \mathbf{\Phi}^t \mathbf{Y}_i - \mathbf{\Phi}^t \mathbf{Y}_j \right\| > \epsilon, \qquad \forall i \ne j \in 1, \ldots, S \ . \tag{3.31}$$

If the invariant set $\mathcal{A}$ is bounded, one can always find a subset $\mathcal{S}$ with a finite number of points. This set is called an $(\epsilon, T)$-separated subset of the invariant set $\mathcal{A}$.

A central function for a given observable $A(\mathbf{X})$ in this formalism is the *topological pressure* which is defined as

$$\mathcal{P}(A) = \lim_{\epsilon \to 0} \lim_{T \to \infty} \frac{1}{2T} \ln \mathcal{Z}(\epsilon, T, A) \ , \tag{3.32}$$

with the partition functional

$$\mathcal{Z}(\epsilon, T, A) = \text{Sup}_{\mathcal{S}} \sum_{\mathbf{Y} \in \mathcal{S}} \exp\left( \int_{-T}^{+T} A(\mathbf{\Phi}^t \mathbf{Y}) \, dt \right) \ , \tag{3.33}$$

where $\mathcal{S}$ is a $(\epsilon, T)$-separated subset of the invariant subset $\mathcal{A}$.

If $B(\mathbf{X})$ is another observable of the system, its average is defined as

$$\begin{aligned}
\langle B \rangle_{\mu_A} \equiv \mu_A(B) &= \frac{d}{d\nu} \mathcal{P}(A + \nu B)|\nu = 0 \\
&= \int B(\mathbf{X}) \mu_A(d\mathbf{X}) \ .
\end{aligned} \tag{3.34}$$

Using the definition of the pressure, we obtain an expression of this *dynamical measure* $\mu_{\mathcal{A}}$

$$\mu_A(d\mathbf{X}) = \lim_{\epsilon \to 0} \lim_{T \to \infty} \mathrm{Sup}_{\mathcal{S}} \sum_{\mathbf{Y} \in \mathcal{S}} \frac{\exp\left(\int_{-T}^{+T} A(\mathbf{\Phi}^t \mathbf{Y})\, dt\right)}{\mathcal{Z}(\epsilon, T, A)} \times \frac{1}{2T} \int_{-T}^{+T} \delta(\mathbf{X} - \mathbf{\Phi}^t \mathbf{Y})\, dt\, d\mathbf{Y}\ . \quad (3.35)$$

Each trajectory of the subset $\mathcal{S}$ is weighted by a Boltzmann-type probability given by

$$\pi_A(\epsilon, T, \mathbf{Y}) = \frac{\exp\left(\int_{-T}^{+T} A(\mathbf{\Phi}^t \mathbf{Y})\, dt\right)}{\mathcal{Z}(\epsilon, T, A)}\ . \quad (3.36)$$

The Kolmogorov-Sinai entropy per unit time with respect to this invariant measure $\mu_A$ is defined by

$$h_{\mathrm{KS}}(\mu_A) = -\lim_{\epsilon \to 0} \lim_{T \to \infty} \frac{1}{2T} \mathrm{Sup}_{\mathcal{S}} \sum_{\mathbf{Y} \in \mathcal{S}} \pi_A(\epsilon, T, \mathbf{Y}) \ln \pi_A(\epsilon, T, \mathbf{Y})\ . \quad (3.37)$$

From Eqs. (3.34) and (3.36), we can deduce the important identity

$$h_{\mathrm{KS}}(\mu_A) = -\mu_A(A) + \mathcal{P}(A)\ . \quad (3.38)$$

An important particular choice for the observable $A(\mathbf{X})$ is the following. $\beta$ being a real parameter, we take

$$A(\mathbf{X}) = -\beta \sum_{\lambda_i > 0} \chi_i(\mathbf{X}) \quad (3.39)$$

where $\chi_i$ are the local stretching rates related to the Lyapunov exponents by (Gaspard, 1998)

$$\lambda_i(\mathbf{X}) = \lim_{t \to \infty} \frac{1}{t} \int_0^t \chi_i(\mathbf{\Phi}^\tau \mathbf{X})\, d\tau\ . \quad (3.40)$$

Using this observable, we observe that, for $\beta > 0$, the probability (3.36) associated with a trajectory is larger for the more stable trajectories. The pressure functional becomes the *pressure function $P(\beta)$*

$$P(\beta) = \mathcal{P}\left[-\beta \sum_{\lambda_i > 0} \chi_i(\mathbf{X})\right] \quad (3.41)$$

which defines an invariant probability measure $\mu_\beta$ depending on the parameter $\beta$. Since the local stretching rates and the Lyapunov exponents are related by Eq. (3.40), and using the time invariance of the measure $\mu_\beta$ we have

$$\mu_\beta(\chi_i) = \mu_\beta(\lambda_i) . \tag{3.42}$$

Therefore, Eq. (3.38) becomes in this case

$$h_{\text{KS}}(\mu_\beta) = \beta \sum_{\lambda_i > 0} \mu_\beta(\lambda_i) + P(\beta) . \tag{3.43}$$

### Closed systems

A closed system is a system in which any trajectory can escape. A time-independent Hamiltonian system presents the *microcanonical measure* $\mu_e(d\mathbf{X})$ as an appropriate invariant measure, which is given by

$$d\mu_e = \mathcal{N}\delta(H - E) \, d\mathbf{q} \, d\mathbf{p} . \tag{3.44}$$

It can be shown that this measure corresponds to the measure associated with the observable $A = -\sum_{\lambda_i > 0} \chi_i(\mathbf{X})$, that is (Gaspard, 1998)

$$\mu_e = \mu_{\beta=1} . \tag{3.45}$$

Furthermore, it can also be shown that, for closed systems, we have $P(\beta = 1) = 0$, so that Eq.(3.38) for $\beta = 1$ becomes

$$h_{\text{KS}}(\mu_e) = \sum_{\lambda_i > 0} \mu_e(\lambda_i) . \tag{3.46}$$

Hence the Pesin's identity (3.30) is recovered.

### Open systems

Contrary to the closed systems, the open systems allow the escape of the trajectories out of the bounded phase-space domain $\mathcal{M}$. Furthermore, in such systems, an escaped trajectory is not allowed to reenter $\mathcal{M}$. The boundaries are therefore considered as absorbing and maintain the system in nonequilibrium. As we will see below the support of the invariant measure that we may here choose is a *fractal repeller*.

The construction of this measure is the following: let us consider a probability measure $\nu_0(d\mathbf{X})$ corresponding to an initial statistical ensemble $\mathbf{X}_0^{(i)}$ on the phase-space domain $\mathcal{M}$. The measure $\nu_0$ is

written as

$$\nu_0(d\mathbf{X}) = \lim_{N_0 \to \infty} \frac{1}{N_0} \sum_{i=1}^{N_0} \delta(\mathbf{X} - \mathbf{X}_0^{(i)}) \, d\mathbf{X} \, . \tag{3.47}$$

Because of the escape, after a time $T$, only $N_T$ points from the initial ensemble are still in $\mathcal{M}$. The ratio of such points is given by

$$\lim_{N_0 \to \infty} \frac{N_T}{N_0} = \int_{\Upsilon_{\mathcal{M}}^{(+)}(T)} \nu_0(d\mathbf{X}) \tag{3.48}$$

where $\Upsilon_{\mathcal{M}}^{(+)}(T)$ is the set of all the initial conditions $\mathbf{X}$ which escape out of $\mathcal{M}$ after a time $T_{\mathcal{M}}^{(+)}(\mathbf{X})$ larger than $T$ (that is, the initial conditions of the trajectories still inside the absorbing boundaries at time $T$)

$$T_{\mathcal{M}}^{(+)}(\mathbf{X}) = \text{Max} \left\{ T > 0 : \mathbf{\Phi}^t \mathbf{X} \in \mathcal{M}, \forall t \in [0, T[ \, \right\} \tag{3.49}$$

$$\Upsilon_T^{(+)}(\mathbf{X}) \equiv \left\{ \mathbf{X} \in \mathcal{M} : T < T_{\mathcal{M}}^{(+)}(\mathbf{X}) \right\} \, . \tag{3.50}$$

The equivalent set of initial conditions $\Upsilon_T^{(-)}(\mathbf{X})$ for backward evolution, is similarly obtained replacing $T$ by $-T$, $\mathbf{\Phi}^t$ and by $\mathbf{\Phi}^{-t}$.

The decay $\frac{N_T}{N_0}$ of the number of trajectories still in $\mathcal{M}$ is exponential since all the trajectories of the repeller are exponentially unstable. The exponential decay is characterized by an *escape rate* given by

$$\gamma = - \lim_{T \to \infty} \frac{1}{T} \ln \nu_0 \left[ \Upsilon_{\mathcal{M}}^{(+)}(T) \right] \, . \tag{3.51}$$

If the system is ergodic, the time average of a dynamical quantity equals the its ensemble average. This is expressed as

$$\mu_{\text{ne}}(A) = \lim_{T \to \infty} \lim_{N_T \to \infty} \frac{1}{N_T} \sum_{i=1}^{N_T} \frac{1}{2T} \int_{-T}^{+T} A(\mathbf{\Phi}^t \mathbf{X}^{(i)}) \, dt$$

$$= \int A(\mathbf{X}) \mu_{\text{ne}}(d\mathbf{X}) \tag{3.52}$$

where $N_T$ is the number of phase-space points remaining in the system during the time interval $] - T, +T[$. This allows us to write $\mu_{\text{ne}}$ as

$$\mu_{\text{ne}}(d\mathbf{X}) = \lim_{T \to \infty} \frac{1}{\nu_0[\Upsilon_{\mathcal{M}}(T)]} \int \nu_0(d\mathbf{y}) I_{\Upsilon_{\mathcal{M}}(T)}(\mathbf{y}) \times \frac{1}{2T} \int_{-T}^{+T} \delta(\mathbf{X} - \mathbf{\Phi}^T \mathbf{y}) \, dt \, d\mathbf{X} \tag{3.53}$$

backward as well as forward evolution in order to get the invariant measure.

It can be shown that $\mu_{\text{ne}}$ corresponds to the invariant measure associated with the observable $A = -\sum_{\lambda_i > 0} \chi_i(\mathbf{X})$ for $\beta = 1$ (Gaspard, 1998)

$$\mu_{\text{ne}} = \mu_{\beta=1} \ . \tag{3.54}$$

The escape rate is related to the topological pressure by

$$P(\beta = 1) = -\gamma \ . \tag{3.55}$$

Thanks to this result, Eq. (3.43) becomes the generalized Pesin's identity available even for open systems. Hence we obtain the so-called *escape-rate formula* (Gaspard, 1998; Eckmann and Ruelle, 1985)

$$\gamma = \sum_{\lambda_i > 0} \mu_{\text{ne}}(\lambda_i) - h_{\text{KS}}(\mu_{\text{ne}}) \tag{3.56}$$

which plays an important role in the escape-rate formalism used in this thesis. A more intuitive derivation of Eq. (3.56) will be given below.

### 3.4.4   Fractal dimensions

As we have seen in section 1.5, we know since the work by Mandelbrot that fractals can be observed everywhere in nature (Mandelbrot, 1975). The famous historical example of fractal appears in the title of Mandelbrot's paper: "How long is the coast of Britain?" (Mandelbrot, 1967). The property of such geometrical objects is that the notion of length (areas, volumes, etc.) loses its sense. In mathematics, fractal objects have been invented in order to study this new geometry, since Cantor (1884), Peano (1890), Koch (1904), Sierpinski (1916), etc. Moreover, their main particularity is their *non-integer dimension*, contrary to usual geometrical objects. For such objects, a more general definition for the dimension is needed.

#### Box-counting dimension

To answer this question, we introduce the notion of box-counting dimension. Let us first consider a simple geometrical object of dimension $D$. Let $N(\epsilon)$ be the number of small $D_e$-dimensional cells of linear size $\epsilon$ needed to cover the object considered. $D_e$ is the embedding dimension. It is an integer chosen large enough to satisfy $D \leq D_e$. $N(\epsilon)$ depends on $\epsilon$ as

$$N(\epsilon) \sim \frac{1}{\epsilon^D} \tag{3.57}$$

where $D$ is the dimension of the object considered. We can therefore define the box-counting dimension or capacity $D$ as

$$D = -\lim_{\epsilon \to 0} \frac{\ln N(\epsilon)}{\ln \epsilon} \tag{3.58}$$

Let us take Koch's curve depicted in section 1.5 as an example. At the beginning, we only need one cell of linear size $\epsilon_0 = 1$ to cover the line. After the first step, the linear length of the cells is multiplied by a factor $a = 1/3$ and the number of cells by $G = 4$. By recurrence, after $k$ steps, we have

$$\epsilon_k = a^k \epsilon_0 \tag{3.59}$$

$$N_k = G^k N_0. \tag{3.60}$$

Eq. (3.59) can be rearranged in order to express $k$ as a function of $\epsilon$

$$k = \frac{\ln \epsilon_k - \ln \epsilon_0}{\ln a} . \tag{3.61}$$

By introducing this result in the logarithm of Eq. (3.60), we obtain

$$\begin{aligned} \ln N_k &= \ln N_0 + k \ln G \\ &= \ln N_0 + \frac{\ln G}{\ln a}(\ln \epsilon_k - \ln \epsilon_0) . \end{aligned} \tag{3.62}$$

By replacing the last results in Eq. (3.58) and by taking the limit $k \to \infty$ corresponding to $\epsilon_k \to 0$, we get the dimension of Koch's curve

$$D = -\lim_{k \to \infty} \frac{\ln N_k}{\ln \epsilon_k} = -\frac{\ln G}{\ln a} = \frac{\ln 4}{\ln 3} = 1.26\ldots \tag{3.63}$$

Let us mention that the general definition of dimension may be applied to usual objects such as the line ($D = 1$), the square ($D = 2$), etc.

**Hausdorff dimension**

The box-counting dimension is determined by covering the fractal with small identical cells and is supposed to be independent of the cell shape. Under these conditions, the limit (3.58) may not be well-defined for very complicated fractals. In this context, it is useful to consider another definition which allows it, that is the so-called *Hausdorff dimension*. In this case, the fractal $A$ is covered by cells $\sigma_k$ of variable diameter $\epsilon_k$, with $\epsilon_k < \epsilon$. By introducing a positive parameter $q$, one defines the following quantity

$$m(q, \epsilon) = \sum_k (\epsilon_k)^q, \qquad \epsilon_k < \epsilon \; . \tag{3.64}$$

In the limit $\epsilon \to 0$, $m(q, \epsilon)$ will vanish for $q > q_0$ and will diverge for $q < q_0$. When $q = q_0$, $m(q, \epsilon)$ may be well-defined and non-vanishing, and $q_0$ is the Hausdorff dimension of $A$

$$D_H = q_0 \; . \tag{3.65}$$

Let us mention that the box-counting dimension coincides with the Hausdorff dimension when it is well-defined.

**Multifractals and generalized dimensions**

The fractals (repellers) with which we deal in this thesis are such that the two previous dimensions do not provide a complete characterization of the scaling properties. Indeed, they may have a self-similar structure which varies from point to point; for instance, because the trajectories do not visit different regions with the same probabilities. Hence it is needed to introduce the concept of *generalized dimensions*. Fractals for which this new concept appears to be non trivial are called *multifractals*. They are the support of a probability measure. Whereas the Hausdorff dimension gives information about the geometry of fractals, the generalized dimensions $D(q)$ take into account the probability distribution on the fractals, and give then additional information.

Let us consider a fractal invariant set $\mathcal{A}$ as the support of an invariant probability distribution $f$. First, let us define dimensions based on the partitioning of phase space into $R$ boxes of equal linear size $\epsilon$ and identical shape. The probability attributed to a box $\sigma_i$ is given by

$$p_i = \int_{\sigma_i} f(\mathbf{X}) \, d\mathbf{X} \; . \tag{3.66}$$

The number of boxes of non-zero probability is denoted by $r$, with $r \leq R$.

Let us define a local quantity $\alpha_i(\epsilon) \equiv \alpha(\epsilon, \mathbf{X})$, for the box $\sigma_i$ of linear length $\epsilon$ centered around $\mathbf{X}$ and of probability $p_i$, by

$$\alpha(\epsilon, \mathbf{X}) = \frac{\ln p_i}{\ln \epsilon} \; . \tag{3.67}$$

We next take the limit $\epsilon \to 0$

$$\alpha(\mathbf{X}) = \lim_{\epsilon \to 0} \alpha(\epsilon, \mathbf{X}) \tag{3.68}$$

to obtain a local scaling exponent or local dimension. The generalized dimensions for multifractals are then defined as

$$D(q) = \lim_{\epsilon \to 0} \frac{1}{\ln \epsilon} \frac{1}{q-1} \ln \sum_{i=1}^{r} p_i^q \; . \tag{3.69}$$

When $q = 0$, we have

$$D(0) = -\lim_{\epsilon} \frac{\ln r}{\ln \epsilon} \tag{3.70}$$

and we recover the box-counting dimension (3.58). For $q = 1$, we get the so-called *information dimension* and is therefore defined as

$$D(1) = \lim_{\epsilon \to 0} \frac{1}{\ln \epsilon} \sum_{i=1}^{r} p_i \ln p_i \; . \tag{3.71}$$

Let us mention that, as for the case of the Hausdorff dimension, we may give a more general definition of the generalized dimensions in terms of boxes of variable linear sizes. We then consider different disjoint cells $\sigma_1, \ldots, \sigma_r$ as previously, and we associated with it a probability $p_i$. We suppose that each cell $\sigma_i$ can be covered by a spherical ball of smallest possible radius $l_i$ with $l_i < l \quad \forall i$. $q$ being a real parameter, we define the following quantity (Halsey et al., 1986)

$$Z(q) = \sum_{i=1}^{r} \frac{p_i^q}{l_i^{(q-1)\xi(q)}} \; . \tag{3.72}$$

The dimension $D(q)$ is given by the value of $\xi(q)$ for which $Z(q)$ does neither diverge nor vanish. $D(0)$ coincides with the Hausdorff dimension defined in Eqs. (3.64) and (3.65).

## 3.5   Thermostated-system approach

In this approach, nonequilibrium systems are defined as systems composed of particles submitted to interparticle forces, to external forces, but also to a fictitious nonHamiltonian force modeling the coupling to some hypothetical thermostat (Evans and Morriss, 1990). For instance, in order to study viscosity, the idea is to reproduce a Couette flow induced by a shearing force (see Fig. 3.2). Hence a velocity gradient is established in the system. However, this gradient leads to considerable viscous heating of the fluid and the energy of the system does not remain constant. To deal with this problem, it is necessary to introduce an internal thermostat – a fictitious frictional force – in order to keep constant the energy. Formally, for a 3-D system, in addition to the shearing force, we have to introduce a term with a thermostating multiplier $\alpha$ (Evans and Morriss, 1990)

$$\begin{aligned}
\frac{d\mathbf{q}_i}{dt} &= \frac{\mathbf{p}_i}{m} + \gamma y_i \\
\frac{d\mathbf{p}_i}{dt} &= \mathbf{F}_i - \gamma p_{yi} - \alpha \mathbf{p}_i \;,
\end{aligned} \tag{3.73}$$

where $\gamma$ is the vector $(\frac{du_x}{dy}, 0)$, and $\mathbf{u}$ is the mean velocity. However, such dynamical systems violate Louville theorem which asserts that the phase-space volumes are presented by the microscopic dynamics. This violation leads to fundamental problems for defining an entropy in nonequilibrium steady states. It thus appears as an artefact of a nonHamiltonian force that the phase-space volume visited by the system decreases in time and is expressed by the non-zero sum of Lyapunov exponents $\sum_{i=1}^{6N} \lambda_i < 0$. This phase-space contraction is introduced in the system through the presence of the thermostat (and the quantity $\alpha$) which takes away the energy given to the system by shearing $\gamma$. It can be showed that the relation between this constant and the sum of Lyapunov exponents is written as

$$\langle 3N\alpha \rangle = -\sum_{i=1}^{6N} \lambda_i \ . \tag{3.74}$$

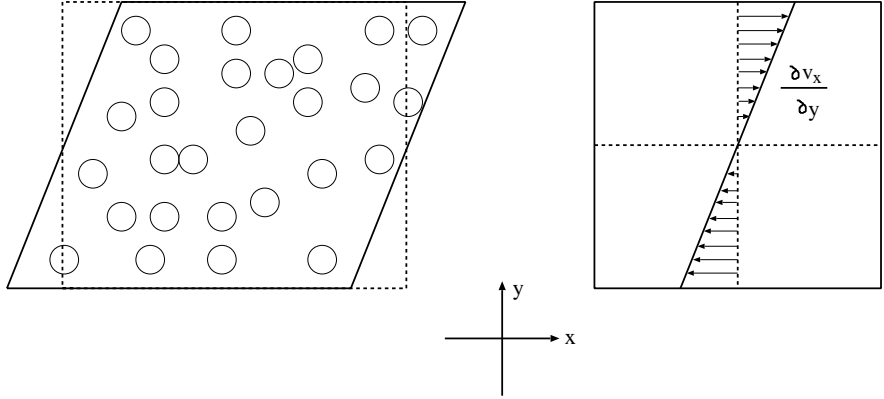

**Figure 3.2**. Illustration of the shearing of the system inducing a velocity gradient $\frac{\partial v_x}{\partial y}$.

The thermostat coupling constant $\alpha$ ensures that the rate of energy produced in the system in the stationary state, due to the work done on the system by the shear forces, is exactly balanced by the energy (heat) removed by the thermostat, so that

$$-P_{xy}V\gamma - 3Nk_BT\alpha = 0 \ . \tag{3.75}$$

where $P_{xy}$ is the $xy$-element of the pressure tensor.

Using that the shear viscosity coefficient $\eta(N, \gamma)$ can be defined from the steady state average $\langle P_{xy} \rangle = -\eta(\gamma)\gamma$, one has

$$\eta(N, \gamma) = \frac{-k_BT}{V\gamma^2} \sum_{i=1}^{6N} \lambda_i(N, \gamma) \ . \tag{3.76}$$

The actual determination of $\eta$ from this relation is hampered by the very large number of exponents in a macroscopic system. An arithmetic mean rule enables one to reduce the sum to the evaluation of the maximum and minimum exponents of the Lyapunov spectrum. Thanks to this pairing rule, Eq. (3.76) becomes for large systems (Evans et al., 1990)

$$\eta(N, \gamma) = \frac{-3nk_BT}{\gamma^2} \left[ \lambda_{max}(\gamma) + \lambda_{min}(\gamma) \right] \ , \tag{3.77}$$

the $N$ dependence of $\eta$ and $\lambda$ disappearing. The shear viscosity occurring in the Navier-Stokes equation is given by $\eta = \lim_{\gamma \to 0} \eta(\gamma)$. This expression relates the Lyapunov exponents to the viscosity coefficient because of the violation of Liouville's theorem by the artificial nonHamiltonian systems. Therefore, this method cannot be used for Hamiltonian systems. Several works have shown that other ways exist to maintain a system out of equilibrium for instance by stochastic boundary conditions or by deterministic scattering (Klages et al., 2000) in which cases the relation (3.77) do not apply. Moreover, violating Liouville's theorem creates problems in defining the entropy for nonequilibrium steady states.

## 3.6   Escape-rate formalism

As for the previous method, the escape-rate formalism introduces nonequilibrium conditions. Here we do not impose an external constraint like a shearing. Instead we open the system in order to generate an escape process. More precisely, we impose absorbing boundary conditions at the statistical level of description, keeping the Hamiltonian character of the equations of motion themselves. The so-called *escape rate* is related to the studied transport coefficient on the one hand, and to the chaotic quantities of the microscopic dynamics on the other hand. The method was first developed in 1990 by Gaspard and Nicolis for the case of diffusion (Gaspard and Nicolis, 1990) and was extended to the other transport processes in 1995 by Dorfman and Gaspard (Dorfman and Gaspard, 1995; Gaspard and Dorfman, 1995). For pedagogical reasons we first will expose the escape-rate formalism for diffusion.

### 3.6.1   Escape-rate formalism and diffusion

Let us take the well-known Lorentz gas (see Fig.3.3) that consists of a particle of mass $m$ moving with energy $E$ among a fixed set of two-dimensional scatterers which is of infinite extent in the $y$-direction, but of finite extent in the $x$-direction, such that the scatterers are confined to the interval

$$-\frac{L}{2} \leq x \leq \frac{L}{2} \ . \tag{3.78}$$

Absorbing walls are placed on the planes at $x = \pm \frac{L}{2}$. The particles, initially introduced at the center of the system, evolve in the system by the successive collisions with the scatterers.

Let us take the spatial distribution function, $p(\mathbf{r}, t)$, of the moving particles. If $L$ is sufficiently large and for long times after some initial time, we expect $p(\mathbf{r}, t)$ to be described by the diffusion equation

$$\frac{\partial p}{\partial t} = D\nabla^2 p \ , \tag{3.79}$$

where $D$ is the diffusion coefficient. The absorbing boundary conditions lead to the condition that $p(\mathbf{r}, t)|_{\text{boundary}} = 0$. Then the probability for the distribution of particles in the $x$-direction has the fom

$$p(x, t) = \sum_{n=1}^{\infty} a_n \sin\left(\frac{\pi n}{L}x\right) \exp\left[-\left(\frac{n\pi}{L}\right)^2 Dt\right] \ , \tag{3.80}$$

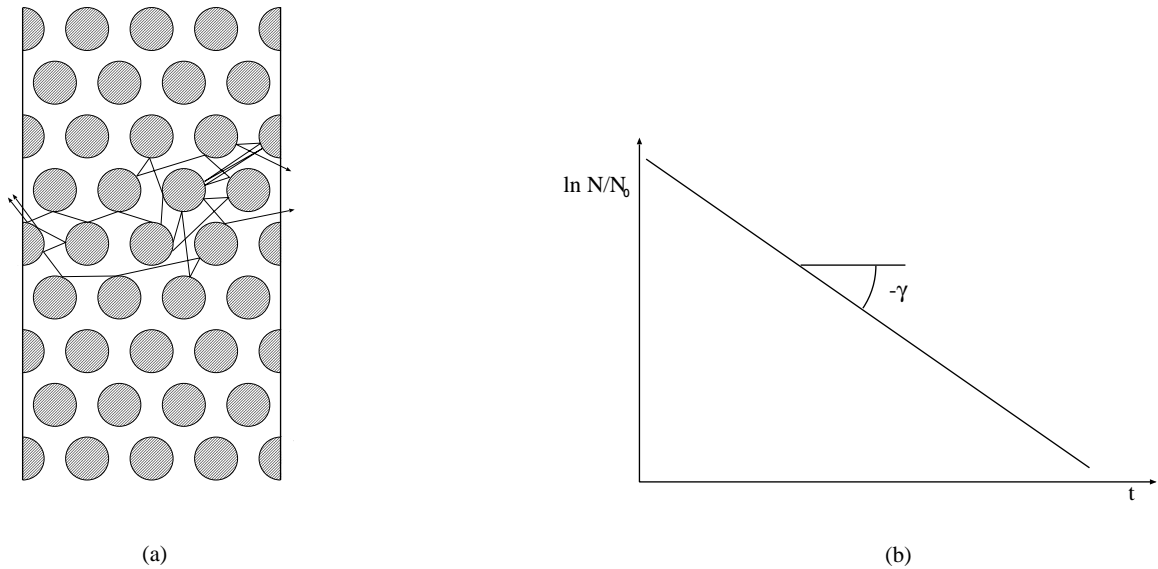

**Figure 3.3**. Lorentz gas with absorbing boundary conditions separated by the distance $L$. (a) Escape process of some particles after diffusion in the scatterer. (b) Exponential decrease in time of the number of particles lying into the limits defined by Eq.(3.78).

where $a_n$ are numerical coefficients fixed by the initial profile of concentration. As time increases, each mode decreases exponentially and vanishes successively, the first ones having the greatest values of $n$. Consequently, for long times, the slowest decaying mode ($n = 1$) describes the escape process and decays as $\exp(-\pi^2 Dt/L^2)$. So for large systems we can define a *macroscopic escape rate* as

$$\gamma_{\mathrm{mac}} = \left(\frac{\pi}{L}\right)^2 D \ . \tag{3.81}$$

In the following we shall call Eq. (3.81) the escape-transport formula.

Let us consider the same process at the microscopic scale. In section 3.4.3, we rigorously introduced the escape rate in the context of the thermodynamic formalism. We here propose a more intuitive derivation of the escape-rate formula (3.56). In open systems such that in Fig. 3.3, particles go out of the boundaries and never go back into the system. But a set of trajectories remain forever (in the future and the past) into the limits. A particle bouncing forever between two scatterers is the simplest example of such trajectories. This set is therefore the best candidate to be the appropriate support for a nonequilibrium invariant measure in order to evaluate the different quantities. This object in phase space presents a particular property: it has a zero Lebesgue measure and is of non-integer dimension. Such an object is called *fractal*. Contrary to the situation in the thermostated systems, this fractal is not an attractor but a *repeller* because trajectories escape from it. We denote this fractal repeller by the symbol $\mathcal{F}_L$ .

In sections 1.5 and 3.4, we saw that the instability of the dynamics characterized by the positive Lyapunov exponents induces a gain of information on the initial conditions of the trajectories in phase space. This information grows exponentially in time and the exponential rate at which information is obtained is measured by the so-called KS entropy, $h_{\text{KS}}$. Let us consider a certain region of phase-space points with a characteristic dimension of the order of $\delta$, which is the error on the observation. The different points in this region are not distinguishable at the initial conditions, but after a certain time $t$, the initial set will be stretched along the unstable directions (which correspond to the positive Lyapunov exponents) to a length of order $\delta \exp\left(t \sum_{\lambda_i > 0} \lambda_i\right)$. Consequently, trajectories emerging from the initial set of points will be separated and we can easily resolve their images in the initial set. In closed systems (without any escape condition) this gain of information therefore is evaluated as

$$\exp(h_{\text{KS}} \, t) = \exp\left(t \sum_{\lambda_i > 0} \lambda_i\right) , \qquad (3.82)$$

which gives us Pesin's theorem (1.17). On the other hand, in open systems, most of the trajectories escape the system because of the absorbing boundaries (3.78). The part of trajectories moving into the limits decays as $\exp(-\gamma t)$, $\gamma$ being the escape rate. When a trajectory escapes the system it can no longer give information on its origin by the aforementioned mechanism of dynamical instability. Accordingly because of the absorbing conditions we loose in time a quantity of information (brought by the instable character of the dynamics) versus the case of closed systems. This phenomenon induces a modification of the Pesin's theorem by the introduction of the escape term

$$\exp(h_{\text{KS}} \, t) = \exp(-\gamma t) \exp\left(t \sum_{\lambda_i > 0} \lambda_i\right) \qquad (3.83)$$

or simply

$$\gamma_{\text{mic}}(\mathcal{F}_L) = \sum_{\lambda_i > 0} \lambda_i(\mathcal{F}_L) - h_{\text{KS}}(\mathcal{F}_L) . \qquad (3.84)$$

Hence, we recover the *escape-rate formula* (3.56). The subscript (mic) specifies that the escape rate is obtained by the microscopic approach.

An equivalent formula can be obtained which involves the partial fractal dimensions of the repeller instead of the KS entropy. Indeed, the fractal character of the repeller is a direct consequence of the escape of trajectories so that the KS entropy is no longer equal to the sum of Lyapunov exponents but to (Young, 1982)

$$h_{\text{KS}} = \sum_{\lambda_i > 0} d_i \, \lambda_i \, , \tag{3.85}$$

where the coefficients are the partial information dimensions[1] of the repeller associated with each unstable direction of corresponding Lyapunov exponent $\lambda_i$ (Eckmann and Ruelle, 1985). These partial dimensions satisfy

$$0 \leq d_i \leq 1 \, , \tag{3.86}$$

so that the KS entropy is in general smaller than the sum of positive Lyapunov exponents. Accordingly, the escape rate can be expressed as

$$\gamma_{\text{mic}}(\mathcal{F}_L) = \left( \sum_{\lambda_i > 0} c_i \, \lambda_i \right)_{\mathcal{F}_L} \tag{3.87}$$

in terms of the partial codimensions defined as

$$c_i \equiv 1 - d_i \, . \tag{3.88}$$

The escape process being the same at the macroscopic and microscopic scales, the identity $\gamma_{\text{mac}} = \gamma_{\text{mic}}$ is obtained in the limit $L \rightarrow \infty$. Consequently we can relate the two levels by combining (3.81) and (3.84) and finally we have

$$D = \lim_{L \rightarrow \infty} \left( \frac{L}{\pi} \right)^2 \left( \sum_{\lambda_i > 0} c_i \, \lambda_i \right)_{\mathcal{F}_L} \, . \tag{3.89}$$

This relation is fundamental: it clearly establishes the link between the microscopic and macroscopic levels, between diffusion as a transport process and the chaotic properties of the underlying microscopic dynamics. This formula allows us to understand how chaos controls transport at the microscopic scale (Gaspard and Baras, 1992, 1995).

---

[1]It is known that the partial information dimension of the repeller can be approximated by the partial Hausdorff dimension if the escape rate is small enough and if Ruelle's topological pressure does not present a discontinuity. This last condition is fulfilled if the system does not undergo a dynamical phase transition. This is the case in the finite-horizon regimes of Sinai's billiard which controls the dynamics of both the Lorentz gas (Gaspard and Baras, 1995) and the two-disk model, as we present in the next chapter. Under these conditions, we can replace the partial information dimension $d_i$ by the partial Hausdorff dimension $d_{Hi}$.

### 3.6.2   The escape-rate formalism for viscosity

In 1995, Dorfman and Gaspard extended the escape-rate formalism to the other transport processes (Dorfman and Gaspard, 1995; Gaspard and Dorfman, 1995), in particular, to the shear viscosity. The problem is quite similar but more abstract. Indeed, the absorbing boundary conditions inducing the escape process is no longer in the physical space but in the space of the Helfand moment associated with the shear viscosity. Indeed, in the previous chapter, we presented the important work realized by Helfand (Helfand, 1960) in which he obtained for example the Helfand moment $G_{xy}(t)$ associated with the shear viscosity. Since this quantity following the Einstein-like relation (2.47), the Helfand moment undergoes a diffusive type of motion in his own space, that is, along the axis of $G_{xy}(t)$. Hence, in the context of the escape-rate formalism, the Helfand moment for viscosity plays the same role as the position for the diffusion.

As for diffusion, the central object of the escape-rate formalism is the fractal repeller composed of the phase-space trajectories for which the Helfand moment fluctuates forever within some interval

$$-\frac{\chi}{2} \leq \tilde{G}_{yx} \leq +\frac{\chi}{2} \; . \tag{3.90}$$

These trajectories are exceptional because the Helfand moment escapes out of this interval for almost all the trajectories. Therefore, the repeller has a vanishing probability measure in the phase space albeit it is typically composed of a non-enumerable set of trajectories. Therefore the repeller typically forms a fractal in the phase space (Gaspard and Dorfman, 1995; Gaspard, 1998).

We set up a first passage problem of the Helfand moment by introducing absorbing boundaries at $\tilde{G}_{yx} = \pm\frac{\chi}{2}$. These absorbing boundaries in the space of variations of the Helfand moment correspond to equivalent absorbing boundaries in the phase space of the system. In the phase space, the absorbing boundaries delimit a domain which contains the fractal repeller. We consider a statistical ensemble of initial conditions taken inside this domain and we run their trajectories. When a trajectory reaches the absorbing boundaries it escapes out of the domain and is thus removed out of the statistical ensemble.

Under the forward time evolution, the remaining trajectories belong to the stable manifolds of the repeller. Under the backward time evolution, the remaining trajectories belong to the unstable manifolds of the repeller. Under both the forward and backward time evolutions, the remaining trajectories belong to the repeller itself which is the intersection of its stable and unstable manifolds (Gaspard and Dorfman, 1995). For a typical chaotic dynamics, almost all trajectories escape out of the domain after some time so that the repeller as well as its stable or unstable manifolds are fractal objects.

These fractals can be generated by allowing the escape of trajectories over a long but finite time interval. Over a finite time, there remains a sizable set of trajectories, which progressively reduces to the fractal as the time interval becomes longer and longer. The number of trajectories in the set decays exponentially with time, the decay being characterized by the escape rate $\gamma$.

The escape rate $\gamma$ can be evaluated by solving the problem of first passage of the Helfand moment at $\tilde{G}_{yx} = \pm\frac{\chi}{2}$ where absorbing boundaries are located. Indeed, the Einstein-Helfand equation (2.47) shows that the Helfand moment performs a diffusive-like random walk. Accordingly, the Helfand moment can be considered as a random variable $g = \tilde{G}_{yx}$ for which the probability density $p(g, t)$ obeys to a diffusion-type equation (Dorfman and Gaspard, 1995)

$$\frac{\partial p}{\partial t} = \eta \, \frac{\partial^2 p}{\partial g^2} \,, \tag{3.91}$$

where the role of the diffusion coefficient is played by the shear viscosity (2.47) itself. At the absorbing boundaries, the probability density must satisfy the absorbing boundary conditions

$$p\left(-\frac{\chi}{2}, t\right) = p\left(+\frac{\chi}{2}, t\right) = 0 \tag{3.92}$$

for all times $t$. The solution of the diffusion-type equation (3.91) with the boundary conditions (3.92) gives us a relation similar to (3.81):

$$\gamma_n = \eta \left(\frac{\pi n}{\chi}\right)^2 \,. \tag{3.93}$$

After a long time, the escape is dominated by the smallest decay rate, $\gamma_1$, which can therefore be identified with the escape rate $\gamma$. In this way, we obtain the escape rate for shear viscosity as a function of $\chi$

$$\gamma = \gamma_1^{(\eta)} = \eta \left(\frac{\pi}{\chi}\right)^2 \,. \tag{3.94}$$

This result is obtained by using the diffusion-type equation (3.91) which is expected to hold if the parameter $\chi$ must be sufficiently large so that the Helfand moment is in a diffusion regime and Eq. (3.91) holds.

The shear viscosity coefficient can thus be obtained from the escape rate which depends on the parameter $\chi$ of separation between the absorbing boundaries as

$$\eta = \lim_{\chi \to \infty} \left(\frac{\chi}{\pi}\right)^2 \gamma(\chi) \,. \tag{3.95}$$

At the microscopic level of description, the escape rate is controlled by the fractal repeller $\mathcal{F}_\chi$ which is composed of all the trajectories satisfying the condition (3.90) under forward and backward time evolutions. As for diffusion, the chaotic quantities like the Lyapunov exponents $\{\lambda_i\}$ and the KS entropy $h_{KS}$ are evaluated with respect to the natural invariant measure of the repeller $\mathcal{F}_\chi$. We thus recover Eq. (3.87) giving the microscopic escape rate in terms of these quantities

$$\gamma(\chi) = \left(\sum_{\lambda_i > 0} c_i \, \lambda_i \right)_{\mathcal{F}_\chi} \,. \tag{3.96}$$

Due to the same arguments as for diffusion, we identify the two escape rates and finally have

$$\eta = \lim_{\chi \to \infty} \left(\frac{\chi}{\pi}\right)^2 \left(\sum_{\lambda_i > 0} c_i \, \lambda_i \right)_{\mathcal{F}_\chi} \,. \tag{3.97}$$

The escape-rate formalism has already been successfully applied to the transport property of diffusion (Gaspard and Baras, 1995) as well as to reaction-diffusion processes (Claus and Gaspard, 2001; Claus et al., 2004). In the present work, we study the properties of viscosity (Viscardy and Gaspard, 2003b).

In the limit $\chi \to \infty$, the Lyapunov exponents reach their equilibrium values $\lambda_{i,\text{eq}}$, while the codimensions vanish typically as $c_i \sim \chi^{-2}$ if transport is normal. If we introduce the quantities

$$a_i \equiv \lim_{\chi \to \infty} \left(\frac{\chi}{\pi}\right)^2 c_i \Big|_{\mathcal{F}_\chi} \,, \tag{3.98}$$

Eq. (3.97) provides a decomposition of the viscosity coefficient on the spectrum of Lyapunov exponents such as

$$\eta = \sum_{\lambda_{i,\text{eq}} > 0} a_i \, \lambda_{i,\text{eq}} \,. \tag{3.99}$$

Typically, the escape is most important in the most unstable direction corresponding to the maximum Lyapunov exponent $\lambda_1$. Therefore, the repeller is more depleted in the most unstable direction and the corresponding partial dimension $d_1$ is lower than the further ones. This reasoning suggests that a

typical behavior is

$$\left(\sum_{\lambda_i>0} c_i \ \lambda_i\right)_{\mathcal{F}_\chi} \simeq (c_1 \ \lambda_1)_{\mathcal{F}_\chi} \ , \tag{3.100}$$

for $\chi \to \infty$ if the maximum Lyapunov exponent $\lambda_1$ is well defined.

In the limit $\chi \to \infty$, the Lyapunov exponents reach their equilibrium values $\lambda_{i,\mathrm{eq}}$, while the codimensions vanish typically as $c_i \sim \chi^{-2}$ if transport is normal. As we will see in chapter 4, this is precisely the case in two-degree-of-freedom systems such as the two-disk model where the chaos-transport formula reduces to Eq. (4.70)

$$\eta = \lim_{\chi\to\infty}\left(\frac{\chi}{\pi}\right)^2 (c_1 \ \lambda)_{\mathcal{F}_\chi} \ , \tag{3.101}$$

where $\lambda$ is the unique positive mean Lyapunov exponent and $c_1$ the corresponding codimension which should be understood as the partial information codimension of the unstable manifolds of the fractal repeller given in terms of the partial information dimension by Young's formula (Young, 1982)

$$c_1 = 1 - d_1 = 1 - \frac{h_{\mathrm{KS}}}{\lambda} \ . \tag{3.102}$$

However, when the number of degrees of freedom is larger, a new difficulty appears. Indeed, it requires the computation of the partial fractal dimensions associated with each positive Lyapunov exponents. Some works has been done on the evaluation of the dimension of fractal repellers (Hunt et al., 1996; Sweet and Ott, 2000). They might be used as a starting point for numerical studies of the escape-rate formalism in many-degree-of-freedom systems.

## 3.7 Hydrodynamic modes

Hydrodynamics describes the macroscopic dynamics of fluids in terms of equations governing the evolution of mass density, fluid velocity, and temperature, such as the Navier-Stokes equations (2.1) and the diffusion equation:

$$\frac{\partial n}{\partial t} = D\nabla^2 n \ . \tag{3.103}$$

Thanks to the kinetic equation developed by Boltzmann nonequilibrium statistical mechanics is able to derive these phenomenological equations, while the Boltzmann equation is itself derived from Liouvillian dynamics, using the *Stosszahlantsatz*. The solutions of Eq.(3.103) called *hydrodynamic*

*modes* are of the form:

$$n_{\mathbf{k}}(\mathbf{r}, t) = \exp(s_{\mathbf{k}}t)\exp(i\mathbf{k} \cdot \mathbf{r}) \,, \qquad (3.104)$$

each modes being characterized by a wavenumber $\mathbf{k}$. The hydrodynamic modes are spatially periodic of wavelength $\lambda = 2\pi/k$ with $k = \|\mathbf{k}\|$. They decay exponentially in time because the corresponding eigenvalues are real and negative

$$s_{\mathbf{k}} = -Dk^2 \,. \qquad (3.105)$$

The paradox between the reversible dynamics at the molecular level and the irreversible macroscopic phenomena that was exposed in section 1.2 can be studied in this context. Indeed the exponential decay of the modes (3.104) toward the thermodynamic equilibrium seems apparently incompatible with the Hamiltonian dynamics governing the motion of atoms and molecules, which is reversible and preserves phase-space volumes. Therefore, until recently the hydrodynamic modes were not described in terms of the Liouvillian dynamics, but at the intermediate level of the kinetic equations. Here, we briefly present a recent method based on the construction of hydrodynamic modes of diffusion in terms of the microscopic deterministic dynamics.

Let us consider the deterministic Liouvillian equation for Hamiltonian systems

$$\frac{\partial f}{\partial t} = \{H, f\} = \hat{L}f \,, \qquad (3.106)$$

and the solution of this equation is

$$f_t(\mathbf{X}) = \hat{P}^t f_0(\mathbf{X}) \qquad (3.107)$$

where $\hat{P}^t = \exp(\hat{L}t)$ is the so-called *Frobenius-Perron operator*.

Boundary conditions are required to solve the Liouville equation. This method does not use the simple periodic boundary conditions. Although we consider here $N$-particle systems periodically extended in position space and forming a lattice $\mathcal{L}^{N}$,[2] such as the Lorentz gas or the multibaker map, the probability density is allowed to extend nonperiodically over the whole lattice so that the periodic boundary conditions does not apply and have to be replaced by the so-called *quasiperiodic boundary conditions*. A Fourier transform must be carried out in position space to reduce the dynamics to the cell at the origin of the lattice ($\mathbf{l} = 0$). A wavenumber $\mathbf{k}$ is introduced which varies continuously in a

---

[2]If the time evolution of the coordinates $\mathbf{\Gamma} = (\mathbf{r}, \mathbf{p}) = (\mathbf{r_1}, ..., \mathbf{r}_N, \mathbf{p_1}, ..., \mathbf{p}_N)$ of the $N$ particles is governed by a first-order equations $\dot{\mathbf{\Gamma}} = \mathbf{F}(\mathbf{\Gamma})$, the vector field $\mathbf{F}$ is therefore symmetric under discrete position translations: $\mathbf{F}(\mathbf{r}, \mathbf{p}) = \mathbf{F}(\mathbf{r} + \mathbf{l}, \mathbf{p})$ with $\mathbf{l} \in \mathcal{L}^{N}$ .

Brillouin zone reciprocal to the lattice

$$f_{t,\mathbf{k}}(\mathbf{r}, \mathbf{p}) = \sum_{\mathbf{l} \in \mathcal{L}^N} \exp(-i\mathbf{k} \cdot \mathbf{l}) f_t(\mathbf{r} + \mathbf{l}, \mathbf{p}) \ . \tag{3.108}$$

In particular the hydrodynamic mode of wavenumber $\mathbf{k}$ is an eigenstate of the operator $\hat{T}_{\mathbf{l}}$ of translation by the lattice vector $\mathbf{l}$

$$\hat{T}_{\mathbf{l}} \Psi_{\mathbf{k}} = \exp(i\mathbf{k} \cdot \mathbf{l}) \Psi_{\mathbf{k}} \ . \tag{3.109}$$

The wavenumber $\mathbf{k}$ characterizes the spatial periodicity of the observables and of the probability densities. Each Fourier component of the probability density evolves differently in time, which requires the introduction of a new Frobenius-Perron operator $\hat{R}_{\mathbf{k}}$ depending explicitly on the wavenumber $\mathbf{k}$. As the operator of translation $\hat{T}_{\mathbf{l}}$ commutes with the Frobenius-Perron operator $R_{\mathbf{k}}$ we can find an eigenstate common to both the spatial translations and the time evolution

$$\hat{R}_{\mathbf{k}} \Psi_{\mathbf{k}} = \exp(s_{\mathbf{k}} t) \Psi_{\mathbf{k}} \ . \tag{3.110}$$

At vanishing wavenumber, we recover the dynamics with periodic boundary conditions which admits an invariant probability measure describing the microcanonical equilibrium state. In contrast, an invariant probability measure no longer exists as soon as the wavenumber is non-vanishing. Instead, we find a complex measure which decays at a rate given by the so-called *Pollicott-Ruelle resonance* $s_{\mathbf{k}}$ (Gaspard, 1998). This measure defines the hydrodynamic mode of wavenumber $\mathbf{k}$ and the associated Pollicott-Ruelle resonance $s_{\mathbf{k}}$ gives the dispersion relation of the hydrodynamic mode (3.105).

Such microscopic hydrodynamic modes present an important difference with the phenomenological hydrodynamics. This difference holds in the fact that they are mathematical distributions or singular measures. The impossibility of constructing eigenstates in terms of functions has its origin in the pointlike character of the deterministic dynamics and in the property of dynamical instability. Therefore, in order to have a representation of the eigenstates we have to consider its cumulative function

$$f_{\mathbf{k}}(\xi) = \int_0^\xi \Psi_{\mathbf{k}}(\mathbf{\Gamma}_{\xi'}) \, d\xi' \tag{3.111}$$

where $\mathbf{\Gamma}_{\xi'}$ is a curve of parameter $\xi$ in the phase space.

The singular property of the eigenstates plays a fundamental role in the understanding of irreversible processes. Indeed it leads to the result that the cumulative function $F_{\mathbf{k}}(\xi)$ is fractal. In the case of diffusion in the periodic Lorentz gas, thanks to the thermodynamic formalism, it has been shown that the diffusion coefficient $D$ is related to the Hausdorff dimension $d_H$ of the cumulative function $F_{\mathbf{k}}(\xi)$ (Gaspard et al., 2001) as

$$D = \lambda \lim_{\mathbf{k} \to 0} \frac{d_H(\mathbf{k}) - 1}{\mathbf{k}^2} \tag{3.112}$$

where $\lambda$ is the positive Lyapunov exponent. Let us mention that a similar study has been done in reactive-diffusion systems (Claus and Gaspard, 2002).

Such processes like diffusion and viscosity are typical irreversible phenomena. This irreversibility can be expressed by the production of entropy. It has been showed that the fractal structure of the diffusive modes plays a crucial role in the positivity of the entropy production in Hamiltonian systems (Gaspard, 1997; Gilbert et al., 2000). Indeed when the distribution functions are smooth, there is no change in the Gibbs entropy and no positive irreversible entropy production. The presence of the singular character therefore appears to be the fundamental element an understanding of the second law of thermodynamics in terms of fractals.

In nonequilibrium systems, stationary states can be obtained by imposing nonequilibrium constraints at the boundaries of an open system. In the case of diffusion, these constraints induce a flux of matter across the system, leading to an irreversible entropy production. An example of such a nonequilibrium steady state can be established in an open Lorentz gas between two particle reservoirs at phase-space densities $p_{\pm}$ separated by a distance $L$. The picture of such a system is similar to the one in Fig.3.3 where the phase-space density corresponding to the reservoir on the left-hand (resp. right-hand) side is $p_-$ (resp. $p_+$). The phase-space density inside the system can only take either the values $p_-$ or $p_+$, according to the reservoir from which the particle enters the system. Let us denote by

$$\mathbf{g} = \frac{p_+ - p_-}{L} \mathbf{e}_x \tag{3.113}$$

the gradient of phase-space concentration in the direction $\mathbf{e}_x$. We therefore can write the invariant density of the nonequilibrium steady state in the form

$$p_{ne}(\mathbf{\Gamma}) = \frac{p_+ + p_-}{2} + \mathbf{g} \cdot \left[ \mathbf{r}(\mathbf{\Gamma}) + \int_0^{T(\mathbf{\Gamma})} \mathbf{v}\left(\mathbf{\Phi}^t \mathbf{\Gamma}\right) dt \right]. \tag{3.114}$$

where $T(\mathbf{\Gamma})$ is the time of the entrance of the particle in the system. We then have $\mathbf{r}\left(\mathbf{\Phi}^{T(\mathbf{\Gamma})} \mathbf{\Gamma}\right) = \pm L/2$.

Finally, we end up with the result that

$$p_{ne}(\boldsymbol{\Gamma}) = p_\pm \ . \tag{3.115}$$

Here, the nonequilibrium situation is complementary to the one encountered with the absorbing boundary conditions introduced in the escape-rate formalism. Indeed, the dynamics takes place on the set of trajectories ingoing from outside the domain $\boldsymbol{\Gamma}$ which is complementary to the fractal repeller appearing in the escape-rate formalism: $\boldsymbol{\Gamma} \setminus \mathcal{F}$. The set of trajectories considered here is therefore also a fractal object.

In the limit where the reservoirs are separated by an arbitrarily large distance $L$, the time of entrance goes to infinity $T(\boldsymbol{\Gamma}) \to \infty$. If we keep constant the density gradient $\mathbf{g}$, we must substract the constant $\frac{p_+ + p_-}{2}$ so that we obtain[3]

$$\Psi_{\mathbf{g}}(\boldsymbol{\Gamma}) = \mathbf{g} \cdot \mathbf{r}(\boldsymbol{\Gamma}) + \int_0^{-\infty} \mathbf{g} \cdot \mathbf{v}\left(\boldsymbol{\Phi}^t \boldsymbol{\Gamma}\right) dt \ . \tag{3.116}$$

The first term describes a mean linear profile of density in the direction of the gradient $\mathbf{g}$. The term can also be derived from phenomenological diffusion equation with flux boundary conditions. However, the second term is new and is the singular part of the steady state, giving the fluctuations around the mean density.

The expression (3.116) is so simple that it suggests a generalization to the other transport processes like the viscosity (Gaspard, 1996). By supposing we have a gradient $\mathbf{g}$ corresponding to the transport coefficient $\alpha$ we can write

$$\Psi_{\mathbf{g}}^{(\alpha)}(\boldsymbol{\Gamma}) = \mathbf{g} \cdot \mathbf{G}^{(\alpha)}(\boldsymbol{\Gamma}) + \mathbf{g} \cdot \int_0^{-\infty} \mathbf{J}^{(\alpha)}\left(\boldsymbol{\Phi}^t \boldsymbol{\Gamma}\right) dt \tag{3.117}$$

where $\mathbf{G}^{(\alpha)}(\boldsymbol{\Gamma})$ is the associated Helfand moment and $\mathbf{J}^{(\alpha)} = d\mathbf{G}^{(\alpha)}/dt$ the associated microscopic current. The representation of the nonequilibrium steady state is a way to observe the fractal character of the modes.

---

[3]Indeed, taking the limit $L \to \infty$, $p_+ - p_- \to \infty$ and the constant $\frac{p_+ + p_-}{2}$ tends also to infinity.

# Chapter 4

# The two-disk model

## 4.1   Introduction

In this chapter, the method exposed in section 2.5.2 to obtain viscosity is applied to the hard-disk fluid. We study in detail the simple model composed of two hard disks in elastic collisions in a domain defined by p.b.c.. Due to the defocusing character of the disks, this model is chaotic. Bunimovich and Spohn (1996) (Bunimovich and Spohn, 1996) have demonstrated that the viscosity already exists in this system with only two particles. The model they studied is defined with p.b.c. in a square geometry. It presents a fluid and a solid phases which are separated by a phase transition. The problems presented by the model in a square geometry are that: (*i*) the viscosity exists only in the solid phase; (*ii*) the viscosity tensor which is of fourth order is *anisotropic* on a square lattice. In the present work, we solve these problems by considering a hexagonal geometry. Indeed, in the hexagonal geometry, the fourth-order viscosity tensor is isotropic and we can proof the existence of viscosity already in the fluid phase (Viscardy and Gaspard, 2003a).

Furthermore, we show that the values of the shear viscosity obtained by our Helfand-moment method are in good agreement with Enskog's theory, already for the fluid of two-hard disks. This shows that transport properties such as viscosity turn out to exist already in small systems at the scale of nanometers.

Before the study of viscosity itself, we present properties of the two-hard-disk model in the two geometries based on the square and hexagonal lattices.

On the other hand, we apply the escape-rate formalism to viscosity. Chaotic properties of the model are studied such as the positive Lyapunov exponent characterizing the instability of the dynamics. The fractal repeller induced by the escape process is depicted for both diffusion and viscosity processes and a comparison is done between both of them. We study the viscosity on the one hand in terms of the escape rate, and, on the other hand, in terms of the chaotic properties such as the Lyapunov exponent and the Hausdorff dimension of the fractal repeller in order to show how microscopic chaos controls the viscosity process (Viscardy and Gaspard, 2003b). Finally, an nonequilibrium steady state corresponding to a velocity gradient is studied and we depict its fractal character by the method of section 3.7.

## 4.2   Description of the two-hard-disk model

In this chapter, we consider a simple model which is composed of two hard disks in elastic col-lisions on a torus. Bunimovich and Spohn (1996) (Bunimovich and Spohn, 1996) have previously studied this model for a square geometry. By periodicity, the system extends to a two-dimensional lattice made of infinitely many images of the two disks. For p. b. c. on a square domain, the infinite images form a square lattice, in which each cell contains two disks (see Fig. 4.1b).

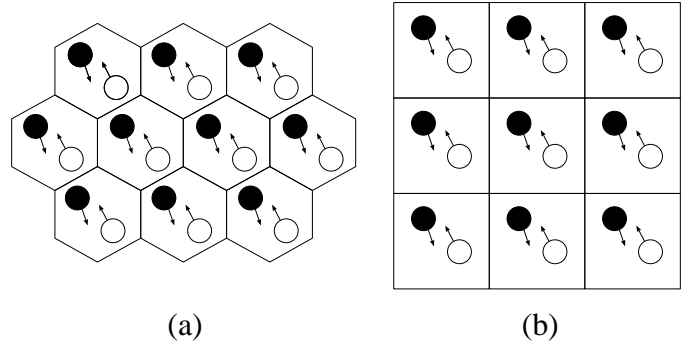

(a)            (b)

**Figure 4.1**. The model of two hard disks: (a) in the hexagonal geometry and (b) in the square geometry.

In the present work, we generalize this model to the hexagonal geometry (see Fig. 4.1a). The possibility of such a model was pointed out by Gaspard (Gaspard, 1998) (1998). The images of each disk now form a triangular lattice. The two disks (the white and the black ones) have the same diameter $\sigma$ and mass $m$. They follow different trajectories. All the black disks move together and all the white ones also move together. The system is periodic and the dynamics of the disks can be reduced to the dynamics in the unit cell or torus.

### 4.2.1 Hexagonal geometry

Let us first introduce some parameters of the system. $L$ is the distance between the centers of two neighboring cells. It also corresponds to the distance between two opposite boundaries of a cell.

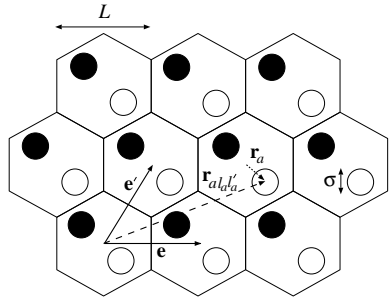

**Figure 4.2**. Basis vector ($\mathbf{e}$ and $\mathbf{e}'$), position vector $\mathbf{r}_a$ of particle $a$ in the cell and the position vector $\mathbf{r}_{a\,l_a\,l'_a}$ in the lattice.

By a linear combination of two vectors

$$\mathbf{e} = (L, 0) \,,$$
$$\mathbf{e}' = \left( \frac{1}{2}L, \frac{\sqrt{3}}{2}L \right) \,, \tag{4.1}$$

we can spot all the cells of the lattice and then localize the center of a disk thanks to

$$\mathbf{r}_{a\, l_a\, l'_a} = \mathbf{r}_a + l_a\, \mathbf{e} + l'_a\, \mathbf{e}' \,, \quad \text{for } a = 1, 2 \,, \tag{4.2}$$

where $l_a$ and $l'_a$ are integer, and $\mathbf{r}_a$ is the position vector of the disk $a$ with respect to the center of the cell (see Fig. 4.2). Therefore, the distance between the two disks is expressed by

$$\| \mathbf{r}_{1\, l_1\, l'_1} - \mathbf{r}_{2\, l_2\, l'_2} \| = \| \underbrace{\mathbf{r}_1 - \mathbf{r}_2}_{\mathbf{r}} + (l_1 - l_2)\, \mathbf{e} + (l'_1 - l'_2)\, \mathbf{e}' \| \,, \tag{4.3}$$

where $\mathbf{r} = \mathbf{r}_1 - \mathbf{r}_2$ is the relative position between both disks. By the minimum image convention, the relative distance $\|\mathbf{r}\|$ should take the smallest value among the infinitely many possible values. Of course, this distance has to be greater than or equal to the disk diameter ($\|\mathbf{r}\| = \|\mathbf{r}_1 - \mathbf{r}_2\| \geq \sigma$). As we have a hard-disk potential, the disks move in a free motion between each collision. Therefore, the equations of motion are written as

$$\frac{d\mathbf{r}_1}{dt} = \frac{\mathbf{p}_1}{m} + \sum_s \Delta\mathbf{r}_1^{(s)}\, \delta(t - t_s) \,,$$
$$\frac{d\mathbf{r}_2}{dt} = \frac{\mathbf{p}_2}{m} + \sum_s \Delta\mathbf{r}_2^{(s)}\, \delta(t - t_s) \,, \tag{4.4}$$

$$\frac{d\mathbf{p}_1}{dt} = \mathbf{F}_1 \,,$$
$$\frac{d\mathbf{p}_2}{dt} = \mathbf{F}_2 \,, \tag{4.5}$$

where $\mathbf{p}_1$ and $\mathbf{p}_2$ are the momenta of the two disks, $\mathbf{F}_1$ and $\mathbf{F}_2$ being the forces applied respectively to the disks 1 and 2. These forces equal zero when $\|\mathbf{r}_1 - \mathbf{r}_2\| > \sigma$ and are infinitely repulsive when $\|\mathbf{r}_1 - \mathbf{r}_2\| = \sigma$. $t_s$ denotes the time of the jump to satisfy the minimum image convention.

At this stage, we can do the following change of variables

$$\mathbf{r} = \mathbf{r}_1 - \mathbf{r}_2 \;,$$
$$\mathbf{R} = \frac{\mathbf{r}_1 + \mathbf{r}_2}{2} \;, \tag{4.6}$$

$$\mathbf{p} = \frac{\mathbf{p}_1 - \mathbf{p}_2}{2} \;,$$
$$\mathbf{P} = \mathbf{p}_1 + \mathbf{p}_2 \;. \tag{4.7}$$

If we introduce the reduced mass $\mu = \frac{m}{2}$, we can write

$$\mu \frac{d\mathbf{r}}{dt} = \mathbf{p} + \sum_s \mu \, \Delta\mathbf{r}^{(s)} \, \delta(t - t_s) = \mu \, \mathbf{v} + \sum_s \mu \, \Delta\mathbf{r}^{(s)} \, \delta(t - t_s) \;, \tag{4.8}$$

$$\frac{d\mathbf{p}}{dt} = \mathbf{F} = \mathbf{F}_1 = -\mathbf{F}_2 \;, \tag{4.9}$$

where $\mathbf{v}$ is the relative velocity and $\Delta\mathbf{r}^{(s)} = \Delta\mathbf{r}_1^{(s)} - \Delta\mathbf{r}_2^{(s)}$. Here we suppose that we are in the reference frame of the mass center (that is $\mathbf{P} = 0$). Accordingly, the energy of the system is reduced to

$$E = \frac{\mathbf{p}^2}{2\mu} \;. \tag{4.10}$$

The interest of this change of variables is to reduce the number of variables. Indeed, the only variables that remain are the relative position and velocity [$\mathbf{r} = (x, y)$ and $\mathbf{v} = (v_x, v_y)$]. We can associate a fictitious pointlike particle with these variables, which moves in a reduced system, known as the *periodic Sinai billiard* (see Fig. 4.3).

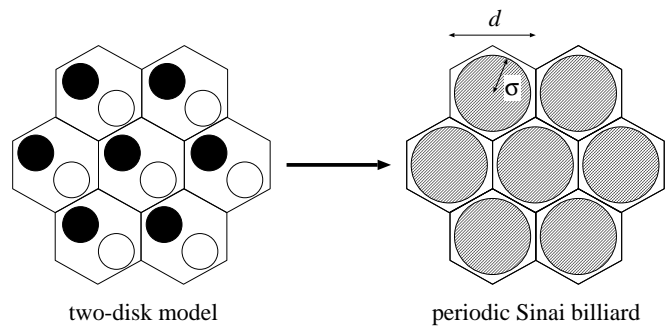

two-disk model          periodic Sinai billiard

**Figure 4.3**. The model of two hard disks in the hexagonal geometry is reduced to the periodic Sinai billiard thanks to a change of variables.

The billiard is also a triangular lattice of hexagonal cells. The size $d$ of these cells is equal to the size of the cells of the model itself ($d = L$). A hard disk is fixed on the center of each cell. Its radius is equal to the diameter $\sigma$ of the two moving disks.

The basis vectors of this lattice are the same as those of the original dynamics (4.4)-(4.5) if we replace $L$ by $d$, which gives us the possibility to spot a cell in the lattice thanks to the vector

$$\mathbf{r}_c = l_c \, \mathbf{e} + l'_c \, \mathbf{e'} \, , \tag{4.11}$$

where $l_c$ and $l'_c$ are integer.

In the Sinai billiard, the system is described by a trajectory in a four-dimensional phase space which are the Cartesian coordinates $(x, y, p_x, p_y)$, or the polar coordinates $(x, y, p_\theta, \theta)$. However, since the energy of the system is conserved, this space is reduced to the three-dimensional space of the variables $(x, y, \theta)$. Furthermore, in hard-ball systems, the topology of the trajectory is independent of the energy level. Therefore, we can study the system on an arbitrary energy level. This energy determines the temperature of the system and is equal to $E = (d/2)(N-1)k_B T = k_B T$ because we have only two degrees of freedom ($d = 2$, $N = 2$). Sinai and Bunimovich have demonstrated that the dynamics in such billiards is ergodic on each energy level (Bunimovich and Sinai, 1980a; Sinai, 1970b; Bunimovich and Sinai, 1980b).

### 4.2.2   Square geometry

The case of the square geometry is similar to the hexagonal one except that the basis vectors are here given by

$$\mathbf{e} = (L, 0) \, ,$$
$$\mathbf{e'} = (0, L) \, , \tag{4.12}$$

where $L$ is the length of a side of the square unit cell which contains two moving disks of diameter $\sigma$. We perform the same change of variables to reduce the dynamics of two hard disks to the one of the fictitious pointlike particle of a Sinai billiard in a square unit cell. Here also, the size $d$ of the cells of the Sinai billiard is the same as for the cells of the two hard disks model: $d = L$.

### 4.2.3   The different dynamical regimes of the model

The physical quantity determining the size of the cell in our model is the density which corre-two-disk model        periodic Sinai billiard sponds to the number of disks per unit volume or, in our case, the number of disks per unit area. Each cell contains two disks. Therefore, the density is $n = \frac{2}{V}$ where $V = \|\mathbf{e} \times \mathbf{e}'\|$ is the area of a cell. In our study, we have chosen that the diameter of the moving disks is equal to the unity: $\sigma = 1$.

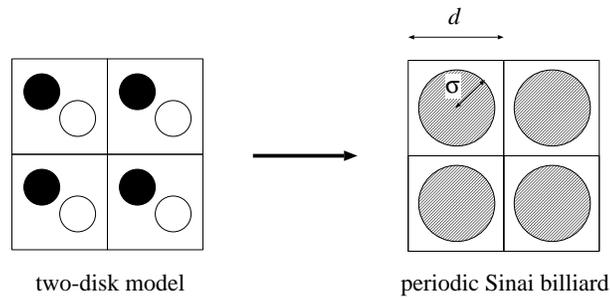

**Figure 4.4.** The model of two hard disks in the square geometry is reduced to the periodic Sinai billiard thanks to a change of variables.

As a function of the density, we observe different dynamical regimes. At low density, the disks are able to move in the whole lattice so that the disks are not localized in bounded phase-space regions. In this case, the billiard may have a finite or an infinite horizon depending on the geometry and on the density. In the opposite, at high density, the disks are so close to each other that they cannot travel across the system and we refer to this regime as the *localized regime*. The critical density between the nonlocalized and localized regimes corresponds to the situation where both disks have a double contact with each other in the configuration shown in Fig. 4.5.

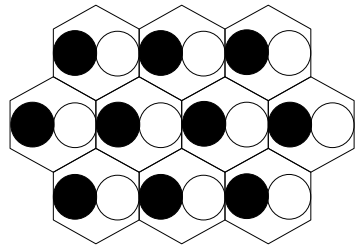

**Figure 4.5.** Hexagonal system at the critical density $n_{\mathrm{cr}}$.

**Hexagonal geometry**

In the hexagonal geometry, the area of the system is $V = \|\mathbf{e} \times \mathbf{e}'\| = \frac{\sqrt{3}}{2}L^2$ and the critical density is equal to

$$n_{\mathrm{cr}} = \frac{\sqrt{3}}{3} \simeq 0.5774 \ , \tag{4.13}$$

even though the maximum density (the *close-packing density*) is

$$n_{\max} = \frac{4\sqrt{3}}{9} \simeq 0.7698 \ . \tag{4.14}$$

At the close-packing density, the system forms a triangular crystal.

In the Sinai billiard, it is well known that there exists different kinds of regimes according to the dynamics of the particles. As a function of the density $n$, we observe three regimes :

- **The infinite-horizon regime:** At the low densities $0 < n < \frac{\sqrt{3}}{4}$, the particles can move in free flight over arbitrarily large distances. In this regime, the self-diffusion coefficient is infinite. (See Fig. 4.6.)

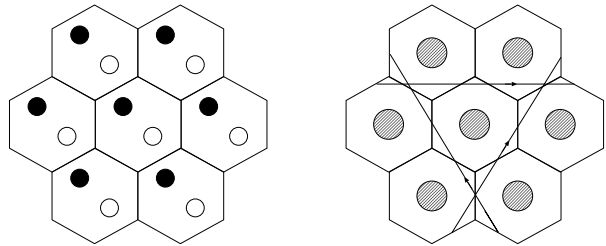

**Figure 4.6**. Typical configuration of the system in the infinite-horizon regime.

- **The finite-horizon regime:** For the intermediate densities $\frac{\sqrt{3}}{4} < n < n_{\mathrm{cr}}$, the free flights between the collisions are always bounded by a finite distance of the order of the interdisk distance $d$. Therefore, the horizon is finite and the self-diffusion coefficient is positive and finite. (See Fig. 4.7.)

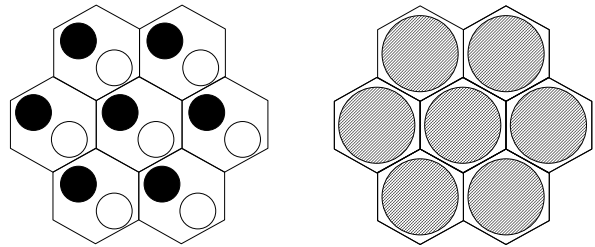

**Figure 4.7**. Typical configuration of the system in the finite-horizon regime.

- **The localized regime:** At the highest densities $n_{\mathrm{cr}} < n < n_{\max}$, the images of the disk overlap each other in the billiard so that the relative motion of the particles is localized in bounded regions. Therefore, the self-diffusion coefficient vanishes. (See Fig. 4.8.)

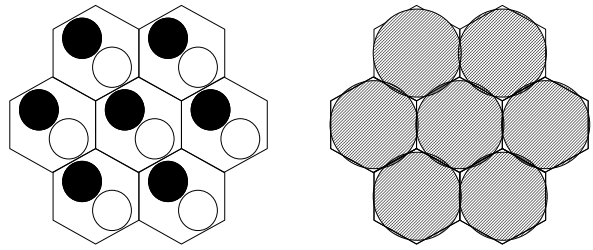

**Figure 4.8**. Typical configuration of the system in the localized regime.

We notice that Figs. 4.6, 4.7 and 4.8 are not depicted at the same scale since the disk diameter is fixed to unity ($\sigma = 1$) and it is the interdisk distance $d$ that varies.

The infinite- and finite-horizon regimes extend over the densities $0 < n < n_{\text{cr}}$. The localized regime corresponds to the densities $n_{\text{cr}} < n < n_{\text{max}}$. The following figure 4.9 shows the different regimes in the hexagonal geometry. The remarkable feature of the hexagonal geometry is that there exists a finite-horizon regime which is not localized, in contrast to the square geometry (see below).

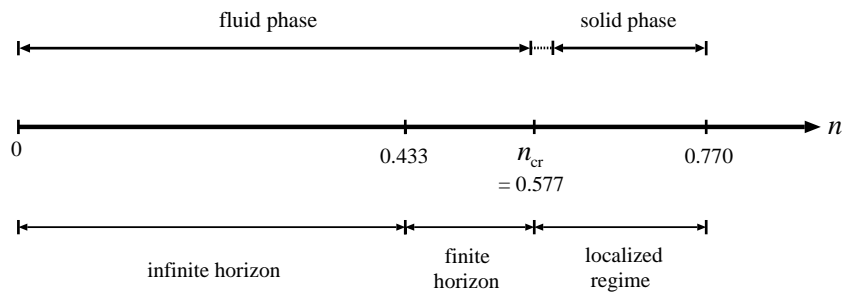

**Figure 4.9**. The different dynamical regimes and thermodynamic phases of the model in the hexagonal geometry versus the density $n$.

### Square geometry

In the square geometry, the volume is $V = \|\mathbf{e} \times \mathbf{e}'\| = L^2$ and the critical density is

$$n_{\text{cr}} = 0.5 \ , \tag{4.15}$$

which is the density of the transition between the infinite-horizon and the localized regimes. The close-packing density is equal to

$$n_{\text{max}} = 1 \ . \tag{4.16}$$

In Fig. 4.10, we have depicted the different regimes in the square geometry. In the square geometry, there also exist nonlocalized and localized regimes, but the horizon is always infinite in the nonlocalized regime. Therefore, it is only in the localized regime that the horizon is finite in the square geometry. This is an important difference with respect to the hexagonal geometry.

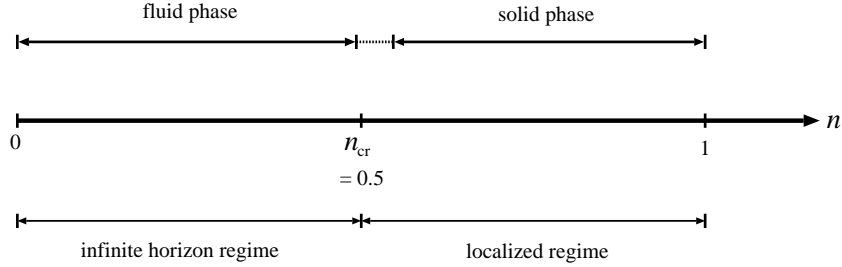

**Figure 4.10**. The different dynamical regimes and thermodynamic phases of the model in the square geometry versus the density $n$.

## 4.3    Properties of the model

### 4.3.1    Mean free path

The *mean free path* $\langle l \rangle$ is the average distance between two successive collisions. It is known that, in two-dimensional billiards, the mean free path is related to the area $\mathcal{A}$ of the billiard and its perimeter $\mathcal{L}$ according to (Machta and Zwanzig, 1983)

$$\langle l \rangle = \frac{\pi \mathcal{A}}{\mathcal{L}} \ . \tag{4.17}$$

In the different regimes, the mean free path is given by

- hexagonal geometry:

    *(i)* $\langle l \rangle = \frac{1}{n} - \frac{\pi}{2}$ ,        $n \leq n_{\mathrm{cr}}$ ,

    *(ii)* $\langle l \rangle = \pi \dfrac{\frac{2}{n} - \pi + 6 \arccos(\frac{1}{\sqrt{\sqrt{3}\,n}}) - \frac{6}{\sqrt{\sqrt{3}\,n}} \sqrt{1 - \frac{1}{\sqrt{3}\,n}}}{2\pi - 12 \arccos(\frac{1}{\sqrt{\sqrt{3}\,n}})}$ ,        $n \geq n_{\mathrm{cr}}$ ;

- square geometry:

    *(i)* $\langle l \rangle = \frac{1}{n} - \frac{\pi}{2}$ ,        $n \leq n_{\mathrm{cr}}$ ,

(*ii*) $\langle l \rangle = \pi \dfrac{\frac{2}{n} - \pi + 4 \arccos(-\frac{1}{\sqrt{2}\,n}) - 2\sqrt{\frac{2}{n}}\,\sqrt{1 - \frac{1}{2\,n}}}{2\,\pi - 8\arccos(-\frac{1}{\sqrt{2}\,n})}$ , $\qquad n \geq n_{\mathrm{cr}}$ .

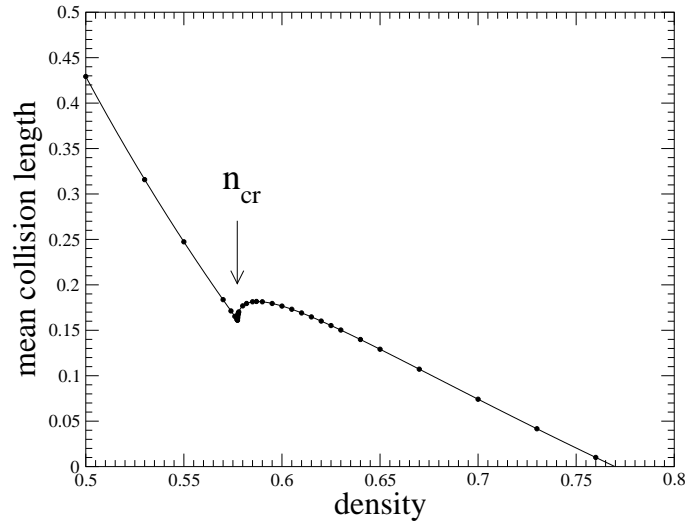

**Figure 4.11**. Theoretical (continuous line) and numerical (dots) values of the mean free path versus the density $n$ in the hexagonal geometry.

We show in Figs. 4.11 and 4.12 the excellent agreement between the above expressions and the values obtained by numerical simulations. The break observed in Figs. 4.11 and 4.12 between the nonlocalized and localized regimes can be explained thanks to Eq. (4.17). Indeed, at the critical density $n_{\mathrm{cr}}$, the disks form a horn. Above criticality, the horn becomes a corner with a finite angle so that the perimeter $\mathcal{L}$ decreases very fast. But, on the other hand, the area $\mathcal{A}$ remains relatively constant. Therefore the ratio $\frac{\mathcal{A}}{\mathcal{L}}$ increases with $n$ until this effect disappears. At higher densities, the mean free path decreases again.

### 4.3.2   Pressure and the different phases of the model

The hydrostatic pressure allows us to interpret the different regimes in terms of thermodynamic phases. The pressure can be calculated in terms of the time average of the Helfand moment as shown in Appendix C. In the two-disk model with $N = 2$ and $d = 2$, the pressure is given by

$$PV = k_{\mathrm{B}}T + R ,\qquad\qquad (4.18)$$

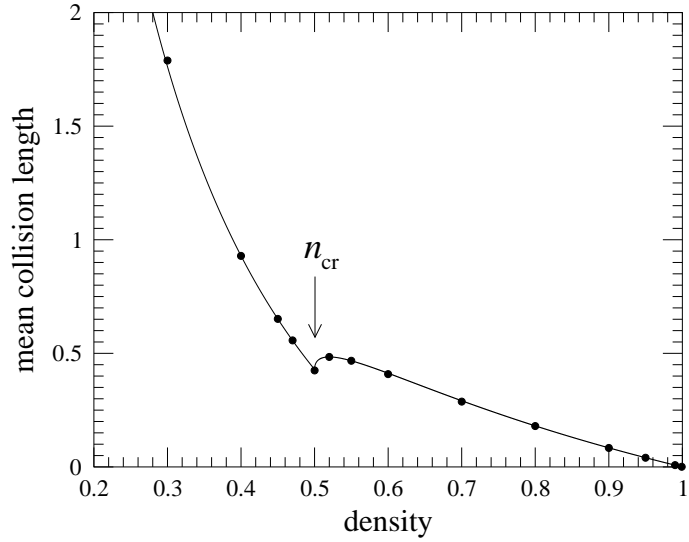

**Figure 4.12**. Theoretical (continuous line) and numerical (dots) values of the mean free path versus the density $n$ in the square geometry.

where the rest can be calculated according to Eq. (C.8) as

$$R = \frac{\langle \Delta \mathbf{p}_1^{(c)} \cdot \mathbf{r}_{12}^{(c)} \rangle}{4 \langle \Delta t_{c-1,c} \rangle} \, , \tag{4.19}$$

where $\langle \Delta t_{c-1,c} \rangle$ is the mean intercollisional time. If we denote by $\phi^{(c)}$ the angle between the velocity at collision and the normal to the disk of the Sinai billiard, the average in the numerator becomes

$$\langle \Delta \mathbf{p}_1^{(c)} \cdot \mathbf{r}_{12}^{(c)} \rangle = m \, v \, \sigma \, \langle \cos \phi^{(c)} \rangle \, , \tag{4.20}$$

$\sigma$ being the diameter of the disks. In the case the total momentum vanishes, the velocity $\mathbf{v}$ of the trajectory in the billiard is related to the relative momentum $\mathbf{p}$, the energy, and the temperature by

$$E = k_\mathrm{B} T = \frac{\mathbf{p}^2}{2\mu} = \frac{\mathbf{p}^2}{m} = \frac{\mu \mathbf{v}^2}{2} = \frac{m \mathbf{v}^2}{4} \, , \tag{4.21}$$

so that $\mathbf{v} = 2\mathbf{p}/m$. At collision, $\sin \phi^{(c)}$ is uniformly distributed in the interval $[-1, +1]$ so that

$$\langle \cos \phi^{(c)} \rangle = \frac{\pi}{4} \, . \tag{4.22}$$

On the other hand, the mean intercollisional time of the billiard is related to the mean free path $\langle l \rangle$ and the speed $v = \|\mathbf{v}\|$ by

$$\langle \Delta t_{c-1,c} \rangle = \frac{\langle l \rangle}{v} \ . \tag{4.23}$$

Gathering the results, we obtain the rest as

$$R = \frac{\pi \, \sigma \, m \, v^2}{16 \, \langle l \rangle} = \frac{\pi \, \sigma}{4 \, \langle l \rangle} \, k_{\mathrm{B}} T \ . \tag{4.24}$$

Accordingly, the hydrostatic pressure of the model is given by

$$PV = k_{\mathrm{B}} T \left( 1 + \frac{\pi \sigma}{4 \langle l \rangle} \right) = k_{\mathrm{B}} T \left( 1 + \frac{\sigma \mathcal{L}}{4 \mathcal{A}} \right) \ . \tag{4.25}$$

In our work, we introduce the reduced pressure defined as

$$P^* \equiv \beta P \, \frac{V}{N} \quad = \quad \frac{PV}{(N-1)k_{\mathrm{B}} T} \tag{4.26}$$

$$= \quad 1 + \frac{\pi \sigma}{4 \langle l \rangle} = 1 + \frac{\sigma \mathcal{L}}{4 \mathcal{A}} \ . \tag{4.27}$$

In Figs. 4.13 and 4.14, the reduced pressure is depicted as a function of the density and we observe the manifestation of a phase transition around the critical density. The hard-ball systems are known to present a fluid-solid phase transition that we here already observe in the two-disk model.

At low density, the fictitious particle of the Sinai billiard can diffuse in the whole lattice. This means that the two disks move over arbitrarily large distances one with respect to the other, which is a feature of a fluid phase. In contrast, at high density, the fictitious particle is trapped between three (or four) disks and its motion is reminiscent of the vibration of atoms in a solid. Of course, it is not really a vibration since the disks bounce in a chaotic motion because of the elastic collisions whereas, in a solid, the atoms have quasi-harmonic oscillations around their equilibrium position. Nevertheless, we are in the presence of a solid phase because the translational invariance is broken. Indeed, the motion is no longer ergodic because the motion now is confined into one among several phase-space domains of the energy shell.

A phase transition occurs between the fluid and solid phases. At the critical density $n_{\mathrm{cr}}$, the pressure has a maximum. Above $n_{\mathrm{cr}}$, the pressure decreases, reaches a minimum at a value $n'_{\mathrm{cr}} > n_{\mathrm{cr}}$, before increasing again. For $n_{\mathrm{cr}} < n < n'_{\mathrm{cr}}$, the compressibility would be negative so that this state would be unstable from a thermodynamic viewpoint. This suggests a Maxwell con-struction to determine a fluid-solid coexistence in the interval of densities $n_F < n < n_S$ with $n_F < n_{cr}$ and $n'_{cr} < n_S$. The values which would delimit this small coexistence interval in a thermodynamic interpretation of the transition would be given by

- hexagonal geometry:   $n_F = 0.57 \pm 0.01$ ,

$\qquad\qquad\qquad\qquad n_S = 0.60 \pm 0.01$ ,                                  (4.28)

and

- square geometry:   $n_F = 0.49 \pm 0.01$ ,

$\qquad\qquad\qquad\quad n_S = 0.55 \pm 0.01$ ,                                  (4.29)

(see Figs. 4.9 and 4.10). In the square geometry, the horizon is infinite in the fluid phase. In the hexagonal geometry, the horizon may also be finite in the fluid phase, which leads to finite viscosity coefficients in the fluid phase of this model as shown in the following.

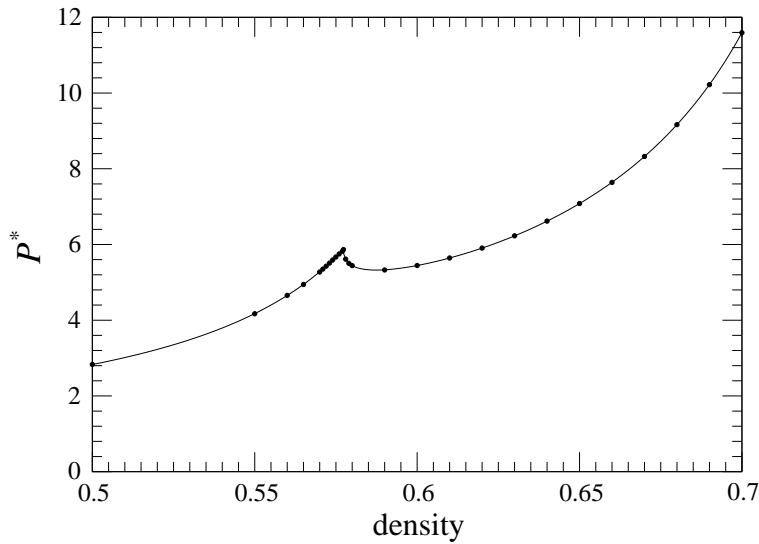

**Figure 4.13**. Theoretical (continuous line) and numerical (dots) values of the reduced pressure $P^*$ versus the density $n$ in the hexagonal geometry.

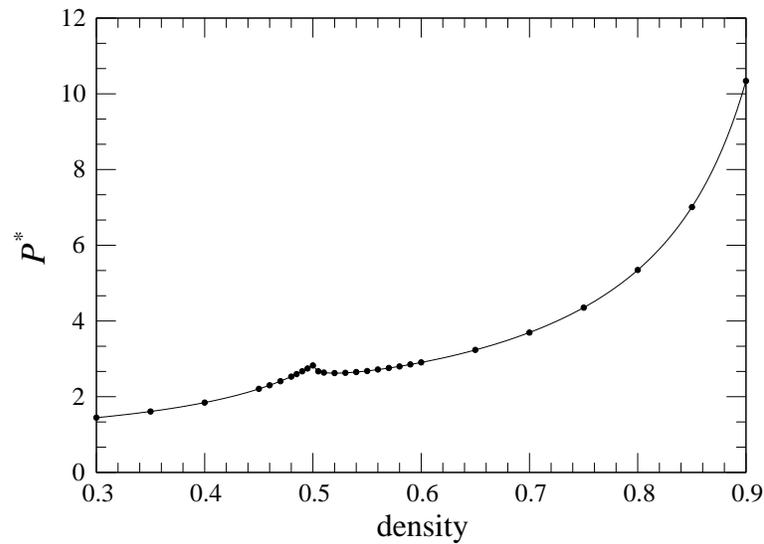

**Figure 4.14**. Theoretical (continuous line) and numerical (dots) values of the reduced pressure $P^*$ versus the density $n$ in the square geometry.

### 4.3.3 Comparison with the pressure in a square box

A comparison between the equation of state in the periodic square system and the one in the square box can give a better understanding of the phase transition. The square box system with hard walls containing two hard disks also presents a phase transition at the density $n = 0.5$. At this density, the disks are trapped in opposite corners of the box, unlike in the fluid phase in which the disks may go from one side to the other one of the square box, as showed in Fig.4.15. Contrary to the periodic system the disks do not cross the boundaries are not reinjected through the opposite boundary.

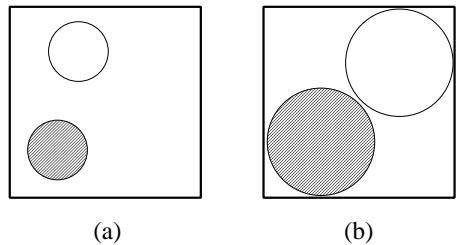

**Figure 4.15**. Two hard disks in a square box. (a) Typical configuration in the fluid phase. The disks may travel in the whole box. (b) Typical configuration in the solid phase in which the disks are constrained to move in opposite corners of the box.

The pressure in such systems have been previously studied by Speedy (1994) and Awazu (2001). We also evaluate this pressure in order to compare with the one in the periodic two-hard-disk system (see Fig. 4.16).

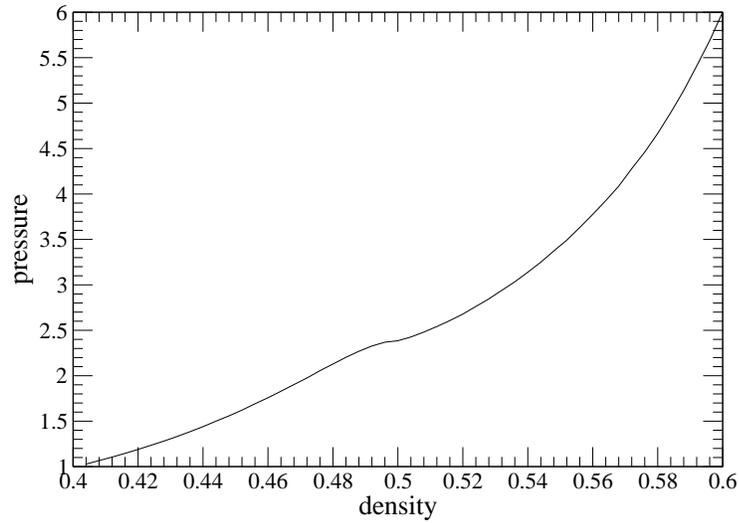

**Figure 4.16**. Pressure versus the density in a square box with hard walls.

Contrary to the equation of state in its periodic analogue, the pressure in the square box is monotonously increasing and only exhibits a little plateau at the critical density. This result leads us to think that the van-der-Waals-type curve obtained for periodic systems in Figs. 4.9 and 4.10 has an origin which is specific to the dynamics of the model.

## 4.4    Viscosity in the two-hard-disk model

### 4.4.1    Symmetry considerations in two-dimensional systems

By symmetry arguments, we can show that most of the elements of the viscosity tensor are either equal or vanish. First, we have

$$\eta_{ij,kl} = \eta_{kl,ij} = \eta_{ji,kl} = \eta_{ij,lk} \; , \tag{4.30}$$

because of the stationarity of the equilibrium average, the reversibility of the microscopic equations, and the fact that $\mathbf{F}(\mathbf{r}_a - \mathbf{r}_b) = \mathbf{F}(\|\mathbf{r}_a - \mathbf{r}_b\|)$ is a central force. Secondly, in our work, the fluid is invariant under rotations by $\varphi = \frac{\pi}{3}$ for the hexagonal geometry and by $\varphi = \frac{\pi}{2}$ for the square one. If we define the viscosity tensor as a linear operator $\hat{\eta}$ acting on matrices A according to $(\hat{\eta}\mathsf{A})_{ij} = \eta_{ij,kl} \; A_{kl}$, our discrete symmetry can be written as

$$\hat{\eta}(\mathsf{R}^{-1}\mathsf{A}\mathsf{R}) = \mathsf{R}^{-1}(\hat{\eta}\mathsf{A})\mathsf{R} \; , \tag{4.31}$$

for all matrices A. R is the rotation matrix

$$R = \begin{pmatrix} \cos\varphi & -\sin\varphi \\ \sin\varphi & \cos\varphi \end{pmatrix} , \tag{4.32}$$

and $\varphi$ is equal to $\frac{\pi}{3}$ or $\frac{\pi}{2}$ respectively for the hexagonal or square systems. Thanks to this symmetry, the only nonvanishing elements are $\eta_{ij,ij} = \eta_{ji,ij}$ and $\eta_{ii,ii} = \eta_{jj,jj}$. Furthermore, for $i \neq j$, $k \neq l$,

$$\eta_{ij,ij} = \eta_{kl,kl} , \qquad \eta_{ii,ii} = \eta_{jj,jj} , \qquad \eta_{ii,jj} = \eta_{kk,ll} . \tag{4.33}$$

Hence, there are in fact only three independent elements: $\eta_{xx,xx}, \eta_{xy,xy}, \eta_{xx,yy}$. On the other hand, for an isotropic system, we can see that

$$\eta = \eta_{xy,xy} , \tag{4.34}$$
$$\zeta = \frac{1}{2} \left( \eta_{xx,xx} + \eta_{xx,yy} \right) . \tag{4.35}$$

The third element $\eta_{xx,yy}$ is in fact a combination of the two other elements

$$\eta_{xx,yy} = \eta_{xx,xx} - 2\,\eta_{xy,xy} . \tag{4.36}$$

### 4.4.2 The Helfand moment in the two-hard-disk model

In our model defined with Eqs. (4.6) and (4.7), the fictitious particle moving in the Sinai billiard always satisfies the minimum image convention

$$|r_i| = |r_{1i} - r_{2i}| \leq \frac{L}{2} \quad \forall i = 1, \dots, d . \tag{4.37}$$

since the particle jumps when $|r_i = \frac{L}{2}|$. Therefore, there is no need to consider a pair composed of one disk with an image of the other one in the two-disk model, and the vector $\mathbf{L}_{b|a}$ (see Eq. (2.45) ) always vanishes. In consequence, the last term in Eq. (2.67) of the Helfand moment for periodic systems proposed in section 2.5.2 does not contribute in the two-hard-disk model studied in this chapter.

In the two-hard-disk model with a vanishing total momentum $\mathbf{P} = \mathbf{0}$, the forces obey $\mathbf{F}_1 = -\mathbf{F}_2 =$

**F** and the microscopic current can be written in relative coordinates as

$$J_{ij} = 2\,\frac{p_i p_j}{m} + F_i\,r_j\,, \tag{4.38}$$

where **r** is the smallest distance between the disks 1 and 2. Following the minimum image convention, the position vector presents discontinuities because of the passages of the relative position through a boundary, after which it is reinjected into the cell at the opposite boundary. We denote the vectors normal to the boundaries of the unit cell by

$$\text{hexagonal geometry}: \begin{cases} \mathbf{c}_1 = \mathbf{a}\,, \\ \mathbf{c}_2 = -\mathbf{a}\,, \\ \mathbf{c}_3 = \mathbf{b}\,, \\ \mathbf{c}_4 = -\mathbf{b}\,, \\ \mathbf{c}_5 = \mathbf{b} - \mathbf{a}\,, \\ \mathbf{c}_6 = \mathbf{a} - \mathbf{b}\,, \end{cases} \tag{4.39}$$

and

$$\text{square geometry}: \begin{cases} \mathbf{c}_1 = \mathbf{a}\,, \\ \mathbf{c}_2 = -\mathbf{a}\,, \\ \mathbf{c}_3 = \mathbf{b}\,, \\ \mathbf{c}_4 = -\mathbf{b}\,. \end{cases} \tag{4.40}$$

In order to satisfy the minimum image convention, the relative position undergoes jumps by vectors which are the vectors normal to the unit cell so that $\Delta \mathbf{r}^{(s)} = -\mathbf{c}_{\omega_s}$ where $\omega_s$ denotes the label of the boundary crossed by the particle at the $s^{\text{th}}$ passage at time $t_s$. In these notations, Hamilton's equations take the form

$$\frac{d\mathbf{r}}{dt} = \frac{2\mathbf{p}}{m} - \sum_s \mathbf{c}_{\omega_s}\,\delta(t - t_s)\,, \tag{4.41}$$

$$\frac{d\mathbf{p}}{dt} = \mathbf{F}\,. \tag{4.42}$$

In this periodic system, the expression for the Helfand moment is given by a reasoning similar to the one leading to Eq. (2.67). We obtain

$$G_{ij}(t) = p_i(t)\, r_j(t) + \sum_s p_i(t_s)\, c_{\omega_s j}\, \theta(t - t_s) \,.$$

(4.43)

Finally, the viscosity coefficients have the expressions

$$\eta_{ij,kl} = \lim_{t \to \infty} \frac{\beta}{2tV} \left( \left\langle \sum_{t_s < t} p_i(t_s)\, c_{\omega_s j} \sum_{t_{s'} < t} p_k(t_{s'})\, c_{\omega_{s'} l} \right\rangle - \left\langle \sum_{t_s < t} p_i(t_s)\, c_{\omega_s j} \right\rangle \left\langle \sum_{t_{s'} < t} p_k(t_{s'})\, c_{\omega_{s'} l} \right\rangle \right) \,.$$

(4.44)

Let us remark that the terms $p_i(t)\, r_j(t)$ do not appear in this relation because they do not contribute to the viscosity coefficients. Indeed, the relative position $\mathbf{r}(t)$ and momentum $\mathbf{p}(t)$ remain bounded in the course of time and their contribution disappears in the limit $t \to \infty$.

In the following, the numerical results are presented in terms of a reduced viscosity tensor which is defined by

$$\eta^*_{ij,kl} \equiv \frac{\eta_{ij,kl}}{2\sqrt{mk_{\mathrm{B}}T}} \,.$$

(4.45)

### 4.4.3   Hexagonal geometry

In the hexagonal geometry the fourth-order tensor of viscosity is isotropic. Indeed, since the system is invariant under rotations by $\frac{\pi}{3}$, we obtain the relation $\eta_{xx,yy} = \eta_{xx,xx} - 2\,\eta_{xy,xy}$ which implies the full rotational invariance of the viscosity tensor. We depict in Figs. 4.17 and 4.18 the results obtained for the reduced viscosities ($\eta^*$, $\zeta^*$) and the relation (4.36) is checked in Fig. 4.19.

In the infinite-horizon regime, the trajectory can present arbitrarily large displacements in the system without undergoing any collision. Accordingly, the variance of the Helfand moment $G_{yx}$ increases faster than linearly as $t\,\log t$, which implies an infinite viscosity coefficient after averaging over an infinite time interval. However, the factor $\log t$ generates a so-weak growth that it does not manifest itself much over the finite time of the simulation. This is the reason why we obtain finite values for the viscosity coefficients in Figs. 4.17 and 4.18. However, these values are only indicative since they should be infinite, strictly speaking (see below).

On the other hand, in the finite-horizon regime, the variance of the Helfand moment has a strictly linear increase in time and the viscosity coefficients are finite and positive. This is the result of a central-limit theorem which holds in the finite-horizon regime of the hexagonal geometry, as can be proved by considerations similar to those developed by Bunimovich and Spohn (1996). We observe in Fig. 4.17 that the viscosity has a diverging singularity at the critical density ($n_{cr} = \frac{\sqrt{3}}{3}$) which corresponds to the fluid-solid phase transition. We shall explain below the origin of this singularity.

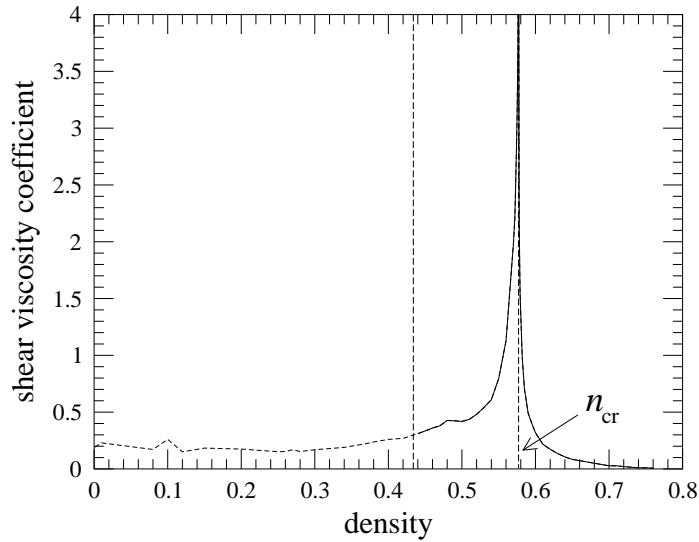

**Figure 4.17**. Shear viscosity coefficient $\eta^*$ versus the density in the hexagonal geometry. The part in dashed line corresponds to the density in which the coefficient would not exist in the limit $t \to \infty$ because the horizon is infinite. The long-dashed vertical lines separate the different regimes: on the left-hand side, the horizon-infinite regime (fluid phase); at the center, the horizon-finite regime (fluid phase); and on the right-hand side, localized regime (solid phase).

Finally, in the localized regime corresponding to the solid phase, the viscosity is finite and positive, and decreases when the density increases until the maximum density.

The results of our Helfand-moment method are compared with Enskog's theory. For a fluid of hard disks of mass $m$ and diameter $\sigma$, Enskog's theory predicts that the shear viscosity is given by (Gass, 1971)

$$\eta = \eta_B \left( \frac{1}{Y} + 2\,y + 3.4916\,Y\,y^2 \right) ,$$

(4.46)

where

$$\eta_B = \frac{1.022}{2\,\sigma} \sqrt{\frac{m\,k_B T}{\pi}} \ .$$

(4.47)

is the Boltzmann value of the shear viscosity, $Y$ is the Enskog factor entering the equation of state as follows

$$P = nk_B T(1 + 2\,y\,Y) .$$

(4.48)

and $y = \pi \sigma^2 n/4$. For the hard-disk fluid, a good approximation of the Enskog factor is given below

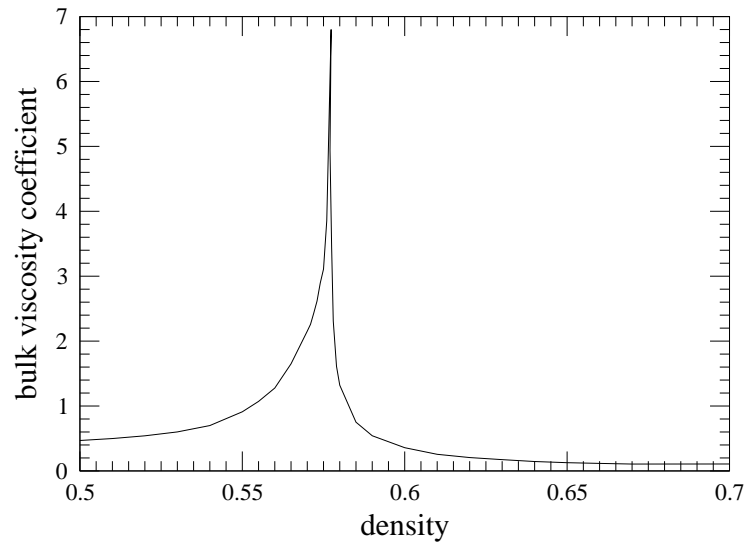

**Figure 4.18**. Bulk viscosity coefficient $\zeta^*$ versus the density in the hexagonal geometry.

the fluid-solid transition by (Barker and Henderson, 1976)

$$Y = \frac{1 - \frac{7}{16}\, y}{(1 - y)^2}\,. \tag{4.49}$$

It is known that the Enskog approximation is not good around the fluid-solid transition and at very high densities.

A remark is here in order. It is known (Alder and Wainwright, 1969) that the viscosity coefficient of the infinite hard-disk fluid is diverging because of long-time tails. However, this divergence is only logarithmic and does not manifest itself in numerical calculations before extremely long times. This explains why the long-time tails do not spoil the agreement between the numerical values and Enskog's theory.

We see in Fig. 4.20 the good agreement between Enskog's theory and the numerical values of our Helfand-moment method at low densities showing the consistency of our results. At high densities, the Enskog's predictions fail because of the hypothesis of low density of this theory. Indeed at high densities, collisions between more than 2 particles are to be taken into account since the probability of such events increases as the density increases.

### 4.4.4 Square geometry

In the square geometry, the fourth-order viscosity tensor is not isotropic. Indeed, the tensor is

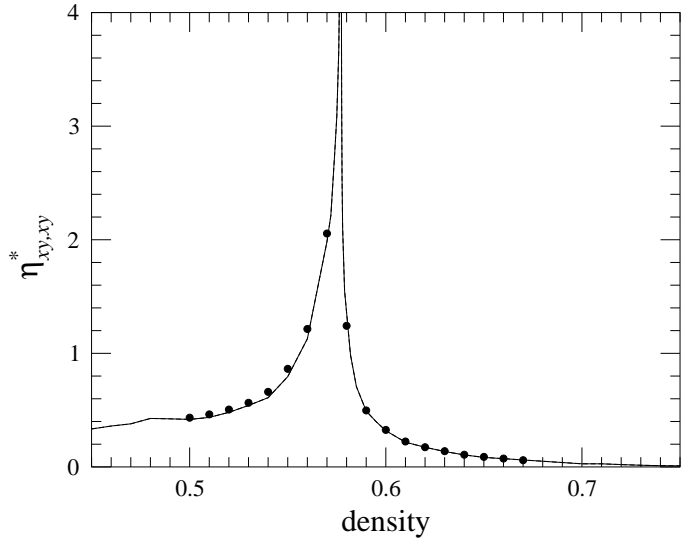

**Figure 4.19**. Tensor element $\eta^*_{xy,xy}$ of shear viscosity versus the density in the hexagonal geometry. The dots represent the results of the relation (4.36): $\eta^*_{xy,xy} = \frac{1}{2}\left(\eta^*_{xx,xx} - \eta^*_{xx,yy}\right)$. The continuous line corresponds to the data of Fig. 4.17.

transformed by the matrix $R_{ij}(\varphi)$ of rotation by an angle $\varphi$ into

$$\eta_{ij,kl}(\varphi) = R_{ii'}(\varphi)\, R_{jj'}(\varphi)\, R_{kk'}(\varphi)\, R_{ll'}(\varphi)\, \eta_{i'j',k'l'}(0) \;. \tag{4.50}$$

For example, if $\varphi = \frac{\pi}{4}$, we have

$$
\begin{aligned}
\eta_{xx,xx}(\tfrac{\pi}{4}) &= \tfrac{1}{2}\left[\eta_{xx,xx}(0) + \eta_{xx,yy}(0) + 2\,\eta_{xy,xy}(0)\right]\;, \\
\eta_{xy,xy}(\tfrac{\pi}{4}) &= \tfrac{1}{2}\left[\eta_{xx,xx}(0) - \eta_{xx,yy}(0)\right]\;, \\
\eta_{xx,yy}(\tfrac{\pi}{4}) &= \tfrac{1}{2}\left[\eta_{xx,xx}(0) + \eta_{xx,yy}(0)\right] - \eta_{xy,xy}(0)\;.
\end{aligned}
\tag{4.51}
$$

Since the system is not isotropic, one more viscosity tensor element is required. For anisotropic systems, the conditions of non-negativity (2.21) no longer apply. Let us consider again the entropy balance equation (2.7)

$$\frac{\partial(\rho s)}{\partial t} + \frac{\partial}{\partial r_j}(\rho s v_j) = \frac{1}{T^2}\kappa\,(\nabla T)^2 + \frac{1}{T}\sigma'_{ij}\frac{\partial v_i}{\partial r_j} \geq 0 \;. \tag{4.52}$$

It implies that

$$\sigma'_{il}\frac{\partial v_i}{\partial r_j} = \eta_{ij,kl}\,\frac{\partial v_k}{\partial r_l}\frac{\partial v_i}{\partial r_j} \geq 0 \;. \tag{4.53}$$

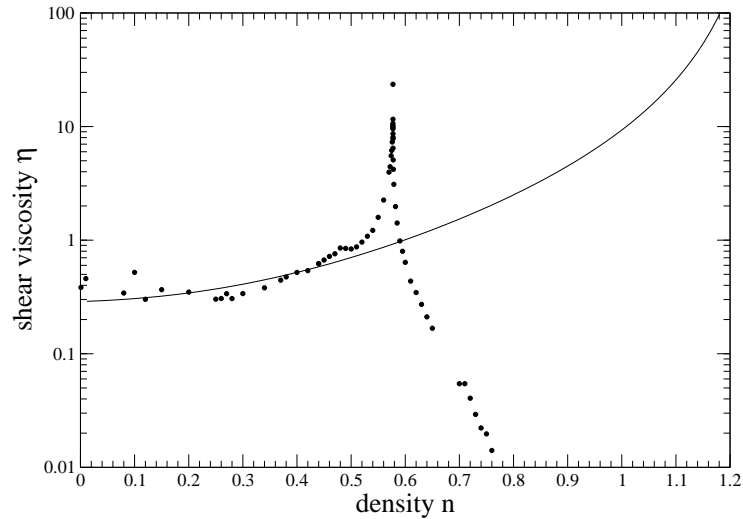

**Figure 4.20**. Comparison between the shear viscosity in the two-hard-disk model(•) and the viscosity obtained by the Enskog's theory (continuous line). The temperature is equal to $T = 1$ and the hard disks are of unit mass and diameter. The data for the two-hard-disk model are the same as in Fig. 4.17 except that we here plot $\eta = 2\eta^*$ instead of $\eta^*$ as in Fig. 4.17.

First, we consider the product $\eta_{ij,kl}\frac{\partial v_k}{\partial r_l}$ by taking into account the relations between the tensor elements obtained above

$$\eta_{ij,kl}\frac{\partial v_k}{\partial r_l} = \begin{pmatrix} \eta_{xx,xx}\frac{\partial v_x}{\partial x} + \eta_{xx,yy}\frac{\partial v_y}{\partial y} & \eta_{xy,xy}\left(\frac{\partial v_x}{\partial y} + \frac{\partial v_y}{\partial x}\right) \\ \eta_{xy,xy}\left(\frac{\partial v_y}{\partial x} + \frac{\partial v_x}{\partial y}\right) & \eta_{xx,xx}\frac{\partial v_y}{\partial y} + \eta_{xx,yy}\frac{\partial v_x}{\partial x} \end{pmatrix} \tag{4.54}$$

Then, by multiplying by the velocity-gradient tensor, we get the sum

$$\begin{aligned} \sigma'_{ij}\frac{\partial v_i}{\partial r_j} &= \eta_{xx,xx}\left(\frac{\partial v_x}{\partial x}\right)^2 + \eta_{xx,yy}\frac{\partial v_x}{\partial x}\frac{\partial v_y}{\partial y} \\ &+ \eta_{xy,xy}\left(\frac{\partial v_x}{\partial y}\right)^2 + \eta_{xy,xy}\frac{\partial v_x}{\partial y}\frac{\partial v_y}{\partial x} \\ &+ \eta_{xy,xy}\left(\frac{\partial v_y}{\partial x}\right)^2 + \eta_{xy,xy}\frac{\partial v_y}{\partial x}\frac{\partial v_x}{\partial y} \\ &+ \eta_{xx,xx}\left(\frac{\partial v_y}{\partial y}\right)^2 + \eta_{xx,yy}\frac{\partial v_x}{\partial x}\frac{\partial v_y}{\partial y} \; . \end{aligned} \tag{4.55}$$

$$\sigma'_{ij}\frac{\partial v_i}{\partial r_j} = \eta_{xx,xx}\left[\left(\frac{\partial v_x}{\partial x}\right)^2 + \left(\frac{\partial v_y}{\partial y}\right)^2\right] + 2\eta_{xx,yy}\frac{\partial v_x}{\partial x}\frac{\partial v_y}{\partial y} + \eta_{xy,xy}\left(\frac{\partial v_x}{\partial y} + \frac{\partial v_y}{\partial x}\right)^2 \geq 0 \; . \tag{4.56}$$

Hence, we obtain the condition of positivity of the element $\eta_{xy,xy}$

$$\eta_{xy,xy} \geq 0 \ . \tag{4.57}$$

Now it has to be shown that

$$\eta_{xx,xx} \left[ \left( \frac{\partial v_x}{\partial x} \right)^2 + \left( \frac{\partial v_y}{\partial y} \right)^2 \right] + \ 2\eta_{xx,yy} \ \frac{\partial v_x}{\partial x} \frac{\partial v_y}{\partial y} \ \geq 0 \tag{4.58}$$

By dividing by $\left( \frac{\partial v_y}{\partial y} \right)^2$ and by defining $\omega \equiv \frac{\partial v_x}{\partial x} \frac{\partial y}{\partial v_y}$, we get

$$\eta_{xx,xx} \ \omega^2 + 2\eta_{xx,yy} \ \omega + \eta_{xx,xx} \geq 0 \ . \tag{4.59}$$

We directly obtain the second condition

$$\eta_{xx,xx} \geq 0 \tag{4.60}$$

in order to ensure the positivity of this equation. Furthermore, the positivity implies that the second-degree equation respects the relation

$$\eta_{xx,yy}^2 - \eta_{xx,xx}^2 \leq 0 \ . \tag{4.61}$$

Hence, the conditions over the elements of the viscosity tensor for the two-disk model in the square geometry are

$$\eta_{xy,xy} \geq 0 \quad ; \quad \eta_{xx,xx} \geq 0 \quad ; \quad \eta_{xx,xx}^2 \geq \eta_{xx,yy}^2 \ . \tag{4.62}$$

Therefore, we have to evaluate the three independent tensor elements $\eta_{xx,xx}, \eta_{xy,xy}, \eta_{xx,yy}$ which are depicted in Figs. 4.21 and 4.22 with respect to two different axis frames: in the first one the axes are parallel to the sides of the square ($\varphi = 0$) and, in the second one, they form an angle of 45 degrees with respect to the lattice ($\varphi = \frac{\pi}{4}$). Figure 4.22 shows that the relations (4.51) are well satisfied between the elements of the viscosity tensor.

We observe that the viscosity tensor element $\eta_{xx,yy}$ is negative. But it does not contradict the conditions imposed by the positivity of the entropy production, conditions that we gave in (4.62).

An important difference with respect to the hexagonal geometry is the absence of a singularity of the viscosity coefficient $\eta_{xy,xy}^*(0)$ at the phase transition in the square geometry. However, such a singularity still appears in the square geometry in the coefficients $\eta^*_{xx,xx}(0)$, $\eta^*_{xx,yy}(0)$, and $\eta^*_{xy,xy}(\frac{\pi}{4})$.

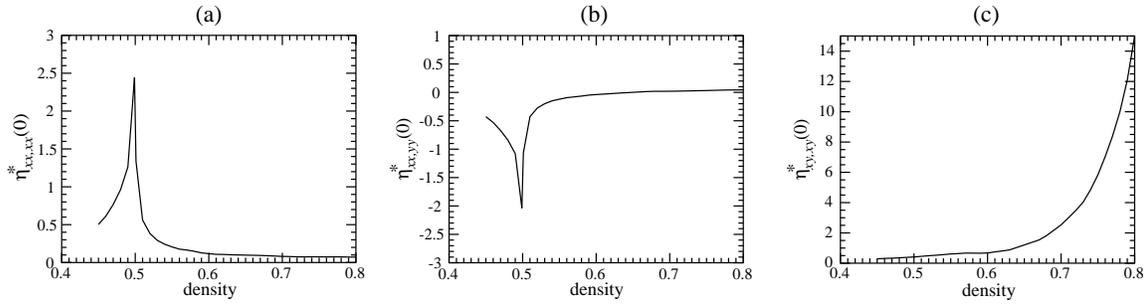

**Figure 4.21**. Square geometry: The three independent tensor elements (a) $\eta^*_{xx,xx}$, (b) $\eta^*_{xx,yy}$, (c) $\eta^*_{xy,xy}$ for $\varphi = 0$.

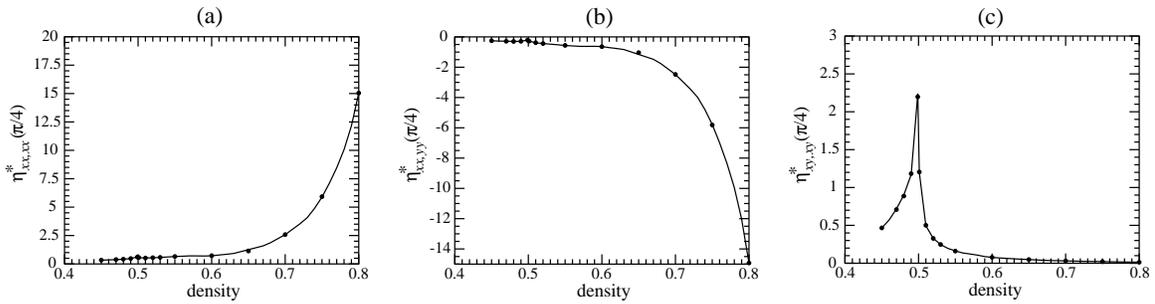

**Figure 4.22**. Square geometry: The three independent tensor elements (a) $\eta^*_{xx,xx}$, (b) $\eta^*_{xx,yy}$, (c) $\eta^*_{xy,xy}$ for $\varphi = \frac{\pi}{4}$. The continuous line corresponds to the results obtained numerically and the dots to the values obtained by the relations (4.51).

Moreover, in the solid phase, the coefficient $\eta^*_{xy,xy}(0)$ increases with the density, as explained here below.

As regards the conditions imposed by the positivity of the entropy production, we see that the the elements $\eta_{xy,xy}$ and $\eta_{xx,xx}$ are both positive for any density. For the last condition, we depict in Fig. 4.23 the relation $\eta^2_{xx,xx} - \eta^2_{xx,yy}$ which is observed to be positive.

### 4.4.5 The $t \log t$ behavior of the variance of the Helfand moment in the infinite regime

In the square geometry, Bunimovich and Spohn have proved a central-limit theorem for viscosity in the localized regime which coincides with the solid phase above the critical density (Bunimovich and Spohn, 1996). In this range of density, the viscosity coefficient is thus guaranteed to be positive and finite.

In the fluid phase, the horizon is infinite and the viscosity is infinite because of a growth as $t \log t$ of the variance of the Helfand moment for a reason similar as in the hexagonal geometry. In Fig.4.24 we compare the behavior of the variance of the Helfand moment divided by time in the two regimes (finite and infinite horizon regimes). In the finite horizon regime, $\langle G_{xy}^2(t)\rangle/t$ tends to a constant value giving a finite viscosity coefficient. On the other hand, this quantity grows continuously in time and quite linearly as a function of $\log t$ as seen in Fig. 4.24. It implies a divergence of the viscosity coefficient in this range of density. In our numerical simulation over a finite time interval, the viscosity takes finite values because the logarithmic growth is very slow.

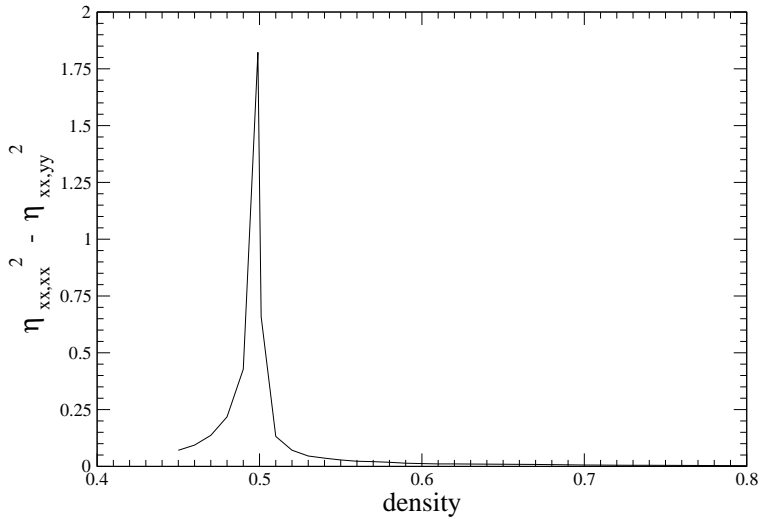

**Figure 4.23**. Square geometry: Condition $\eta_{xx,xx}^2 - \eta_{xx,yy}^2$. The positivity of this relation ensures the positivity of the entropy production.

### 4.4.6 Explanation of the numerical observations

**Solid phase**

The behavior of the viscosity tensor is clearly different in the two geometries. In this section, we explain these differences by comparing the topology of the trajectories in both geometries, since these trajectories form the basis of the evolution of the Helfand moment. More precisely, we will compare the behavior of $\eta_{xy,xy}^*$ between the hexagonal and square geometries for $\varphi = 0$. This viscosity coefficient is given by

$$\eta_{xy,xy}^* = \eta_{yx,yx}^* \sim \frac{\left\langle G_{yx}(t)^2\right\rangle}{t}, \qquad \left\langle G_{yx}(t)\right\rangle = 0, \qquad G_{yx} \sim \sum_s v_y(t_s)\, c_{\omega_s,x}, \qquad (4.63)$$

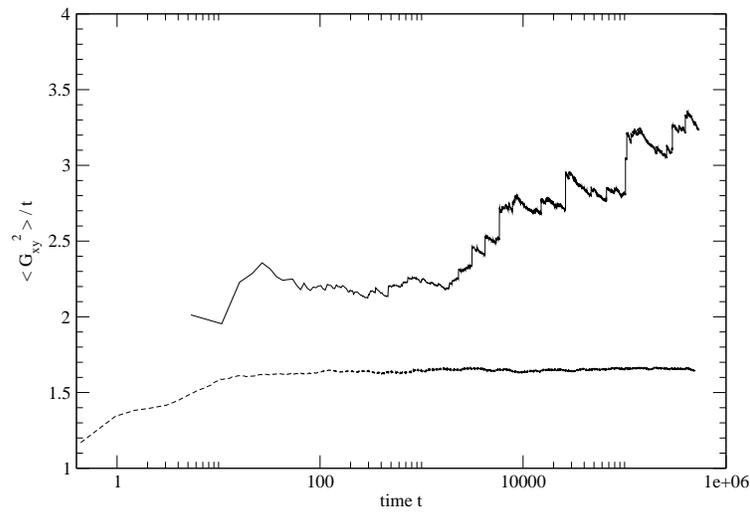

**Figure 4.24**. Variance of the Helfand moment divided by time $\langle G_{xy}^2(t)\rangle/t$ versus log $t$ in the hexagonal geometry. Comparison of the evolution of this quantity between the finite (dashed line) and the infinite (continuous line) horizon regime.

where $v_y(t_s)$ is the $y$-component of the velocity at the time $t_s$ of the jump.

When the density tends to the closed-packing density, the accessible domain of the particles tends to a perfect triangle in the hexagonal geometry. On the other hand, in the square geometry, it tends to a perfect square. This difference is at the origin of the different behaviors of the $\eta_{xy,xy}^*$ in both lattices.

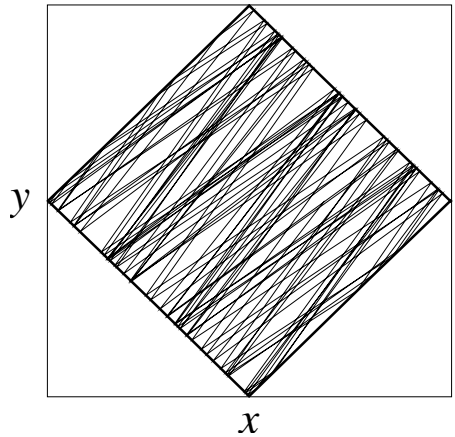

**Figure 4.25**. Part of a typical trajectory in the square geometry when the density tends to the closed-packing density.

## Square geometry

First, let us consider the case of the square geometry. In Fig. 4.25, we depict a typical trajectory of the fictitious particle moving in the Sinai billiard. We observe that this trajectory presents a regular motion between two opposite "walls" (these walls are made of parts of the fixed hard disks in the billiard). At the limit where the billiard is a perfect square, the trajectories will bounce back and forth in a regular motion. Indeed the square billiard is an *integrable system*.

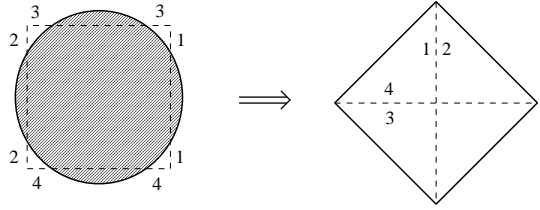

**Figure 4.26**. Geometry and notation for the boundaries in the case of the square geometry at high density.

As we have seen before, the evolution of the Helfand moment along the trajectories is determined by the passages through the boundaries (see Fig. 4.26). Both horizontal boundaries (3 and 4) do not contribute to the evolution of $G_{yx}$ since the $x$-component of the normal vectors to these boundaries equals zero. Therefore, only the passages through the vertical boundaries contribute to the Helfand moment in the square geometry.

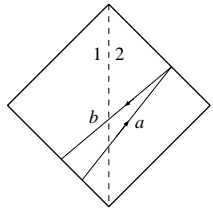

**Figure 4.27**. Part of a typical trajectory in the square geometry at high density.

To understand the behavior of the Helfand moment, let us take a small part of the typical trajectory drawn in Fig. 4.25 (see Fig. 4.27). First, let us consider the part denoted by the letter $a$ in Fig. 4.27. This one crosses the boundary in the direction $1 \rightarrow 2$, which means that $c_{\omega_s x}$ is positive (since $c_{1x} = \frac{d}{2}$). On the other hand, the $y$-component of the velocity, $v_y$, is also positive. Therefore, the contribution of the small part $a$ to the evolution of $G_{yx}$ is positive.

Now, let us take the part of the trajectory denoted $b$ in Fig. 4.27. In this case, the particle crosses the boundary in the direction $2 \rightarrow 1$ and $c_{2x}$ is negative. Since $v_y$ is also negative, the product of these two quantities is positive, and so at each successive crossings of the boundary $1 - 2$. Consequently, we obtain a sum of positive terms and the Helfand moment quickly increases along a trajectory as the one of Fig. 4.25.

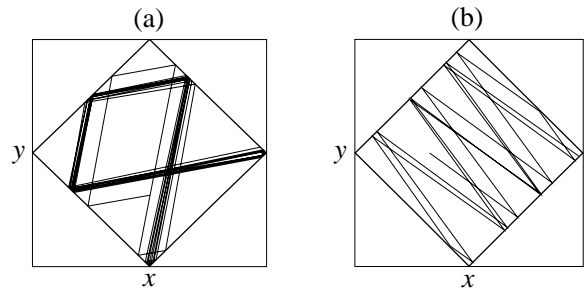

**Figure 4.28**. Square geometry at high density: The trajectory is depicted (a) during a transient regime before (b) another regime with most bounces on the two other opposite walls.

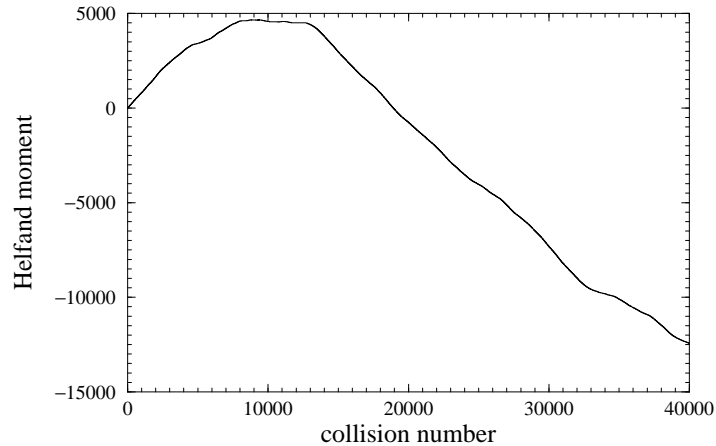

**Figure 4.29**. Evolution of the Helfand moment along a typical trajectory in the square geometry at high density.

However, the square is not perfect and the walls are still slightly convex. Therefore, after a certain time, the trajectory shown in Fig. 4.25 goes into a transient regime shown on the left-hand side of Fig. 4.28 before another regime in which the particle collides most often the two other walls (see the right-hand side of Fig. 4.28).

With the same reasoning as before, we conclude that the contributions are negative in this new regime and the Helfand moment decreases during a long time interval.

The evolution of the Helfand moment along the whole trajectory is depicted in Fig. 4.29 where we observe the succession of the three types of regimes which we have described here above. We notice that the nearly constant part corresponds to the transient regime.

The larger is the density the more perfect is the square and the longer the trajectory remains in a particular regime. Therefore, the Helfand moment can have larger and larger variations, which implies an increase of the coefficient $\eta^*_{xy,xy}(0)$ of shear viscosity with density.

**Hexagonal geometry**

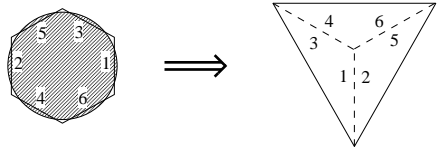

**Figure 4.30**. Geometry and notation for the boundaries in the case of the hexagonal geometry at high density.

In the hexagonal geometry (see Fig. 4.30), the trajectories present another behavior. We show in Fig. 4.30 a typical trajectory in this geometry with a density larger than the critical density. We observe that the trajectory visits the whole billiard in different directions and therefore goes into very different velocities. Accordingly, the particle crosses the boundaries with random values of its velocity in contrast to its behavior in the square geometry. Consequently, the quantity $c_{\omega_s,x}$ can be positive at a particular crossing and negative at the next one. Hence the Helfand moment cannot increase or decrease over long periods as in the square geometry (see Fig. 4.31). This explains qualitatively why, in the solid phase, the coefficient $\eta^*_{xy,xy}(0) = \eta^*$ is much smaller in the hexagonal geometry than in the square one.

In the square geometry with $\varphi = \frac{\pi}{4}$, the same arguments as in the hexagonal case explain the decrease of $\eta^*_{xy,xy}(\frac{\pi}{4})$ at high density. By the relations between the different elements of the viscosity tensor, we can also understand the behavior of the other elements in both geometries.

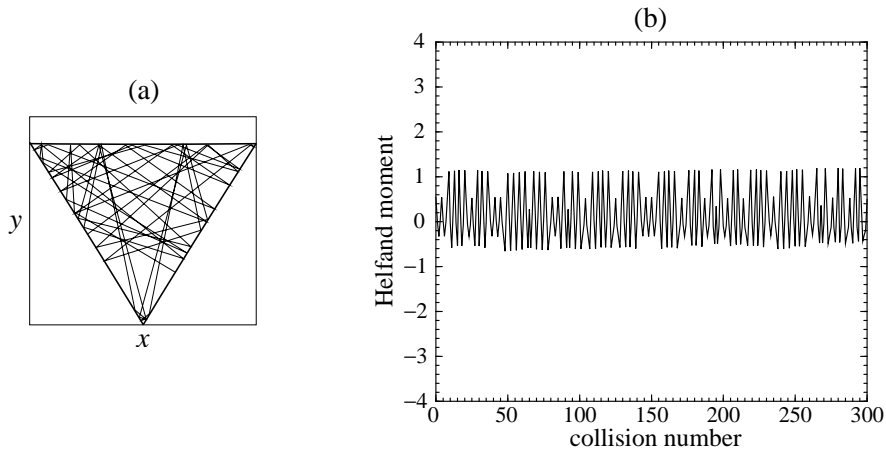

**Figure 4.31**. Hexagonal geometry at high density: (a) Part of a typical trajectory when the density tends to the closed-packing density. (b) Evolution of the Helfand moment along this typical trajectory.

**Fluid-solid phase transition**

In both the hexagonal and square geometries, the two-disk model presents a phase transition. This transition is reminiscent of the fluid-solid phase transition in the many-disk system where the viscosity coefficient is also singular. In this regard, the two-disk model can contribute to the understanding of the changes in the transport properties across the fluid-solid phase transition.

We first explain why $\eta^*_{xy,xy}$ presents a diverging singularity at the critical density in the hexagonal geometry and not in the square geometry for $\varphi = 0$. Here again, we compare the topology of the trajectories in both geometries and the way in which the Helfand moment evolves along these trajectories. At densities close to the critical density, both geometries present what we call *traps*.

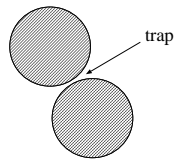

**Figure 4.32**. Example of traps in which the particles can enter and remain a long time.

Figure 4.32 shows an example of a trap. These traps are particular regions of the billiard where the particle can remain during a long time interval. Figure 4.33 depicts typical examples of a particle moving in such traps. When the particle travels out of the traps, the Helfand moment does not increase quickly in both geometries. Therefore, it is the presence of the traps which is at the origin of the difference between both geometries.

In the square geometry, the traps do not influence the evolution of $G_{yx}$. Indeed, as we have already mentioned here above, the passages through the horizontal boundary $3 - 4$ do not contribute since $c_{3x} = c_{4x} = 0$ (see Fig. 4.26 for the definitions of the boundaries in the square geometry). Therefore, the horizontal traps around these boundaries do not contribute. There remains the vertical traps. When a particle bounces for a long time in one of these traps, $c_{1x}$ and $c_{2x}$ are not vanishing, but the velocity $v_y$ is almost equal to zero so that the vertical traps does not contribute much either. This implies that both kinds of traps contribute very slightly to the evolution of the Helfand moment. To conclude the Helfand moment diffuse in the same way as for the other densities and the coefficient $\eta^*_{xy,xy}(0)$ does not present any divergence at the fluid-solid transition in the square geometry.

On the other hand, in the hexagonal geometry, the traps along the boundaries making an angle of 30° with respect to the horizontal are very important for the evolution of $G_{yx}$, whereas the vertical traps do not participate significantly. Figure 4.34 shows a typical diffusion of the Helfand moment. We observe in Fig. 4.34 the presence of jumps which correspond to the passages in the traps like the one drawn on the left-hand side of Fig. 4.33. Because of these jumps, the Helfand moment quickly diffuses. Furthermore, the importance of these traps in the hexagonal geometry can also be understood by comparing the behavior of the Helfand moment as a function of time at densities below and above the critical one $n_{\text{cr}}$.

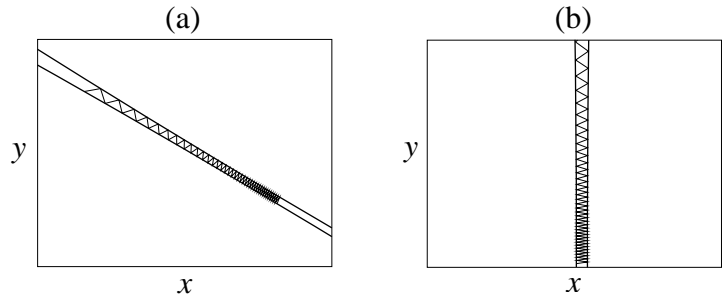

**Figure 4.33**. Particle trapped between two disks very close to each other in the hexagonal geometry. The line joining their centers either (a) forms an angle with the horizontal or (b) is horizontal.

We illustrate this point in Fig. 4.35 where we observe that there are no more jumps above the critical density. Therefore, $G_{yx}$ does not vary much contrary to the case of densities just below $n_{\text{cr}}$. Above criticality, the size of the traps decreases so quickly that the contribution of these traps decreases and, thus, the viscosity coefficient $\eta^*_{xy,xy} = \eta^*$ also decreases. By these arguments, we have an explanation for the diverging singularity of the shear viscosity at the phase transition in the hexagonal geometry.

This results show that, at a fluid-solid phase transition the viscosity coefficients may depend sensitively on the geometry of the lattice of the solid phase in formation.

### 4.4.7 Viscosity by the method of Alder *et al.*

We have also verified numerically that our method of calculation of the viscosity based on the Helfand moment (2.67) gives the same values as the method of Alder *et al.* based on the expression (2.58) (Alder et al., 1970). In the two-disk system, this expression reduces to

$$G_{ij}(t) = \sum_c \left[ 2 \, \frac{p_i \, p_j}{m} \, \Delta t_{c-1,c} + \Delta p_i^{(c)} \, r_j(t_c) \, \theta(t - t_c) \right]. \tag{4.64}$$

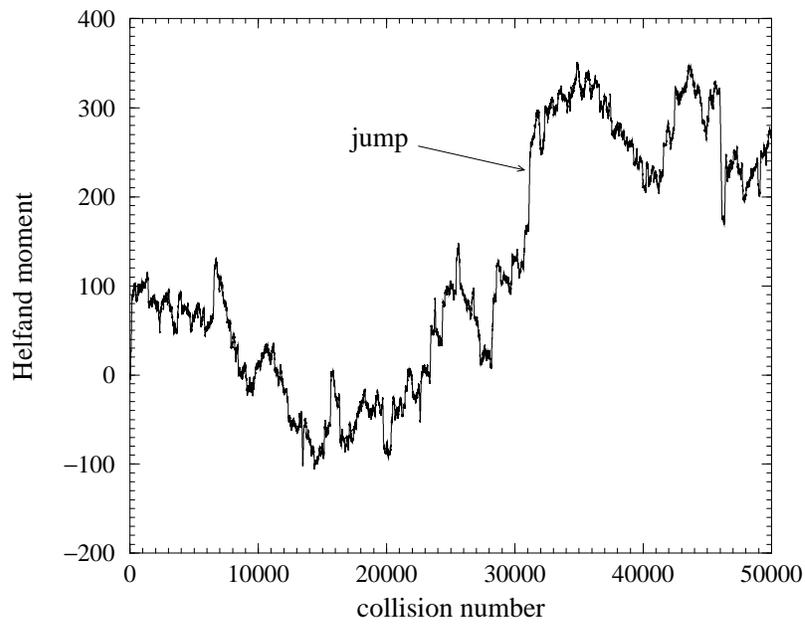

**Figure 4.34**. Helfand moment in the hexagonal geometry evaluated along a particular trajectory at a density tending to the critical density.

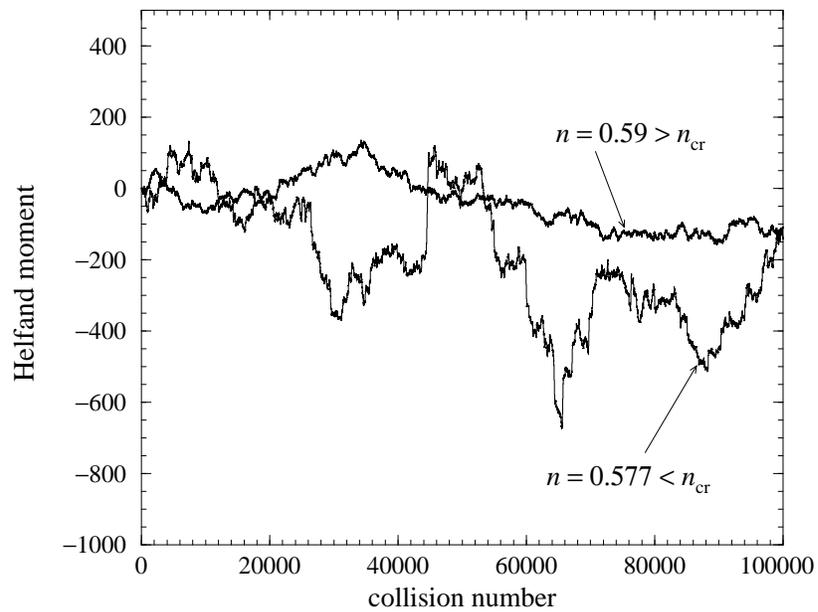

**Figure 4.35**. Comparison of the evolution of the Helfand moment for two different densities separated by the critical density in the hexagonal geometry.

As shown in Fig. 4.36 for the shear viscosity in the hexagonal geometry, there is an excellent agreement between the values obtained by both methods, which confirms the exact equivalence of both methods.

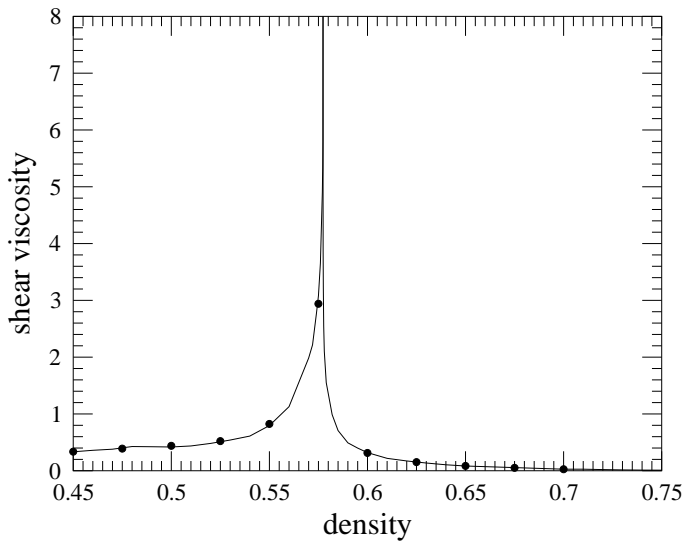

**Figure 4.36**. Shear viscosity $\eta^*$ in the hexagonal geometry calculated by our Helfand moment (2.67) (continuous line) and the one of Alder *et al.* (dots).

## 4.5   The escape-rate formalism and the fractal repeller

In this section, our purpose is to display the fractal repeller associated with viscosity in the two-disk model and to compare it with the fractal repeller of diffusion in the Lorentz gas in order to show that they are different and therefore specific to each transport property.

### 4.5.1   Shear viscosity in the two-disk model

In section 4.4, we have considered the two-disk model in the hexagonal geometry which we shall use in the following. We showed that the dynamics reduces to a Sinai billiard in the center-of-mass frame and that the Helfand moment with $N = 2$ is then given by

$$\tilde{G}_{yx}(t) = \sqrt{\frac{\beta}{V}} \left[ x(t) \, p_y(t) - \sum_s \Delta x^{(s)} \, p_y^{(s)} \, \theta(t - t_s) \right] , \tag{4.65}$$

where $(x, y)$ are the coordinates of the relative position of both disks and $(p_x, p_y)$ the canonically conjugated relative momentum. The jumps happen when the trajectory of Sinai's billiard crosses the hexagonal boundary. If the trajectory crosses the side of label $\omega$ the trajectory is reinjected at the opposite side so that the jump in position is given by the lattice vector $\Delta \mathbf{r}^{(s)} = -\mathbf{c}_\omega^{(s)}$ corresponding to the side $\omega$.

A fractal repeller is defined by considering all the trajectories such that their Helfand moment satisfies the conditions

$$-\frac{\chi}{2} \leq \tilde{G}_{yx} \leq +\frac{\chi}{2} \,, \tag{4.66}$$

where the parameter $\chi$ should be large enough. The stable manifolds of the fractal repeller can be visualized by plotting the initial conditions of trajectories satisfying the conditions (4.66) over a long time interval extending forward in time. These initial conditions are taken on the disk of Sinai's billiard. The initial conditions are specified by the angle $\theta$ of the initial position and the angle $\phi$ that the initial velocity makes with a vector which is normal to the disk at the initial position (see Fig. 4.37). The initial conditions are plotted in the Birkhoff coordinates ($\theta$, $\sin \phi$).

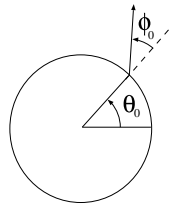

**Figure 4.37**. Initial conditions of the particules in the Sinai billiard.

Figure 4.38 depicts such the fractal composed of the stable manifolds of the repeller for viscosity in the two-disk model. We provide evidence that the set is fractal by zooming successively on it in Figs. 4.39 and 4.40, where the self-similarity of the repeller clearly appears.

Let us take a section across the repeller in Fig. 4.38 at $\theta_0 = \pi/4$. Taking the escape time of the corresponding trajectory, we have obtained the escape-time function depicted in Fig. 4.41. The time for the trajectory to escape out of the phase-space region corresponding to the interval (4.66) is infinite if the trajectory belongs to the stable manifold of a trajectory of the repeller. Indeed, this trajectory is then asymptotic to a trajectory which does not escape. Accordingly, the escape-time function has vertical asymptotes on the stable manifolds of the repeller. Since the repeller is fractal the vertical asymptotes are not enumerable, which explains the behavior in Fig. 4.41.

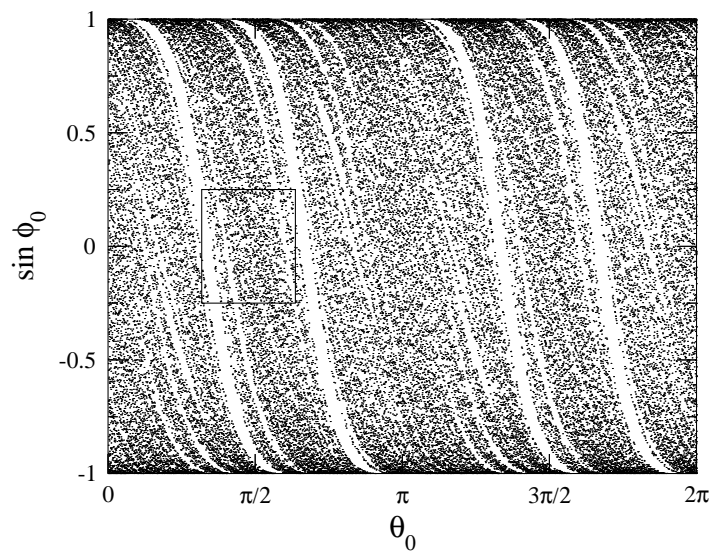

**Figure 4.38**. Fractal repeller associated with viscosity in the hexagonal geometry with absorbing boundaries at $\chi = 2.70$. The density is $n = (2/V) = 0.45$.

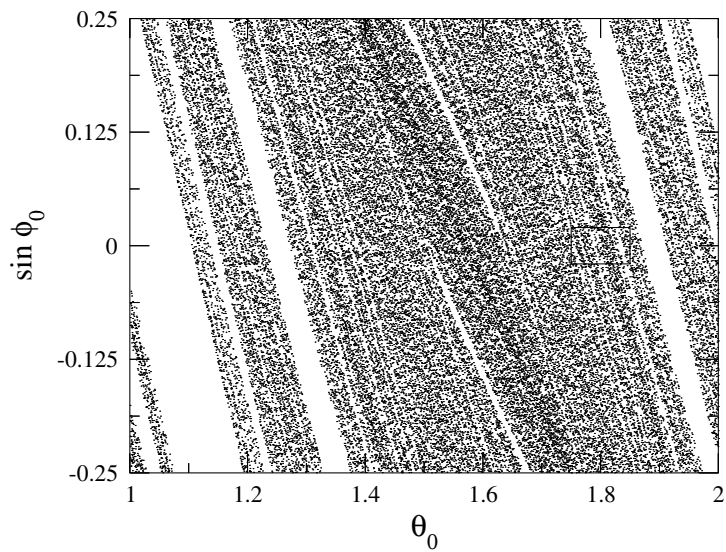

**Figure 4.39**. Enlarging of the domain in to the square in Fig. 4.38.

### 4.5.2  Diffusion in the Lorentz gas

Diffusion of a tracer particle in the hard-disk periodic Lorentz gas has been studied with the escape-rate formalism by Gaspard and Baras (1995). In this Lorentz gas, the tracer particle undergoes elastic collisions on hard disks forming a triangular lattice. In a unit cell of the lattice, the dynamics also reduces to Sinai's billiard. The energy of the tracer particle is conserved as well as the phase-space volumes. Sinai and Bunimovich have proved that the dynamics is ergodic and mixing and that the diffusion coefficient is positive and finite in the finite-horizon regime (Bunimovich and Sinai, 1980b). For diffusion, the associated Helfand moment is simply given by one of the coordinates $(x, y)$ of position of the tracer particle (Dorfman and Gaspard, 1995). An escape process is associated with diffusion by setting up a problem of first passage of the tracer particle at some absorbing boundaries. If we consider the $x$-coordinates, the tracer particle does not escape as long as the following condition is satisfied

$$-\frac{R}{2} \leq x \leq +\frac{R}{2} \, . \tag{4.67}$$

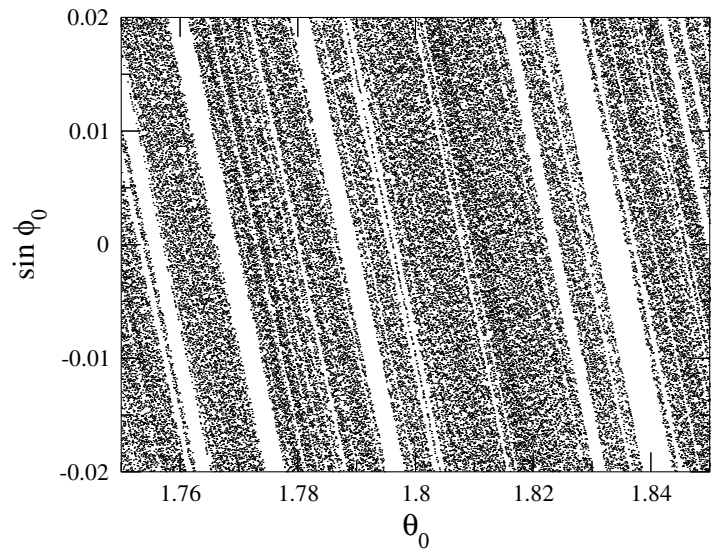

**Figure 4.40**. Enlarging of the domain into the square in Fig. 4.39.

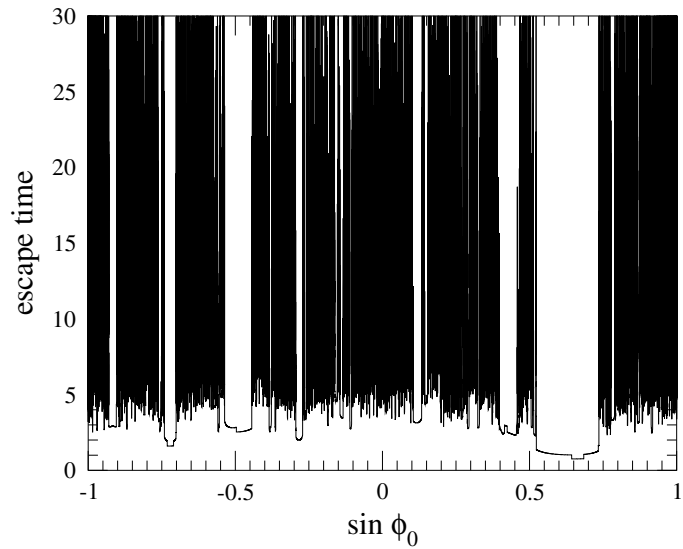

**Figure 4.41**. Escape-time function for viscosity versus $\sin\phi_0$ ($\theta_0 = \pi/4$). This function corresponds to a section in Fig. 4.38 along a vertical line at $\theta_0 = \pi/4$.

The absorbing boundary conditions are therefore defined at $x = \pm \frac{R}{2}$. With these absorbing boundaries, the system is called an *open Lorentz gas* (Gaspard and Baras, 1995).

The trajectories trapped within the interval (4.67) form a fractal repeller as shown by Gaspard and Baras (1995). In order to compare with the fractal repeller of viscosity, we can plot the fractal repeller of diffusion in a similar way as here above for viscosity.

Here again, we plot all the initial conditions of trajectories remaining within the interval (4.67) over a long forward time interval. These initial conditions are plotted in the same Birkhoff coordinates $(\theta, \sin \phi)$ of a disk around the coordinate $x \simeq 0$ in the Lorentz gas. The set of the selected initial conditions depicts the stable manifolds of the fractal repeller. We successively zoom on this fractal in Figs. 4.43 and 4.44, which provides evidence of its self-similarity. As a consequence, the repeller is also fractal. The fractal dimension of the repeller is related to the diffusion coefficient of the Lorentz gas and its Lyapunov exponent, as shown by Gaspard and Nicolis (Gaspard and Nicolis, 1990) (1990), and Gaspard and Baras (1995).

### 4.5.3   Comparison between diffusion and viscosity

We point out that the two fractal repellers associated respectively with diffusion (see Fig. 4.42) and viscosity (see Fig. 4.38) are different. Indeed, although the global structure is similar, the trajectories belonging to the different repellers are not the same.

To convince us of this difference, we take some examples of trajectories. In Fig. 4.45, we have a periodic trajectory bouncing between two disks in the billiard. This trajectory belongs to the repeller associated with diffusion since the position $x$ is bounded and satisfies (4.67). However, the viscosity Helfand moment of this trajectory does not satisfy the condition (4.66) so that it does not belong to the repeller of shear viscosity. With Eq. (4.65), we see that, in one direction, both $\Delta x^{(s)}$ and $p_y^{(s)}$ are positive. Therefore, the contribution at this passage is positive for the Helfand moment. In the other direction, both $\Delta x^{(s)}$ and $p_y^{(s)}$ are negative but the product $\Delta x^{(s)} p_y^{(s)}$ is also positive. Accordingly, the Helfand moment increases forever on this trajectory which, therefore, does not belong to the repeller associated with shear viscosity.

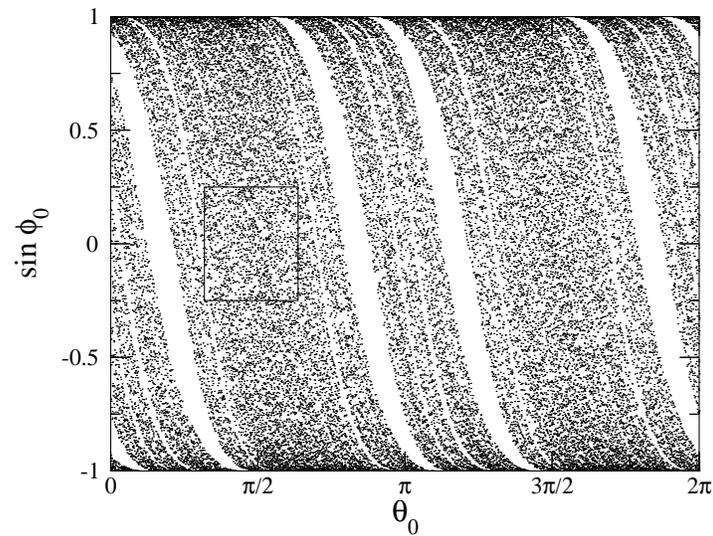

**Figure 4.42**. Fractal repeller associated with diffusion in the hexagonal geometry with absorbing boundaries at $R = 4$. The density of hard disks is $n = 0.45$.

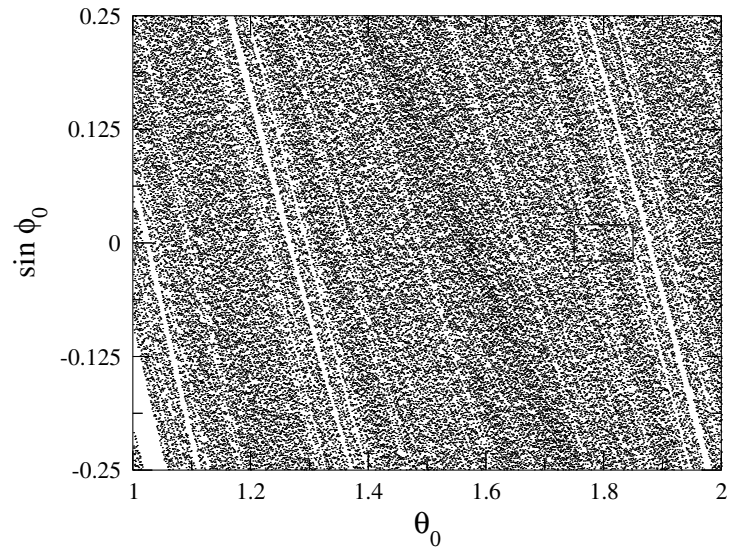

**Figure 4.43**. Enlarging of the domain into the square in Fig. 4.42.

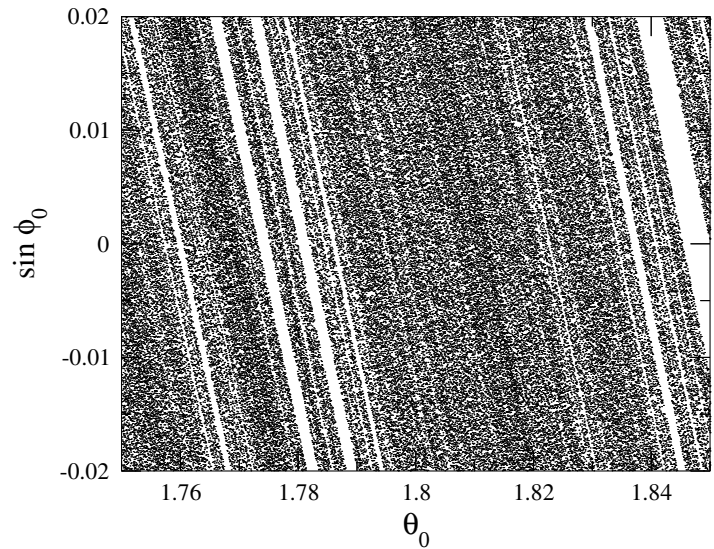

**Figure 4.44**. Enlarging of the domain into the square in Fig. 4.43.

On the other hand, we can observe the opposite case. Figure 4.46 depicts an example of trajectory escaping from the interval (4.67) although its Helfand moment of viscosity remains between the absorbing boundary conditions (4.66).

The repellers associated with different transport properties are therefore different.

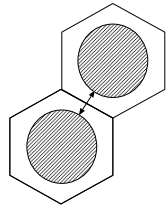

**Figure 4.45**. Periodic trajectory belonging to the fractal repeller associated with diffusion but not to the one associated with viscosity.

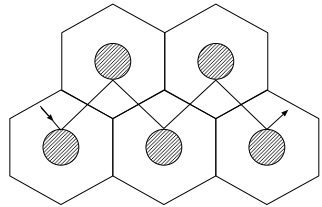

**Figure 4.46**. Typical trajectory which moves through the whole system but which has a Helfand moment that remains close to zero. This trajectory belongs to the repeller of viscosity but not to the one of diffusion.

### 4.5.4 Escape rate and viscosity

In this subsection, we show that the shear viscosity can be obtained from the escape rate of the repeller by using the escape-transport formula (3.95). We consider a sequence of repellers with larger and larger values of the parameter $\chi$. The escape rate $\gamma(\chi)$ is numerically evaluated for each repeller by computing the decay of the number $\mathcal{N}(t)$ of trajectories having their Helfand moment still within the interval (4.66) at current time and by extracting the escape rate from the exponential decay. The escape rate is observed to behave as $\gamma(\chi) \sim \chi^{-2}$ and the shear viscosity coefficient is then obtained with Eq. (3.95).

Figures 4.47 and 4.48 depict the viscosity directly computed from the escape rate and compared with the values obtained by the Einstein-Helfand formula in section 4.4, respectively in the hexagonal and square geometries. We observe in Figs. 4.47 and 4.48 the excellent agreement between both methods.

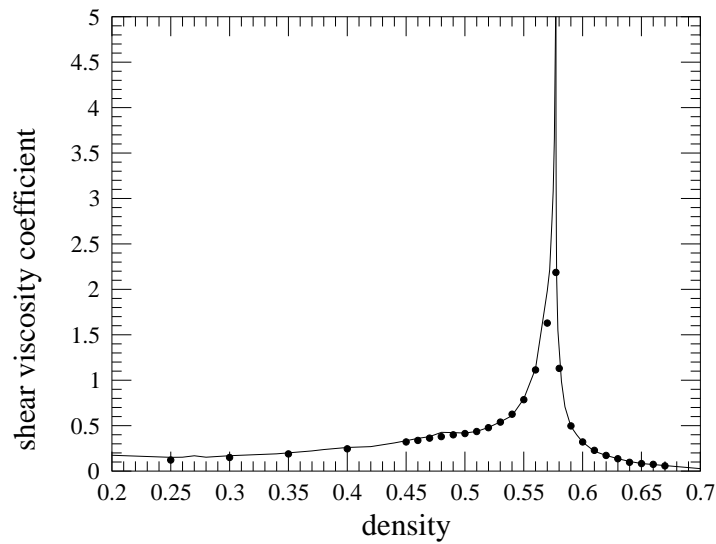

**Figure 4.47**. Comparison between two methods of calculating the shear viscosity coefficient $\eta^* = \eta^*_{xy,xy}$ in the hexagonal geometry: the Einstein-Helfand formula (continuous line) and the escape-transport formula (3.95) with $\chi = 60\sqrt{n}$ (dots).

## 4.6 Viscosity from the chaotic and fractal properties of the repeller

In this section, we compute the shear viscosity coefficient in terms of the chaotic and fractal properties of the repeller by using the chaos-transport formula (3.97) which relates the viscosity to the Lyapunov exponent and the Hausdorff codimension of the repeller of viscosity.

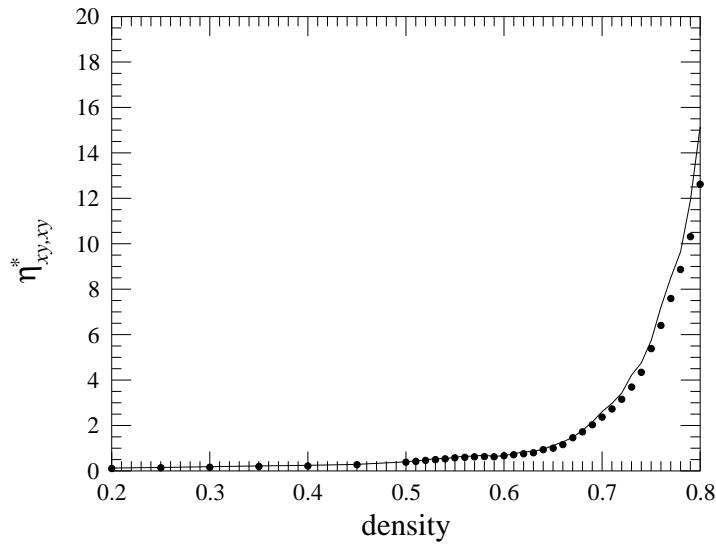

**Figure 4.48**. Comparison between two methods of calculating the shear viscosity coefficient $\eta^*_{xy,xy}$ in the square geometry: the Einstein-Helfand formula (continuous line) and the escape-transport formula (3.95) with $\chi = 45\sqrt{n}$ for density $n < 0.66$, $\chi = 100\sqrt{n}$ for $0.67 < n < 0.75$, and $\chi = 150\sqrt{n}$ for $0.76 < n$ (dots).

### 4.6.1  Lyapunov exponent

In Sinai's billiard which controls the reduced dynamics of the two-disk model, the elastic collisions between the disks are defocusing. This induces a dynamical instability of the trajectories which is characterized by the Lyapunov exponents. These exponents are the rates of exponential separations between a reference orbit and infinitesimally close orbits. Since the dynamics of Sinai's billiard is symplectic and volume-preserving in the four-dimensional phase space, the spectrum of Lyapunov exponents is $(+\lambda, 0, 0, -\lambda)$ so that their sum is vanishing. One of the Lyapunov exponents vanishes because of the absence of exponential separation in the direction of the flow. Another one corresponding to the direction perpendicular to the energy shell equals zero because of energy conservation.

There exists a method to calculate the positive Lyapunov exponent by considering the motion of a front of particles accompanying the reference particle and issued from the same initial position but with different initial velocities (Sinai, 1970a). Because the dynamics is defocusing, this front is expanding. Locally on the reference orbit $\Gamma_t$, the front (called the unstable horocycle) is characterized by a curvature $\kappa_u(\Gamma_t)$ or, equivalently, by its radius of curvature $1/\kappa_u(\Gamma_t)$. Thanks to this method explained in detail by Gaspard and Baras (Gaspard, 1998; Gaspard and Baras, 1995) (1995), we have computed the positive Lyapunov exponent as a function of the density of the system (in the hexagonal and square geometries). The equilibrium values of the Lyapunov exponent are obtained by running a trajectory in Sinai's billiard without absorbing boundaries and by averaging over a long time interval. The resulting numerical values are depicted in Figs. 4.49 and 4.50.

In the chaos-transport formula (3.97), the Lyapunov exponent has to be evaluated for the trajectories belonging to the fractal repeller. The statistical average is here carried out for the natural invariant probability measure concentrated on the fractal repeller. This invariant measure defines a nonequilibrium state for the motion. As aforementioned, the natural invariant measure is generated by the dynamics itself. Accordingly, the Lyapunov exponent is numerically computed by averaging over a statistical ensemble of trajectories which has not yet escaped after a long but finite time. This ensemble can be as large as wished by increasing the number of initial conditions. In this way, we can calculate the nonequilibrium values of the Lyapunov exponent.

In Table 4.1, we present a comparison between the equilibrium Lyapunov exponent $\lambda_{eq}$ without absorbing boundary conditions (as depicted in Fig. 4.49 and 4.50 and the nonequilibrium Lyapunov exponent $\lambda_{neq}(\chi)$ evaluated over a nonequilibrium measure which has the fractal repeller as support. The difference between these exponents is small and of the order of the escape rate, in agreement with the results of van Beijeren et al. (2000) for the disordered Lorentz gas.

| $n$ | $\lambda_{eq}$ | $\lambda_{neq}$ | $\gamma$ | $h_{KS}$ | $c_I = \gamma/\lambda_{neq}$ | $c_H$ |
|------|------|------|------|------|------|------|
| 0.40 | 1.5156 | 1.5163 | 0.0017 | 1.5146 | 0.0011 | 0.0011 |
| 0.50 | 2.3519 | 2.3539 | 0.0023 | 2.3516 | 0.00098 | 0.00092 |
| 0.60 | 3.7258 | 3.7249 | 0.0015 | 3.7234 | 0.00040 | - |

**Table 4.1.** Values of the characteristic quantities of chaos for different densities $n$ in the hexagonal system: $\lambda_{eq}$ is the equilibrium Lyapunov exponent for the closed system. The following quantities characterize the fractal repeller for viscosity with $\chi = 60\sqrt{n}$: $\lambda_{neq}$ is the nonequilibrium Lyapunov exponent of the repeller, $h_{KS}$ its KS entropy (calculated with the relation $h_{KS} = \lambda_{neq} - \gamma$), $\gamma$ its escape rate, $c_I$ its partial information codimension, and $c_H$ its partial Hausdorff codimension.

## 4.6.2   Hausdorff dimension and viscosity

As we have seen in the previous chapter, the viscosity may be evaluated in terms of quantities of

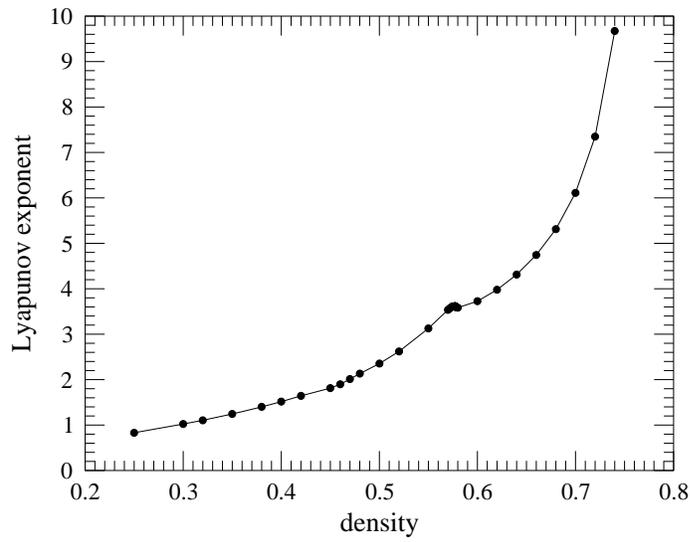

**Figure 4.49**. Equilibrium Lyapunov exponent versus density in the hexagonal geometry.

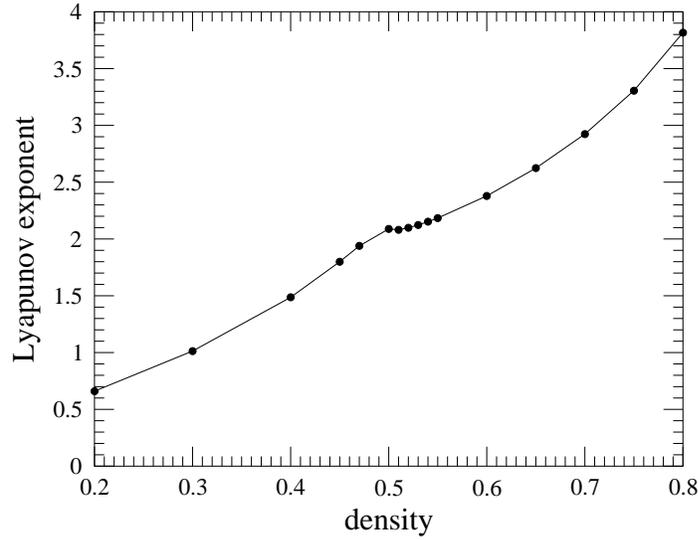

**Figure 4.50**. Equilibrium Lyapunov exponent versus density in the square geometry.

the microscopic chaos through the relation (3.97)

$$\eta = \lim_{\chi \to \infty} \left( \frac{\chi}{\pi} \right)^2 \left( \sum_{\lambda_i > 0} c_i \, \lambda_i \right)_{\overline{\mathcal{F}}_\chi} .$$
(4.68)

which reduces to

$$\eta = \lim_{\chi \to \infty} \left( \frac{\chi}{\pi} \right)^2 (\lambda \, c_I)_{\overline{\mathcal{F}}_\chi}$$
(4.69)

for a two-degree-of-freedom system.

In the particular case of the present system, only one Lyapunov exponent $\lambda$ is strictly positive. Furthermore, it has been shown that the information dimension $d_I = 1 - c_I$ tends to the Hausdorff dimension $d_H = 1 - c_H$ when the parameter $\chi$ determining the escape process tends to infinity (Gaspard, 1998; Gaspard and Baras, 1995). Numerical evaluations obtained for both dimensions are given in Table I and their very small difference confirms that the multifractal character of the repeller is very small and this last tends to a uniform fractal. Hence, the information dimension may be substituted by the Hausdorff dimension. Finally, Eq. (4.68) can be rewritten as

$$\eta = \lim_{\chi \to \infty} \left( \frac{\chi}{\pi} \right)^2 (\lambda \, c_H)_{\mathcal{F}_\chi} \ . \tag{4.70}$$

In order to determine the viscosity by the chaos-transport formula (4.70), we need to determine the partial Hausdorff codimension $c_H$ of the fractal repeller. The corresponding dimension $d_H = 1 - c_H$ is the Hausdorff dimension of the vertical asymptotes of the escape-time function depicted in Fig. 4.41. Its values range in the interval $0 \le d_H \le 1$.

The Hausdorff codimension can be obtained using the following numerical algorithm developed by the group of Maryland (McDonald et al., 1985). We consider an ensemble of pairs of trajectories starting from initial conditions $\phi_0$ differing in a value $\epsilon$. The time taken by the trajectories to escape out of the system is given by the escape-time function in Fig. 4.41. The pair is said to be *uncertain* if the trajectories and their Helfand moment present at least one of the following conditions: (i) both trajectories follow paths that differ by the successive passages through the cell boundaries, that is, if we associate to each trajectory a symbolic sequence $(\omega_1, \omega_2, ...)$ which gives the labels of the cell boundaries across which the successive passages occur, and both sequences are different; (ii) one of both trajectories has its Helfand moment which reaches the fixed absorbing boundaries (4.66) when the Helfand moment of the other one still remains within these limits. If the pair does not present one of these conditions it is called *certain*. The fraction $f(\epsilon)$ of uncertain pairs in the initial ensemble is known to behave as the power

$$f(\epsilon) \sim \epsilon^{c_H} \ , \tag{4.71}$$

giving the Hausdorff codimension as its exponent. Derivations of this result can be found elsewhere (Ott, 1993; Claus et al., 2004; McDonald et al., 1985). This method has been already used by Gaspard and Baras (1995), Claus and Gaspard (2001), and Claus et al. (2004).

We have here applied the Maryland algorithm to obtain the Hausdorff codimension of the fractal repeller of viscosity. Table I compares the partial Hausdorff codimension with the partial information dimension in particular cases. We observe that both codimensions take very close values as expected.

By varying the parameter $\chi$, we have obtained the shear viscosity thanks to the chaos-transport formula (3.97). These values are plotted in Figs. 4.51 and 4.52 for the hexagonal and square geometries, respectively. We consider the shear viscosity coefficient $\eta^*$ in the hexagonal geometry and the element $\eta^*_{xy,xy}$ of the viscosity tensor in the square geometry. The values obtained with the chaos-transport formula (3.97) (Viscardy and Gaspard, 2003b) are compared with the values obtained by the escape-transport formula (3.95) (Viscardy and Gaspard, 2003b) and those by the Einstein-Helfand formula (2.47) obtained in section 4.4 (Viscardy and Gaspard, 2003a). The agreement between the three formulas is excellent, which confirms the theoretical results.

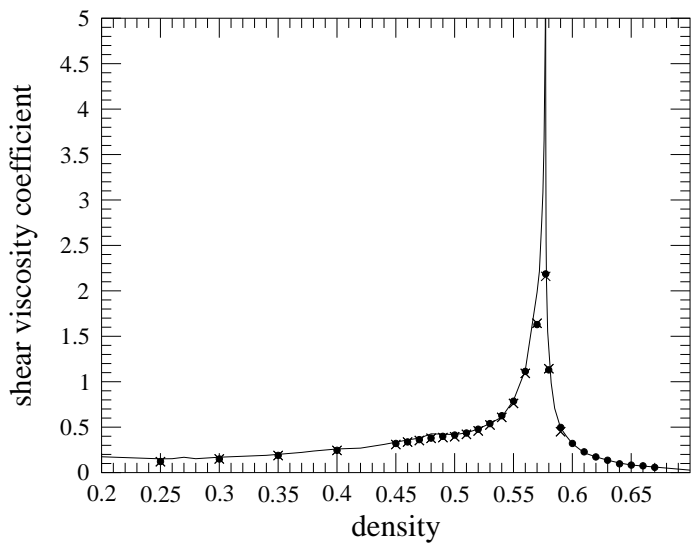

**Figure 4.51**. Comparison between the three methods calculating the shear viscosity coefficient $\eta^*$ in the hexagonal geometry: the Einstein-Helfand formula (2.47) (continuous line), the escape-transport formula (3.95) (dots), and the chaos-transport formula (3.97) (crosses) with $\chi = 60 \sqrt{n}$.

## 4.7  Nonequilibrium steady state

In section 3.7, we presented the construction of the hydrodynamic modes in terms of the microscopic dynamics. We showed that the singular character of the hydrodynamic modes due to the pointlike character of the deterministic dynamics and the property of dynamical instability induces the addition of a term to the nonequilibrium steady state usually obtained by the phenomenological equations. In the case of viscosity, we have

$$\Psi^{(\eta)}_g(\mathbf{\Gamma}) = g\, G^{(\eta)}(\mathbf{\Gamma}) + g \int_0^{-\infty} J^{(\eta)}\left(\mathbf{\Phi}^t\mathbf{\Gamma}\right) dt \qquad (4.72)$$

where $G^{(\eta)}(\mathbf{\Gamma})$ is the Helfand moment and $J^{(\eta)} = dG^{(\eta)}/dt$ the associated microscopic current. By resolving the integral, Eq.(4.72) formally becomes

$$\Psi^{(\eta)}_g(\mathbf{\Gamma}) = \lim_{t\to\infty} g\, G(\mathbf{\Phi}^{-t}\mathbf{\Gamma}) \ . \qquad (4.73)$$

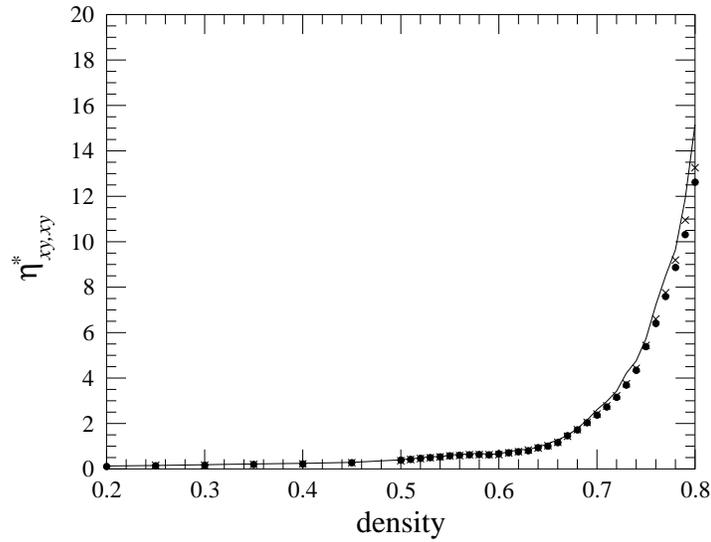

**Figure 4.52.** Comparison between the three methods calculating $\eta^*_{xy,xy}$ in the square geometry: the Einstein-Helfand formula (2.47) (continuous line), the escape-transport formula (3.95) (dots), and the chaos-transport formula (3.97) (crosses) with $\chi = 45\sqrt{n}$ for density $n < 0.66$, $\chi = 100\sqrt{n}$ for $0.67 < n < 0.75$, and $\chi = 150\sqrt{n}$ for $0.76 < n$.

Here we want to obtain a representation of such a nonequilibrium steady state corresponding to a velocity gradient. Because of its singular character the steady state can be represented by its cumulative function. Here we are concerned with the two-disk model. The dynamics of the two disks is reduced to the one of a pointlike particle moving in the periodic Sinai billiard with the angle of initial position $\theta$. Here we choose $\theta$ as the variable of the cumulative function $\mathcal{I}(\theta)$

$$\begin{aligned} \mathcal{I}(\theta) &= \int_0^\theta \Psi(\theta') d\theta' \\ &= \lim_{t\to\infty} g \int_0^\theta G(\mathbf{\Phi}^{-t}\theta') d\theta' = gT(\theta) \ . \end{aligned} \qquad (4.74)$$

Numerically, $T(\theta)$ is obtained by taking several initial conditions $\theta$ distributed homogeneously on the fixed disk with a fixed velocity angle and by evaluating for each trajectory the Helfand moment after 30 collisions. By summing over $\theta$ from $\theta = 0$ to $\theta = \theta^*$ the corresponding values of the Helfand moment we obtain the cumulative function $T(\theta^*)$. In Fig. 4.7 we plot $T(\theta)$ in function of $\theta$. The self-similar character of the fractal is showed by zooming a domain of the complete cumulative function.

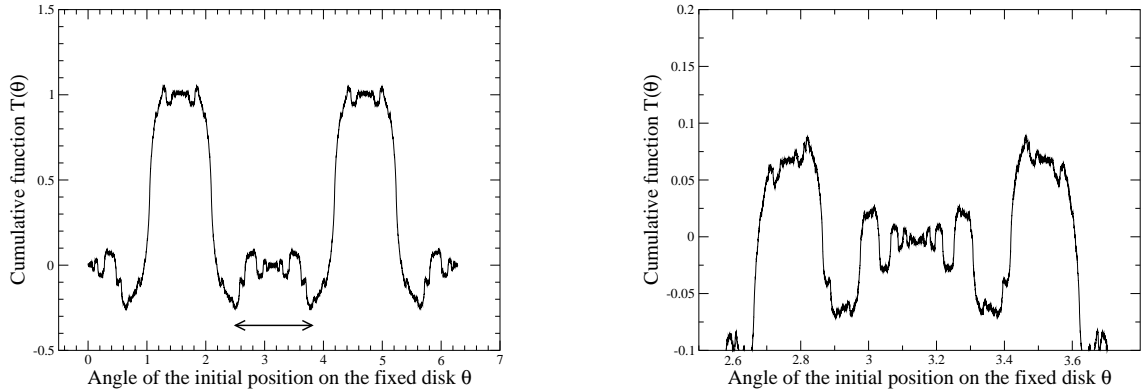

**Figure 4.53**. On the left-hand side: Fractal curve of the cumulative function of a nonequilibrium steady state corresponding to a velocity gradient in the two-hard-disk model. On the right-hand side: Zoom of the domain underlined in the figure on the left. The self-similar character clearly appears and is the signature of the fractality of the cumulative function.

## 4.8   Conclusions

In this chapter we have studied viscosity in the simplest model. This model composed by two hard disks has already been investigated in the square geometry by Bunimovich and Spohn (1996). Here, we have generalized this model to the hexagonal geometry. First, we show that the fourth-order viscosity tensor is isotropic in the hexagonal geometry although it is not in the square geometry. Secondly, we show that the viscosity can be positive and finite in the fluid phase of the hexagonal geometry, although it is always infinite in the fluid phase of the square geometry. The reason is that the horizon of the Sinai billiard driving the dynamics of the two-disk model is always infinite in the fluid phase of the square geometry although there is a regime with a finite horizon in the fluid phase of the hexagonal geometry. In an infinite-horizon regime, the viscosity becomes infinite so that, from a physical point of view, the proof of the existence of a positive and finite viscosity coefficient strictly holds in the hexagonal two-disk model. In the solid phase, the transport coefficients acquire a different meaning because the spontaneous breaking of translational invariance modifies the structure of the hydrodynamic modes and the viscosity coefficient should be reinterpreted in terms of the damping coefficients of the transverse sound modes and of the diffusive modes (Martin et al., 1972; Fleming III and Cohen, 1975; Kirkpatrick et al., 1990). The remarkable result is that the two-disk systems already gives the shear viscosity in quantitative agreement with Enskog's theory at moderate densities.

The two-disk model presents a phase transition between a fluid and a solid phase. This transition is reminiscent of the fluid-solid transition in the system composed of many disks. Indeed, the transition manifests itself in the hydrostatic pressure in a very similar way as in the many-particle system. The hydrostatic pressure can be directly related to the mean free path in the two-disk model and we can thus explain the manifestation of the transition on the pressure in terms of the behavior of the mean free path near the transition. In this simple model, the transition can be understood as a geometric property of the dynamical system. Indeed, the trajectories are unbounded in the fluid phase albeit there remain localized in bounded domains in the solid phase where ergodicity is broken. The fluid-solid transition also manifests itself as a diverging singularity in the viscosity in the two-disk model. We have here shown that this singularity in the viscosity versus the density may depend sensitively on the geometry of the lattice of the solid phase in formation.

The escape-rate formalism implies the appearance of a fractal repeller associated with viscosity. We have numerically generated the fractal repeller associated with viscosity in this model and we have provided evidence for its fractal character. Using the chaos-transport formula of the escape-rate formalism, we have been able to evaluate the shear viscosity from the positive maximum Lyapunov exponent and the Hausdorff codimension of the fractal repeller of viscosity. The values obtained by using the chaos-transport formula for shear viscosity have been compared with the values obtained by other methods based on the Einstein-Helfand formula, which is equivalent to the Green-Kubo formula as shown in Appendix B. An excellent agreement has been observed between the different methods. This agreement brings an important support to the escape-rate formalism as a method to establish a connection between the transport properties – here of viscosity – and the underlying microscopic chaotic dynamics. The agreement therefore confirms the theoretical results of the escape-rate formalism (Gaspard and Nicolis, 1990; Dorfman and Gaspard, 1995).

Finally, we have studied a nonequilibrium steady state corresponding to a velocity gradient. The singular character of such a distribution plays a fundamental role in irreversible processes like viscosity. We have evaluated the cumulative function of this steady state which clearly exhibits a self-similar character.

# Chapter 5

# The $N$-particle systems

In this chapter, we study the properties of *N*-particle systems in two and three dimensions with periodic boundary conditions with emphasis on the *N*-hard-ball systems. The main purpose of this chapter is to extend the study realized in the previous one to the *N*-hard-ball systems. After presenting the equations ruling the dynamics of the system, the equation of state is evaluated and the well-known fluid-solid phase transition is put in evidence. It is shown that the dynamics of *N* hard balls is extremely sensitive to its initial conditions. This dynamical chaos is characterized by a full spectrum of positive Lyapunov exponents. Then viscosity is calculated with the Helfand-moment method for systems of *N* hard balls with $N > 2$ increasing. Finally, we apply our method described in section 2.5.2 for calculating viscosity in *Lennard-Jones fluids*. Furthermore, we propose a similar Helfand-moment method for the *thermal conductivity*. For both transport properties, a comparison is done with our own results obtained by the Green-Kubo method as well as the ones found in the literature (Viscardy et al., 2005).

## 5.1    Molecular dynamics in hard-ball systems

The simulation of the dynamics of hard balls is easier than particles with potentials like the Lennard-Jones potential. Indeed, the integration of Newton's equation of motion may be divided into two steps: the first one considering the *free flight* between the successive binary collisions; the second one calculating the change of the velocities of the bouncing particles.

1. *Free flight between binary collisions:*

   Between two successive collisions $n-1$ and $n$, the particles do not experience any force. Mathematically, it is expressed as

   $$t_{n-1} \to t_n : \begin{cases} \mathbf{r}_a^{(-)}(t_n) = \mathbf{r}_a^{(+)}(t_{n-1}) + (t_n - t_{n-1}) \, \mathbf{v}_a^{(+)}(t_{n-1}) \\ \mathbf{v}_a^{(-)} = \mathbf{v}_a^{(+)}(t_{n-1}) \, . \end{cases} \tag{5.1}$$

2. *Binary collision:*

   On the other hand, the $n^{\text{th}}$ collision between two particles *a* and *b* is elastic and instantaneous.

It means that, while the position of the colliding particles remains constant

$$t_n: \quad \mathbf{r}_a^{(+)} = \mathbf{r}_a^{(-)} \ , \tag{5.2}$$

the change of velocities is given by the law of geometric optics

$$t_n: \begin{cases} \mathbf{v}_a^{(+)} = \mathbf{v}_a^{(-)} - \left( \epsilon_{ab} \cdot \mathbf{v}_{ab}^{(-)} \right) \epsilon_{ab} \\ \mathbf{v}_b^{(+)} = \mathbf{v}_b^{(-)} + \left( \epsilon_{ab} \cdot \mathbf{v}_{ab}^{(-)} \right) \epsilon_{ab} \\ \mathbf{v}_k^{(+)} = \mathbf{v}_k^{(-)} \qquad \forall \quad k \neq a, b \end{cases} \tag{5.3}$$

where $\mathbf{v}^{(-)}$ and $\mathbf{v}^{(+)}$ indicate respectively the velocities before and after the collision. The unit vector $\epsilon_{ab}$ joins the centers of the $a^{\text{th}}$ and $b^{\text{th}}$ balls at the collision given by

$$\epsilon_{ab} = \frac{\mathbf{r}_a^{(\pm)} - \mathbf{r}_b^{(\pm)}}{\sigma} \tag{5.4}$$

with $\mathbf{r}^{(\pm)}$ the position at the collision, and $\sigma$ the diameter of the particles. Further, we define the relative velocity vector as

$$\mathbf{v}_{ab}^{(-)} = \mathbf{v}_a^{(-)} - \mathbf{v}_b^{(-)} \ . \tag{5.5}$$

As a matter of fact, the method used for the molecular dynamics of hard balls goes back to Alder and Wainwright in the late fifties (Alder and Wainwright, 1959) who proposed a general algorithm initially for systems of particles with a square-well potential.

## 5.2 Equation of state

### 5.2.1 Introduction

Thanks to the development of the mechanical philosophy during the 17th century, scientists started to reject Aristotle's physics. In particular, as an explanation of the suction of water in a tube due to the action of a pump, the so-called *nature's abhorrence of a vacuum* was replaced by the idea that the suction of the air is rather the result of the mechanical pressure of the air surrounding the pump. In this new philosophical context were developed a lot of works on pressure of gases realized, e.g., by Torricelli (the well-known *Torricelli's tube* or *barometer*) or Pascal (showing that the air pressure decreases with the altitude). But the culmination of these studies was the work of Boyle who established a clear relation between the pressure and the volume of a vessel expressed by the well-known *Boyle's law PV* = const. And later, in 1787, Charles (1746-1823) showed that the volume increases linearly with temperature, thanks to which the important law, that is, the today well-known *ideal gas law*[1]

$$PV = Nk_B T \ , \tag{5.6}$$

could be obtained. This consisted of the first *equation of state* relating the quantities characterizing the gas[2].

Rapidly, if this relation was verified for gases at low densities, its validity was no longer observed when the density increases. According to Regnault (1810-1878),

> Boyle already thought to observe that from 4 atmospheres, the air compressed less than it should according to the [Boyle's] law (from Regnault (1846)).

Regnault himself is famous for his experimental work devoted to this property (Regnault, 1846). Among scientists having derived an equation of state introducing a correction to the ideal gas law is the the physicist Ritter (1801-1862) who obtained theoretically in 1846 such a theoretical expression (Ritter, 1846).

However, the starting point for the modern theories devoted to the equation of state is the work of Clausius (1870) on the *virial theorem*. This theorem states that the mean kinetic energy $\langle E_c \rangle$ in a system of *N* material points is equal to minus the half of the mean value of the so-called *virial* $\langle \sum_{a=1}^{N} \mathbf{F}_a \cdot \mathbf{r}_a \rangle$

$$\langle E_c \rangle = -\frac{1}{2} \sum_{a=1}^{N} \langle \mathbf{F}_a \cdot \mathbf{r}_a \rangle \tag{5.7}$$

where $\mathbf{r}_a$ is the position of the particles *a* whereas $\mathbf{F}_a$ is the interaction force acting on the particles *a*. By considering separately the internal (interaction between particles) and the external (interaction with the container) parts of the virial, Clausius arrived at the equation that is today written as

$$PV = N \left[ k_B T + \frac{1}{6} \sum_{a=2}^{N} \langle \mathbf{r}_{1a} \cdot \mathbf{F}_{1a} \rangle \right] \tag{5.8}$$

where $\mathbf{r}_{1a}$ and $\mathbf{F}_{1a}$ are respectively the relative position and the force between the particle 1 and *a*.

---

[1] We write here the ideal gas law in modern terminology.

[2] For a general overview of the development of equation of state in the kinetic theory of gases, see (Brush, 1976; Brush, 1961).

The first who attempted to apply the virial theorem was van der Waals (1873). He obtained the henceforth called the *van der Waals equation of state*

$$\left(P + \frac{a}{\hat{v}^2}\right)(\hat{v} - b) = \sum mv^2/3 = RT \tag{5.9}$$

where $a/\hat{v}^2$ represent the contribution of the interatomic forces, $\hat{v}$ the volume per mole, and $b$ firstly an adjustable parameter, which was later equal to four times the volume of the particles.

In 1896, van der Waals tried to evaluate the next correction term by extending his original method and by writing the equation of state as a series in power of the density $n$ (van der Waals, 1896). It then gives one

$$P = nRT \left(1 + b_0 \, n + \frac{5}{8} \, b_0^2 \, n^2 + \dots \right) \tag{5.10}$$

where $b_0$ is the so-called *second virial coefficient*, $B_3 = \frac{5}{8} \, b_0^2 \, n^2$ being the third coefficient[3]. For hard-disk systems, it is to Tonks (Tonks, 1936) that we owe the first calculation of $B_3$. Boltzmann (1899) and Jäger (1899) already obtained the fourth virial coefficient $B_4$ for hard-sphere systems whereas one had to wait for Rowlinson (Rowlinson, 1964) and simultaneously Hemmer (Hemmer, 1964) for its evaluation for hard disks.

All the previous coefficients have been calculated analytically. However, the calculation of the next ones needed computers. Numerical computations started to be available since the fifties. In particular, the fifth virial coefficient $B_5$ was obtained by Metropolis *et al.* (1953) for hard spheres and by the Rosenbluths (1954) for hard disks. During the following decade were computed $B_6$ as well as $B_7$ by Ree and Hoover (1964a, 1964b, 1967) for hard disks and hard spheres. For the next virial coefficients, their evaluations were accomplished only in 1993 by Janse van Rensburg (1993) for $B_8$. And finally, it is during the last and current years that the ninth (Kolafa et al., 2004; Labík et al., 2005) and the tenth (Clisby and McCoy, 2005) were computed. Hence the equation of state with the first terms in the virial expansion can be written (in a different way than previously) as

$$P = nk_B T \sum_{k=1}^{10} B_k \, (b_0 \, n)^{k-1} + \dots \tag{5.11}$$

where $b_0 = \frac{\pi \sigma^2}{2}$ for a two-dimensional system ($d = 2$) or $b_0 = \frac{2\pi \sigma^3}{3}$ for $d = 3$. The factors $B_k$

---

[3]Remark that, originally, van der Waals obtained $\frac{15}{32}$ instead of $\frac{5}{8}$. It is to Boltzmann (1896) and independently Jäger (1896) that we owe this correction done the same year.

| $B_k$ | $d = 2$ | $d = 3$ |
|---|---|---|
| $B_1$ | 1 | 1 |
| $B_2$ | 1 | 1 |
| $B_3$ | 0.782004 | 0.625 |
| $B_4$ | 0.53223180 | 0.2869495 |
| $B_5$ | 0.33355604 | 0.110252 |
| $B_6$ | 0.1988425 | 0.03888198 |
| $B_7$ | 0.1148728 | 0.01302354 |
| $B_8$ | 0.0649930 | 0.0041832 |
| $B_9$ | 0.0362193 | 0.0013094 |
| $B_{10}$ | 0.0199537 | 0.0004035 |

**Table 5.1**. Numerical values of the first ten virial coefficients for $d = 2$ and $d = 3$; values given in Ref. (Clisby and McCoy, 2005).

recalculated by Clisby and McCoy (2005) are given in Table 5.1.

### 5.2.2   The *N*-hard-disk model

By using molecular dynamics for systems composed of *N* particles with hard-sphere potentials submitted to periodic boundary conditions, we compute the pressure and depict in Fig. 5.1 the equation of state at constant temperature ($k_B T = 1$), unit diameter $\sigma$ and unit mass *m* for the case of hard disks ($d = 2$). In this figure, we compare the numerical results with the "analytical" virial expansion given by Eq. (5.11). For low and intermediate densities, the agreement is excellent.

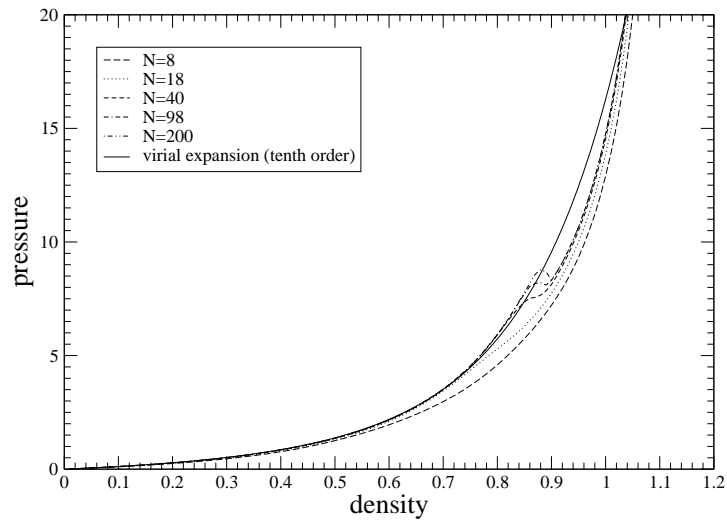

**Figure 5.1**. Pressure versus the density in the *N*-hard-disk model. The numerical data are given for $N = 8, 18, 40, 98, 400$ and are compared with the virial expansion (tenth order) for the fluid phase (continuous line).

It is well-known that hard-disk systems present a fluid-solid phase transition, as already observed in this thesis for the two-hard-disk model (see section 4.3.2). The equation of state is of van-der-Waals type. Due to the unstable character of the local decrease of pressure as a function of the density at the phase transition, thermodynamics imposes a Maxwell construction in order to avoid this instabilility and to determine the coexistence of fluid and solid phases in the range of densities $0.87 < n < 0.91$.

Today, the study of granular media presents more and more interest, not only at the theoretical level, but also for industrial perspectives (e.g. the pharmaceutical and agro-alimentary industries). For experiments on earth, gravity plays an important role on such systems and implies at equilibrium a pressure profile along the vertical axis. Recently, Luding and coworkers developed a *global equation of state* for such systems in order to study this pressure gradient (Luding, 1996; Luding and Strauß, 2001; Luding, 2002) which is of interest for our purposes. Indeed, This equation is derived for *non-dissipative*[4] hard-disks for all the densities. The idea is to combine a term for the pressure at low density with a term giving the pressure at high density. Hence the global equation of state may be written as

$$\frac{P}{Nk_BT} - 1 = P_f^* + m(v)(P_s^* - P_f^*) \tag{5.12}$$

where $v = \frac{\pi\sigma^2}{4}n$, and $m(v)$ is an empirical merging function

$$m(v) = \frac{1}{1 + \exp\left(\frac{v - v_c}{m_0}\right)}, \tag{5.13}$$

$v_c = 0.7006$ being the phase-transition density, and $m_0 = 0.0011$, both obtained numerically. For low densities $v \ll v_c$, Eq. (5.12) is reduced to the pressure given by the virial expansion. Luding considered the expression

$$P_f = 2v\left(\frac{1 - 7v/16}{(1 - v)^2} - \frac{v^3/16}{8(1 - v)^4}\right). \tag{5.14}$$

At the opposite, for high densities $v \gg v_c$, the *free-volume theory* allows one to write the high-density pressure $P_s$ as

$$P_s = \frac{c_0}{v_{\max} - v}h_3(v_{\max} - v) - 1 \tag{5.15}$$

where $c_0 = 1.8137$ is obtained numerically, and $h_3(x) = 1 + c_1x + c_3x^3$ is a fit polynomial with $c_1 = -0.04$ and $c_3 = 3.25$. $v_{\max}$ is the maximum packing density for the triangular lattice in which the hard disks are ordered at high densities. Consequently, we have got an expression of the equation of state for hard-disk systems, covering the whole range of density. In Fig. 5.2 we compare the results obtained in Fig. 5.1 to the analytical global equation of state (5.12). Our numerical data are in very good agreement with Eq. (5.12).

---

[4]In reality, the granular media, composed of *macroscopic* solid entities, present dissipation due to collision between particles. Here, the coefficient of restitution (or inelasticity) is equal to the unity, vanishing the dissipation.

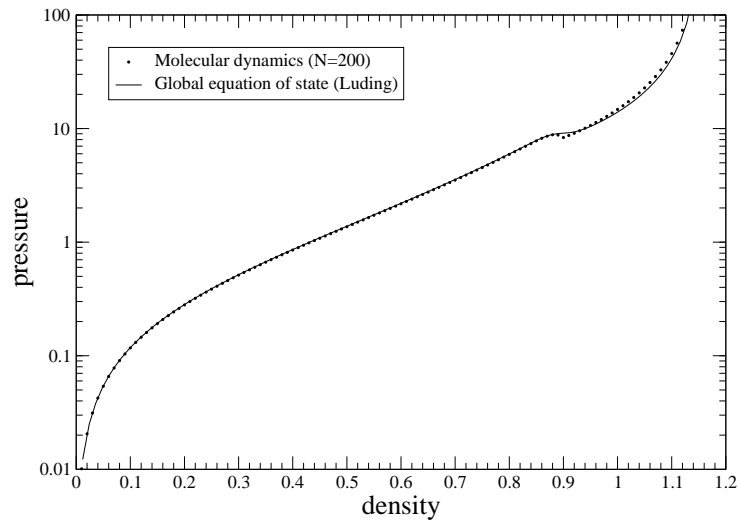

**Figure 5.2**. Pressure in *N*-hard-disk systems versus the density. Comparison between numerical data obtained by molecular dynamics for *N* = 200 and the *global equation of state* (5.12) given by Luding (Luding, 1996; Luding and Strauß, 2001; Luding, 2002).

As we have seen in section 1.6 the hard-sphere system has allowed physicists to explain and to predict important phenomena. However, in particular for the *real* equation of state, this model rapidly fails with increasing density. If the equation of state for hard-ball systems reproduces qualitatively the fluid-solid transition, the comparison with experimental data shows that the hard-ball model remains valid for dilute and moderately dense system. In this context, an impressive experimental work has been recently done by Brunner *et al.* (2003) measuring the equation of state of a system of colloidal particles confined in two dimensions. The hard-sphere potential of interaction between the charged colloids is assured by the addition of a salt. The particles are maintained on a plane thanks to a widened beam of laser directed from above into the sample cell, exerting a vertical pressure on the suspension. By computing pair correlation functions versus the density, they have been able to evaluate experimentally the equation of state of hard-disk system and obtained a good agreement with theoretical predictions (see Fig. 5.3).

### 5.2.3   The $N$-hard-sphere model

We also evaluate the pressure in function of the density in the $N$-hard sphere model. A van der Waals-type curve is also obtained. In Fig.5.4 is depicted the numerical results. The latter are compared to the pressure expressed by the virial expansion (5.11) for $d = 3$.

For high densities, we consider the theory developed by Hall (1970). According to Hall, the pressure in the solid phase of hard spheres may be expressed as

$$P = nk_BTz \tag{5.16}$$

where the factor $z$ is written as

$$
\begin{aligned}
z \;=\; & \frac{12 - 3\beta}{\beta} + 2.557696 + 0.1253077\beta + 0.1762393\beta^2 \\
& -1.053308\beta^3 + 2.818621\beta^4 - 2.921934\beta^5 + 1.118413\beta^6 \;.
\end{aligned}
\tag{5.17}
$$

$\beta$ is defined as $\beta = 4\left(1 - \frac{n}{n_0}\right)$ and $n_0$ is the density at the close packing ($n_0 = \frac{\sqrt{2}}{N\sigma^3}$). The comparison in Fig. 5.4 with the numerical computation of the equation of state in the high-density range shows

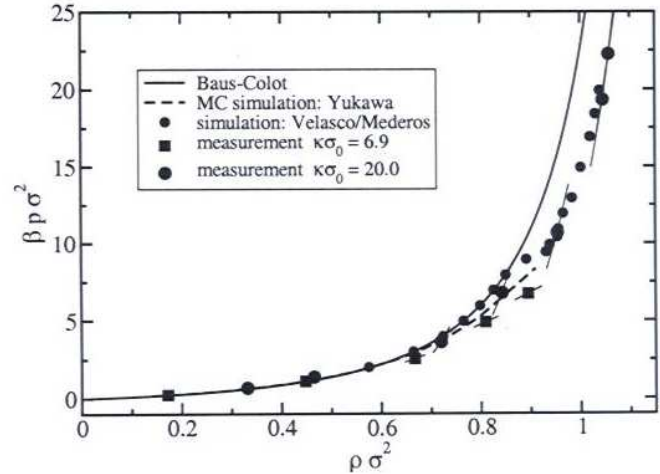

**Figure 5.3**. Experimental equation of state in a two-dimensional system composed of screened charged polystyrene colloids. Theoretical predictions for the fluid branch (Baus/Colot (Baus and Colot, 1986; Baus and Colot, 1987)) and the solid branch (Velasco/Mederos (Velasco and Mederos, 1997)) are compared with experimental data, measured in high-salt (filled circles) and low-salt (filled squares) colloidal suspensions. Error bars (solid short lines attached to the filled symbols) are inclined for reasons explained in the text. Monte Carlo data for a Yukawa fluid (dashed line) are provided to interpret the $\kappa\sigma_0 = 6.9$ measurement. Let us point out that $\kappa^{-1}$ is the screening length, $\sigma_0$ the actual particle diameter, $\sigma$ the effective hard-core diameter, and $\rho$ the particle density. This figure is taken from Ref. (Brunner et al., 2003).

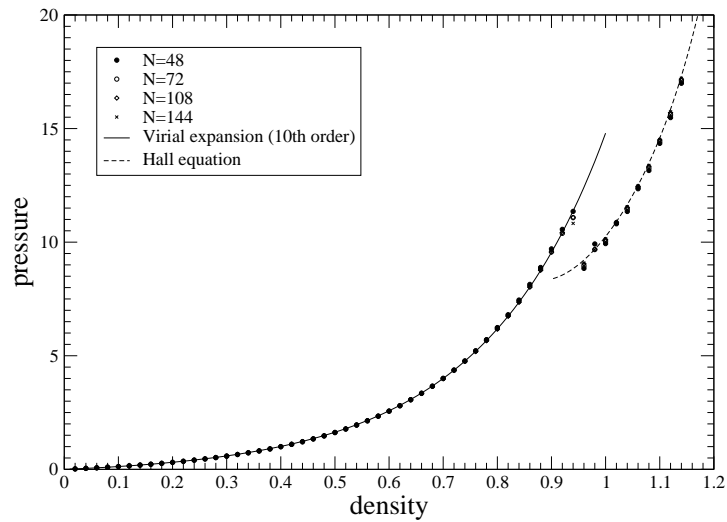

**Figure 5.4**. Pressure versus the density in the N-hard-sphere model. The numerical data are given for $N = 48, 72, 108, 144$ and are compared with theoretical predictions: the virial expansion (tenth order) for the fluid phase (continuous line); Hall equation for the solid phase (dashed line).

an excellent agreement.

As for hard-disk systems, a decade ago, an experimental group invented a method for measuring the equation of state considering colloidal particles behaving like true hard spheres (Rutgers et al., 1996). Indeed, in this experiment, polystyrene spheres are suspended in water. Electrostatic repulsion between particles stabilizes the suspension and an electrolyte in the concentration is added in order to screen the electrostatic interaction. Consequently the potential of interaction between the colloidal particles becomes precisely the hard-sphere potential.

The method they used to calculate the equation of state is based on an idea developed by Perrin in 1910. It consists in measuring the osmotic pressure of pollen particles by observing their sediments presenting a density profile. Considering different colloidal particles of various diameters, the experimental results are depicted in Fig. 5.5 and are compared to theoretical preditions for low and high densities[5]. The excellent agreement punctuates this beautiful experiment in the sense that the experimental conditions have been adapted in order to be compared to the theory whereas the latter is originally a relatively strong simplification of the reality.

---

[5]Here, the Carnahan-Starling equation of state (Carnahan and Starling, 1969) is considered for the fluid phase, whereas an approximation of the Hall equation is chosen for the high-density range.

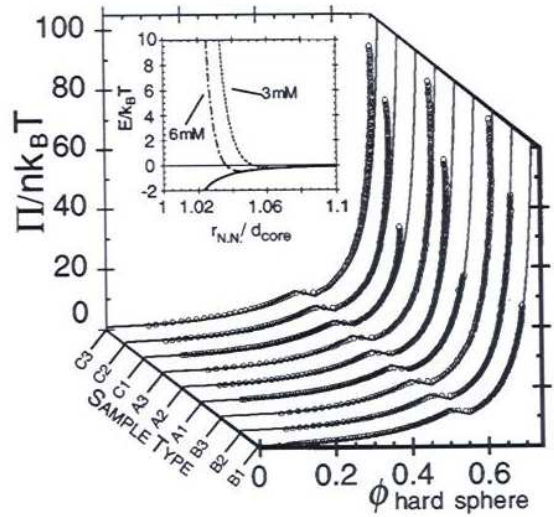

**Figure 5.5**. Experimental equation of state in a system composed of screened charged polystyrene colloids. Nine different samples are considered (three different diameters $\sigma$ and three different electrolytic solutions). The continuous line is the theoretical results, whereas the circles ○ correspond to the experimental data. The figure which is inside depicts the interparticle potential energy. The solid line gives the van der Waals attraction. The dotted lines correspond to the sum of van der Waals and electrostatic repulsion, marked with added electrolyte concentrations. This figure is taken from Rutgers et al. (1996).

## 5.3   Lyapunov spectrum

### 5.3.1   Introduction

The dynamics of hard-ball systems is well known to be chaotic. As we did in section 4.6.1, the Lyapunov exponents characterize this property. Since the relationship between statistical mechanics and the theory of dynamical systems is studied, the calculation of the Lyapunov spectrum in various systems has been done. The simpliest one is the well-known Sinai billiard or Lorentz gas for which the evaluation of the Lyapunov spectrum has been several times investigated in the past by Gaspard and Baras (1995), and Dellago and Posch (1995), as well as in section 4.6.1 of this thesis, since the dynamics of two disks with periodic boundary conditions may be reduced to the one of a particle moving in the Sinai billiard.

Here, we are interested in systems composed of $N$ hard balls. The Lyapunov spectrum is no longer composed of four exponents with only one positive exponent but of $2dN$ where $d$ is the dimension of the system. For each quantity conserved by the equations of motion one Lyapunov exponent vanishes. We therefore have $2d + 2$ vanishing exponents due to the conservation of the total energy, and the total momentum, the uniform motion of the center of mass, and the last one being associated with the displacement in the flow direction. The first numerical study of chaotic properties goes back to the end of the eighties with Posch and Hoover (1989) who considered soft repulsive interaction between particles. However, since hard-ball systems have played and still are playing an important role in the understanding of problems in statistical mechanics, their chaotic properties have already been studied numerically several times (Posch and Hoover, 1989; C. Dellago and Hoover, 1996; Dellago and Posch, 1996; Dellago and Posch, 1997; Milanovic et al., 1998; Posch and Hirschl, 2000), and more recently in (McNamara and Mareschal, 2001b; Mareschal and McNamara, 2004; de Wijn and van Beijeren, 2004; de Wijn, 2004; de Wijn, 2005; Forster, 2002; Forster et al., 2004; Posch and Forster, 2004; Eckmann et al., 2005). Extension to granular media has also been done (McNamara and Mareschal, 2001a).

### 5.3.2 Numerical method

To begin with the description of the numerical method, let us write the definition of the Lyapunov exponents

$$\lambda = \lim_{t \to \infty} \frac{1}{t} \ln \frac{\|\delta \mathbf{X}_t\|}{\|\delta \mathbf{X}_0\|} . \qquad (5.18)$$

Their computation is realized by calculating the evolution in time of the initial perturbation $\delta \mathbf{X}_0$ where $\delta \mathbf{X} = \{\delta \mathbf{r}_a, \delta \mathbf{v}_a\}_{a=1}^N$. The time evolution of this perturbation, basically consisting in a Taylor series of Eqs. (5.1)-(5.3), is given in Refs. (C. Dellago and Hoover, 1996; Gaspard and van Beijeren, 704).

1. *Free flight between binary collisions:*

$$t_{n-1} \to t_n : \begin{cases} \delta \mathbf{r}_a^{(-)}(t_n) = \delta \mathbf{r}_a^{(+)}(t_{n-1}) + (t_n - t_{n-1}) \, \delta \mathbf{v}_a^{(+)}(t_{n-1}) \\ \delta \mathbf{v}_a^{(-)}(t_n) = \delta \mathbf{v}_a^{(+)}(t_{n-1}) . \end{cases} \qquad (5.19)$$

2. *Binary collision:*

$$t_n : \begin{cases} \delta \mathbf{r}_a^{(+)} = \delta \mathbf{r}_a^{(-)} - (\epsilon_{ab} \cdot \delta \mathbf{r}_{ab}^{(-)}) \, \epsilon_{ab} \\ \delta \mathbf{r}_b^{(+)} = \delta \mathbf{r}_b^{(-)} + (\epsilon_{ab} \cdot \delta \mathbf{r}_{ab}^{(-)}) \, \epsilon_{ab} \\ \delta \mathbf{r}_k^{(+)} = \delta \mathbf{r}_k^{(-)} \qquad \forall \quad k \neq a, b \end{cases} \qquad (5.20)$$

$$t_n : \begin{cases} \delta\mathbf{v}_a^{(+)} = \delta\mathbf{v}_a^{(-)} - \left[ (\epsilon_{ab} \cdot \delta\mathbf{v}_{ab}^{(-)}) \, \epsilon_{ab} + (\delta\epsilon_{ab} \cdot \mathbf{v}_{ab}^{(-)}) \, \epsilon_{ab} + (\epsilon_{ab} \cdot \mathbf{v}_{ab}^{(-)}) \, \epsilon_{ab} \right] \\ \delta\mathbf{v}_b^{(+)} = \delta\mathbf{v}_b^{(-)} + \left[ (\epsilon_{ab} \cdot \delta\mathbf{v}_{ab}^{(-)}) \, \epsilon_{ab} + (\delta\epsilon_{ab} \cdot \mathbf{v}_{ab}^{(-)}) \, \epsilon_{ab} + (\epsilon_{ab} \cdot \mathbf{v}_{ab}^{(-)}) \, \epsilon_{ab} \right] \\ \delta\mathbf{v}_k^{(+)} = \delta\mathbf{v}_k^{(-)} \qquad \forall \quad k \neq a, b \end{cases} \tag{5.21}$$

where $\delta\epsilon_{ab}$ is defined as

$$\delta\epsilon_{ab} = \frac{1}{\sigma} \left( \delta\mathbf{r}_{ab}^{(-)} - \mathbf{v}_{ab}^{(-)} \frac{\epsilon_{ab} \cdot \delta\mathbf{r}_{ab}^{(-)}}{\epsilon_{ab} \cdot \mathbf{v}_{ab}^{(-)}} \right) \tag{5.22}$$

and

$$\begin{aligned} \delta\mathbf{r}_{ab}^{(-)} &= \delta\mathbf{r}_a^{(-)} - \delta\mathbf{r}_b^{(-)} \\ \delta\mathbf{v}_{ab}^{(-)} &= \delta\mathbf{v}_a^{(-)} - \delta\mathbf{v}_b^{(-)} \end{aligned} \tag{5.23}$$

### 5.3.3 Numerical results

Below, in Figs. 5.6 and 5.7 the Lyapunov spectra are depicted for systems of hard disks ($N = 40$) and of hard spheres ($N = 48$). We observe that the positive and negative branches of the Lyapunov spectrum are equal in magnitude which is the consequence of the pairing rule for Hamiltonian systems (Eckmann and Ruelle, 1985). In the hard-disk systems, it confirms that we have got six vanishing exponents, whereas eight exponents equal zero in the hard-sphere systems.

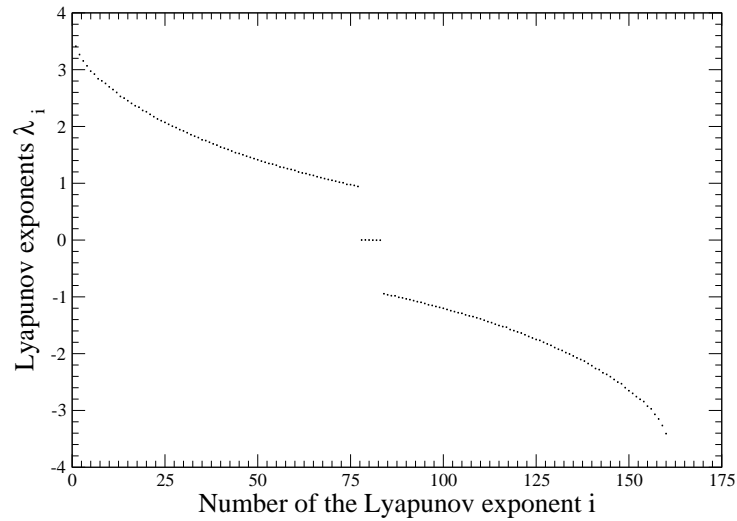

**Figure 5.6.** Lyapunov spectrum of a system composed of 40 hard disks (density $n = 0.5$). The exponents are sorted out in the decreasing order of the values of the exponents.

A decade ago, it has been observed that the shape of the Lyapunov spectra in fluid and solid phases are different (Dellago and Posch, 1996; Dellago and Posch, 1997). Indeed, the decreasing of the Lyapunov spectrum in the fluid phase is more abrupt than in the solid phase. We depict a comparison of the spectra for $n = 0.5$ (fluid phase) and $n = 1.1$ (solid phase) for a system composed of 40 hard disks and of 48 hard spheres in Fig. 5.8.

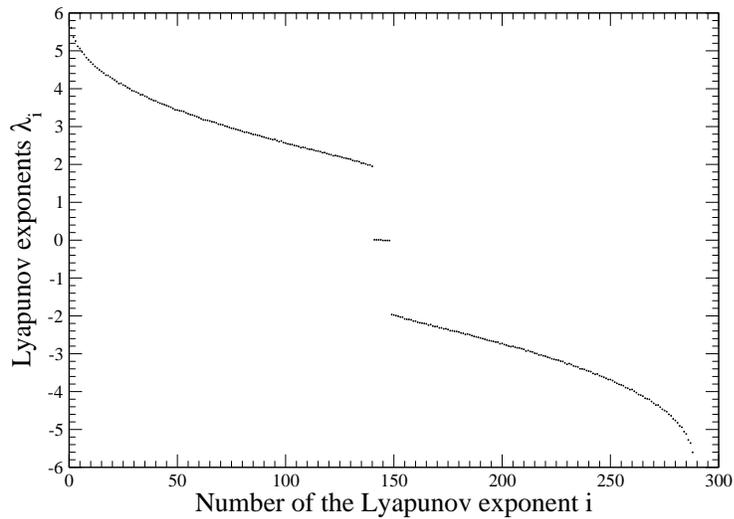

**Figure 5.7**. Lyapunov spectrum of a system composed of 48 hard spheres (density $n = 0.5$). The exponents are sorted out in the decreasing order of the values of the exponents.

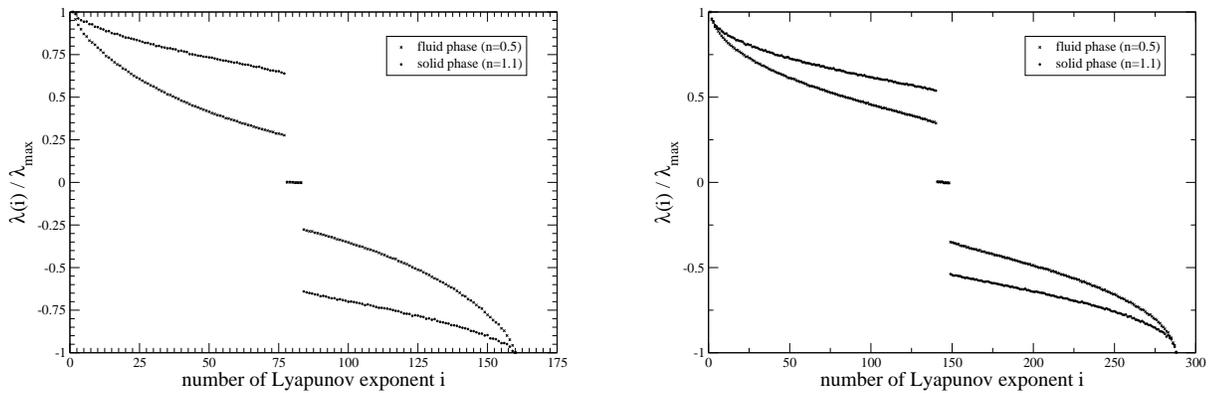

**Figure 5.8**. Comparison of the Lyapunov spectra in the fluid phase ($n = 0.5$) and in the solid phase ($n = 1.1$).On the left-hand side: system composed of 40 hard disks. On the right-hand side: system composed of 48 hard spheres.

By considering the maximum positive Lyapunov exponent of the whole spectrum (see Figs. 5.9 and 5.10), we can see the influence of the phase transition on the chaotic properties of the microscopic dynamics. In the case of hard-disk systems, we observe that the effect of the phase transition clearly appears only with increasing value of *N*, in much the same way as it is observed for the equation of state in Fig. 5.1. However, as shown by Dellago and Posch (1997), the drop is caused by the decrease of the collision rate at the transition and does not appear for the Lyapunov exponent per collision.

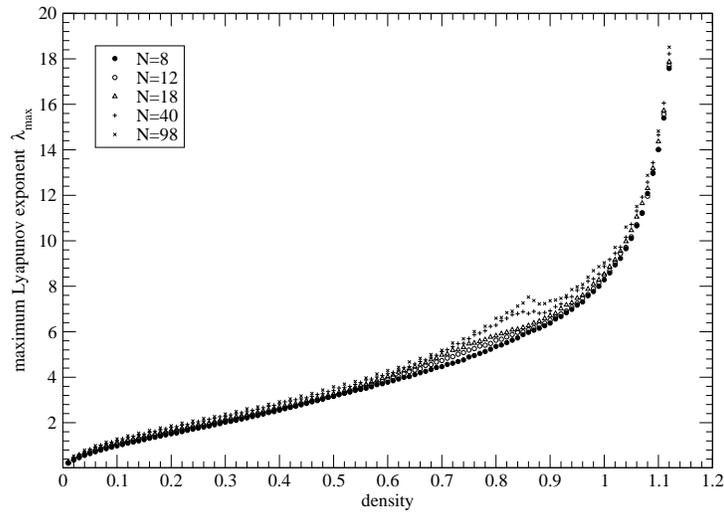

**Figure 5.9**. Maximum Lyapunov exponent versus density in $N$-hard-disk systems ($N = 8, 12, 18, 40, 98$).

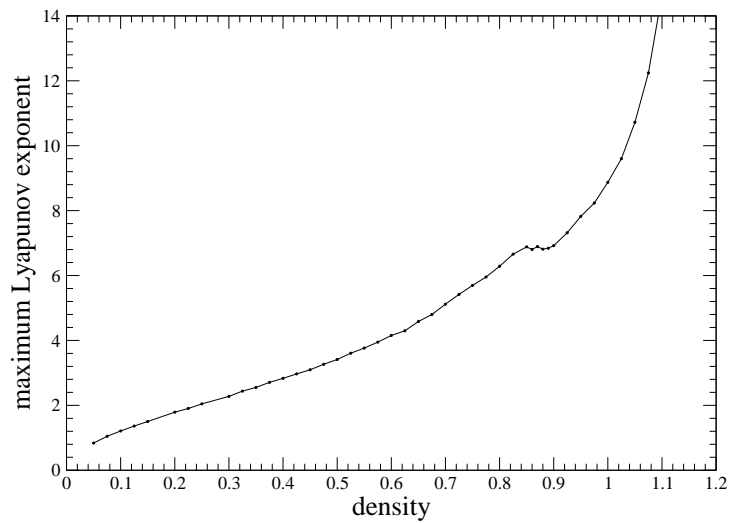

**Figure 5.10**. Maximum Lyapunov exponent versus density in a system composed of 40 hard disks. The influence of the phase transition on the Lyapunov exponent appears clearly.

## 5.4    Viscosity in *N*-hard-particle systems

### 5.4.1    The hydrodynamic modes

In the continuity of the previous chapter, this section is devoted to the viscosity in systems composed of $N$ hard disks or hard spheres ($N > 2$). The first studies of viscosity in such systems go back to the beginning of the seventies with Alder, Gass and Wainwright (Alder et al., 1970). As discussed in section 2.5.1, they used the Helfand relation instead of the usual Green-Kubo formula, since problems appear in the latter due to the singularity of the potential of interaction.

Firstly, let us consider a fluid in three dimensions. Let us rewrite in another form the hydrodynamic equations by adding the equation for the temperature $T$ obtained from the conservation of the energy, in addition to the equations for the mass density $\rho$ and the momentum density $\{\rho v_i\}$ (Balescu, 1975; Résibois and Leener, 1977)

$$
\begin{aligned}
\frac{\partial \rho}{\partial t} &= -\frac{\partial}{\partial r_j}\left(\rho v_j\right) \\[2mm]
\frac{\partial v_i}{\partial t} &= -v_j\frac{\partial v_i}{\partial r_j} - \frac{1}{\rho}\left(\frac{\partial P}{\partial \rho}\right)_T\frac{\partial \rho}{\partial r_i} - \frac{1}{\rho}\left(\frac{\partial P}{\partial T}\right)_\rho\frac{\partial T}{\partial r_i} + \frac{1}{\rho}\eta\frac{\partial^2 v_i}{\partial r_j\partial r_j} + \frac{1}{\rho}\left(\zeta + \frac{1}{3}\eta\right)\frac{\partial^2 v_j}{\partial r_i\partial r_j} \\[2mm]
\frac{\partial T}{\partial t} &= -\frac{T}{\rho c_V}\left(\frac{\partial P}{\partial T}\right)_\rho\frac{\partial v_j}{\partial r_j} - v_j\frac{\partial T}{\partial r_j} + \frac{\kappa}{\rho c_V}\frac{\partial^2 T}{\partial r_j\partial r_j} \\[2mm]
&\quad + \frac{\eta}{\rho c_V}\left(\frac{\partial v_j}{\partial r_i} + \frac{\partial v_i}{\partial r_j}\right)\frac{\partial v_j}{\partial r_i} + \frac{1}{\rho c_V}\left(\zeta - \frac{2}{3}\eta\right)\left(\frac{\partial v_j}{\partial r_j}\right)^2
\end{aligned}
\tag{5.24}
$$

where we have expanded the pressure $P$ in terms of the mass density $\rho$ and temperature $T$. If we consider the system to be close to equilibrium, the quantities $\rho$, $\{v_i\}$ and $T$ may be rewritten as

$$
\begin{aligned}
\rho(\mathbf{r}, t) &= \rho_{\text{eq}} + \delta\rho(\mathbf{r}, t) \\[1mm]
\mathbf{v}(\mathbf{r}, t) &= \mathbf{v}(\mathbf{r}, t) \\[1mm]
T(\mathbf{r}, t) &= T_{\text{eq}} + \delta T(\mathbf{r}, t) \ .
\end{aligned}
\tag{5.25}
$$

By replacing $\rho$, $\{v_i\}$ and $T$ by these expressions in Eqs. (5.24) and keeping only the linear terms, we

get

$$
\begin{aligned}
\frac{\partial}{\partial t}\,\delta\rho &= -\rho_{\mathrm{eq}}\,\frac{\partial v_j}{\partial r_j} \\
\frac{\partial v_i}{\partial t} &= -\frac{1}{\rho_{\mathrm{eq}}}\left(\frac{\partial P}{\partial \rho}\right)_T \frac{\partial\,\delta\rho}{\partial r_i} - \frac{1}{\rho_{\mathrm{eq}}}\left(\frac{\partial P}{\partial T}\right)_\rho \frac{\partial\,\delta T}{\partial r_i} \\
&\quad + \frac{\eta}{\rho}\frac{\partial^2 v_i}{\partial r_j\partial r_j} + \frac{1}{\rho}\left(\zeta + \frac{1}{3}\eta\right)\frac{\partial^2 v_j}{\partial r_i\partial r_j} \\
\frac{\partial}{\partial t}\,\delta T &= -\frac{T}{\rho c_V}\left(\frac{\partial P}{\partial T}\right)_\rho \frac{\partial v_j}{\partial r_j} + \frac{\kappa}{\rho c_V}\frac{\partial^2\,\delta T}{\partial r_j\partial r_j}\;.
\end{aligned}
\tag{5.26}
$$

This set of linear equations admits solutions of the form $b(\mathbf{r},t) = \exp\left(i\mathbf{q}\cdot\mathbf{r} + \lambda t\right)b_{\mathbf{q}}$ where $b$ is one of the above quantities. Therefore, Eqs. (5.27) become

$$
\begin{aligned}
\lambda\,\delta\rho_{\mathbf{q}} + i\rho_{\mathrm{eq}}q_j v_{\mathbf{q}j} &= 0 \\
\lambda v_{\mathbf{q}i} + i\frac{1}{\rho}\left(\frac{\partial P}{\partial \rho}\right)_T q_i\delta\rho_{\mathbf{q}} + i\frac{1}{\rho}\left(\frac{\partial P}{\partial T}\right)_\rho q_i\delta T_{\mathbf{q}} + \frac{\eta}{\rho}q^2 v_{\mathbf{q}i} + \frac{1}{\rho}\left(\zeta + \frac{1}{3}\eta\right)q_i q_j v_{\mathbf{q}j} &= 0 \\
\lambda\,\delta T_{\mathbf{q}} + i\frac{T}{\rho c_V}\left(\frac{\partial P}{\partial T}\right)_\rho q_j v_{\mathbf{q}j} + \frac{\kappa}{\rho c_V}q^2\,\delta T_{\mathbf{q}} &= 0\;.
\end{aligned}
\tag{5.27}
$$

By vanishing the characteristic determinant of this set of equations where we choose a particular direction to the vector $\mathbf{q}$, for instance the $x$-direction $\mathbf{q} = q\mathbf{1}_x$, one obtains the five roots associated with the *hydrodynamic modes*. The general solution of the set of equations (5.27) is therefore the superposition of five independent hydrodynamic modes, each describing a motion among the five variables. The roots are given in Table 5.2 and schematically represented in Fig. 5.11 where the sound speed $v_s = \sqrt{\frac{c_P}{c_V}\left(\frac{\partial P}{\partial \rho}\right)_T}$ is introduced. Hence we observe that the different modes are associated with different physical processes. The first two are associated with the propagation of sound, in which the two viscosity coefficients $\eta$ and $\zeta$ appear. Two degenerate modes describe the dissipation due to the shear viscosity $\eta$. And the fifth mode is associated with the heat diffusion through the presence of the heat conductivity coefficient $\kappa$.

| Sound mode | $\lambda_1 = iv_s q - \frac{1}{2\rho}\left[\left(\frac{1}{c_V} - \frac{1}{c_P}\right)\kappa + \frac{4}{3}\eta + \zeta\right]q^2$ |
|---|---|
| Sound mode | $\lambda_2 = -iv_s q - \frac{1}{2\rho}\left[\left(\frac{1}{c_V} - \frac{1}{c_P}\right)\kappa + \frac{4}{3}\eta + \zeta\right]q^2$ |
| Shear mode | $\lambda_3 = -\frac{\eta}{\rho}q^2$ |
| Shear mode | $\lambda_4 = -\frac{\eta}{\rho}q^2$ |
| Thermal mode | $\lambda_5 = -\frac{\kappa}{\rho c_P}q^2$ |

**Table 5.2**. The dispersion relations for the hydrodynamic modes in fluids.

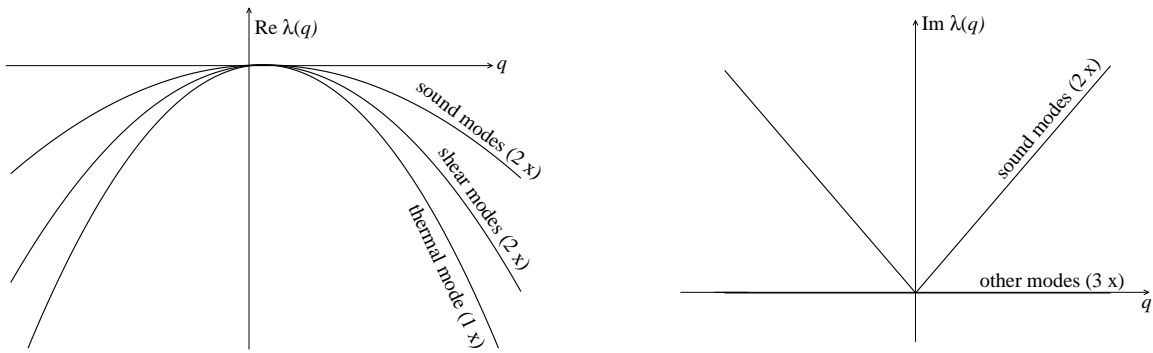

**Figure 5.11.** Schematic representation of the dispersion relation of the five modes in fluids: 2 degenerate sound modes; two degenerate shear modes; one thermal modes. On the left: the real part. On the right: the imaginary part.

In the particular case of the two viscosity coefficients, we observe that the shear viscosity coefficient $\eta$ appears alone in the two degenerate shear modes whereas the bulk viscosity $\zeta$ contributes to the damping of sound. This is precisely what we saw in section 1.3 in the experimental evidence and measuring of this transport coefficient.

In solids, the problem of the hydrodynamics received a particular interest since the seventies (Martin et al., 1972; Fleming III and Cohen, 1975). Such media sustain transport as well as in fluids. However, an important property appears since the continuous translation symmetries are henceforth broken in the solid state. Due to the *Goldstone theorem*, a new model corresponds to each broken symmetry (Martin et al., 1972). We therefore have eight independent modes in a three-dimensional system. Martin *et al.* showed that the *distorsion vector* (Landau and Lifschitz, 1986) must be treated as a hydrodynamic variable in addition to the conserved variables. It is well known that a crystal at equilibrium exhibits vacancies. In this context, Fleming *et al.* showed that diffusion of vacancies in solids has to be considered and appears precisely in an additional hydrodynamic mode (Fleming III and Cohen, 1975).

When the hydrodynamic equations for solids are solved, it is observed that the shear modes existing in fluids disappear. Indeed, the structure of the hydrodynamic modes is changed and we obtain two degenerate longitudinal sound modes, four degenerate transversal sound modes, one vacancy diffu-sion mode, and one thermal mode, as schematically depicted in Fig. 5.12. Hence the two shear modes in fluids no longer exist and the shear viscosity appears only in the sound modes. Since shear viscosity does no longer control the relaxation to equilibrium in shear modes but is mixed with other transport coefficients, it should be reinterpreted by considering the damping coefficients of the hydrodynamic modes in solids.

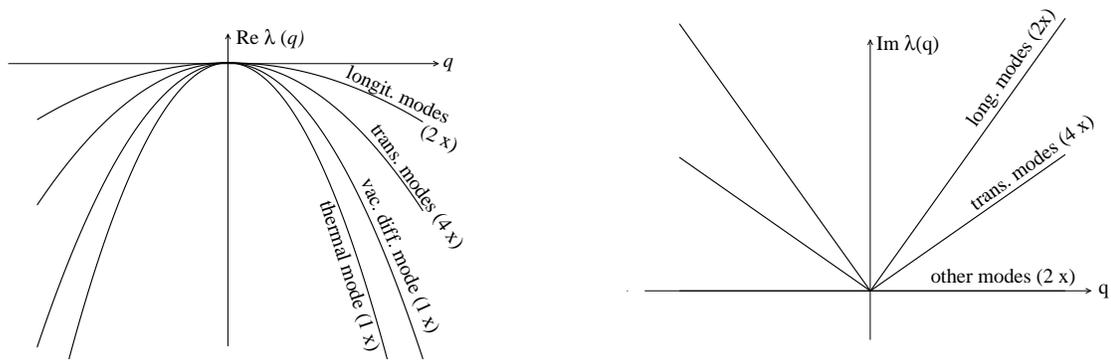

**Figure 5.12**. Schematic representation of the dispersion relation of the eight modes in solids: two degenerate longitudinal-sound modes; four degenerate transversal-sound modes; one vacancy diffusion mode; one thermal mode. On the left: the real part. On the right: the imaginary part.

### 5.4.2  Numerical results

Here, the viscosity is evaluated by considering unit mass ($m = 1$), diameter ($\sigma = 1$) and temperature ($T = 1$). The method of Alder *et al.* based on Eq. (2.58) is used. In the hard-disk model (see Fig. 5.13), we compute it for $N = 8, 40, 98, 200$. A comparison is done with the Enskog viscosity (4.46) already used in section 4.4

$$\eta = \eta_B \left( \frac{1}{Y} + 2\,y + 3.4916\,Y\,y^2 \right) , \tag{5.28}$$

where the Boltzmann viscosity is written as

$$\eta_B = \frac{1.022}{2\,\sigma} \sqrt{\frac{m\,k_B T}{\pi}} , \tag{5.29}$$

and $Y$ is the Enskog factor entering the equation of state as follows

$$P = n k_B T (1 + 2\,y\,Y) \tag{5.30}$$

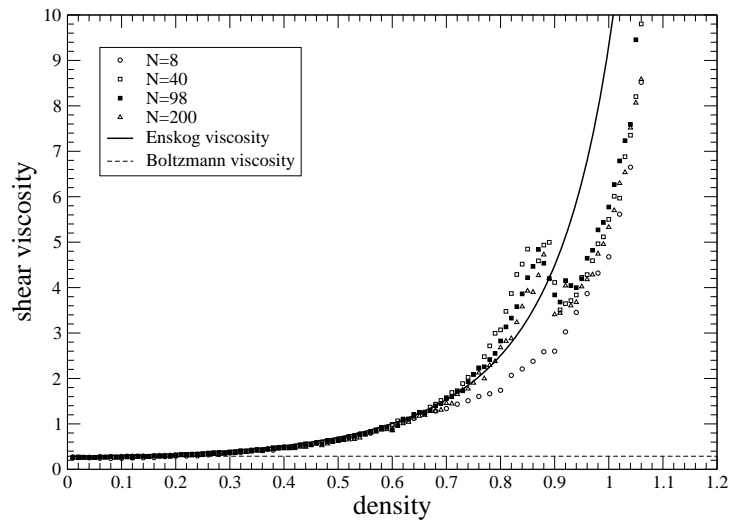

**Figure 5.13**. Shear viscosity in the *N*-disk model versus the density. Numerical results obtained by molecular dynamics ($N = 8, 40, 98, 200$) are compared to the Boltzmann viscosity (dashed line) for dilute-fluid range and to the Enskog viscosity (continuous line) for moderately dense fluid.

where $y = \pi\sigma^2 n/4$ and

$$Y = \frac{1 - \frac{7}{16} y}{(1 - y)^2} \, . \tag{5.31}$$

In Fig. 5.14, we zoom the range of small and moderate densities. We observe a good agreement with

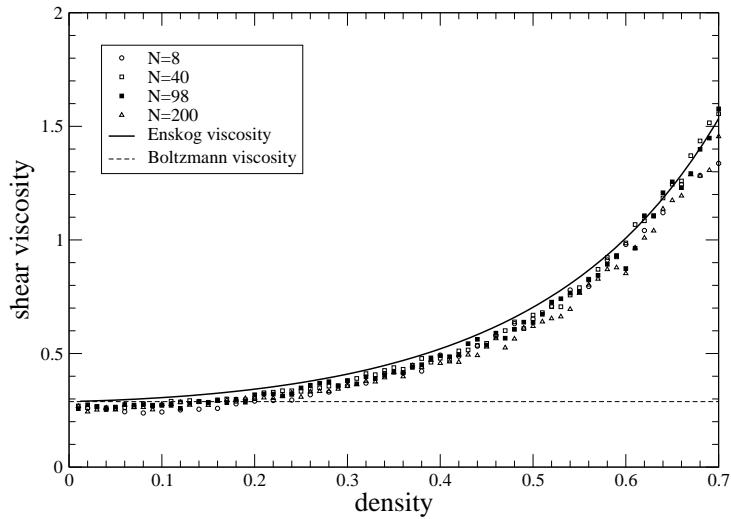

**Figure 5.14**. Zoom of Fig. 5.13 on the range of small and moderate densities.

Enskog's theory.

In the hard-sphere systems with $N = 32, 48, 72, 108$, viscosity is evaluated and depicted in Fig.

5.15. For the dilute and moderately dense gases (see Fig. 5.16), the results are compared with the

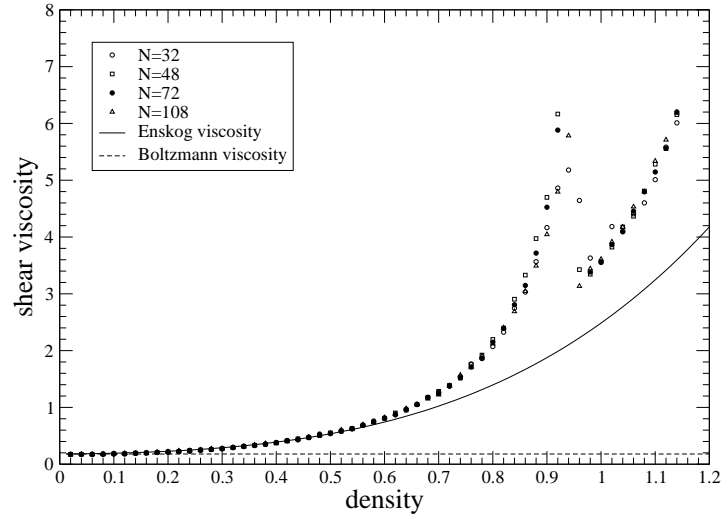

**Figure 5.15**. Shear viscosity in the *N*-hard-sphere model versus the density. Numerical results obtained by molecular dynamics ($N = 32, 48, 72, 108$) are compared to the Boltzmann viscosity (dashed line) for dilute-fluid range of density and to the Enskog viscosity (continuous line) for moderately dense fluid.

Enskog viscosity (1.14) previously written in section 1.4 as

$$\eta = \eta_B \left( \frac{1}{\chi} + \frac{4}{5} \, b_0 n + 0.7614 \, b_0^2 n^2 \chi \right) \tag{5.32}$$

with the Enskog factor expressed as

$$\chi = 1 + \frac{5}{8} b_0 n + 0.2869 \, (b_0 n)^2 \, , \tag{5.33}$$

$b_0 = \frac{2\pi\sigma^3}{3}$, and the Boltzmann viscosity (1.13) is given by

$$\eta_B = 1.0162 \, \frac{5}{16 \, \sigma^2} \, \sqrt{\frac{mk_BT}{\pi}} \, . \tag{5.34}$$

We observe that the viscosity is sensible to the phase transition. Indeed, whereas it increases monotonously in the dilute and moderately dense gases, near the phase transition, it decreases before increasing again in solid phase. The agreement is very good.

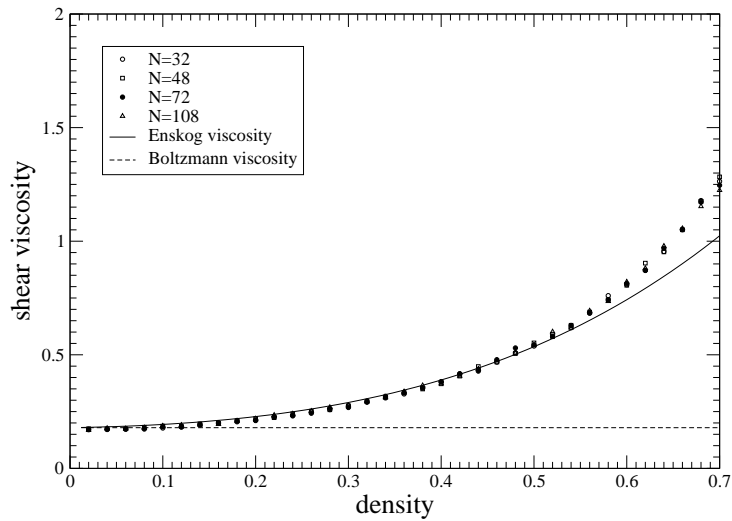

**Figure 5.16**. Zoom of Fig. 5.15 on the range of small and moderate densities.

## 5.5 Transport coefficients in the Lennard-Jones fluid

In chapter 2, we have proposed a new method for calculating the viscosity coefficients. It consists in an adaptation of the Helfand formula for periodic systems. In the literature, such a technique was considered to be unusable (Erpenbeck, 1995; Allen et al., 1994; Allen, 1993). The present section is devoted to the computation of the shear viscosity in a Lennard-Jones fluid by our Helfand-moment method based on Eq. (2.67). Furthermore, an analog derivation is given for the thermal conductivity. Its calculation by molecular dynamics is also achieved.

### 5.5.1 Integration algorithm

In 1967, Verlet proposed a powerful algorithm which involves the calculation of the current positions in terms of the positions at the two last time steps (Verlet, 1967). One of the advantages of this method is that it is of the third order, even if such terms are not computed. On the other hand, in the *Verlet method*, the computation of the velocities is not required. Since the velocities are necessary for the evaluation of the energy (one of the best test to verify that a molecular-dynamics simulation is proceeding correctly) as well as of the quantities of interest in this thesis, we have used the so-called *velocity Verlet scheme* (Swope et al., 1982). The positions and velocities are given by the following

algorithm

$$r_{ai}(t + \Delta t) = r_{ai}(t) + v_{ai}(t)\Delta t + \frac{1}{2}\alpha_{ai}(t)\Delta t^2$$
$$v_{ai}\left(t + \frac{\Delta t}{2}\right) = v_{ai}(t) + \frac{1}{2}\alpha_{ai}(t)\Delta t \ . \tag{5.35}$$

The acceleration $\alpha$ is simply obtained by the evaluation of the interparticle force with the other particles

$$\alpha_{ai}(t + \Delta t) = \frac{1}{m_a}\sum_{b \neq a} F_{abi} \ . \tag{5.36}$$

The velocities are finally recomputed as

$$v_{ai}(t + \Delta t) = v_{ai}\left(t + \frac{\Delta t}{2}\right) + \frac{1}{2}\alpha_{ai}(t + \Delta t)\Delta t \ . \tag{5.37}$$

In this section, we are dealing with systems of particles interacting through the Lennard-Jones potential. For numerical reasons (see section 2.4), the potential is truncated in a certain distance called the *cutoff distance*, and is then written as in Eq. (2.44)

$$V(r_{ab}) = \begin{cases} 4\epsilon\left[\left(\frac{\sigma}{r_{ab}}\right)^{12} - \left(\frac{\sigma}{r_{ab}}\right)^6\right] & r_{ab} \leq r_{\text{cutoff}} \\ 0 & r_{ab} > r_{\text{cutoff}} \end{cases} \tag{5.38}$$

where $\mathbf{r}_{ab} = \mathbf{r}_a - \mathbf{r}_b - \mathbf{L}_{b|a}$ defined in section 2.4 for systems with periodic boundary conditions.

In the two following sections devoted to the results obtained for the shear viscosity and the thermal conductivity, we consider the reduced quantities: $T^* = k_B T/\epsilon$, $\rho^* = \rho\sigma^3$, $t^* = t\sqrt{\epsilon/m}/\sigma$, $r^* = r/\sigma$ and $\eta^* = \eta\sigma^2/\sqrt{m\epsilon}$. Furthermore, we choose a cutoff radius $r^* = 2.5$.

### 5.5.2 Shear viscosity

Shear viscosity is probably the transport coefficient which has been most studied in numerical calculations. The principal method applied for this purpose in soft-potential systems is the standard *Green-Kubo* formula (2.23) that Levesque *et al.* already used in the seventies for computing viscosity in the Lennard-Jones liquid near its triple point (Levesque et al., 1973), and that was used several times since then (Erpenbeck, 1988; Heyes, 1988). Besides, nonequilibrium methods simulating a Couette flow were also used since the seventies (Ashurst and Hoover, 1973; Ashurst and Hoover, 1975; Hoover et al., 1980). More recently, different works have employed the Einstein-like formula for viscosity in periodic systems (Hess and Evans, 2001; Hess et al., 2003; Meier et al., 2004). However, instead of considering the variance of the Helfand moment (as proposed in chapter 2), they simply computed the integral of the current.

In this section, we study the shear viscosity by the Einstein-like formula in systems with periodic boundary conditions with our expression (2.67) for the Helfand moment

$$G_{ij}(t) = \sum_{a=1}^{N} p_{ai}(t)\, r_{aj}(t) - \sum_{a=1}^{N} \sum_{s} p_{ai}^{(s)}\, \Delta r_{aj}^{(s)}\, \theta(t - t_s) - \frac{1}{2} \sum_{a=1}^{N} \sum_{b \neq a} \int_0^t d\tau\, F_i(\mathbf{r}_a - \mathbf{r}_b - \mathbf{L}_{b|a})\, L_{b|aj}\ . \quad (5.39)$$

In order to test our Helfand-moment method for Lennard-Jones systems, we consider a state point ($T^* = 0.722$, $\rho^* = 0.8442$) close to the triple point. In Table 5.3, we compare the results obtained by our method with those obtained by the standard Green-Kubo method.

| Number of particles $N$ | Temperature $T^*$ | Shear viscosity $\eta_H^*$ | Shear viscosity $\eta_{GK}^*$ |
|---|---|---|---|
| 108 | 0.71627 | 3.05 | 2.98 |
| 256 | 0.72229 | 3.18 | 3.15 |

**Table 5.3**. Simulation data for the shear viscosity at the state point ($T^* = 0.722$, $\rho^* = 0.8442$) close to the triple point of the Lennard-Jones fluid. The cutoff radius is $r^* = 2.5$. We compare the results $\eta_H^*$ obtained by our method with our own data $\eta_{GK}^*$ given by the Green-Kubo method.

We observe that our method is in good agreement with the data obtained by the standard Green-Kubo method as well as with Erpenbeck (1988): $\eta^* = 3.345$ given by the extrapolation of the data for the viscosity to infinite system size ($N^{-1} \to 0$). We may therefore conclude that our Helfand-moment method based on Eq. (5.39) is valid.

### 5.5.3 Thermal conductivity

**Helfand-moment method**

Since the first numerical studies of transport properties, several works have been devoted in particular to the thermal conductivity $\kappa$. As for viscosity, the Green-Kubo method has been a standard technique used in soft-potential systems like Lennard-Jones systems (Voegelsang et al., 1987; Sharma et al., 1995). On the other hand, thermal conductivity have been often computed through nonequilibrium molecular dynamics (Evans, 1982; Massobrio and Ciccotti, 1984; Evans, 1986; Paolini et al., 1986).

We here propose a *Helfand-moment method* for the thermal conductivity in systems with periodic boundary conditions. As for viscosity, Helfand proposed a quantity associated with thermal conductivity $\kappa$ in order to express the transport coefficient in terms of the variance of this new quantity written as

$$G_i^{(\kappa)}(t) = \sum_{a=1}^{N} r_{ai} \left( E_a - \langle E_a \rangle \right) \tag{5.40}$$

where the energy of the particle $a$ is defined as $E_a = \frac{p_a^2}{2m} + \sum_{b \neq a} \frac{1}{2} V_{ab}$.

However, the periodicity of the dynamics affects the expression of the so-called *Helfand moment*. Following the same derivation than for viscosity in section 2.5.2, $G_i^{(\kappa)}(t)$ must be modified by adding a quantity $I_i(t)$. The time derivative of the modified Helfand moment is

$$\frac{dG_i^{(\kappa)}(t)}{dt} = \sum_{a=1}^{N} \frac{dr_{ai}}{dt} \left( E_a - \langle E_a \rangle \right) + \sum_{a=1}^{N} r_{ai} \frac{dE_a}{dt} + \frac{dI(t)}{dt} \ . \tag{5.41}$$

The time derivative of $E_a$ is given by

$$\begin{aligned}
\frac{dE_a}{dt} &= \frac{\mathbf{p}_a}{m} \cdot \frac{d\mathbf{p}_a}{dt} + \frac{1}{2} \sum_{b \neq a} \frac{\partial V_{ab}}{\partial \mathbf{r}_{ab}} \cdot \frac{\mathbf{p}_a - \mathbf{p}_b}{m} \\
&= \sum_{b \neq a} \frac{\mathbf{p}_a}{m} \cdot \mathbf{F}(\mathbf{r}_a - \mathbf{r}_b - \mathbf{L}_{b|a}) - \frac{1}{2} \sum_{b \neq a} \frac{\mathbf{p}_a - \mathbf{p}_b}{m} \cdot \mathbf{F}(\mathbf{r}_a - \mathbf{r}_b - \mathbf{L}_{b|a}) \\
&= \frac{1}{2} \sum_{b \neq a} \frac{\mathbf{p}_a + \mathbf{p}_b}{m} \cdot \mathbf{F}(\mathbf{r}_a - \mathbf{r}_b - \mathbf{L}_{b|a})
\end{aligned} \tag{5.42}$$

where $\mathbf{L}_{b|a}$ has been defined in Eq. (2.45). We notice that there is here no jump in position to consider because $\frac{d\mathbf{r}_{ab}}{dt}$ concerns a relative position $\mathbf{r}_{ab} = \mathbf{r}_a - \mathbf{r}_b - \mathbf{L}_{b|a}$ which satisfies the minimum image convention within the range of the force. Symmetrizing the second term in Eq. (5.41), it is hence written as

$$
\begin{aligned}
\sum_{a=1}^{N} r_{ai} \frac{dE_a}{dt} &= \frac{1}{2} \sum_{b \neq a} r_{ai} \frac{\mathbf{p}_a + \mathbf{p}_b}{m} \cdot \mathbf{F}(\mathbf{r}_a - \mathbf{r}_b - \mathbf{L}_{b|a}) \\
&= \frac{1}{4} \sum_{a=1}^{N} \sum_{b \neq a} r_{ai} \frac{\mathbf{p}_a + \mathbf{p}_b}{m} \cdot \mathbf{F}(\mathbf{r}_a - \mathbf{r}_b - \mathbf{L}_{b|a}) \\
&\quad + \frac{1}{4} \sum_{b=1}^{N} \sum_{a \neq b} r_{bi} \frac{\mathbf{p}_a + \mathbf{p}_b}{m} \cdot \mathbf{F}(\mathbf{r}_b - \mathbf{r}_a - \mathbf{L}_{a|b})
\end{aligned} \tag{5.43}
$$

$$
\begin{aligned}
\sum_{a=1}^{N} r_{ai} \frac{dE_a}{dt} &= \frac{1}{4} \sum_{a=1}^{N} \sum_{b \neq a} (r_{ai} - r_{bi} - L_{b|ai}) \frac{\mathbf{p}_a + \mathbf{p}_b}{m} \cdot \mathbf{F}(\mathbf{r}_a - \mathbf{r}_b - \mathbf{L}_{b|a}) \\
&\quad + \frac{1}{4} \sum_{a=1}^{N} \sum_{b \neq a} L_{b|ai} \frac{\mathbf{p}_a + \mathbf{p}_b}{m} \cdot \mathbf{F}(\mathbf{r}_a - \mathbf{r}_b - \mathbf{L}_{b|a})
\end{aligned} \tag{5.44}
$$

where $\mathbf{F}(\mathbf{r}_a - \mathbf{r}_b - \mathbf{L}_{b|a}) = -\mathbf{F}(\mathbf{r}_b - \mathbf{r}_a - \mathbf{L}_{a|b})$ according to Newton's third law. Substituting this expression in Eq. (5.41), where $\frac{dr_{ai}}{dt}$ is given by modified Newton's equations of motion (2.46), we get the expression

$$
\begin{aligned}
\frac{dG_i^{(\kappa)}(t)}{dt} &= \sum_{a=1}^{N} \frac{p_{ai}}{m} (E_a - \langle E_a \rangle) + \sum_{a=1}^{N} \sum_{s} \Delta r_{ai}^{(s)} (E_a - \langle E_a \rangle) \delta(t - t_s) \\
&\quad + \frac{1}{4} \sum_{a=1}^{N} \sum_{b \neq a} (r_{ai} - r_{bi} - L_{b|ai}) \frac{\mathbf{p}_a + \mathbf{p}_b}{m} \cdot \mathbf{F}(\mathbf{r}_a - \mathbf{r}_b - \mathbf{L}_{b|a}) \\
&\quad + \frac{1}{4} \sum_{a=1}^{N} \sum_{b \neq a} L_{b|ai} \frac{\mathbf{p}_a + \mathbf{p}_b}{m} \cdot \mathbf{F}(\mathbf{r}_a - \mathbf{r}_b - \mathbf{L}_{b|a}) + \frac{dI(t)}{dt} \ .
\end{aligned} \tag{5.45}
$$

On the other hand, it is well known that the current for the thermal conductivity $J_i^{(\kappa)}$ is given by

$$
J_i^{(\kappa)}(t) = \sum_{a=1}^{N} \frac{p_{ai}}{m} (E_a - \langle E_a \rangle) + \frac{1}{4} \sum_{a=1}^{N} \sum_{b \neq a} (r_{ai} - r_{bi} - L_{b|ai}) \frac{\mathbf{p}_a + \mathbf{p}_b}{m} \cdot \mathbf{F}(\mathbf{r}_a - \mathbf{r}_b - \mathbf{L}_{b|a}) \ . \tag{5.46}
$$

Since $\frac{dG_i^{(\kappa)}(t)}{dt} = J_i^{(\kappa)}(t)$ by definition, we then obtain that

$$\frac{dI(t)}{dt} = -\sum_{a=1}^{N} \sum_{s} \Delta r_{ai}^{(s)} \left(E_a - \langle E_a \rangle\right) \delta(t - t_s) - \frac{1}{4} \sum_{a=1}^{N} \sum_{b \neq a} L_{b|ai} \frac{\mathbf{p}_a + \mathbf{p}_b}{m} \cdot \mathbf{F}(\mathbf{r}_a - \mathbf{r}_b - \mathbf{L}_{b|a}) \ . \quad (5.47)$$

Finally, the quantity to be added to the usual Helfand moment (5.40) is

$$I(t) = -\sum_{a=1}^{N} \sum_{s} \Delta r_{ai}^{(s)} \left(E_a - \langle E_a \rangle\right) \theta(t - t_s) - \frac{1}{4} \sum_{a=1}^{N} \sum_{b \neq a} \int_0^t d\tau \, L_{b|ai} \frac{\mathbf{p}_a + \mathbf{p}_b}{m} \cdot \mathbf{F}(\mathbf{r}_a - \mathbf{r}_b - \mathbf{L}_{b|a}) \ . \quad (5.48)$$

We conclude that the good expression for the Helfand moment which should be used in periodic systems is

$$\begin{aligned} G_i^{(\kappa)}(t) &= \sum_{a=1}^{N} r_{ai} \left(E_a - \langle E_a \rangle\right) - \sum_{a=1}^{N} \sum_{s} \Delta r_{ai}^{(s)} \left(E_a - \langle E_a \rangle\right) \theta(t - t_s) \\ &\quad - \frac{1}{4} \sum_{a=1}^{N} \sum_{b \neq a} \int_0^t d\tau \, L_{b|ai} \frac{\mathbf{p}_a + \mathbf{p}_b}{m} \cdot \mathbf{F}(\mathbf{r}_a - \mathbf{r}_b - \mathbf{L}_{b|a}) \ . \end{aligned} \quad (5.49)$$

rather than merely Eq. (5.40)

### Numerical results

As in the case of the shear viscosity, we here calculate the thermal conductivity for a state point ($T^* = 0.722$, $\rho^* = 0.8442$) close to the triple point in the Lennard-Jones fluid. In Table 5.4, we give the results $\kappa_H^*$ obtained by our Helfand-moment method for systems with periodic boundary conditions, which is based on the Helfand moment written in Eq. (5.40). We compare these results to those that we obtained by the standard Green-Kubo method ($\kappa_{GK}^*$). Both methods are in good agreement between themselves and also with the data found in literature: in the same conditions, Massobrio and Ciccotti obtained $\kappa^* = 6.87$ (Massobrio and Ciccotti, 1984). We may therefore conclude that our Helfand-moment method based on Eq. (5.49) is valid.

| Number of particles $N$ | Temperature $T^*$ | Thermal conductivity $\kappa_H^*$ | Thermal conductivity $\kappa_{GK}^*$ |
|---|---|---|---|
| 108 | 0.71627 | 6.84 | 6.72 |
| 256 | 0.72229 | 6.59 | 6.54 |

**Table 5.4**. Simulation data for the thermal conductivity at the state point ($T^* = 0.722$, $\rho^* = 0.8442$) close to the triple point of the Lennard-Jones fluid. The cutoff radius is $r^* = 2.5$. We compare the results $\kappa_H^*$ obtained by our method with our own data $\kappa_{GK}^*$ given by the Green-Kubo method.

## 5.6   Conclusions

The main goal of this chapter has been to extend the study done in the chapter 3 to the dynamical and nonequilibrium properties of *N*-particle systems in two and three dimensions for $N > 2$.

First, the equation of state of hard-disk and hard-sphere systems have been computed. A phase transition of fluid-solid type has been observed and an excellent agreement has been obtained with the virial expansion to the tenth order (Clisby and McCoy, 2005) for the fluid phase. Furthermore, for the hard-disk systems, a comparison has also been done with a global equation of state valid for both phases (Luding, 1996; Luding and Strauß, 2001; Luding, 2002). The agreement is also excellent.

It is well known that the dynamics of hard balls is chaotic. Contrary to the case of the two-hard-disk model, the Lyapunov spectrum is composed of several positive exponents. In this chapter, we have computed this spectrum in the hard-disk and hard-sphere systems.

The viscosity has been computed in these hard-ball systems and the agreement with Enskog's theory is very good for dilute and moderately dense gases. The viscosity has been obtained for both phases. However, as mentioned in section 5.4, the structure of the hydrodynamic modes is modified in solids due to the broken symmetries. In particular, the shear modes existing in fluids disappear and the viscosity therefore does no longer control the relaxation of such modes. On the other hand, this transport coefficient only appears in sound modes. Hence it suggests to reinterpret the role of viscosity in terms of damping coefficients in the hydrodynamic modes in solids.

The study of the relationship established by the escape-rate formalism is being the most important perspective of this thesis in the context of the nonequilibrium statistical mechanics and its relationships with the dynamical systems theory. The main problem is how to compute the partial fractal dimensions associated with the positive Lyapunov exponents. The works done by Hunt et al. (1996) and Sweet and Ott (2000) should be a starting point for establishing such an algorithm. On the other hand, this problem does not concern only the viscosity but also the other transport properties, such as the coefficients of diffusion, heat conductivity and charge conductivity.

**Chapter 6**

# Conclusions and perspectives

This thesis has been mainly devoted to the viscosity properties in hard-ball systems. We proposed a method, based on Helfand's formula, which may be applied to any systems with periodic boundary conditions. As an irreversible process, the study of the viscosity properties and their relationships with the microscopic chaos gives one a better understanding of the emergence of irreversibility from the underlying reversible microscopic dynamics.

## 6.1   Overview of the results

In this work, we have dealt with periodic systems. In this context, we have proposed a new method for the computation of the viscosity (Viscardy and Gaspard, 2003a). This method is based on the Helfand formula (Helfand, 1960) in which the viscosity coefficient is expressed in terms of the variance of the so-called *Helfand moment*

$$\eta_{ij,kl} = \lim_{t\to\infty} \frac{\beta}{2tV} \left[ \langle G_{ij}(t)G_{kl}(t)\rangle - \langle G_{ij}(t)\rangle\langle G_{kl}(t)\rangle \right] \ . \tag{6.1}$$

By analogy with diffusion, this quantity evolves in the space of the Helfand moment as a Brownian particle in the physical space.

Previously, the main method to calculate the transport coefficients was the Green-Kubo method implying the computation of integrals of time autocorrelation functions (Green, 1951; Green, 1960; Kubo, 1957; Mori, 1958a). Besides this, Alder *et al.* (1970) proposed an expression for the transport coefficients in the particular case of periodic systems composed of hard balls. More precisely, this method is an intermediate between the Helfand and the Green-Kubo methods.

Here, we have derived a general formula for the viscosity coefficients which may be applied to periodic systems with any potential of interaction between the particles. This derivation implies the addition of a term to the original expression of the Helfand moment appearing in the Einstein-like equation for the viscosity, this added term taking into account the periodicity of the system (or more precisely, the usual *minimum-image convention*), and hence, we have obtained our expression (2.67)

$$G_{ij}(t) = \sum_{a=1}^{N} p_{ai}(t) \, r_{aj}(t) - \sum_{a=1}^{N} \sum_{s} p_{ai}^{(s)} \, \Delta r_{aj}^{(s)} \, \theta(t - t_s) - \frac{1}{2} \sum_{a=1}^{N} \sum_{b \neq a} \int_{0}^{t} d\tau \, F_i(\mathbf{r}_a - \mathbf{r}_b - \mathbf{L}_{b|a}) \, L_{b|aj} \, . \quad (6.2)$$

We have shown in Appendix B that our method is completely equivalent to the Green-Kubo formula. Furthermore, this new method presents interesting advantages. First, it may be used for any periodic systems of particles, contrary to the Alder *et al*'s method which is restricted to the hard-ball potential. Secondly, the Helfand-moment method is numerically more efficient than the Green-Kubo method. Indeed, in the Helfand-moment method, the computation of the viscosity is realized through the accumulation during the simulation of the variance of the Helfand moment $G_{ij}(t)$ instead of the computation of the time integral of an autocorrelation function.

We have also discussed the validity of our method with respect to the literature of this topic. We first have remarked that the origin of the so-called *McQuarrie formula* for shear viscosity (McQuarrie, 2000) could be due to a simple typing error. Whereas the original Helfand-moment formula implies the evaluation of viscosity through a *collective* quantity, that is the so-called Helfand moment (2.52), this formula reduced the computation to the evolution of *a single particle* by a simple typographic exchange of a sum and a bracket[1]. However, it is easy to be convinced that a typing error is at the origin of this. Furthermore, the discussions in the literature (Chialvo and Debenedetti, 1991; Chialvo et al., 1993; Allen et al., 1994; Allen, 1994) about this formula showed that the latter is not equivalent to the original Helfand expression.

As regards our method, a more important problem has been discussed in the literature (Erpenbeck, 1995; Allen et al., 1994; Allen, 1993) in which it seemed to be concluded that an expression for the Helfand moment for periodic systems cannot be derived. In this context, we think that our present work overcomes the previous difficulties and therefore can have a particular interest in the future for its simplicity, efficiency and generality.

The main model we have used is the simplest model in which viscosity has been proved to exist (Bunimovich and Spohn, 1996). It is composed of only two hard disks per cell. Since the dynamics of the disks is the same in the different cell, only one cell may be considered if we impose periodic boundary conditions. On the other hand, it is well known that the dynamics of these two disks may be reduced to the one of a pointlike particle moving in the well-known *Sinai billiard*.

---

[1] Eq. (3.13) in Helfand (1960)'s paper and Eq. (21-304) in McQuarrie (1976)'s book.

We have considered two different lattices: the hexagonal and the square geometries. Both of them present two ranges of density separated by a critical density. The two phases may be called the *fluid* and the *solid* phases. Previously, Bunimovich and Spohn considered only the square geometry. In this thesis, we have extended the model to the hexagonal geometry. This presents some advantages compared to the square case. Indeed, the viscosity is proved to exist in the fluid phase (in the finite-horizon regime) as well as in the solid phase, whereas viscosity is found only in the latter. Furthermore, we have shown that the four-order viscosity tensor is *isotropic* on the hexagonal lattice, contrary to the situation on the square lattice. This implies that the viscosity tensor may be reduced to the usual two viscosity coefficients: the *shear viscosity* $\eta$ and the *bulk viscosity* $\zeta$. On the other hand, the three elements $\eta_{xx,xx}$, $\eta_{xx,yy}$ and $\eta_{xy,xy}$ of the viscosity tensor have to be taken into account and do depend on the chosen direction.

First, we have studied some properties of the model. By considering the treatment of the diffusion in the Lorentz gas by Machta and Zwanzig (1983), analytic expressions for the mean free path has been obtained for both geometries and thermodynamic regimes and have been compared with numerical results. This agreement has been shown to be excellent. Furthermore, this quantity put in evidence the phase transition at the critical density in both geometries. We have also studied the pressure, which is the usual quantity considered for showing a phase transition. Since it may be expressed in terms of the mean free path in such systems, pressure might be obtained analytically and these results agree very well with the numerical data obtained by simulation. The pressure presents a maximum at the critical density, and thereafter, decreases until reaching a minimum. Then it increases again monotonously in the solid phase. It thus involves a negative compressibility in the critical density range so that the system is thermodynamically unstable. It is here usual to consider the Maxwell construction which leads to the appearance of a coexistence between the fluid and the solid phases. Nevertheless, by making a comparison with the pressure in a square box defined by hard walls (system in which the same "phases" also exist), it has been shown that the origin of such a van-der-Waals-type curve is due to the particular dynamics observed in the periodic version of the two-hard-disk model.

We have studied the viscosity properties in the two-hard-disk model in both geometries. To do that, we have used our method (6.2) based on the Helfand formula. In the two-hard-disk model, the last term vanishes. In the hexagonal lattice, we have computed the coefficients of shear and bulk viscosities and have shown that the hexagonal lattice is isotropic by checking the relation $\eta_{xy,xy} = \frac{1}{2}\left(\eta_{xx,xx} - \eta_{xx,yy}\right)$ implying the dependence of the third viscosity tensor element on the two others. As previously recalled, our method is equivalent to the Alder *et al*'s method in periodic systems of hard balls. We have successfully checked this equivalence between both methods. Finally we compared the calculated shear viscosity with the Enskog viscosity. It is astonishing to observe the agreement, although the system we consider is composed of only two particles. In the square lattice, the three elements $\eta_{xx,xx}$, $\eta_{xx,yy}$ and $\eta_{xy,xy}$ are independent of each other. However, according to the direction in which the viscosity tensor is defined, the elements change in magnitude and become a linear combination of the elements obtained when another referential is considered (here, after a rotation of an angle $\phi$). We have therefore computed these viscosity tensor elements and have observed that the data obtained for $\phi = \pi/4$ agree very well with the values calculated by linear combinations of the elements for $\phi = 0$.

In the hexagonal lattice, it has been found that the shear viscosity presents a divergence at the phase transition, and decreases monotonously in the solid phase whereas the bulk viscosity increases monotonously and diverges in the solid phase, at the close-packing density. In the square lattice, according to the chosen referential, the three elements present or do not present divergence at the phase transition or close-packing density. We have then given a qualitative explanation of these different behaviors of the viscosity properties in terms of the evolution of the Helfand moment along the trajectories drawn by the pointlike particle in the Sinai billiard.

One of the main purposes of this thesis is to study the relationships between the viscosity as a transport coefficient characterizing a macroscopic process, and chaotic properties of the microscopic dynamics. A decade ago, such a relationship has been established by Dorfman and Gaspard (Dorfman and Gaspard, 1995; Gaspard and Dorfman, 1995) in the context of the *escape-rate formalism* (Gaspard, 1998; Gaspard and Nicolis, 1990). This formalism introduces nonequilibrium conditions on the system without changing the Hamiltonian character of the microscopic dynamics, contrary to the thermostated-system approach. In the case of viscosity, this nonequilibrium conditions are introduced by imposing absorbing limits on the evolution of the Helfand moment, which involve the escape process of trajectories characterized by an *escape rate*. This escape-rate formalism establishes a relationship between this quantity and the viscosity. We have then first studied numerically this relationship in the two-hard-disk model (Viscardy and Gaspard, 2003b) and observed an excellent agreement with the data obtained by other methods.

In the two-hard-disk model reduced to the Sinai billiard, only one Lyapunov exponent is found to be positive. Since this quantity is a measure of the chaotic character of the dynamics, we have also evaluated it and have observed that this exponent is also sensible to the phase transition.

With time, almost all trajectories escape, except some of them composing a *fractal repeller* which is characterized by a noninteger dimension. In the case of the two-hard-disk model, the escape-rate formalism proposes a formula (4.68) expressing the viscosity in terms of the positive Lyapunov exponent and the partial fractal dimension of the repeller, more precisely its *Hausdorff dimension*. This quantity has been computed by applying the so-called Maryland algorithm (McDonald et al., 1985). Hence viscosity may be obtained by the calculation of chaotic quantities (Viscardy and Gaspard, 2003b) and these new data agree very well with the previous ones computed by the Helfand-moment method and by the escape-rate approach. Hence we have confirmed the previous theoretical results of the escape-rate formalism and the relationship between viscosity as a transport property and quantities of the underlying microscopic chaotic dynamics.

In the last chapter, we have started the extension to systems composed of $N$ particles in two and in three dimensions. First, we have calculated the equation of state in the $N$-hard-disk and the $N$-hard-sphere systems. In both systems, a fluid-solid phase transition is observed. A comparison with analytical data obtained by the virial expansion to the tenth order (Clisby and McCoy, 2005) for the fluid phase and the agreement has been found excellent. For the $N$-hard-disk system, we have also compared our results to the semi-empirical *global equation of state* for both phases (Luding, 1996; Luding and Strauß, 2001; Luding, 2002). They agree very well. In the case of the $N$-hard-sphere systems, we compared our results to the theoretical expressions for the equation of state for the solid state obtained by Hall (1970), and we obtained a very good agreement.

Since viscosity is the transport coefficient of interest in this thesis, we have studied the viscosity properties in such systems. The results obtained for dilute and moderately dense gases agree very well with the predictions made by Enskog. These results appear to be a support for the application of the escape-rate formalism for system with many degrees of freedom. Indeed, when we increase the number of particles and/or dimensions, more than one Lyapunov exponent is found to be positive. it therefore consists in a new challenge for the application of the escape-rate formula for transport coefficient, here for viscosity.

Finally, we have applied our method based on the Helfand-moment formula for the computation of shear viscosity in Lennard-Jones fluid near the triple point. Furthermore, we have proposed a similar method for the thermal conductivity. Indeed, the Einstein-like expression for the coefficient of thermal conductivity and the associated Helfand moment $G_i^{(\kappa)}(t)$ are given by

$$
\begin{aligned}
\kappa &= \frac{1}{V k_B T^2} \lim_{t \to \infty} \frac{\left\langle G_i^{(\kappa)\,2} \right\rangle}{2\,t} \\
G_i^{(\kappa)}(t) &= \sum_{a=1}^{N} r_{ai} \left( E_a - \langle E_a \rangle \right) - \sum_{a=1}^{N} \sum_{s} \Delta r_{ai}^{(s)} \left( E_a - \langle E_a \rangle \right) \theta(t - t_s) \\
&\quad - \frac{1}{4} \sum_{a=1}^{N} \sum_{b \neq a} \int_0^t d\tau \, L_{b|ai} \frac{\mathbf{p}_a + \mathbf{p}_b}{m} \cdot \mathbf{F}(\mathbf{r}_a - \mathbf{r}_b - \mathbf{L}_{b|a}) \; .
\end{aligned}
\tag{6.3}
$$

We have compared our results for the shear viscosity and the thermal conductivity with our own Green-Kubo data and they are both in good agreement. In consequence, we can conclude that our Helfand-moment method for periodic systems is valid.

## 6.2 Perspectives

Besides the well-known *Green-Kubo* formula, other methods have been developed for the computation of the transport coefficients, in particular the viscosity. The method we have proposed in chapter 2 may be applied in periodic systems with any potential of interaction. In the literature, the *Lennard-Jones* potential of interaction is the most used, especially when the validity of a new method has to be confirmed. In this context, the perspective of this thesis is to extend the present results for the hard-ball systems to systems with more realistic potentials. Our method presents different advan-tages compared to the Green-Kubo, in particular its application to hard-ball systems and its simplicity. Indeed, the computation is realized by the accumulation during the simulation of a quantity instead of the integral of an autocorrelation function. Since that, we think that our method could be interesting in the future.

In the context of the relationship between transport processes and the chaotic properties of the underlying microscopic dynamics, as the escape-rate formalism have established, we are confronted to a challenge when we wish to increase the number of degrees of freedom. Indeed, the advantage of the two-hard-disk model is that the dynamics of the two disks may be reduced to the one of a pointlike particle moving in the Sinai billiard. We therefore deal with only one positive Lyapunov exponent. Hence the sum in the escape-rate formula for viscosity is reduced to only one term, that $\lambda \, c_H$. When the system is composed of more than two disks, other positive exponents are added and it becomes necessary to consider partial fractal dimensions associated with the different positive Lyapunov exponents. Hence another perspective would be the development of an algorithm evaluating these quantities. The works done by Hunt et al. (1996) as well as by Sweet and Ott (2000) might be a good start of such a purpose. This question does not concern only viscosity but also the other transport processes (diffusion and heat conductivity). Thanks to the expression of the Helfand moment for the heat conductivity that we propose in Eq. (6.3) in addition to the one used in this thesis for the viscosity, the escape-rate formalism can be considered in the study of the relationship between different transport coefficients and the chaotic properties of the microscopic dynamics.

As shown in the last chapter, the structure of the hydrodynamic modes in the solid phase is different than in the fluid phase. Indeed, according to the Goldstone theorem, one hydrodynamic mode associated with each broken symmetry has to be added to the five modes existing in the fluid phase. Furthermore, in a solid phase, the shear modes disappear and the viscosity is transformed into a damping rate for the sound modes. It implies that the viscosity no longer characterizes the relaxation to equilibrium of shear modes. Hence it implies the necessity to give a new interpretation of the viscosity studied in this thesis.

# Appendix A

# Microscopic derivation of the viscosity tensor

In this Appendix, we provide a short microscopic derivation of the viscosity tensor (2.42). First, we need the balance equation for the local conservation of momentum. If we define the density of momentum as

$$g_i(\mathbf{r}) = \sum_{a=1}^{N} p_{ai} \, \delta(\mathbf{r} - \mathbf{r}_a) \, , \tag{A.1}$$

the balance equation is

$$\partial_t \, g_i + \partial_j \, \tau_{ij} = 0 \, , \tag{A.2}$$

with $\partial_j = \partial/\partial r_j$. The microscopic momentum current density is given by

$$\begin{aligned}
\tau_{ij} = &\sum_{a=1}^{N} \frac{p_{ai} \, p_{aj}}{m} \, \delta(\mathbf{r} - \mathbf{r}_a) \\
&+ \frac{1}{2} \sum_{a \neq b=1}^{N} F_i(\mathbf{r}_a - \mathbf{r}_b) \int_0^1 d\lambda \, \frac{dr_{abj}}{d\lambda} \, \delta \left[ \mathbf{r} - \mathbf{r}_{ab}(\lambda) \right] \, ,
\end{aligned} \tag{A.3}$$

where $\mathbf{r}_{ab}(\lambda)$ is the parametric equation of a curve joining the particles $a$ and $b$: $\mathbf{r}_{ab}(0) = \mathbf{r}_b$ and $\mathbf{r}_{ab}(1) = \mathbf{r}_a$.

The microscopic current associated with viscosity is defined by integrating the momentum current density over the volume $V$:

$$J_{ij} = \int_V \tau_{ij}(\mathbf{r}) \, d\mathbf{r} \;, \tag{A.4}$$

which is given by Eq. (2.41). We notice that the hydrostatic pressure is given at equilibrium by

$$\langle J_{ij} \rangle_{\text{eq}} = P \, V \, \delta_{ij} \;, \tag{A.5}$$

if second-order tensors are isotropic in the system of interest.

We suppose that, at the initial time, the fluid is close to the equilibrium and described by the following nonequilibrium phase-space distribution:

$$\begin{aligned} \mathcal{P}(\Gamma) &= \mathcal{P}_{\text{eq}}(\Gamma) \left[ 1 + \beta \int \mathbf{g}(\mathbf{r}) \cdot \mathbf{v}(\mathbf{r}) \, d\mathbf{r} \right] \\ &= \mathcal{P}_{\text{eq}}(\Gamma) \left[ 1 + \beta \sum_{a=1}^{N} \mathbf{p}_a \cdot \mathbf{v}(\mathbf{r}_a) \right] \;, \end{aligned} \tag{A.6}$$

where $\mathcal{P}_{\text{eq}}$ is the equilibrium distribution and $\beta$ is a normalization constant such that

$$\langle p_{ai} \, p_{bj} \rangle_{\text{eq}} = \frac{m}{\beta} \, \delta_{ij} \, \delta_{ab} \;. \tag{A.7}$$

In the microcanonical state, we have that

$$\beta = \frac{1}{k_{\text{B}} T} \, \frac{N}{N-1} \;. \tag{A.8}$$

The aforementioned distribution describes a fluid with a macroscopic velocity field $\mathbf{v}(\mathbf{r})$ since the nonequilibrium average of the momentum density can easily be shown to be given by

$$\langle \mathbf{g}(\mathbf{r}) \rangle_{\text{noneq}} = \rho_{\text{eq}} \, \mathbf{v}(\mathbf{r}) \;, \tag{A.9}$$

where

$$\rho_{\text{eq}} = m \, \frac{N}{V} \;, \tag{A.10}$$

is the mass density at equilibrium.

The time evolution of the probability density (A.6) is ruled by Liouville's operator given by the Poisson bracket with the Hamiltonian $\hat{L} = \{H, \cdot\}$ or the pseudo-Liouville operator in the case of hard-ball dynamics. This operator has the effect of replacing the phase-space coordinates $\Gamma$ by $\Gamma(-t)$

$$
\begin{aligned}
\mathcal{P}_t &= \mathrm{e}^{\hat{L}t} \mathcal{P}_0 = \mathcal{P}_{\mathrm{eq}}(\Gamma) \left[ 1 + \beta \int \mathrm{e}^{\hat{L}t} \mathbf{g}(\mathbf{r}) \cdot \mathbf{v}(\mathbf{r}) \, d\mathbf{r} \right] \\
&= \mathcal{P}_{\mathrm{eq}}(\Gamma) \left[ 1 + \beta \sum_{a=1}^{N} \mathbf{p}_a(-t) \cdot \mathbf{v} \left[ \mathbf{r}_a(-t) \right] \right] .
\end{aligned}
\tag{A.11}
$$

Alternatively, we known that the time evolution of the momentum density is given by Eq. (A.2). In this case, the momentum density should be considered as an observable so that the solution of Eq. (A.2) is

$$
\mathbf{g}(\mathbf{r}, t) = \mathrm{e}^{-\hat{L}t} \mathbf{g}(\mathbf{r}, 0) ,
\tag{A.12}
$$

so that

$$
\mathrm{e}^{\hat{L}t} \mathbf{g}(\mathbf{r}) = \mathbf{g}(\mathbf{r}, -t) ,
\tag{A.13}
$$

is solution of the equation

$$
\partial_t g_i = \partial_j \tau_{ij} .
\tag{A.14}
$$

Integrating both sides over time we get

$$
g_i(\mathbf{r}, -t) = g_i(\mathbf{r}, 0) + \int_0^t dt' \, \partial_j \tau_{ij}(t') .
\tag{A.15}
$$

Close to equilibrium, we may consider the time evolution of deviations with respect to the equilibrium. We neglect terms which are quadratic in the deviations such as the velocity field itself. The time evolution of these deviations is obtained by considering the nonequilibrium average of the balance equation (A.2) for the deviations:

$$
\partial_t \langle \delta g_i \rangle_{\mathrm{noneq}} + \partial_j \langle \delta \tau_{ij} \rangle_{\mathrm{noneq}} = 0 ,
\tag{A.16}
$$

with $\delta \tau_{ij} = \tau_{ij} - \langle \tau_{ij} \rangle_{\mathrm{eq}}$. The nonequilibrium average of the deviation of the momentum current density

is given by

$$
\begin{aligned}
\langle \delta\tau_{ij}(\mathbf{r}) \rangle_{\text{noneq}} &= \int \delta\tau_{ij}(\mathbf{r}) \, \mathcal{P}(\Gamma, t) \, d\Gamma \\
&= \beta \int d\mathbf{r}' \langle \delta\tau_{ij}(\mathbf{r}) \, g_k(\mathbf{r}', -t) \rangle_{\text{eq}} \, v_k(\mathbf{r}') \, .
\end{aligned}
\tag{A.17}
$$

We use Eq. (A.15) to transform the average as

$$
\begin{aligned}
\langle \delta\tau_{ij}(\mathbf{r}) \, g_k(\mathbf{r}', -t) \rangle_{\text{eq}} &= \langle \delta\tau_{ij}(\mathbf{r}) \, g_k(\mathbf{r}', 0) \rangle_{\text{eq}} + \\
&\quad \int_0^t dt' \, \langle \delta\tau_{ij}(\mathbf{r}, 0) \, \partial_l' \delta\tau_{kl}(\mathbf{r}', t') \rangle_{\text{eq}} \, ,
\end{aligned}
\tag{A.18}
$$

where we have used the property that $\partial_l' \langle \tau_{kl} \rangle_{\text{eq}} = 0$ because the equilibrium state is spatially uniform. We notice that the first term in the right-hand side of Eq. (A.18) vanishes because the equilibrium average of an odd power of particle momenta vanishes. After an integration by part over the velocity field, Eq. (A.17) becomes

$$
\begin{aligned}
\langle \delta\tau_{ij}(\mathbf{r}) \rangle_{\text{noneq}} &= -\beta \int d\mathbf{r}' \int_0^t dt' \, \langle \delta\tau_{ij}(\mathbf{r}, 0) \, \delta\tau_{kl}(\mathbf{r}', t') \rangle_{\text{eq}} \, \partial_l' v_k(\mathbf{r}') \\
&= -\eta_{ij,kl} \, \partial_l v_k(\mathbf{r}) \, ,
\end{aligned}
\tag{A.19}
$$

where the identification with the viscosity tensor is carried out in the limit $t \to \infty$ by

$$
\eta_{ij,kl} \, \delta(\mathbf{r} - \mathbf{r}') = \beta \int_0^\infty dt' \, \langle \delta\tau_{ij}(\mathbf{r}, 0) \, \delta\tau_{kl}(\mathbf{r}', t') \rangle_{\text{eq}} \, .
\tag{A.20}
$$

Taking the double volume integral $\int_V d\mathbf{r} \int_V d\mathbf{r}'$ of both sides of Eq. (A.20) and dividing by the volume $V$, we obtain the viscosity tensor as

$$
\eta_{ij,kl} = \frac{\beta}{V} \int_0^\infty dt \, \langle \delta J_{ij}(0) \, \delta J_{kl}(t) \rangle_{\text{eq}} \, ,
\tag{A.21}
$$

with

$$
\delta J_{ij}(t) = \int_V d\mathbf{r} \, \delta\tau_{ij}(\mathbf{r}, t) = J_{ij}(t) - \langle J_{ij} \rangle_{\text{eq}} \, ,
\tag{A.22}
$$

which is Eq.(2.42). Q.E.D.

# Appendix B

# Proof of the equivalence between Green-Kubo and Einstein-Helfand formulas

Our aim is here to deduce the Green-Kubo formula (2.42) from the Einstein-Helfand formula (2.51), proving the equivalence between both formulas under the condition that the Helfand moment is defined by Eq. (2.54) as the time integral of the microscopic current (2.41) and the further condition that the time auto-correlation functions decrease fast enough.

We start from the Einstein-Helfand formula (2.51) with

$$\delta G_{ij}(t) = \int_0^t \delta J_{ij}(\tau) \, d\tau \, ,$$

$$(B.1)$$

$\delta J_{ij}$ being defined by Eq. (A.22) and supposing for simplicity that $\delta G_{ij}(0) = 0$. Accordingly, we have

successively from Eq. (2.51) that

$$
\begin{aligned}
\eta_{ij,kl} &= \lim_{T\to\infty} \frac{\beta}{2TV} \langle \delta G_{ij}(T)\delta G_{kl}(T)\rangle \\
&= \lim_{T\to\infty} \frac{\beta}{2TV} \int_0^T dt_1 \int_0^T dt_2 \, \langle \delta J_{ij}(t_1)\delta J_{kl}(t_2)\rangle \\
&= \lim_{T\to\infty} \frac{\beta}{2TV} \int_{-T}^{+T} dt \int_{|t|/2}^{T-|t|/2} d\tau \, \langle \delta J_{ij}(0)\delta J_{kl}(t)\rangle \\
&= \lim_{T\to\infty} \frac{\beta}{2V} \int_{-T}^{+T} dt \left(1 - \frac{|t|}{T}\right) \langle \delta J_{ij}(0)\delta J_{kl}(t)\rangle \\
&= \frac{\beta}{V} \int_0^{+\infty} dt \, \langle \delta J_{ij}(0)\delta J_{kl}(t)\rangle \, ,
\end{aligned}
\tag{B.2}
$$

where we have performed the change of integration variables

$$
\begin{aligned}
t &= t_2 - t_1 \, , \\
\tau &= \frac{t_1 + t_2}{2} \, ,
\end{aligned}
\tag{B.3}
$$

and supposed that

$$
\lim_{T\to\infty} \frac{1}{T} \int_{-T}^{+T} dt \, |t| \, \langle \delta J_{ij}(0)\delta J_{kl}(t)\rangle = 0 \, ,
\tag{B.4}
$$

which requires that the time autocorrelation functions decrease faster than $|t|^{-1-\epsilon}$ with $\epsilon > 0$. Q.E.D.

# Appendix C

# Pressure and Helfand moment

The hydrostatic pressure at equilibrium is given as the mean value of the momentum current density, i.e., as the mean value of the same microscopic current entering the Green-Kubo relation:

$$P_{ij}V = \int_V \langle \tau_{ij} \rangle_{\text{eq}} \, d\mathbf{r} = \langle J_{ij} \rangle_{\text{eq}} \,. \tag{C.1}$$

The average over the thermodynamic equilibrium state can be replaced by a time average:

$$P_{ij}V = \langle J_{ij} \rangle_{\text{eq}} = \lim_{t \to \infty} \frac{1}{t} \int_0^t d\tau \, J_{ij} \,. \tag{C.2}$$

We can here introduce the Helfand moment to obtain the hydrostatic pressure from the Helfand moment as:

$$P_{ij}V = \lim_{t \to \infty} \frac{1}{t} \left[ G_{ij}(t) - G_{ij}(0) \right] \,. \tag{C.3}$$

In the microcanonical equilibrium state we have that

$$\langle p_{ai} \, p_{aj} \rangle_{\text{eq}} = m \, k_{\text{B}} T \, \frac{N-1}{N} \, \delta_{ij} \,. \tag{C.4}$$

If we assume that the system is isotropic, $P_{ij} = P \, \delta_{ij}$ and we obtain

$$PV = (N-1)k_{\text{B}}T + R \,, \tag{C.5}$$

where the rest $R$ provides the corrections to the law of perfect gases in dense systems. By using Eqs.

(2.67) and (2.58), the virial can be computed alternatively by

$$R \;=\; \left\langle \frac{1}{2d} \sum_{a \neq b=1}^{N} \mathbf{F}(\mathbf{r}_{ab}) \cdot \mathbf{r}_{ab} \right\rangle_{\text{eq}} \tag{C.6}$$

$$=\; \lim_{t \to \infty} \frac{-1}{td} \sum_{s} \sum_{a=1}^{N} \mathbf{p}_{a}^{(s)} \cdot \Delta \mathbf{r}_{a}^{(s)} \, \theta(t - t_s) \tag{C.7}$$

$$=\; \lim_{t \to \infty} \frac{1}{td} \sum_{c} \Delta \mathbf{p}_{a}^{(c)} \cdot \mathbf{r}_{ab}^{(c)} \, \theta(t - t_c) \tag{C.8}$$

where $d$ is the dimension, $\mathbf{r}_{ab} = \mathbf{r}_a - \mathbf{r}_b$, $t_s$ are the times of jumps to satisfy the minimum image convention, while the last expression only holds for hard-ball systems, $t_c$ are the collision times, and $\mathbf{r}_{ab}^{(c)} = \mathbf{r}_a(t_c) - \mathbf{r}_b(t_c)$.